\documentclass[prx,twocolumn]{revtex4-1}

\usepackage{amssymb} 
\usepackage{amsmath}

\usepackage{amsthm}
\usepackage{graphicx}
\usepackage{dcolumn}
\usepackage{bm}
\usepackage{comment}
\usepackage{tabularx} 
\usepackage{mathrsfs}
\usepackage{tensor}
\usepackage{xr}
\usepackage{xcolor}
\newtheorem*{theorem*}{Theorem}
\newtheorem*{prop*}{Proposition}

\begin{document}


\title{Bulk~topology~of~line-nodal~structures~protected~by~space~group~symmetries~in~class~AI}

\author{Adrien Bouhon}%
\email{adrien.bouhon@physics.uu.se}
\author{Annica M. Black-Schaffer}
\affiliation{%
Department of Physics and Astronomy, Uppsala University, Box 516, SE-751 21 Uppsala, Sweden 
}%

\date{\today}

\begin{abstract}

We give an exhaustive characterization of the topology of band structures in class AI, using nonsymmorphic space group 33 ($Pna2_1$) as a representative example where a great variety of symmetry protected line-nodal structures can be formed. We start with the topological classification of all line-nodal structures given through the combinatorics of valence irreducible representations (IRREPs) at a few high-symmetry points (HSPs) at a fixed filling. We decompose the total topology of nodal valence band bundles through the local topology of elementary (i.e.~inseparable) nodal structures and the global topology that constrains distinct elementary nodal elements over the Brillouin zone (BZ). Generalizing from the cases of simple point nodes and simple nodal lines (NLs), we argue that the local topology of every elementary nodal structure is characterized by a set of poloidal-toroidal charges, one monopole, and one thread charge (when threading the BZ torus), while the global topology only allows pairs of nontrivial monopole and thread charges. We show that all these charges are given in terms of symmetry protected topological invariants, defined through quantized Wilson loop phases over symmetry constrained momentum loops, which we derive entirely algebraically from the valence IRREPs at the HSPs. In particular, we find highly connected line-nodal structures, line-nodal monopole pairs, and line-nodal thread pairs, that are all protected by the unitary crystalline symmetries only. Furthermore, we show symmetry preserving topological Lifshitz transitions through which independent NLs can be connected, disconnected, or linked. Our work constitutes a heuristic approach to the systematic topological classification and characterization of all momentum space line-nodal structures protected by space group symmetries in class AI. 

\end{abstract}

\maketitle


\section{Introduction}

The problem of classifying and characterizing topological phases of matter, taking into account the symmetries of the system, is playing a very central role in the most recent developments of solid state theory. Since the prediction of topological insulators protected by time reversal symmetry (TRS) \cite{KaneMele_Z2, FuKane_Z2, FuKaneMele_Z2}, the scope of topological classification of gapped quantum phases with symmetries has been extended in many directions. Symmetry protected topological phases, including both insulators and superconductors but only considering the noninteracting case with short-range entanglement \cite{Wen_zoo_top}, have been classified in the Altland-Zirnbauer tenfold-way \cite{Schnyder08, Kitaev, RyuSchnyder_10ways} for TRS, particle-hole symmetry, and chiral symmetry. The common conceptual framework for further classification in momentum space of fully gapped phases and of stable nodal phases was early advocated by Volovik \cite{Volovik} and has since then also been integrated within the tenfold-way \cite{Top_defect_tenfold,ZhaoWang_FS,ChiuSchnyder_reflection}. However, even though the classification of topological semimetals protected by all the symmetries present in crystalline materials has been the subject of intense activity in the last few years it is still far from complete. 

The aim of this work is to provide a fully systematic scheme for the classification of topological crystalline bulk nodal phases of which nonsymmorphic space groups constitute the largest subset. Very early on, Michel and Zak \cite{Zak1,Zak2,Zak3,Zak4} showed that nonsymmorphic space groups host band permutations under reciprocal lattice translations that must be accompanied by symmetry protected band crossings. Higher dimensional {\it essential} degeneracies at special high symmetry points, lines, or planes of the BZ boundary in nonsymmorphic space groups is a well-known and direct consequence of the existence of projective representations of the crystal symmetries acting on the Bloch eigenfunctions at those momenta \cite{BradCrack}. What Michel and Zak also showed is that the combination of several non-commuting nonsymmorphic symmetries naturally leads to symmetry protected band crossings beyond essential degeneracies, i.e.~band crossing points (lines) that are entirely free to move on high-symmetry lines (planes), but cannot be globally removed from the BZ. This has very recently been revived in the modern context of topological semimetals \cite{Vishwanath1,YoungKane_simple,Kane_nonsym,Vishwanath2,WiederKane_double,Vishwanath3,Thomas_line,Solouyanov_triplepoints,Kane_2D_SOC,ZhaoSchnyder_1D,Furusaki_line,GB_line,GelBoBoBa_II,Furusaki_multiplescrews,Murakami_NLsym,Vanderbilt_WeylPoints,Sato_3,1611.07925}. 

In this work we consider as an important case study the nonsymmorphic space group 33 or $Pna2_1$ in class AI (denoted SG33-AI) \cite{Schnyder08,Kitaev}, i.e.~assuming TRS and neglecting spin-orbit coupling, such that spin is unimportant. Due to the combination of one screw and two glide symmetries that do not commute one with an other, SG33 allows a great variety of symmetry protected NLs formed by the crossing of bands over the whole BZ and is thus a very illustrative example. To be specific, we call a nodal structure the locus of band-crossing momenta over the BZ at a fixed filling, i.e.~for a fixed number of valence states and constant over the BZ. Since all the essential degeneracies are well-known and found in the tables of space group representations, see e.g.~Ref.~\cite{BradCrack,Bilbao}, we here focus exclusively on the nonessential symmetry protected band crossings. 

We perform a completely exhaustive characterization of the bulk topology for any band structure in SG33-AI that supports symmetry protected nodal structures. This is achieved first through a topological classification of all band structures, only given a fixed total number of bands and entirely in terms of combinatoric sets of the valence band irreducible representations (IRREPs) at only a few \textit{active} high-symmetry points (HSPs). Then, for each topological class of band structures, we determine the corresponding nodal structure imposed by the compatibility relations and band permutation rules over the whole BZ. In general we find \textit{composed} nodal structures that can be split into indivisible \textit{elementary} nodal structures. 
We also decompose the total topology of nodal valence band bundles in the local topology of elementary nodal structures and the global topology that constrains the local topology of distinct nodal elements over the BZ. 
Generalizing from the cases of simple point nodes and a simple NL, we argue that the local topology of every elementary nodal structure is characterized by a set of poloidal-toroidal charges, one monopole, and one thread charge. We show that the global topology originating from the geometry of the BZ itself imposes that only pairs of nontrivial monopole and thread charges can be realized.
We then show that all the local charges are given in terms of symmetry protected topological invariants obtained from quantized Wilson loop phases over symmetry constrained momentum loops. We show that all these topological invariants can be computed entirely algebraically knowing only the valence IRREPs at the HSPs. At the core of our approach is thus the combination of space group representation theory \cite{BradCrack,Bilbao} with the Wilson loop techniques \cite{Bernevig1, Bernevig_point_groups,Alex1,Alex_noSOC_noTRS,Kane_line,AlexBernevig_berryphase,Bernevig3,Alex0,Alex2} and it is therefore also straightforwardly extendable to any other space group.

For the SG33-AI we find highly connected line-nodal structures characterized by multiple poloidal charges, line-nodal monopole pairs, and line-nodal thread pairs that are protected by the unitary crystalline symmetries only. This notably extends previous line-nodal monopoles discussions, where one anti-unitary symmetry was crucial \cite{Fu_line_node_monopole,Agterberg_BdGsurface,Bzdusek_mulitnodes,Moroz_nodalsurface}. We also show that symmetry preserving topological Lifshitz transitions are easily achievable, through which independent NLs can be connected, disconnected, or even linked in which case also nontrivial toroidal charges are present. 
Furthermore, we find that no topological insulator phases are allowed in SG33-AI, which makes this topological classification and characterization of the nodal band structures exhaustive in this system.  

The philosophy of the initial part of our work is similar to several proposals that have been made towards a systematic topological classification of band structures for a given space group, taking as an input the set of valence IRREPs at relevant HSPs and combined with the compatibility relations over the BZ \cite{Slager_SG_pfaffian,Liu_coreps, Liu_reps, Slager_K, BBS_1,ShiozakiWallPaper_K, PoVishWata_symmetry_indic,Graph_theory,Band_connectivity}. An extension of this has also been recently proposed \cite{TopQuantChem,Alex_BlochOsc, Cano_EBR,Po_FragileTop}, also considered in \cite{ShiozakiWallPaper_K}, showing that the knowledge of the sub-lattice degrees of freedom composing the bands is sometimes needed in order to distinguish topologically inequivalent band structures. 
A complementary task is however to systematically give the algebraic computation of the symmetry protected topological invariants in terms of the valence IRREPs. This was initiated by Fu and Kane \cite{FuKane_inv} who derived the algebraic expression of the $\mathbb{Z}_2$ invariants of centrosymmetric topological insulators in terms of inversion symmetry eigenvalues. Later Chern numbers were also shown to be computable algebraically from point symmetry eigenvalues \cite{Bernevig_point_groups}, see also Refs.~\cite{TeoFuKane_mirror_Z2, Alex1, Alex_noSOC_noTRS, AlexBernevig_berryphase, Bernevig3,Liu_coreps, Liu_reps,Alex0}. 
In this work we combine and also extend these two disparate sets of previous works when we combine a topological classification of the nodal band structures with algebraic computation of all the symmetry protected topological invariants characterizing the nodal structure topology. 

The remainder of the article is organized in the following way. 
In Section \ref{SG33} we present the topological classification of all symmetry protected line-nodal structures for SG33-AI. This is done through the combinatorics of valence IRREP sets constrained by all the compatibility relations and band permutations over the BZ. 
In Section \ref{top_meta} we present the problem of the topological characterization of nodal structures in a formal way using a homotopy approach. This leads to the definitions of the local poloidal-toroidal charges and the monopole and thread charges of any elementary nodal structure. 
In Section \ref{top_inv} we derive all necessary symmetry protected topological invariants for SG33-AI, given as quantized Wilson loop phases over symmetry-constrained momentum loops and derived algebraically directly from the valence IRREPs. 
In Sections \ref{four_top} and \ref{eight_top} we discuss in detail all the symmetry protected nodal structures obtained from four and eight bands, respectively, at half-filling. We show explicitly that all the local charges of a nodal structure (poloidal-toroidal, monopole, and thread) are determined by the previously defined symmetry protected topological invariants. In Section \ref{Lifshitz} we also discuss symmetry preserving topological Lifshitz transitions through which independent NLs can be connected, disconnected, or even linked. 
Finally, in Section \ref{discussion} we give a few brief concluding remarks.

\section{Symmetry protected line-nodal structures for SG33-AI}\label{SG33}
We start with presenting the topological classification of all band structures, for fixed total band number $N$ and valence band number $N_v$, in terms of combinatoric sets of valence IRREPs for SG33-AI. We show that only the few \textit{active} HSPs $\{\Gamma,\text{Z},\text{S},\text{R}\}$ is needed, see also Refs.~\cite{Slager_K,BBS_1}. For each band structure class and fixed valence band number we determine the corresponding nodal structure imposed by the compatibility relations and band permutation rules over the BZ.

\subsection{Building blocks of any band structure} 
The nonsymmorphic space group SG33 (P$na2_1$) \footnote{We use the parameterization of the International Tables for Crystallography \cite{ITC}.} exhibits a great variety of symmetry-protected line-nodal structures. SG33 is composed of a primitive orthorhombic Bravais lattice and three nonsymmorphic symmetries; one screw axis and two glide planes, that take the form $\{g\vert \boldsymbol{\tau}_g\}$ with the point group elements $g\in C_{2v} = \{E,C_{2z},m_y,m_x\}$ and the fractional translations $ \boldsymbol{\tau}_z = \boldsymbol{a}_3/2$, $ \boldsymbol{\tau}_y = (\boldsymbol{a}_1+\boldsymbol{a}_2)/2$, $ \boldsymbol{\tau}_x = (\boldsymbol{a}_1+\boldsymbol{a}_2+\boldsymbol{a}_3)/2$, where $\{\boldsymbol{a}_i\}_{i=1,2,3}$ are the primitive lattice vectors with their duals in reciprocal space written $\{\boldsymbol{b}_i\}_{i=1,2,3}$.  
In this work we also keep within the AI class, thus assuming TRS and neglect spin-orbit coupling. 

\begin{figure}[htb]
\centering
\begin{tabular}{cc} 
	\includegraphics[width=0.5\linewidth]{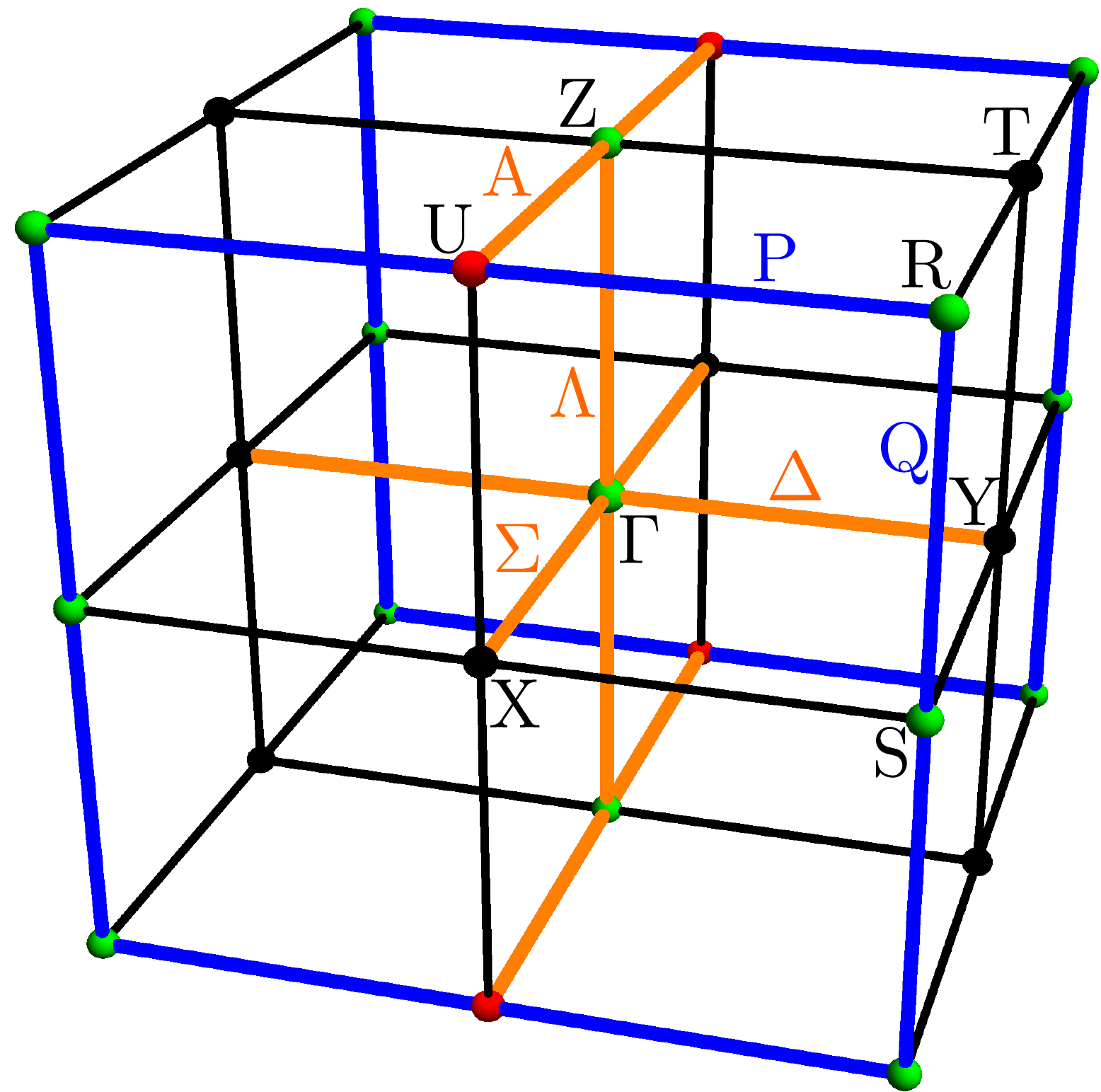} &
	\includegraphics[width=0.35\linewidth]{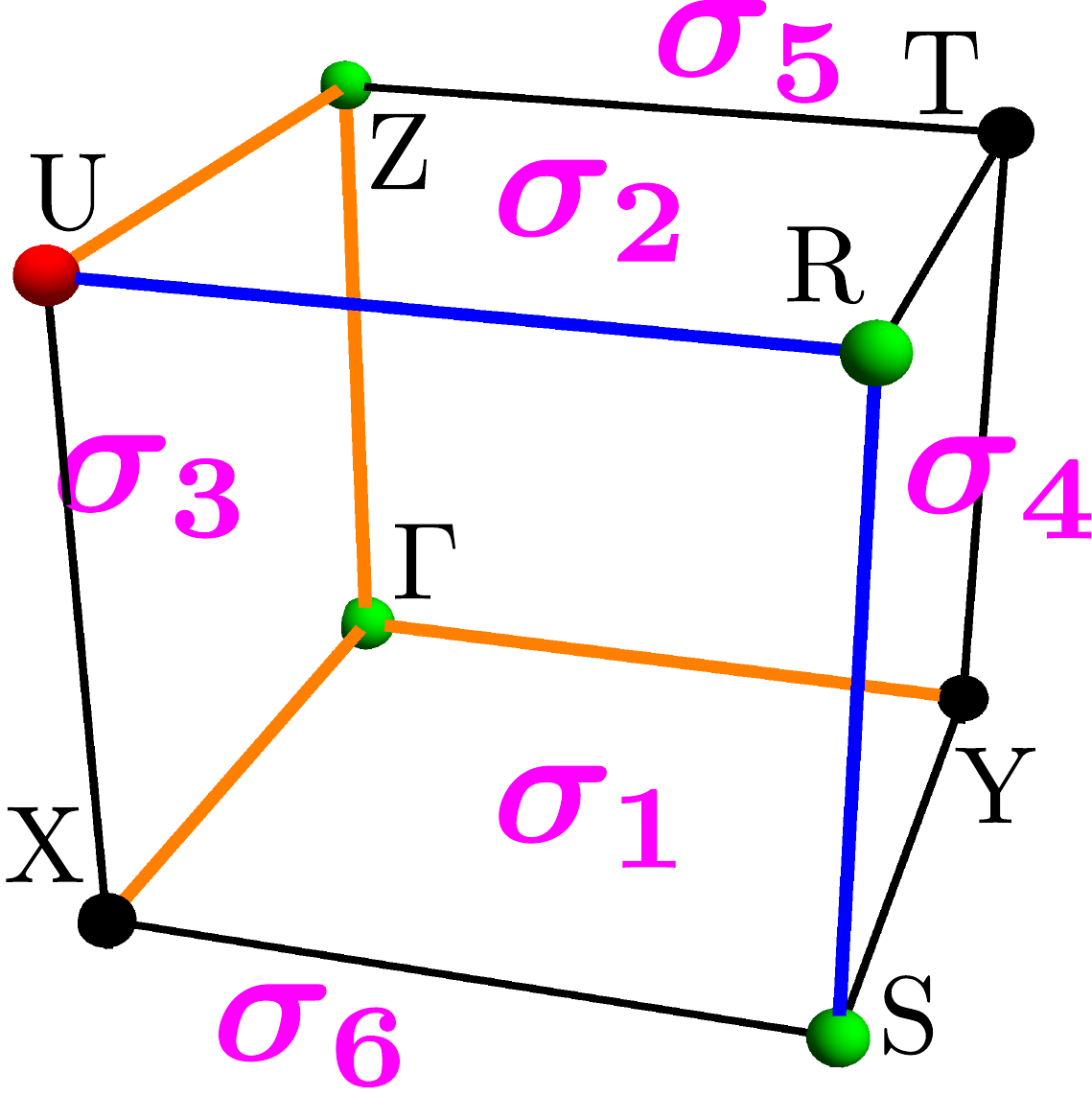}\\
	(a) & (b)  
\end{tabular}
\begin{tabular}{c} 
	\includegraphics[width=0.59\linewidth]{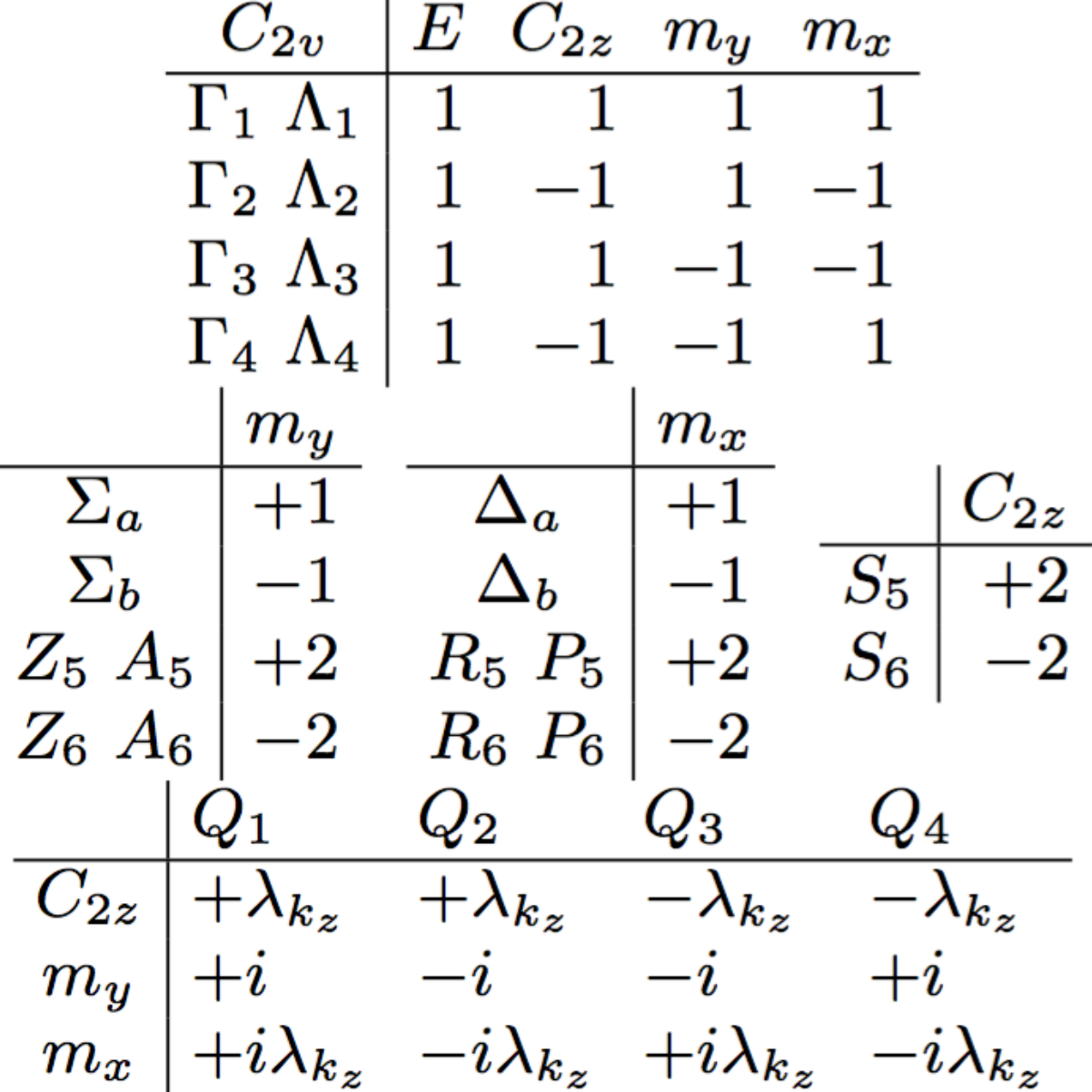} \\
	(c) \\
	\includegraphics[width=0.99\linewidth]{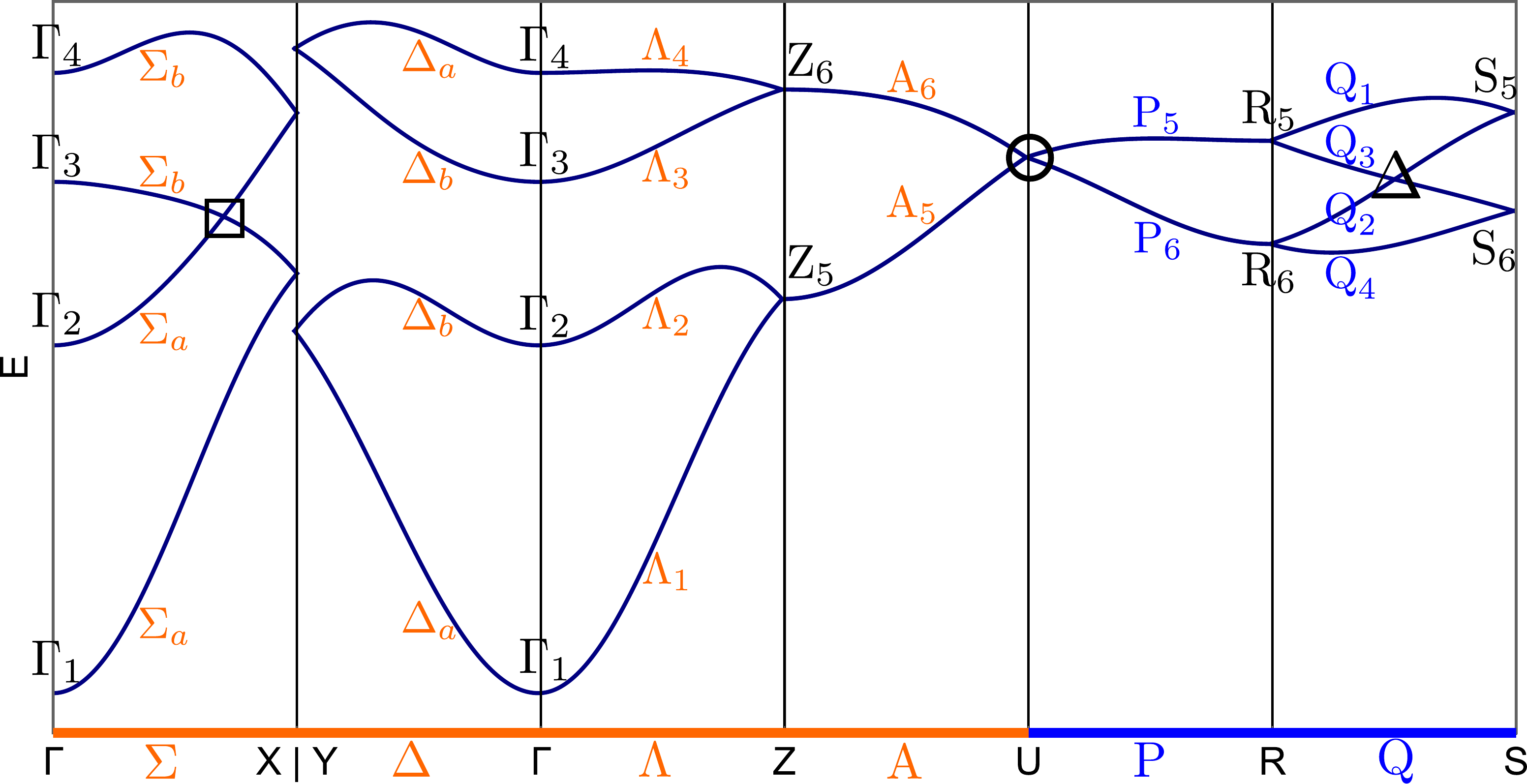} \\
	(d)
\end{tabular}
\caption{\label{fig_SG33} (a) First BZ for SG33 with the HSPs and HSLs highlighted. (b) The six planes of the fundamental domain, $\sigma_i$ $i=1,\dots,6$. (c) All relevant character tables of the IRREPs of HSPs and HSLs, with $\lambda_{k_z} = \mathrm{e}^{-i  (\pi/2) k_z}$. (d) Electronic band structure of a four-band subspace obtained from a tight-binding model for SG33-AI. Band crossing points part of nonessential NLs are marked with $\square$ for NLs on a $m_y$-invariant plane, $\triangle$ for NLs on a $m_x$-invariant plane, and by circle $\bigcirc$ for two connected NLs with one on a $m_y$-invariant plane and one on the perpendicular $m_x$-invariant plane. Axis and label colors correspond to the line colors in (a).
}
\end{figure}
Analyzing the topology of the band structures the different high-symmetry regions in the BZ play a crucial role. Figure~\ref{fig_SG33}(a) shows the HSPs and high-symmetry lines while Fig.~\ref{fig_SG33}(b) labels the planes ($\sigma_{1,\dots,6}$) of the fundamental domain.
Since SG33 has a single Wyckoff position with no symmetry, the set of all IRREPs at $\Gamma$ must split into $n\in \mathbb{N}$ copies of the four one-dimensional (1D) IRREPs of $C_{2v}$ $\{\Gamma_1, \Gamma_2 , \Gamma_3 , \Gamma_4 \}$ defined by the character table in Fig.~\ref{fig_SG33}(c). Likewise, the set of all 2D IRREPs at the HSPs $ p \in \{\text{Z},\text{S},\text{R}\}$ of the BZ boundary splits into $N$ copies of the two projective IRREPs $\{p_{5},p_{6}\}$, also defined in Fig.~\ref{fig_SG33}(c) \cite{BradCrack, Bilbao}. Importantly, the remaining HSPs $\{\text{X},\text{Y},\text{T},\text{U}\}$ allow only a unique IRREP, fourfold degenerate at U and twofold degenerate at the other points. We call the HSPs that allow inequivalent IRREPs, i.e.~$\{\Gamma,\text{Z},\text{S},\text{R}\}$ {\it active}, while those that only allow a single IRREP, i.e.~$\{\text{X},\text{Y},\text{T},\text{U}\}$ are then inactive. 

It is convenient to split the high-symmetry lines (HSLs) into three subsets as shown in Fig.~\ref{fig_SG33}(a): the active lines $\mathcal{B}_{\Gamma} = \{\Sigma,\Delta, \Lambda,\text{A} \}$ (orange) and $\mathcal{B}_{\text{R}}=\{\text{P},\text{Q}\}$ (blue), and the remaining inactive lines $\mathcal{B}_{\Gamma\text{-}\text{R}} $ (black). Since only a unique 2D IRREP is allowed on each high-symmetry line of $\mathcal{B}_{\Gamma\text{-}\text{R}}$ \cite{BradCrack,Bilbao} symmetry protected band crossings are excluded on that region of the BZ. Also, $\mathcal{B}_{\Gamma}$ and $\mathcal{B}_{\text{R}}$ are \textit{symmetry independent} since there is no constraint between their IRREPs. Indeed, the compatibility relations map the IRREPs of $\mathcal{B}_{\Gamma}$ and $\mathcal{B}_{\text{R}}$ to the single IRREP of the intermediary region $\mathcal{B}_{\Gamma\mathrm{-}\text{R}}$. 

Furthermore, time reversal symmetry $\mathcal{T}$ combined with the nonsymmorphic space symmetry group $\mathcal{G}$ leads to the enlarged group of symmetries $\mathcal{G}\times \{E,\mathcal{T}\}$. Since the spectrum is invariant under complex conjugation the band structure is symmetric under the effective $k$-space point symmetry group $C_{2v}\times \{E, I \} = D_{2h}$ where $I$ is the space inversion. Indeed, combining any unitary point symmetry $g\in\{E,C_{2z},m_y,m_x\}$ with time reversal, we obtain the anti-unitary symmetry $\mathcal{T}*g = a \mathcal{K}$ with the unitary transformation $a = gI \in \{I,m_z,C_{2y},C_{2x}\}$ and the complex conjugation $\mathcal{K}$. In the following we implicitly consider the extended group of $k$-space point symmetries $g\in D_{2h}$ satisfied by the band structure. Note that while the effect of the anti-unitary symmetries on the spectrum is trivial, it is not on the eigenvectors as they are in general complex. This explains some of the essential degeneracies at HSPs and HSLs of SG33-AI \cite{BradCrack,Bilbao}. 
We use a short notation for the point unitary symmetries $ g\in\{1,z,y,x\} = \{E, C_{2z},m_y,m_x\}= C_{2v}$, but keep the full symbols for the effective band structure symmetries coming from TRS, i.e.~ $g\in\{I,m_z,C_{2y},C_{2x}\}$.

It can be shown that {\it any} $4N$-band structure for SG33-AI can be reconstructed by hand from the list of energ- ordered IRREPs at the active HSPs $\{\Gamma, Z,S,R\}$, making use of (i) the band permutation rules of Table \ref{permutation_SG33}, and (ii) the compatibility relations from the HSPs into the HSLs defined by the character tables of Fig.~\ref{fig_SG33}(c) and Table \ref{permutation_SG33}. The guiding rules (i) and (ii) are illustrated in the four band example in Fig.~\ref{fig_SG33}(d). Since the proof is in complete analogy to the one given in detail for another space group in Ref.~\cite{BBS_1} we do not repeat it here. However, we still discuss in detail two examples, one with four bands and one with eight bands. The generalization to $4N$ bands is straightforward. Also, we concentrate exclusively on the symmetry-protected band crossings that go beyond the essential degeneracies \footnote{An essential degeneracy happens on the whole connected subspace of the BZ that is characterized by the same little co-group. Therefore, these include band crossing points at HSPs, band crossing lines at HSLs, and band crossing planes at high-symmetry planes.}, since the latter are systematically listed in space group representation tables, see e.g.~ Ref.~\cite{BradCrack,Bilbao}. 

{\def\arraystretch{1.3}  
\begin{table}[t]
\caption{\label{permutation_SG33} Band permutation rules in $\mathcal{B}_{\Gamma}$ and $\mathcal{B}_{\text{R}}$, as well as compatibility relations from S and R onto $\text{Q}\subset \mathcal{B}_{\text{R}}$. The permutation rule $\mathcal{P}_{\overline{\text{p} \boldsymbol{b}_i} \parallel \text{L}} = (p_1 p_2)$ reads as the permutation between the two bands with IRREPs $p_1$ and $p_2$ at the HSP $\text{p}$ under a reciprocal lattice translation $\boldsymbol{b}_i$ along the HSL L.
}
 \begin{tabular*}{\linewidth}{X @{\extracolsep{\fill}} c c c c}
 \hline
 \hline  
    $\mathcal{P}_{\overline{\Gamma\boldsymbol{b}_1}\parallel \Sigma}$  & $\mathcal{P}_{\overline{\Gamma\boldsymbol{b}_2}\parallel \Delta}$ & $\mathcal{P}_{\overline{\Gamma\boldsymbol{b}_3}\parallel \Lambda}$ & $\mathcal{P}_{\overline{\text{Z}\boldsymbol{b}_1}\parallel \text{A}}$  \\
 \hline
  $\begin{array}{c}   (13)(24) \\  (14)(23) \end{array} $ & 
 		$\begin{array}{c}  (12)(34) \\  (13)(24)  \end{array}$ & 
		$ (12)(34) $  & $(56)$ \\
 \hline
 \hline
  \end{tabular*}
 \begin{tabular*}{\linewidth}{X @{\extracolsep{\fill}} c c c}
    $\mathcal{P}_{\overline{\text{R}\boldsymbol{b}_2}\parallel \text{P}}$
    & $\Gamma^{\text{S}} \rightarrow \Gamma^{\text{Q}}$ 
    & $\Gamma^{\text{R}} \rightarrow \Gamma^{\text{Q}}$ \\ 
 \hline
  $   (56) $ 
	& $\begin{array}{c} S_5 \rightarrow Q_1 \oplus Q_2 \\ 
	S_6 \rightarrow Q_3 \oplus Q_4 \end{array}$ & 
	$\begin{array}{c} R_5 \rightarrow Q_1 \oplus Q_3 \\ 
	R_6 \rightarrow Q_2 \oplus Q_4 \end{array}$ \\ 
 \hline
 \hline
 \end{tabular*}
\end{table}
}

In the following we arbitrarily split the band structures of $N$-band subspaces into $N_v$ valence (i.e.~occupied) states and $N-N_v$ conduction (unoccupied) states and fix $N_v$ over the whole BZ, where we refer to $N_v = N/2$ as half-filling. We then characterize the symmetry protected nodal structures that result from the crossings between valence and conduction bands. Strictly speaking this would require a $\boldsymbol{k}$-dependent Fermi level, but, as we argue in Section \ref{top_meta}, this is the appropriate conceptual construction in order the characterize the topology of any band nodal structure. The relationship between this abstract construction and a proper Fermi surface, i.e.~the iso-energy cross-section of a band structure, is briefly addressed in Section \ref{discussion}.

\subsection{Four-band subspace}\label{four}

\subsubsection{Compatibility relations and symmetry protected band crossings}
We start with the case of a four-band subspace, i.e.~four bands separated from the other bands by an energy band gap from above and from below. Fig.~\ref{fig_SG33}(d) shows one numerical example obtained from a tight-binding model, where we included enough terms such that all artificial degeneracies are lifted. In the following we show in detail how the compatibility relations and the band permutation rules over the different regions of the BZ lead to the band structure of Fig.~\ref{fig_SG33}(d).

Let us start with the HSLs of $\mathcal{B}_{\Gamma}$. The compatibility relations from the IRREPs of the HSP $\Gamma$ to the IRREPs of the HSL $\Sigma$ are given by $\Gamma_1,\Gamma_2 \rightarrow \Sigma_a$ and $\Gamma_3,\Gamma_4 \rightarrow \Sigma_b$, i.e.~for $\boldsymbol{k}\in\Sigma$ (with the little co-group $\overline{G}^{\Sigma} = \{E,m_y\}$) we have $^{\{m_y \vert \boldsymbol{\tau}_y\}} \vert \psi_{\Gamma_j \rightarrow \Sigma_\nu } , \boldsymbol{k} \rangle = \lambda_{\nu}(\boldsymbol{k}) \vert \psi_{\Gamma_j \rightarrow \Sigma_\nu } , \boldsymbol{k} \rangle$ with the eigenvalues $\lambda_{\nu}(\boldsymbol{k}) = s_{\nu} e^{-i \boldsymbol{k}\cdot \boldsymbol{\tau}_y }$ for $\nu=a,b$ chosen such that $s_{a} = +1$ and $s_{b} = -1$ according to the character tables of Fig.~\ref{fig_SG33}(c). Overall, Fig.~\ref{fig_SG33}(d) realizes the permutation $(13)(24)$ along the HSL $\Sigma$. Because of the little co-group of the HSP X on the BZ boundary, $\overline{G}^{X} = C_{2v}$, the symmetry protected crossings between the bands $\Gamma_1$ and $\Gamma_3$ and between the bands $\Gamma_2$ and $\Gamma_4$ must occur at X itself. Actually, the doubly degenerated points on X are part of lines of double degeneracy over the whole HSLs $\overline{XS}$ and $\overline{XU}$. Since these are well known essential degeneracies listed in e.g.~Ref.~\cite{BradCrack,Bilbao}, we do not further discuss them. The discussion is similar along the HSL $\Delta$ which has the little co-group $\overline{G}^{\Delta} = \{E,m_x\}$ and the band permutation $(12)(34)$. 

The HSL $\Lambda$, with the little co-group $\overline{G}^{\Lambda} = C_{2v}$, only admits the band permutation $(12)(34)$, as seen in Table~\ref{permutation_SG33}. Also, there are two distinct doubly degenerate IRREPs at Z: $Z_5$, giving the crossing point along $\Lambda$ of the bands $\Gamma_1 \rightarrow \Lambda_1$ and $\Gamma_2\rightarrow \Lambda_2$, and $Z_6$, giving the crossing point of the bands $\Gamma_3 \rightarrow\Lambda_3$ and $\Gamma_4\rightarrow\Lambda_4$, see Fig.~\ref{fig_SG33}(d). Interestingly, the combination of the three band permutations along $\Sigma$, $\Delta$ and $\Lambda$ aways lead to an unavoidable pair of band crossing points (it forms a pair because one point has its image under $C_{2z}$ or $m_x$). The crossing points are set only by the energy ordering of the IRREPs at $\Gamma$ and at Z for the four band case. For the example in Fig.~\ref{fig_SG33}(d) it occurs along $\Sigma$, and marked by a small square.
Contrary to essential degeneracies, these crossings are free to move along the high-symmetry line $\Sigma$. Since, given the energy ordering of IRREPs at $\Gamma$, the crossing must happen between the bands $\Gamma_2\rightarrow\Sigma_a$ and $\Gamma_3\rightarrow\Sigma_b$ that have distinct $m_y$-symmetry characters, the crossing is necessarily protected by symmetry. Actually, the whole plane at $k_x = 0$, $\sigma_5$, is invariant under $m_y$, i.e.~it has the same little co-group as the line $\Sigma$ ($\Sigma \subset \sigma_5$). It thus follows that the pair of crossing points on $\Sigma$ are never gapped when we move away from $\Sigma$ within $\sigma_5$. Therefore, there must be a pair of NLs on $\sigma_5$ crossing the line $\Sigma$. 

The HSL Q is also special because of the compatibility relations that hold between the two 2D IRREPs at S ($S_5$, $S_6$) and at R ($S_5$, $S_6$), see Table~\ref{permutation_SG33}. They guarantee the crossing point on Q marked by a triangle in Fig.~\ref{fig_SG33}(d). Depending on the relative energy ordering of the IRREPs at S and R, the crossing happens between the bands $Q_2$ and $Q_3$ or between the bands $Q_1$ and $Q_4$. In both cases, the crossing point must continue over the $m_x$-invariant plane $\sigma_6$ leading to a (nonessential) NL on the BZ boundary. Because of the fourfold degeneracy at U, i.e.~a \textit{double} point node, the NLs on $\sigma_3$ and on $\sigma_6$ must cross at U, as marked by a circle in Fig.~\ref{fig_SG33}(d). Note that U is both the crossing point of these two nonessential NLs on the planes $\sigma_3$ and $\sigma_6$ and of two essential NLs (on HSLs A and P).
 
In Fig.~\ref{fig_SG33}(d) and throughout this whole work, we use the following convention: square ($\square$) marks a point node that is part of a nonessential NL belonging to a $m_y$-invariant high-symmetry plane, triangle ($\triangle$) marks a point node that is part of a nonessential NL belonging to a $m_x$-invariant high-symmetry plane, and circle ($\bigcirc$) marks a point node that is the section of two nonessential NLs, one belonging to a $m_y$-invariant high-symmetry plane and one belonging to a $m_x$-invariant high-symmetry plane.

\subsubsection{IRREP combinatorics and nodal structures}
\begin{figure}[htb]
\centering
\begin{tabular}{ccc} 
	\includegraphics[width=0.4\linewidth]{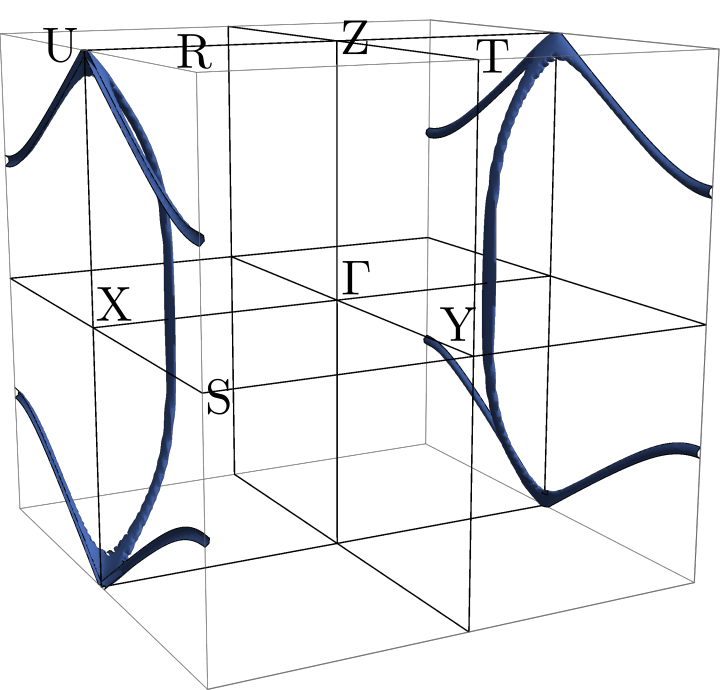}  \\
	$(\Gamma_1,\Gamma_2,\Gamma_3,\Gamma_4)$ $(\text{Z}_5,\text{Z}_6)$ \\
	(a) 
\end{tabular} \\
\begin{tabular}{cc} 
	\includegraphics[width=0.4\linewidth]{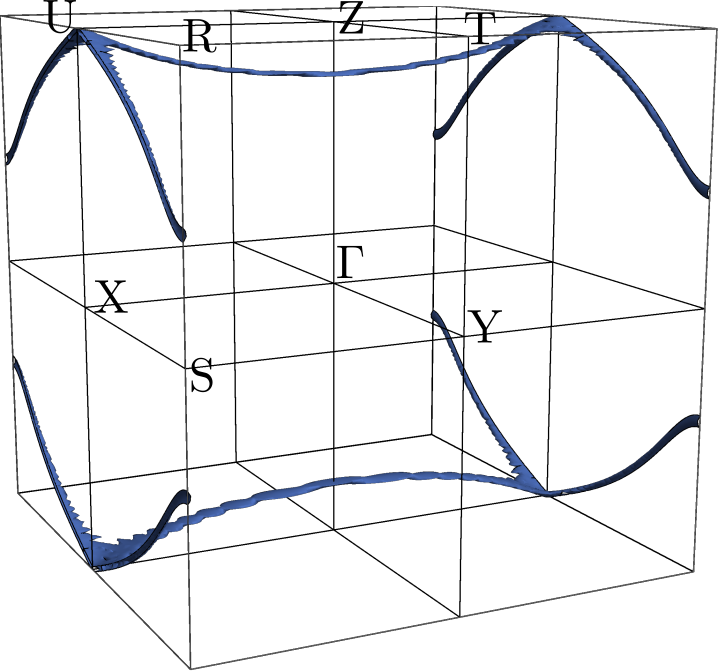} &
	\includegraphics[width=0.4\linewidth]{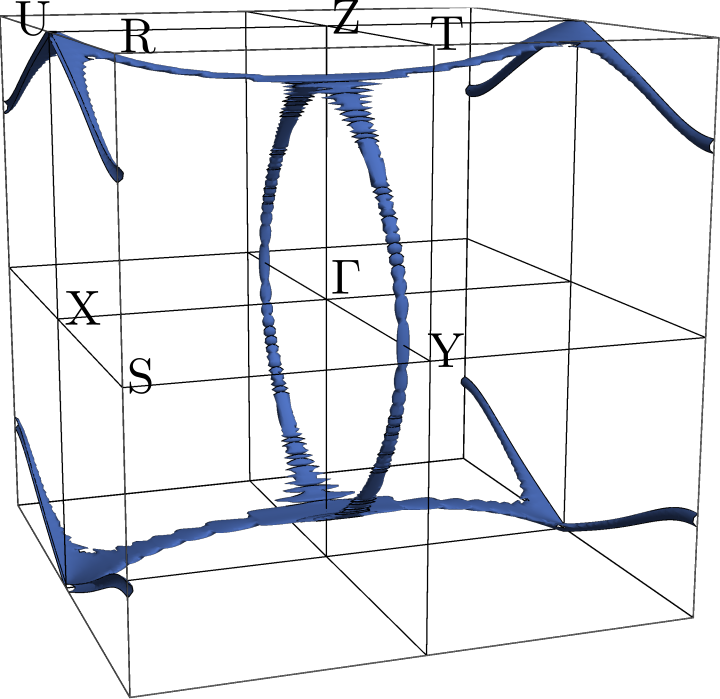} \\
	$(\Gamma_1,\Gamma_3,\Gamma_2,\Gamma_4)$ $(\text{Z}_5, \text{Z}_6)$ & 
	$(\Gamma_1,\Gamma_4,\Gamma_2,\Gamma_3)$ $(\text{Z}_5, \text{Z}_6)$\\
	(b) & (c) \\
	\includegraphics[width=0.4\linewidth]{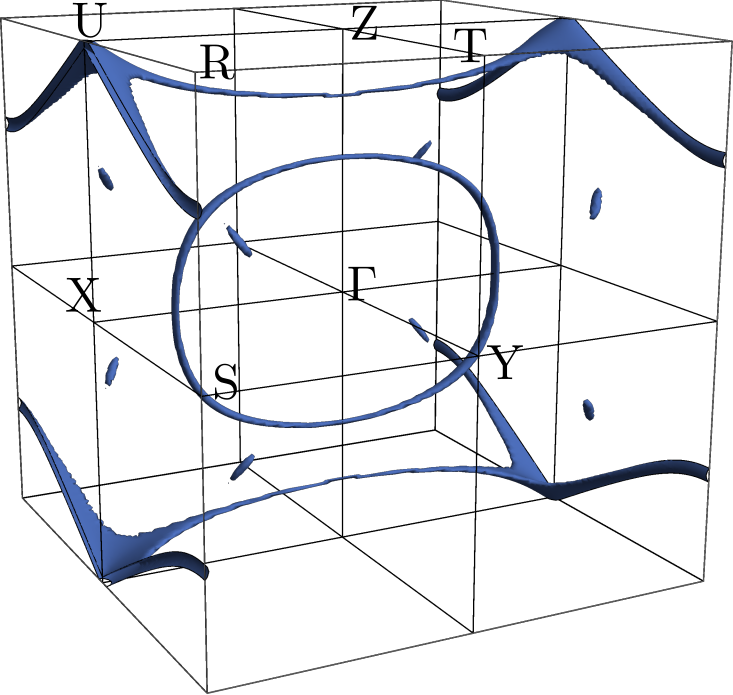} &
	\includegraphics[width=0.4\linewidth]{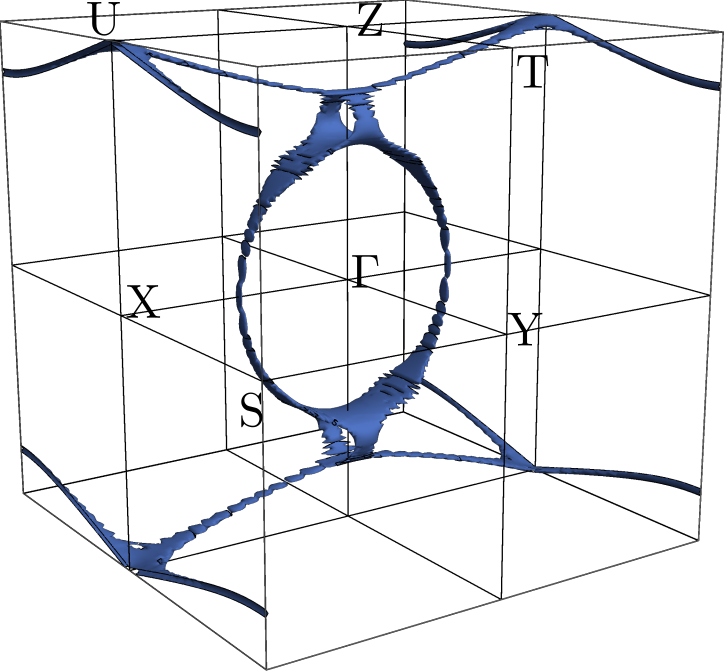} \\
	$(\Gamma_1,\Gamma_2,\Gamma_3,\Gamma_4)$ $(\text{Z}_6,\text{Z}_5)$ & 
	$(\Gamma_2,\Gamma_1,\Gamma_3,\Gamma_4)$ $(\text{Z}_6,\text{Z}_5)$ \\
	(d) & (e)
\end{tabular} 
\caption{\label{fig_SG33_4B_lines} Line-nodal structures of a four-band subspace at half-filling. Each case is determined by the energy ordered list of IRREPs at the HSPs $\{\Gamma,\text{Z}\}$, written for increasing energies. Fig.~\ref{fig_SG33_4B_lines}(a) corresponds to the band structure of Fig.~\ref{fig_SG33}(d).}
\end{figure}

In Fig.~\ref{fig_SG33_4B_lines}(a) we show the position in $k$-space of the NLs occurring at half-filling over the whole BZ. From the discussion it is clear that the nodal structure is fully determined by the energy ordering of the IRREPs at only the $\Gamma$ and Z points for four bands. Exhausting the combinatoric possibilities of these IRREPs ordering, we show in Fig.~\ref{fig_SG33_4B_lines} all the inequivalent line-nodal structures \footnote{All nodal structures of Fig.~\ref{fig_SG33_4B_lines} actually come from an eight-band tight-binding model that splits into two four-band subspaces separated by an energy band gap. It is easy to show that any four-by-four tight-binding model with SG33-AI (corresponding to four sub-lattice sites with a single electronic orbital per site) has an accidental fourfold degenerate NL along the whole HSL $\text{P}$. We therefore introduce more bands in order to avoid this artificial degeneracy.}.

Fig.~\ref{fig_SG33_4B_lines}(a), which corresponds to the band structure of Fig.~\ref{fig_SG33}(d), is formed by splitting the IRREPs at $\Gamma$ into the groups $\{\Gamma_1,\Gamma_2\}$ (valence) and $\{\Gamma_3,\Gamma_4\}$ (conduction) and by setting $E_{Z_5}< E_{Z_6}$. 
Now considering all the other ways of ordering the IRREPs at $\Gamma$ and Z, leads to the remaining inequivalent symmetry protected line-nodal structures in Fig.~\ref{fig_SG33_4B_lines}. Splitting the IRREPs into $\{\Gamma_1,\Gamma_3\}$ and $\{\Gamma_2,\Gamma_4\}$ gives the NLs of Fig.~\ref{fig_SG33_4B_lines}(b). Here one NL on the $m_y$-invariant plane $\sigma_3$ (and its image under $m_z$) crosses the line $\Lambda$ (compared to $\Sigma$ in Fig.~\ref{fig_SG33_4B_lines}(a)), while the nonessential NL on the BZ boundary remains. Notably, this case is actually independent of the IRREP ordering at Z.  
Instead splitting the IRREPs into $\{\Gamma_1,\Gamma_4\}$ and $\{\Gamma_2,\Gamma_3\}$ gives the NLs of Fig.~\ref{fig_SG33_4B_lines}(c) with the NL pair on $\sigma_3$ of Fig.~\ref{fig_SG33_4B_lines}(b) but now also connected by another NL on the perpendicular plane $\sigma_5$ with the two connection points on $\Lambda$. These connection points are \textit{simple} point nodes, i.e.~only twofold degenerate. The NL on the BZ boundary remains. Similarly to the case Fig.~\ref{fig_SG33_4B_lines}(b), the nodal structure is independent of the IRREP ordering at Z.  

Taking again the IRREPs splitting at $\Gamma$ in Fig.~\ref{fig_SG33_4B_lines}(a), but now exchanging the IRREPs at Z, we get Fig.~\ref{fig_SG33_4B_lines}(d). This results in two new crossing point pairs on $\Lambda$ since the two valence  bands at $\Gamma$ ($\{\Gamma_1,\Gamma_2\}$) must now instead connect with the conduction band at Z ($Z_{5}$). These crossings are only protected on the $m_y$-invariant plane $\sigma_3$ leading to two independent NLs coexisting on $\sigma_3$. We also note the presence of eight accidental isolated point-nodes inside the BZ, which we  discuss in details in Section \ref{four_top} when we address the local topology of the nodal structures. 
Finally, reusing the IRREP ordering of Fig.~\ref{fig_SG33_4B_lines}(d) but now exchanging the relative ordering of the two valence states at $\Gamma$, i.e.~setting $E_{\Gamma_2} < E_{\Gamma_1}$, the bands $\{\Gamma_1\rightarrow\Lambda_1,\Gamma_3\rightarrow\Lambda_3\}$ and $\{\Gamma_2\rightarrow\Lambda_2,\Gamma_4\rightarrow\Lambda_4\}$ must cross on $\Lambda$. These crossings are protected on both the $m_x$-invariant plane $\sigma_5$ and the $m_y$-invariant plane $\sigma_3$. This explains the higher connectivity of the nodal structure in Fig.~\ref{fig_SG33_4B_lines}(e), where a small extra NL connects the two independent NLs of $\sigma_3$.

\subsection{Eight-band subspace}\label{eight}
We have already seen above that band structures of a four-band subspace with SG33 must be connected, i.e.~there cannot be any energy band gap splitting the four bands into smaller unconnected subspaces. Beyond the minimum four-band connectivity, we show here that piled-up four-band structures can also be connected in a non-trivial way, leading to $4N$ fully connected bands. We give here a detailed study of the eight-band subspace case from which it is easy to generalize to an arbitrary number of bands. 

\begin{figure}[htb]
\centering
\begin{tabular}{c} 
	\includegraphics[width=0.99\linewidth]{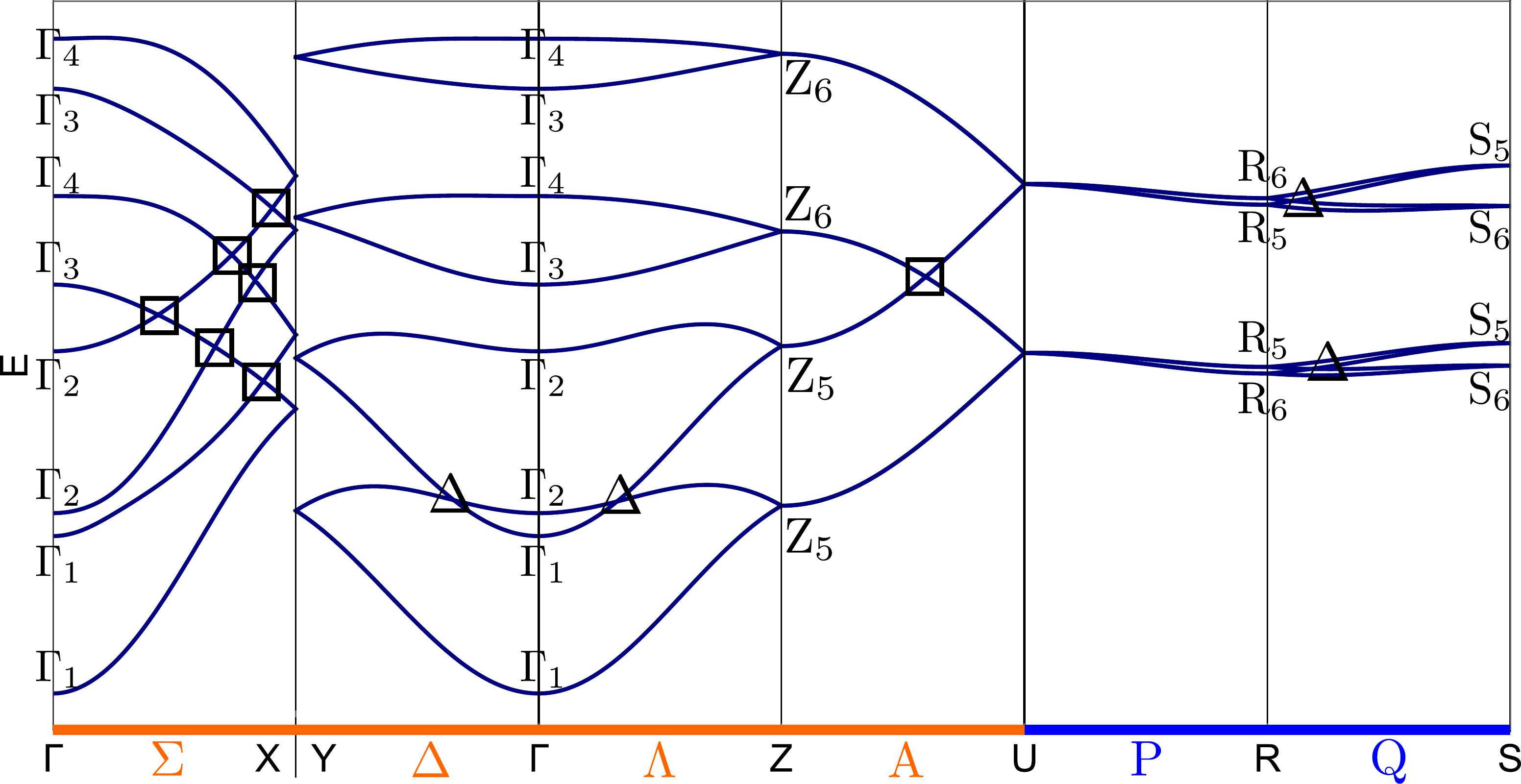} 
\end{tabular}
\caption{\label{fig_SG33_8bands} Electronic band structure of an eight-band subspace obtained from a tight-binding model for SG33-AI. The global band topology is fully determined by the energy-ordered list of IRREPs at the HSPs $\{\Gamma,\text{Z},\text{S},\text{R}\}$. Band crossing points that are part of nonessential NLs are marked with $\square$, $\triangle$ and $\bigcirc$ following the convention in Fig.~\ref{fig_SG33}(d).
}
\end{figure}
\begin{figure*}[t]
\centering
\begin{tabular}{l} 
\hline
\hline
\begin{tabular}{ccc|ccc} 
	&$ (\Gamma_{II} , \text{Z}_I ) $& &
	 &$ (\Gamma_{III} , \text{Z}_{I} ) $ 
	  & \\
\hline
	\includegraphics[width=0.16\linewidth]{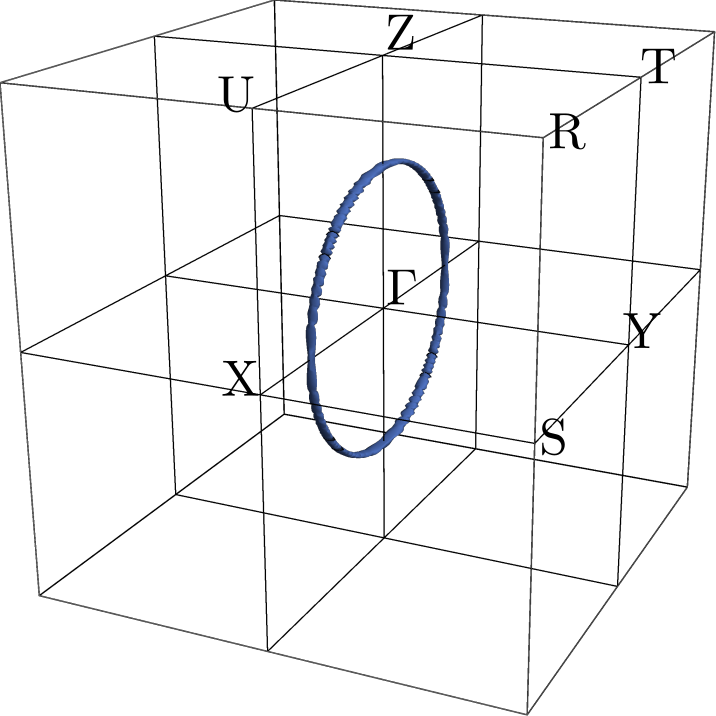} & 		
	\includegraphics[width=0.16\linewidth]{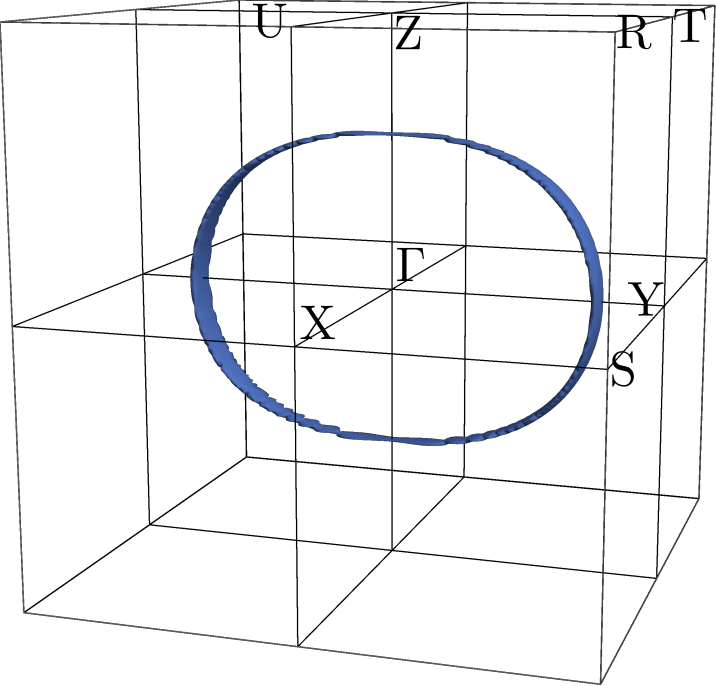} & 
	\includegraphics[width=0.16\linewidth]{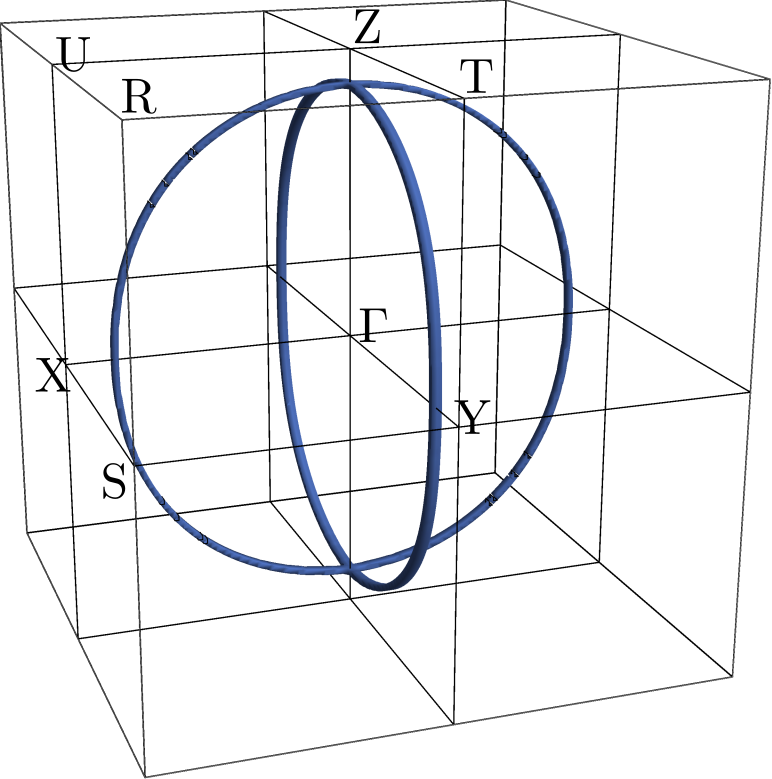} &
	\includegraphics[width=0.16\linewidth]{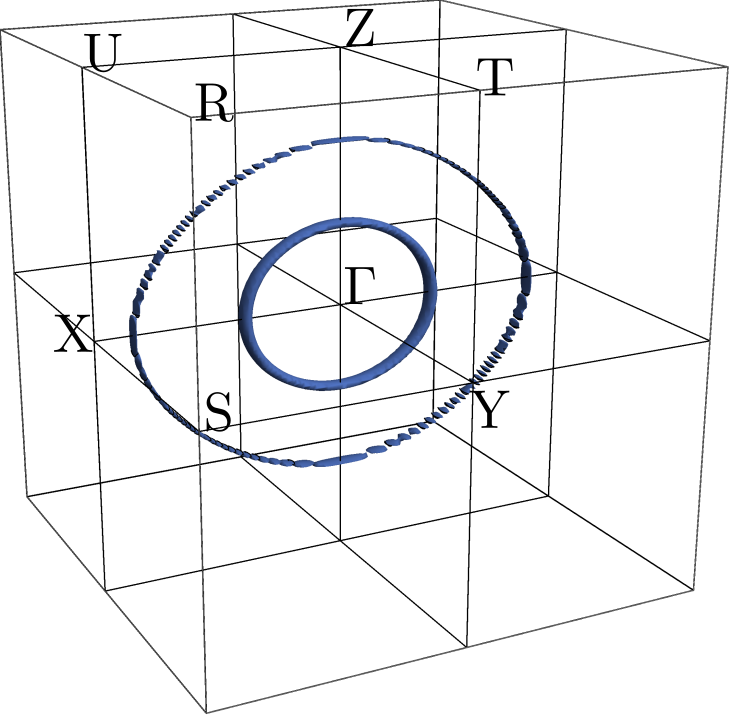} & 
	\includegraphics[width=0.16\linewidth]{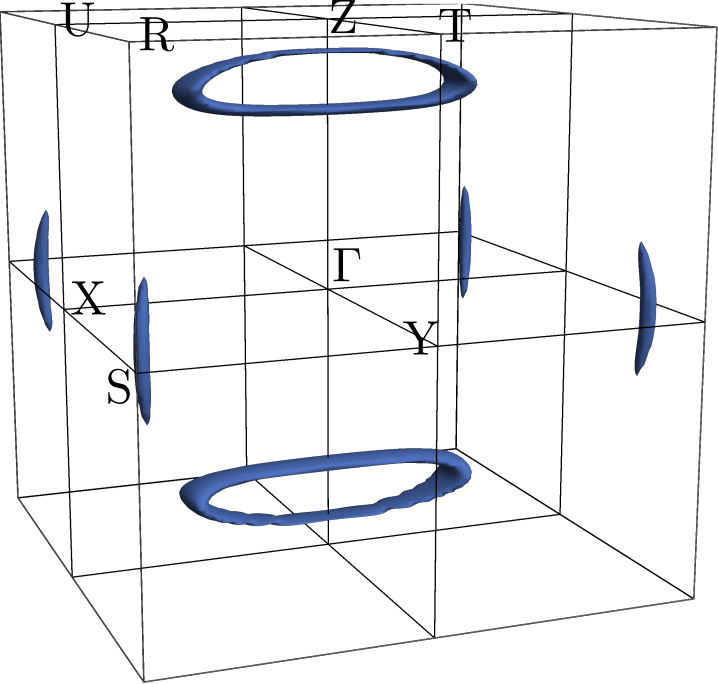} &
	 \includegraphics[width=0.16\linewidth]{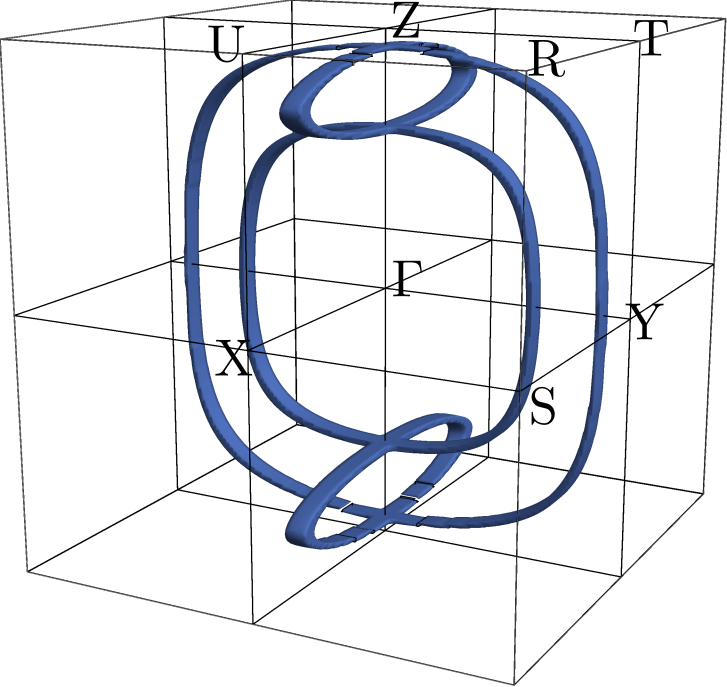} \\
	$\{\Gamma_1,\Gamma_1,\Gamma_2,\Gamma_3 \}$ & 
	$\{\Gamma_1,\Gamma_1,\Gamma_3,\Gamma_4 \}$ & 
	$\{\Gamma_1,\Gamma_1,\Gamma_2,\Gamma_4 \}$ &
	$\{\Gamma_1,\Gamma_1,\Gamma_2,\Gamma_2 \}$ & 
	$\{\Gamma_1,\Gamma_1,\Gamma_3,\Gamma_3 \}$ &
	$\{\Gamma_1,\Gamma_1,\Gamma_4,\Gamma_4 \}$ \\
	$\{Z_5,Z_6\}$ & $\{Z_5,Z_6\}$ & $\{Z_5,Z_6\}$ &
	 $\{Z_5,Z_6\}$ & $\{Z_5,Z_6\}$ & $\{Z_5,Z_6\}$ \\
	 (a) & (b) & (c) & (d) & (e) & (f) 
\end{tabular} \\
\hline
\hline
\begin{tabular}{c|ccccc} 
	$ (\Gamma_{I} , \text{Z}_{II}  ) $ & &  &  $ (\Gamma_{II} , \text{Z}_{II}  ) $
	&  \\
\hline
	 \includegraphics[width=0.16\linewidth]{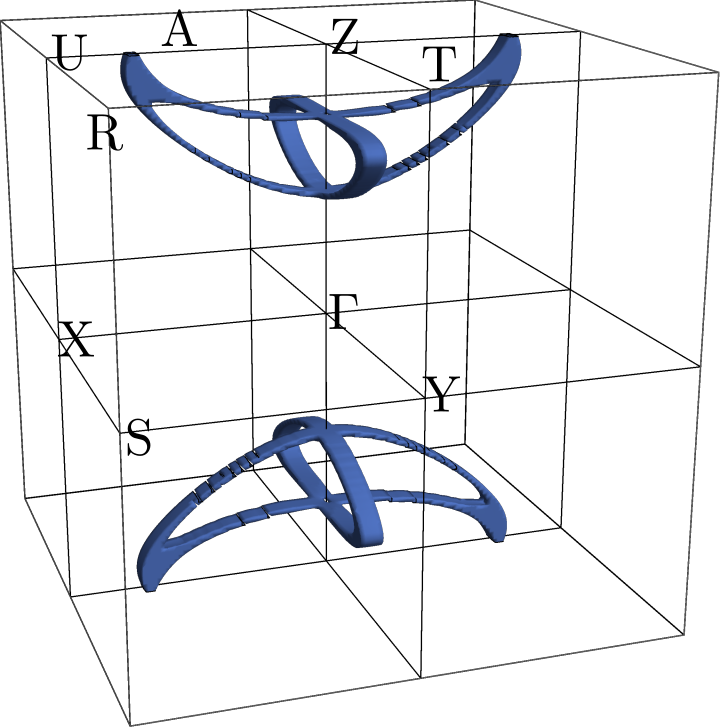} &
	 \includegraphics[width=0.16\linewidth]{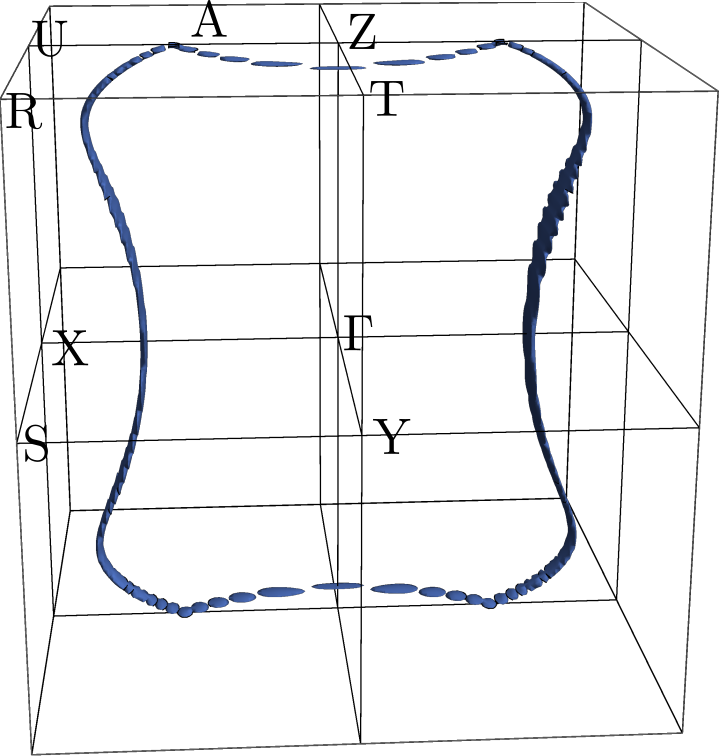} & 
	 \includegraphics[width=0.16\linewidth]{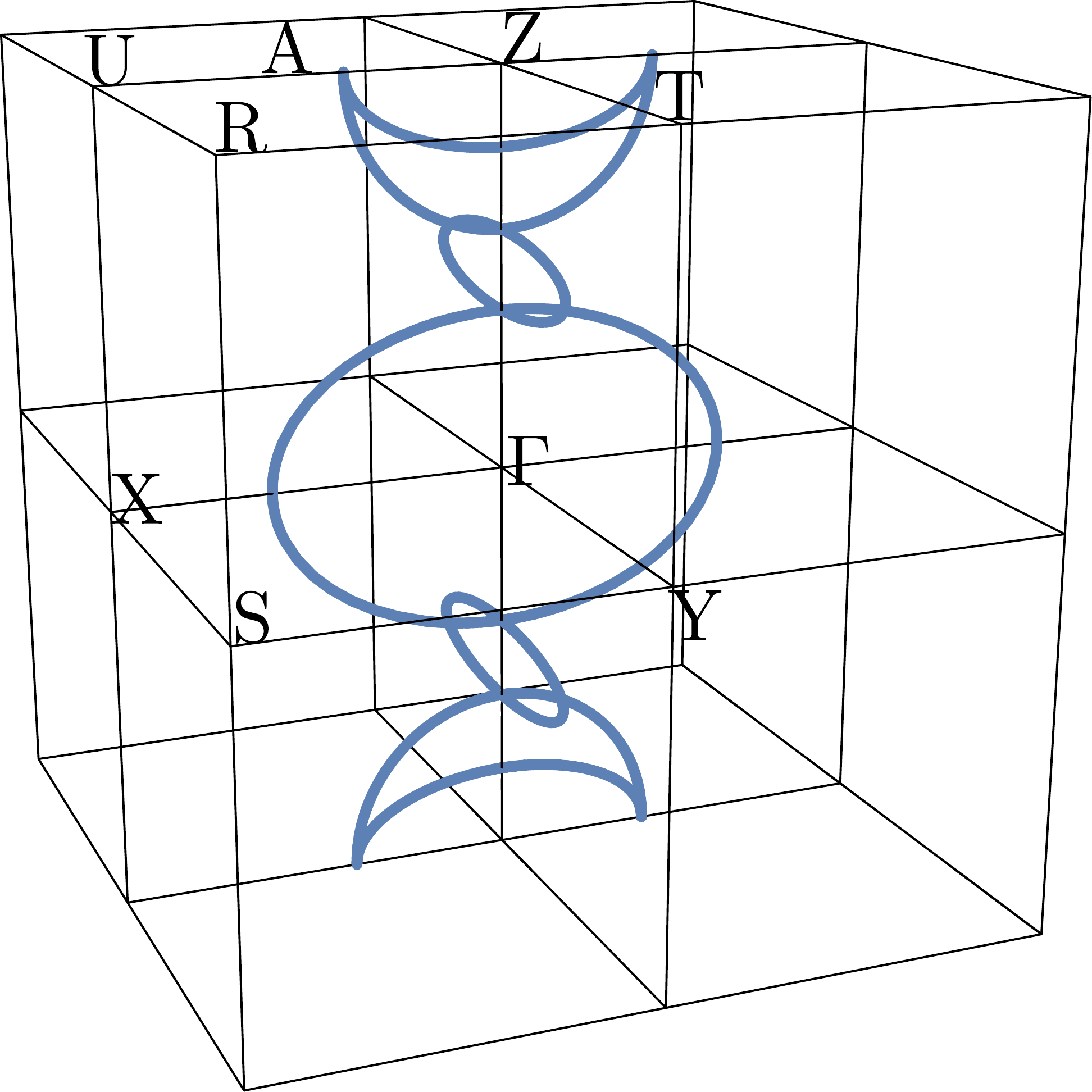}&
	 \includegraphics[width=0.16\linewidth]{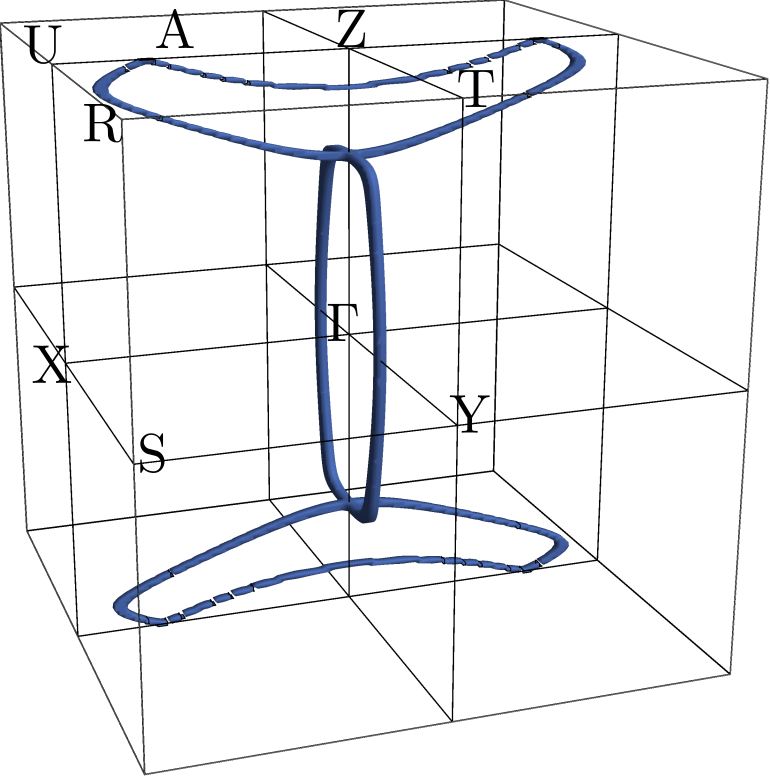} &
	\includegraphics[width=0.16\linewidth]{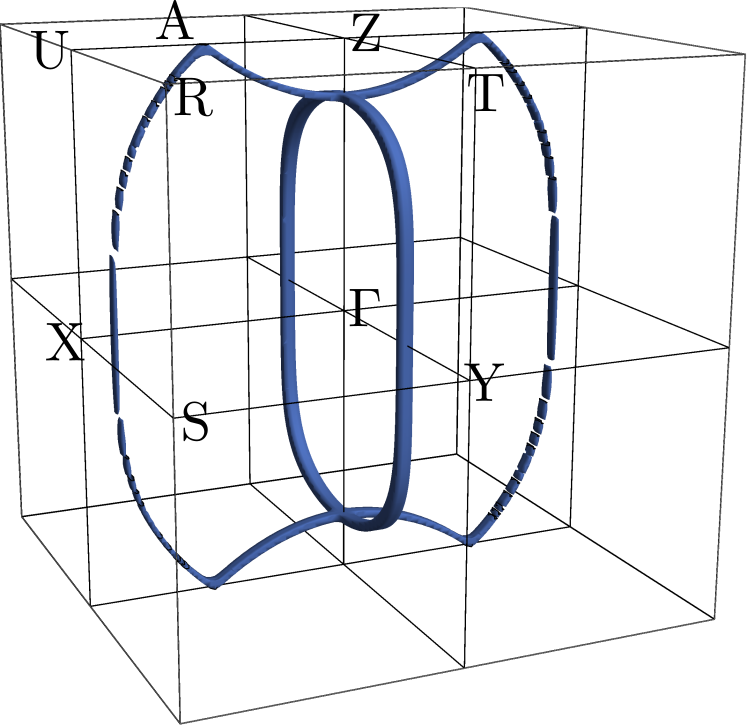} & 
	\includegraphics[width=0.16\linewidth]{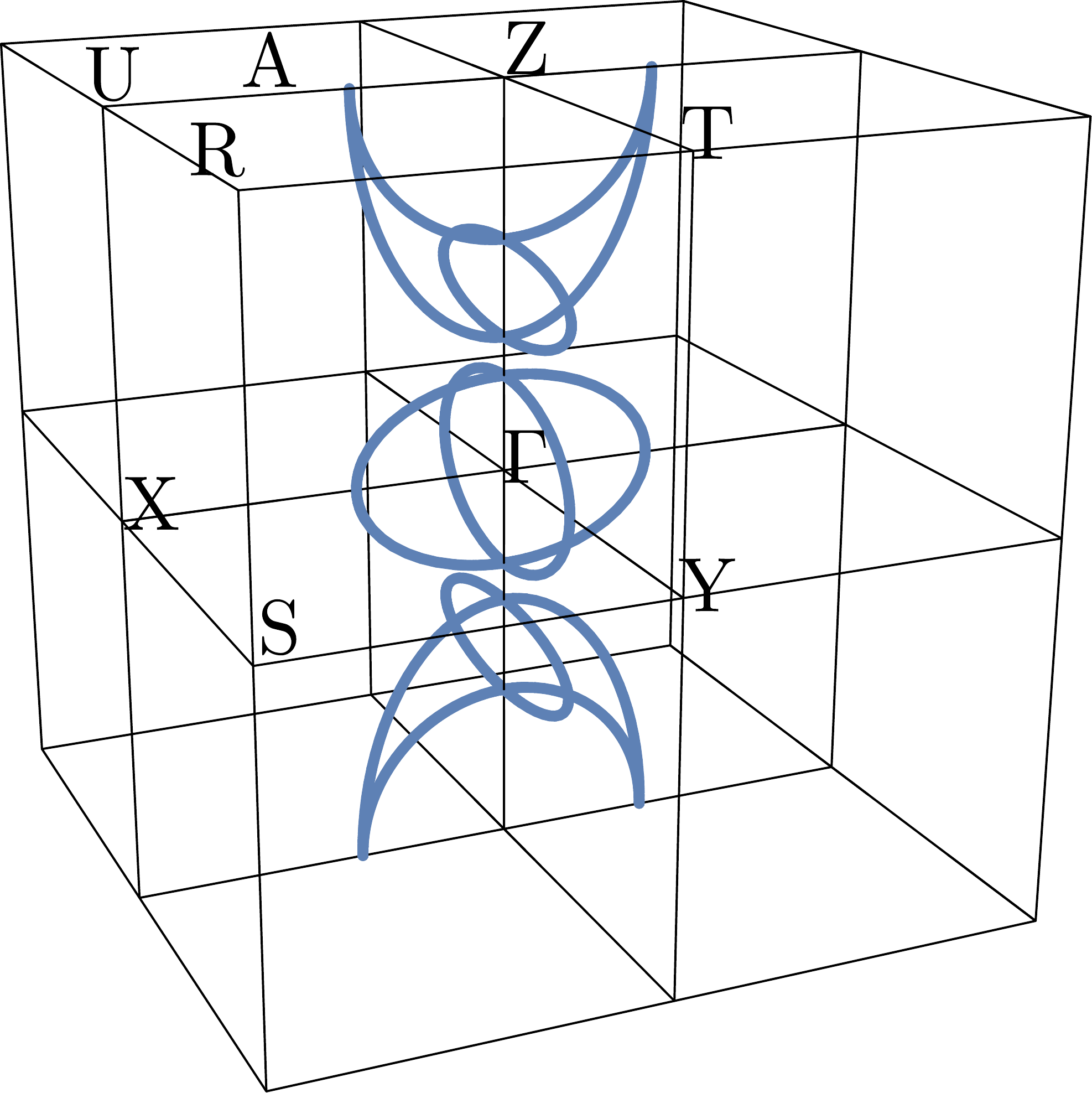} \\
	$\{\Gamma_1,\Gamma_2,\Gamma_3,\Gamma_4  \}$ & 
	$\{\Gamma_1,\Gamma_1,\Gamma_2,\Gamma_3 \}$ & 
	$\{\Gamma_1,\Gamma_1,\Gamma_2,\Gamma_3 \}$ &
	 $\{\Gamma_1,\Gamma_1,\Gamma_3,\Gamma_4 \}$ &
	$\{\Gamma_1,\Gamma_1,\Gamma_2,\Gamma_4 \}$ & 
	 $\{\Gamma_1,\Gamma_1,\Gamma_2,\Gamma_4 \}$  \\
	 $\{Z_i,Z_i\}_{i=5\mathrm{or}6}$  &
	 $\{\text{Z}_5,\text{Z}_5\}$  &  
	 $\{\text{Z}_6,\text{Z}_6\}$ &
	 $\{Z_i,Z_i\}_{i=5\mathrm{or}6}$ &
	 $\{\text{Z}_5,\text{Z}_5\}$ &
	 $\{\text{Z}_6,\text{Z}_6\}$ \\
	  (g) & (h) & (i) & (j) & (k) & (l) 
\end{tabular} \\
\hline
\hline
\begin{tabular}{cccc} 
	   & $ (\Gamma_{III} , \text{Z}_{II} ) $ &  &  \\
\hline 
	 \includegraphics[width=0.16\linewidth]{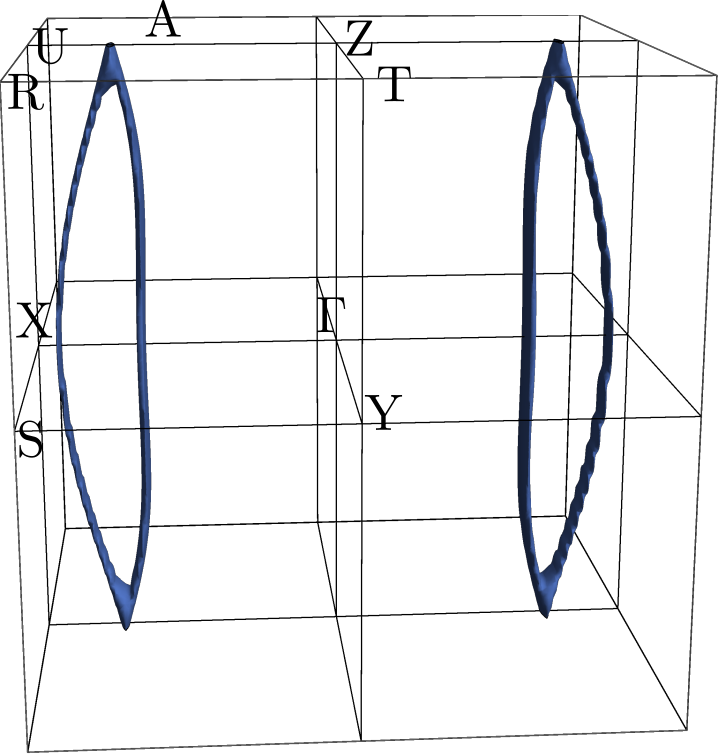} & 
	 \includegraphics[width=0.16\linewidth]{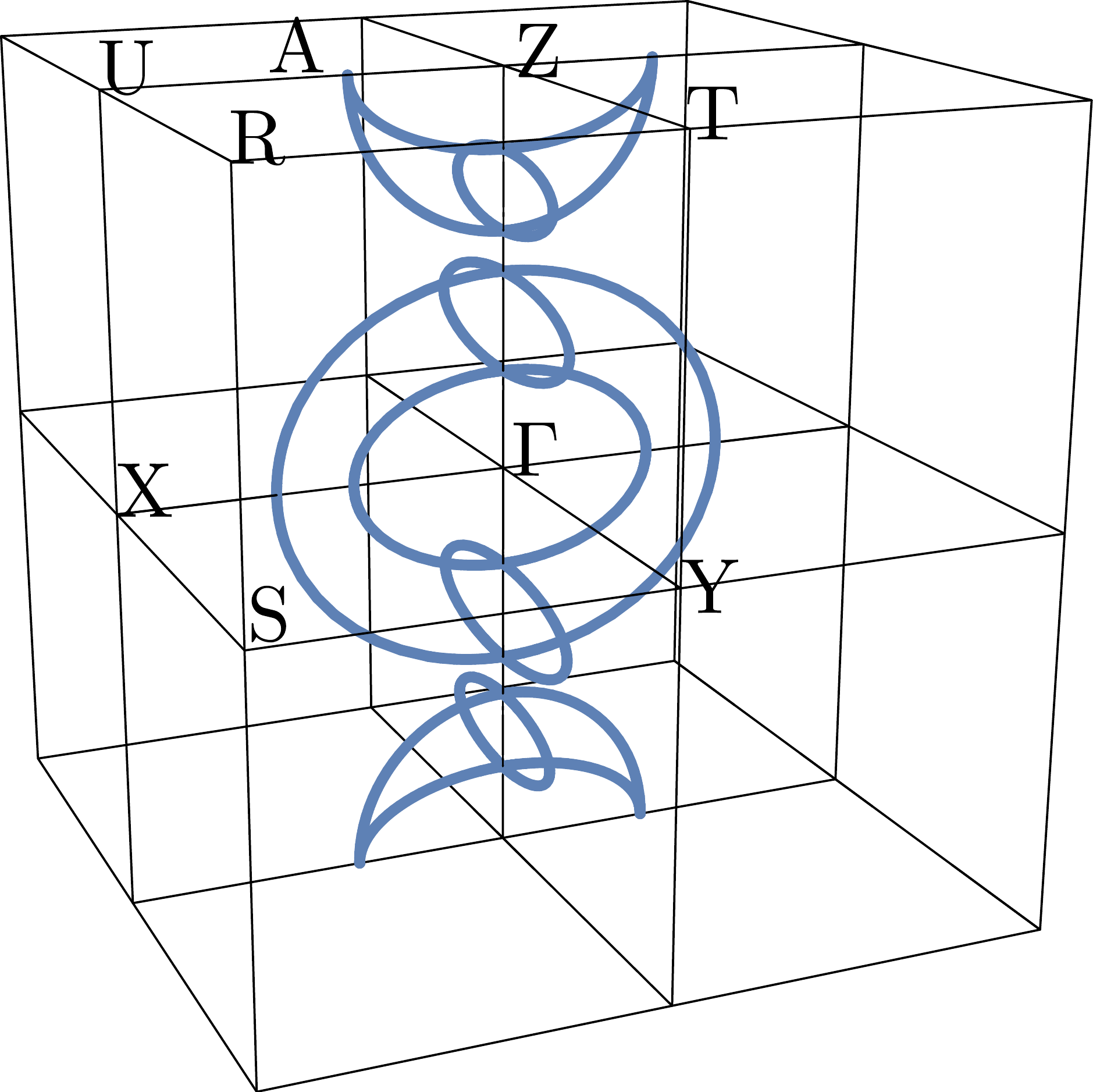} &
	 \includegraphics[width=0.16\linewidth]{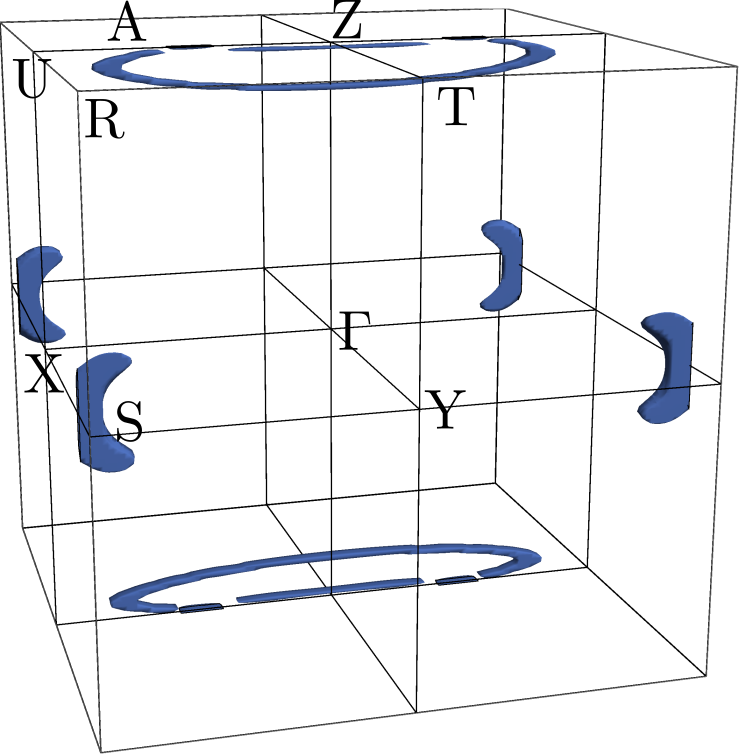} &
	 \includegraphics[width=0.16\linewidth]{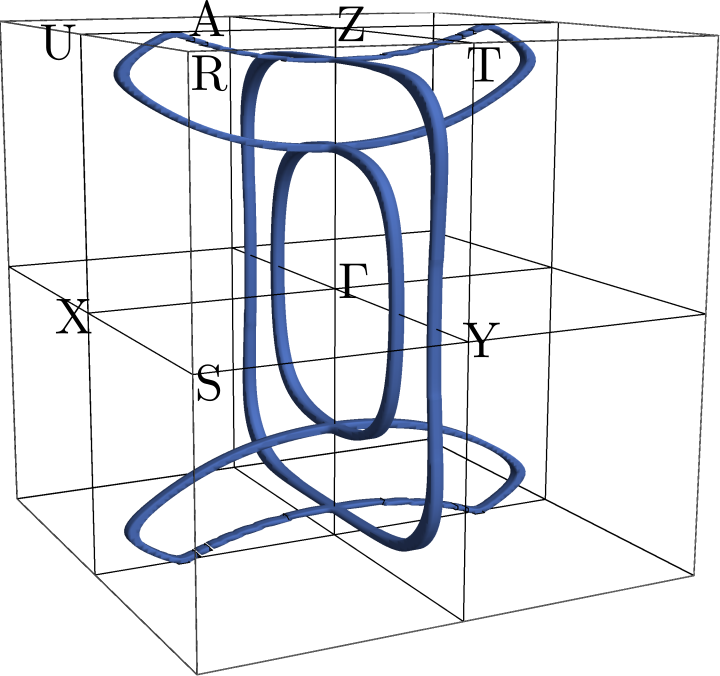}   \\
	$\{\Gamma_1,\Gamma_1,\Gamma_2,\Gamma_2 \}$ &
	$\{\Gamma_1,\Gamma_1,\Gamma_2,\Gamma_2 \}$ & 
	$\{\Gamma_1,\Gamma_1,\Gamma_3,\Gamma_3 \}$ &
	$\{\Gamma_1,\Gamma_1,\Gamma_4,\Gamma_4 \}$  \\
	 $\{Z_5,Z_5\}$ &
	 $\{Z_6,Z_6\}$ & 
	  $\{Z_i,Z_i\}_{i=5\mathrm{or}6}$ & 
	  $\{Z_i,Z_i\}_{i=5\mathrm{or}6}$ \\
	   (m) & (n) & (o) & (p) 
\end{tabular} \\
\hline
\hline
\begin{tabular}{c|c|c|c|c|c}
 $ ( \text{S}_{II} , \text{R}_I ) $ & $ ( \text{S}_I , \text{R}_{II} ) $ & $ (\text{S}_{II} , \text{R}_{II} ) $ &
 $(\Gamma_{III},\text{Z}_{I},\text{S}_{II},\text{R}_{I})$ & 
	$(\Gamma_{II},\text{Z}_{II},\text{S}_{II},\text{R}_{I})$ & 
	$(\Gamma_{III},\text{Z}_{II},\text{S}_{II},\text{R}_{I})$ \\
 \hline
	 \includegraphics[width=0.16\linewidth]{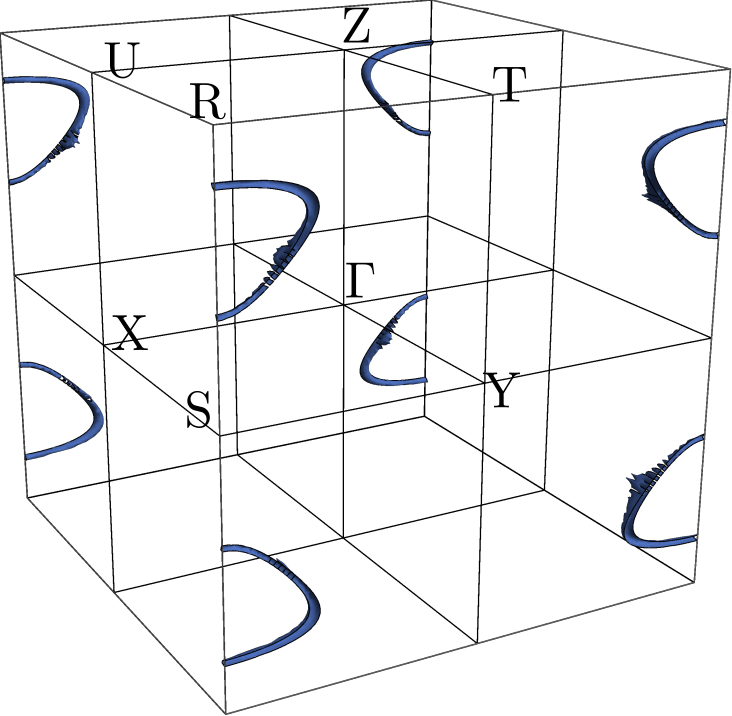} & 
	 \includegraphics[width=0.16\linewidth]{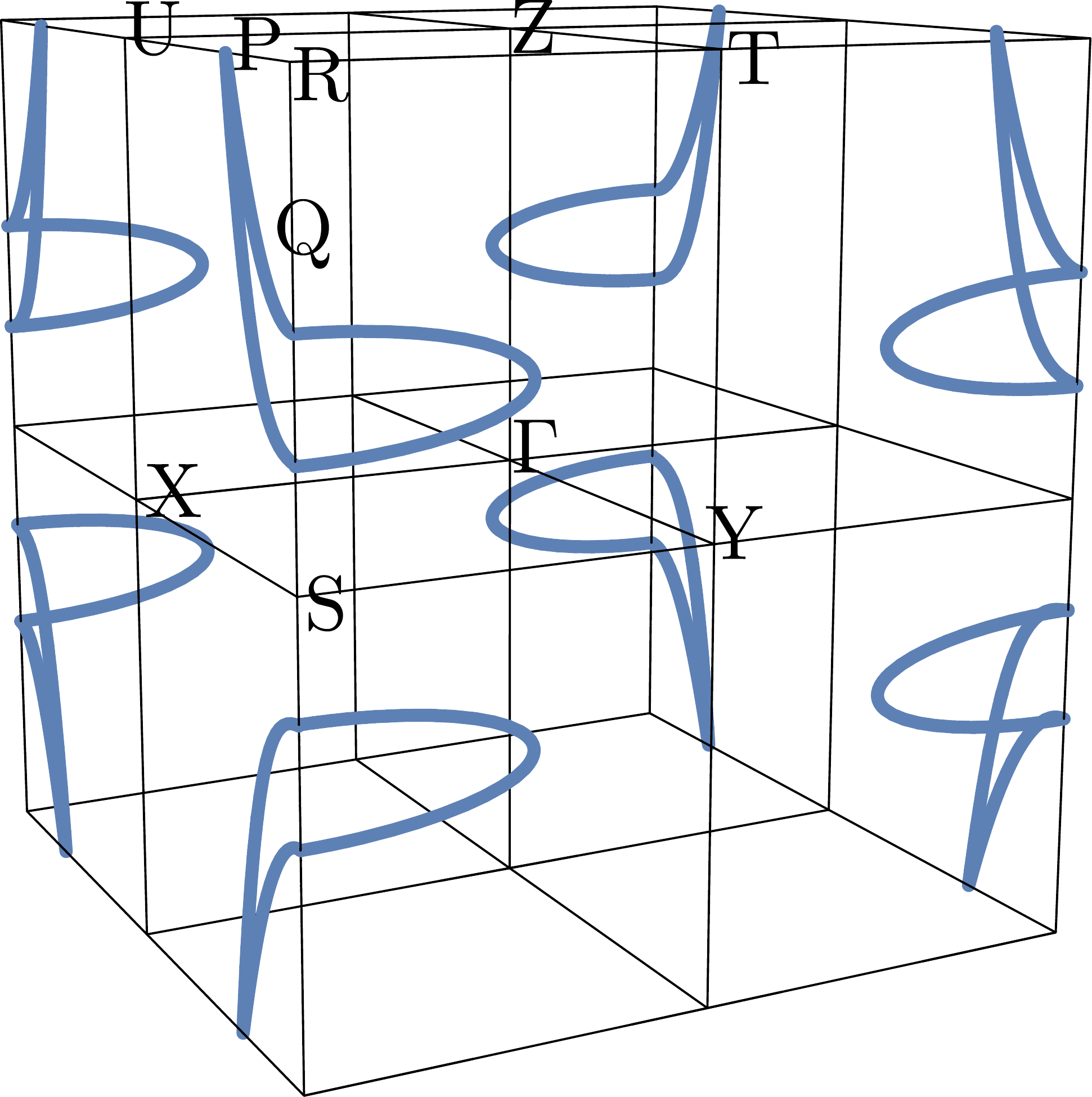} & 
	 \includegraphics[width=0.16\linewidth]{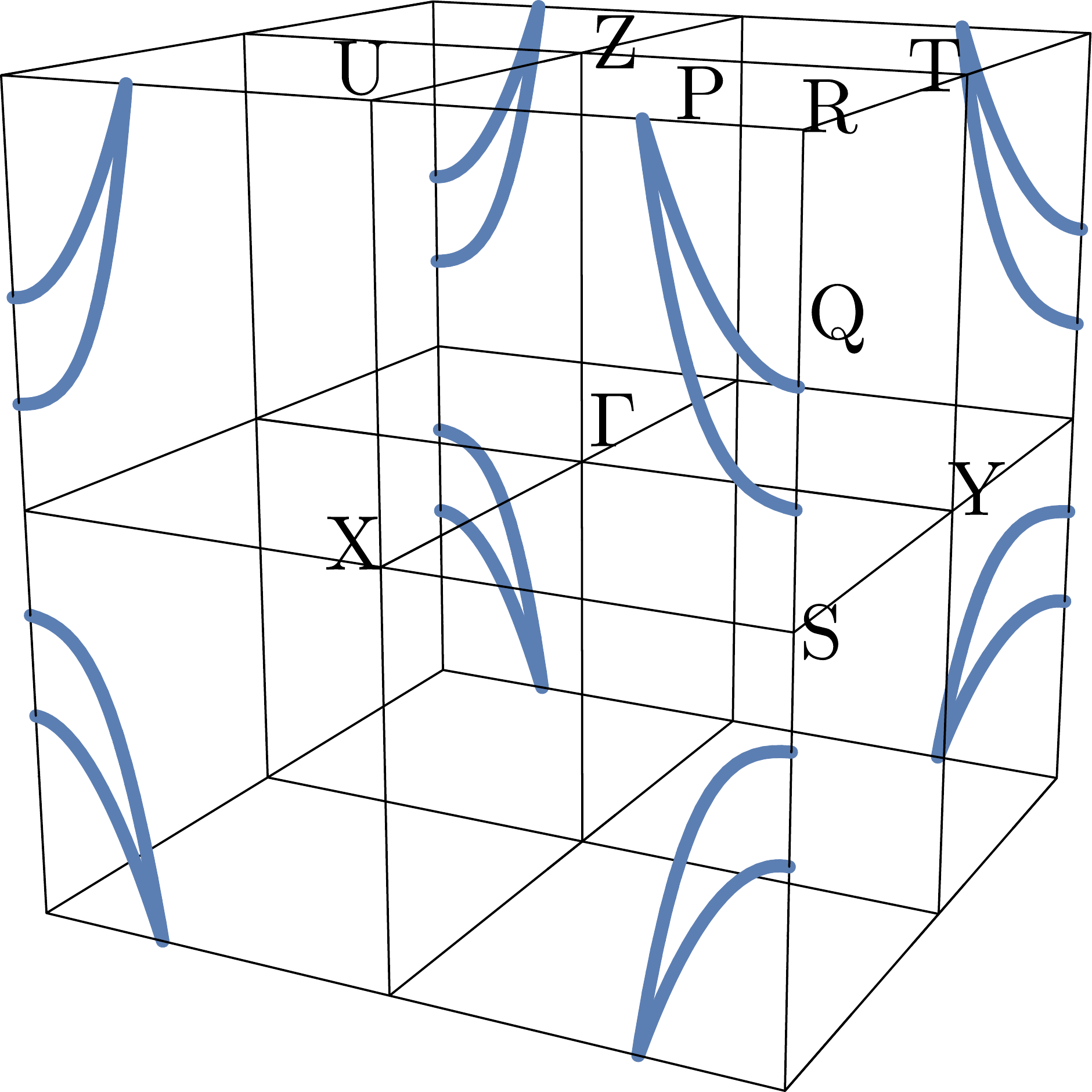} &
	\includegraphics[width=0.16\linewidth]{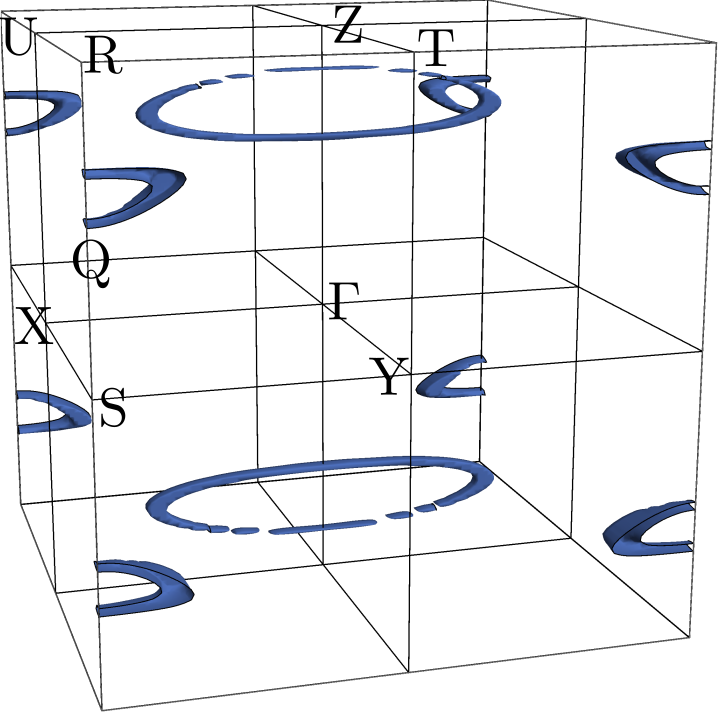} &
	\includegraphics[width=0.16\linewidth]{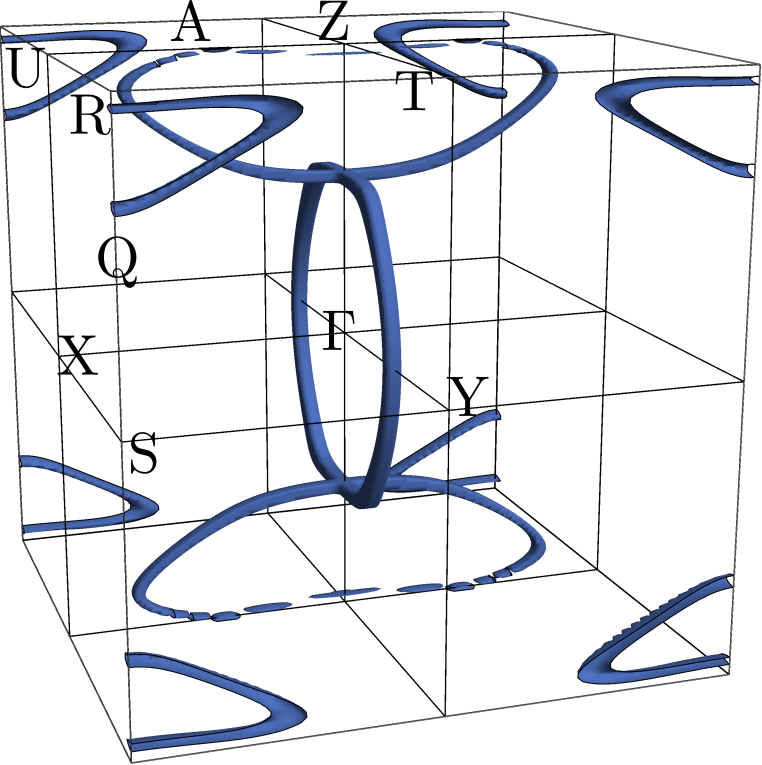} &
	\includegraphics[width=0.16\linewidth]{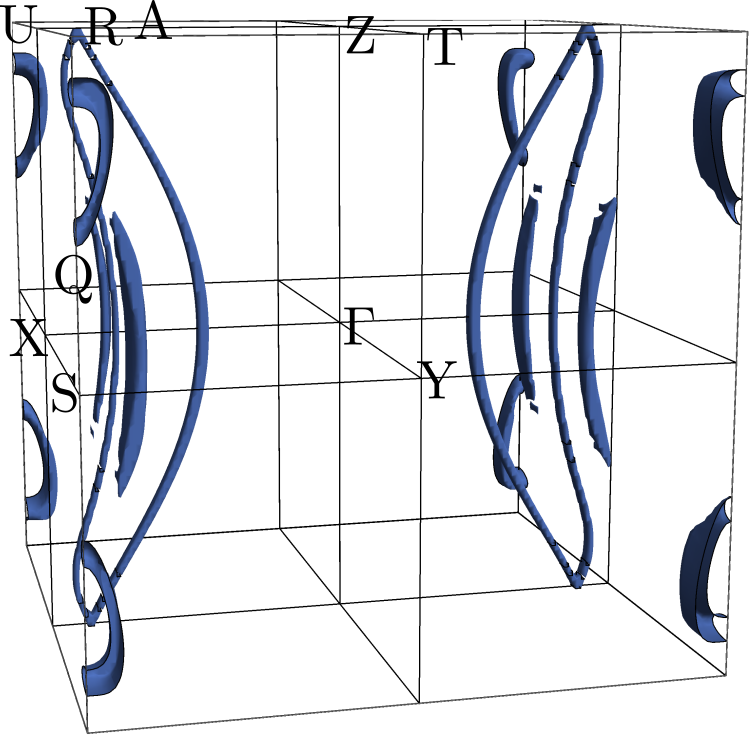}  \\	
	$\{\text{S}_i , \text{S}_i\}_{i=5,6}$ & 
	$\{\text{S}_5 , \text{S}_6\}$ & 
	$\{\text{S}_i , \text{S}_i\}_{i=5,6}$ &	$\{\Gamma_1,\Gamma_1,\Gamma_3,\Gamma_3\}$ & 
	$\{\Gamma_1,\Gamma_1,\Gamma_3,\Gamma_4\}$ & 
	$\{\Gamma_1,\Gamma_1,\Gamma_2,\Gamma_2\}$ \\
	  $\{\text{R}_6 , \text{R}_5\}$ & 
	  $\{\text{R}_i , \text{R}_i\}_{i=5,6}$ & 
	  $\{\text{R}_i , \text{R}_i\}_{i=5,6}$ &
	  $\{S_i,S_i\}_{i=5,6}$ & 
	$\{Z_i,Z_i\}$,$\{S_i,S_i\}_{i=5,6}$ & 
	$\{Z_i,Z_i\}$,$\{S_i,S_i\}_{i=5,6}$\\
	(q) & (r) & (s) & (t) & (u) & (v)
\end{tabular} \\
\hline
\hline
\end{tabular}
\caption{\label{fig_SG33_8B_lines} Line-nodal structures of an eight-band subspace at half-filling for the HSP classes  according to Table \ref{point_classes}. (a-p) Line-nodal structures for the five nontrivial HSP classes $(\Gamma_{\alpha},\text{Z}_{\beta})$ of the region $\mathcal{B}_{\Gamma}$. (q-s) Line-nodal structures for the three nontrivial classes $(\text{S}_{\beta'},\text{R}_{\beta''})$ of the region $\mathcal{B}_{\text{R}}$. (t,u,v) Examples of combined HSP classes of the regions $\mathcal{B}_{\Gamma}$ and $\mathcal{B}_{\text{R}}$ showing additivity of the NLs: (t)=(e)+(q), (u)=(j)+(q), and (v)=(m)+(q).
} 
\end{figure*}

Figure~\ref{fig_SG33_8bands} shows an example of a band structure for SG33-AI with eight bands non-trivially connected. The nonessential symmetry protected band crossings along the HSLs are marked with squares, triangles and circles following the convention established for Fig.\ref{fig_SG33}(d). Similarly to the four-band case we now extract all line-nodal structures at half-filling for an eight-band subspace.
In Fig.~\ref{fig_SG33_8B_lines}(m) we show the line-nodal structure corresponding to Fig.~\ref{fig_SG33_8bands}. The two simple band crossings on the line $\Sigma$ at half-filling (middle squares in Fig.~\ref{fig_SG33_8bands}), are parts of NLs within the plane $\sigma_3$. The simple NLs merge two-by-two at double band crossing points on the HSL A. These double point nodes, marked by a square in Fig.~\ref{fig_SG33_8bands} along the HSL A, are formed by the crossing of the two doubly degenerate bands $Z_5\rightarrow A_5$ and $Z_6\rightarrow A_6$. 

Generally for the eight band case, we find the line-nodal structure fully determined by the energy ordered lists of IRREPs at the active HSPs $\{\Gamma,\text{Z},\text{S},\text{R}\}$. This is a result from combining the compatibility relations and the band permutation rules of Fig.~\ref{fig_SG33}(c) and Table~\ref{permutation_SG33}. 
In order to explore the space of all topologically inequivalent eight-band structures with SG33-AI it is convenient to coarse-grain the IRREP combinatorics into valence high-symmetry point classes $\Gamma_{I,II,III} ,~ \text{Z}_{I,II} ,~ \text{S}_{I,II}$, and $\text{R}_{I,II}$, defined through the sets of valence IRREPs at the corresponding HSPs given in Table~\ref{point_classes}. In total 24 global band classes $ ( \Gamma_{\alpha} , \text{Z}_{\beta} , \text{S}_{\beta'} , \text{R}_{\beta''} )$, with $\alpha=I,II,III$ and $\beta,\beta',\beta'' = I,II$, can be formed. Since the crossings on $\mathcal{B}_{\Gamma}$ and $\mathcal{B}_{\text{R}}$ are independent, we can split the classes into the two subregion classes $(\Gamma_{\alpha}, \text{Z}_{\beta})$ for $\mathcal{B}_{\Gamma}$ combined with the trivial, i.e.~fully gapped, $(\text{S}_{I},\text{R}_{I})$ class and $(\text{S}_{\beta'},\text{R}_{\beta''})$ for $\mathcal{B}_{\text{R}}$ combined with the trivial $(\Gamma_I,\text{Z}_I)$. This leaves $5+3=8$ nontrivial cases. All the other global band classes can be straightforwardly obtained by combining the two subregion classes. 
{\def\arraystretch{1.3}  
\begin{table}[hb]
\caption{\label{point_classes} HSP classes of an eight-band subspace at half-filling for SG33-AI, separated into the symmetry independent regions $\mathcal{B}_{\Gamma}$ and $\mathcal{B}_{\text{R}}$. Each class is defined by the set of valence IRREPs at the corresponding HSP. 
}
\begin{tabular*}{\linewidth}{c @{\extracolsep{\fill}} c c }
\hline
\hline
 \begin{tabular}{l}   
	$\Gamma_I = \left\{\Gamma_1,\Gamma_2,\Gamma_3,\Gamma_4\right\}$ \\
	$\Gamma_{II} = \left\{ \Gamma_j,\Gamma_j, \Gamma_k,\Gamma_{l}  \right\}_{j\neq k \neq l}$ \\
	$\Gamma_{III} = \left\{\Gamma_j,\Gamma_j,\Gamma_k,\Gamma_k \right\}_{j\neq k}$ 
 \end{tabular}  &  
  \begin{tabular}{l}  
	$\text{p}_{I}  = \left\{ p_{5} , p_{6} \right\}$  \\
	$\text{p}_{II}  = \left\{ p_{j} , p_{j} \right\}_{j=5,6}$  \\
	 \hline
	$\text{p} \in \{ \text{Z} , \text{S} , \text{R} \}$ 
 \end{tabular} \\
 \hline
  \hline
\end{tabular*}
\end{table} 
}

In Fig.~\ref{fig_SG33_8B_lines}(a-p) we show all the nodal structures for the five nontrivial HSP classes of $\mathcal{B}_{\Gamma}$ and in Fig.~\ref{fig_SG33_8B_lines}(q-s) the three nontrivial classes of $\mathcal{B}_{\text{R}}$. As seen, each class can be further split according to the explicit set of valence IRREPs possibly leading to qualitatively different nodal structures within the same global band class \footnote{Most of Fig.~\ref{fig_SG33_8B_lines} is obtained numerically from a generic eight-band tight-binding model, but a few cases (i,l,n,r,s) are drawn by hand as they would require an eight-band subspace imbedded in additional bands in order to lift artificial symmetries.}. Still, as we will see in Section \ref{top_inv}, the different cases among the same global HSP class share similar topological features.
For these nodal structures, every NL is in general twofold degenerate (i.e.~two bands crossing), except the NL crossing the HSLs $\text{A}$ or $\text{P}$ leading to a fourfold degenerate point since the IRREPs $A_{5,6}$ and $P_{5,6}$ are doubly degenerate. 
We note that depending on the explicit ordering of the IRREPs within a given set of valence or conduction IRREPs, some cases allow the presence of a small extra NL connecting two other independent NLs, as illustrated for example in Fig.~\ref{fig_SG33_8B_lines}(f). This is true for the cases (d,f,g,i,l,n,r,). We discuss in Section \ref{Lifshitz} the topological Lifshitz transformations involved in such connections of independent NLs.
We also note that the quadruple point nodes on the plane $\sigma_1$ seen in Fig.~\ref{fig_SG33_8B_lines}(e), (o) and (v) are accidental, similarly to those of Fig.~\ref{fig_SG33_4B_lines}(d). Their cigar-like shape is a mere numerical effect.

\subsection{Additivity of nodal lines}\label{additivity}
We can also show that any nodal structure of a HSP class $(\Gamma_{\alpha},\text{Z}_{\alpha'})$ for $\mathcal{B}_{\Gamma}$, is additive with any nodal structure of a HSP class $(\Gamma_{\beta},\text{Z}_{\beta'})$ for $\mathcal{B}_{\text{R}}$. Three examples of this is shown in Fig.~\ref{fig_SG33_8B_lines}(t-v). 
The case Fig.~\ref{fig_SG33_8B_lines}(t) combines Fig.~\ref{fig_SG33_8B_lines}(e), belonging to classes $(\Gamma_{III},\text{Z}_{I})$ and $(\text{S}_{I},\text{R}_{I})$, with Fig.~\ref{fig_SG33_8B_lines}(q), belonging to classes $(\Gamma_{I},\text{Z}_{I})$ and $(\text{S}_{II},\text{R}_{I})$, leading to the global class $(\Gamma_{III},\text{Z}_{I},\text{S}_{II},\text{R}_{I})$. The resulting nodal structure is given as the superposition of the NLs of Figs.~\ref{fig_SG33_8B_lines}(e) and \ref{fig_SG33_8B_lines}(q), which we write simply as (t)=(e)+(q). For Fig.~\ref{fig_SG33_8B_lines}(u) we have (u)=(j)+(q) while Fig.~\ref{fig_SG33_8B_lines}(v), shows that (v)=(m)+(q). 
The same can be done by combining any case from Fig.~\ref{fig_SG33_8B_lines}(a-p) $\in \mathcal{B}_{\Gamma}$ with any case from Fig.~\ref{fig_SG33_8B_lines}(q,r,s)$\in \mathcal{B}_{\text{R}}$. This leads to a total of 48 qualitatively distinct nodal structures among the 24 inequivalent global HSP classes listed in Table~\ref{point_classes}. 
This additivity of NLs when combing a HSP class from $\mathcal{B}_{\Gamma}$ with one of $\mathcal{B}_{\text{R}}$ is quite remarkable and very different from the case of symmetry protected point-nodes. In the latter case, the Nielsen-Ninomiya theorem constrains the global nodal structure through the requirement of  total charge cancellation over the whole BZ. Thus point nodes are still related even though they belong to symmetry independent BZ regions, see e.g.~Ref.~\cite{BBS_1}.

\subsection{Topological classification} 
We end this section by noting that the combinatorial classification of band structures we have presented above is related to the topological classification of insulating phases proposed earlier in Refs.~\cite{Slager_K,ShiozakiWallPaper_K,PoVishWata_symmetry_indic} and independently also in Ref.~\cite{BBS_1} for the classification of symmetry protected point nodes (see also the earlier contributions \cite{Liu_coreps,Liu_reps} emphasizing a classification scheme based on band IRREPs, and the recent works \cite{Graph_theory,Band_connectivity} presenting the systematic mapping of band combinatorics to a graph theory problem). 
While Ref.~\cite{Slager_K,ShiozakiWallPaper_K} gave the mapping from the band combinatorics to K-theory without TRS and in the spinless case (i.e.~the spinless class A of the Altland-Zirnbauer tenfold symmetry classes \cite{Schnyder08}), Ref.~\cite{PoVishWata_symmetry_indic} proposed to extend this approach to other Altland-Zirnbauer classes among which is the class AI we consider here. While it has been argued recently in Refs.~\cite{TopQuantChem,ShiozakiWallPaper_K, Alex_BlochOsc, Cano_EBR} that the knowledge of the sub-lattice degrees of freedom (the Wyckoff's positions) composing the bands is sometimes needed in order to distinguish topologically inequivalent band structures, we show in Section \ref{discussion} that this is not relevant in our case, see also Ref.~\cite{Po_FragileTop}. This justifies our choice to label our IRREPs-based combinatorial classification of band structures as \textit{topological}. 

Complementary to this task of topological classification of band structures and their symmetry-protected nodal structures is the topological characterization of the nodal structures themselves in terms of their local and global topologies. These are determined by the local topological charges reflecting the stability of the nodal elements (points, lines) of a nodal structure and the global topology constraining the local charges of different coexisting elementary nodal structures. We solve this topological characterization in the next section.

\section{Local and global topology of crystalline-nodal structures}\label{top_meta}
Up to now we have unfolded the classification of band structures and their induced line-nodal structures for SG33-AI in terms of HSP IRREPs combinatorics. We now want to determine the topology of the line-nodal structures. To that end, we need to step back a bit and formalize the problem within vector bundle theory, which is the natural conceptual framework for Bloch band theory. We show in this section that the local topology of any nodal structure is characterized through symmetry constrained homotopy groups corresponding to poloidal-toroidal and monopole charges, and with a global topology given in terms of constraints among the local charges. 

In our context the non-interacting crystalline system is modeled in terms of a Bloch Hamiltonian operator $\mathcal{H}({\boldsymbol{k}}) = \sum_{\mu,\nu=1}^{N} \vert c_{\mu},\boldsymbol{k}\rangle H_{\mu\nu}(\boldsymbol{k}) \langle c_{\nu},\boldsymbol{k} \vert$ with $H$ a $N\times N$ complex Hermitian matrix defined at every point of the BZ ($\boldsymbol{k}\in$ BZ $\cong \mathbb{T}^3$) and given a separable Hilbert space $\mathscr{H} = \bigoplus_{\boldsymbol{k}} \mathscr{H}_{\boldsymbol{k}}$ on which $\mathcal{H}(\boldsymbol{k})$ acts \footnote{This is obtained through the Fourier transform of an appropriate basis set modeling the physical degrees of freedom. In particular, we assume real space functions that have the periodicity of the sub-lattice sites belonging to the same Wyckoff position and not only the periodicity of the Bravais lattice. This choice of a trivializing reference section of the total Bloch bundle has been shown to be the more physically relevant for studying parallel transports, see Refs.~\cite{FruchartCarpentier1, Berry_connection_Moore}.}.
The spectrum is given by the set of eigenvalues $\mathrm{eig}\{\mathcal{H}(\boldsymbol{k})\} = \{E_1(\boldsymbol{k}),E_2(\boldsymbol{k}),\dots,E_N(\boldsymbol{k})\}$ with the Bloch-eigenstates $\vert \psi_n,\boldsymbol{k}\rangle =  \vert c_{\nu},\boldsymbol{k}\rangle \breve{U}_{\nu n}(\boldsymbol{k}) \in \mathscr{H}_{\boldsymbol{k}}$. We can order the spectrum in energy as $E_{n_1}\leq E_{n_2}$ for $n_1<n_2$, $n_i = 1,2,\dots,N$. In the following we refer to a simplified picture by considering the vector space $V_{\boldsymbol{k}}$ spanned by the (orthonormal) eigenvectors $[\breve{U}(\boldsymbol{k})]_n = \vert \breve{U}_n, \boldsymbol{k} \rangle$ of the matrix Hamiltonian $H({\boldsymbol{k}})$, given for instance by a tight-binding model, with $\breve{U}(\boldsymbol{k}) \in U(N)$ the diagonalization matrix. The eigenvalues vary continuously with $\boldsymbol{k}$ and form the energy bands. When some bands cross, say $E_{N_v}(\boldsymbol{k}_i)=E_{N_v+1}(\boldsymbol{k}_i)$, we can split the bands away from $\boldsymbol{k}_i$ into a valence subspace with the eigenvalues $\{E_1,\dots,E_{N_v}\}$ and a conduction subspace with the eigenvalues $\{E_{N_v+1},\dots,E_{N}\}$. The locus of degeneracies for a given $N_v$, called $L_{v}=\{\boldsymbol{k}\vert E_{N_v}(\boldsymbol{k})=E_{N_v+1}(\boldsymbol{k}), ~\boldsymbol{k}\in \mathrm{BZ}\}$, defines a nodal structure in $k$-space. Note that more than two bands can be degenerated at a fixed $N_v$. Fixing $N_v$ over the whole BZ, we aim to characterize the topology of the nodal structures that are enforced by the symmetries of the system. 

In general, the nodal structure (i.e.~the locus of degeneracies) is the union of several disconnected elements $L_{v} = \bigcup_i L_{v}^{(i)}$ where each nodal component $L_{v}^{(i)}$ cannot be split without breaking it. Let us write the punctured BZ over which the spectrum is gapped $B_v =\mathbb{T}^3 \backslash L_{v} $. Therefore, the vector space spanned by the eigenvectors over the base space $B_v$ naturally splits into the valence and the conduction vector subspaces as $V_{\boldsymbol{k}} = V_{v,\boldsymbol{k}} \bigoplus V_{c,\boldsymbol{k}}$ for every point of the base space $\boldsymbol{k} \in B_v$. We can then define the valence (vector) bundle $ \mathcal{E}_{v} = \bigcup_{\boldsymbol{k}\in B_v} V_{v,\boldsymbol{k}} $ that is a sub-bundle of the total Bloch bundle $\mathcal{E}_B = \bigcup_{\boldsymbol{k}\in \mathbb{T}^3} V_{\boldsymbol{k}}$. While the total Bloch bundle is trivialized through the appropriate Fourier-transform basis \cite{FruchartCarpentier1,Berry_connection_Moore} the valence sub-bundle can be nontrivial as we will see. 

The vector space $V_{v,\boldsymbol{k}}$ defines a point in the space of all rank $N_v$ vector subspaces of $V_{\boldsymbol{k}}$, i.e.~by definition a point of the Grassmannian $ Gr_{N_{v}}(\mathbb{C}^N) $ \footnote{\unexpanded{The Grassmanian can also be defined as the space of valence projector matrices $P_{v,\boldsymbol{k}} = \sum_{n=1}^{N_v}  \vert \breve{U}_n, \boldsymbol{k} \rangle \langle \breve{U}_n, \boldsymbol{k} \vert$, which takes a vector in $V_{\boldsymbol{k}}$ and gives a vector in $V_{v,\boldsymbol{k}}$ such that $\mathrm{Ran}~P_{v,\boldsymbol{k}} = V_{v,\boldsymbol{k}}$.}}.
It follows that the topological classification of the valence bundles is given by $[\mathcal{E}_v] \simeq [B_v , Gr_{N_{v}}(\mathbb{C}^N)]$. Here $[\mathcal{E}_v]$ is the set of equivalence classes under bundle isomorphism and $[B_v , Gr_{N_{v}}(\mathbb{C}^N)]$ is the set of homotopy equivalence classes of the continuous maps $\Phi : \boldsymbol{k} \mapsto V_{v,\boldsymbol{k}}$ (indirectly given through the Hamiltonian matrices $H(\boldsymbol{k})$ which determine the set of valence eigenvectors) from the base space $B_v$ to the classifying space $Gr_{N_{v}}(\mathbb{C}^N)$ \footnote{This follows from the valence bundle always be obtainable as the pullback of the tautological valence bundle $\mathcal{F}_v = \bigcup_{P_v \in Gr_{N_{v}}(\mathbb{C}^N)} \mathrm{Ran}~P_v  $ by the continuous map $\Phi:B_v \rightarrow Gr_{N_{v}}(\mathbb{C}^N)$, i.e.~$\mathcal{E}_v = \Phi^*\mathcal{F}_v$, and noting the invariance of bundle isomorphism classes under homotopy \cite{NittisGomiAI}.}. 
In principle, this construction gives all the topological invariants that characterize the topology of the valance bundle. When the system is constrained to satisfy symmetries it restricts the classifying space $\mathcal{C}_v \subsetneq Gr_{N_{v}}(\mathbb{C}^N)$ possibly leading to the existence of symmetry protected topological invariants. Here, we want to take into account the space symmetry group $\mathcal{G}$ of the system within the class AI, i.e.~we consider the enlarged group of symmetries $\mathcal{G}\times \{E,\mathcal{T}\}$ with TRS and we neglect the spin degrees of freedom. We write the corresponding classifying space as $ \mathcal{C}^{\mathcal{G}_{\mathrm{AI}}}_v$. While a direct computation of the homotopy classes $[B_v ,  \mathcal{C}^{\mathcal{G}_{\mathrm{AI}}}_v]$ is difficult, we present a heuristic solution to the problem combining space group representation theory and Wilson loop techniques. However, before to proceeding, we first decompose the problem of the total topology of the valence bundle into the local topology of the nodal structures and the global topology constraining it. 

We have seen in the previous sections that the combined glide and screw symmetries in SG33-AI lead to a rich variety of symmetry protected nodal structures, sometimes even rather complex nodal structures $L_v$ and punctured base spaces $B_v$. We call a nodal structure \textit{composed} when it can be separated into smaller disconnected nodal structures where each can be surrounded by a closed surface. We call a nodal structure is \textit{elementary} when it cannot be further subdivided by pulling apart the unconnected elements without breaking it. Let us write $\boldsymbol{L}_{v}^{(i)} $ for the surrounding surface of an elementary component $L_{v}^{(i)}$, obtained by inflating it \footnote{\unexpanded{Strictly speaking $\boldsymbol{L}_{v}^{(i)}$ is the manifold obtained as the retract of $(\mathbb{T}^3 \backslash L_{v}^{(i)}) \cap D_{v,i}$, where $D_{v,i}$ is an open disk covering $L_{v}^{(i)}$ \cite{MathaiThiang_Weyls_I}. It is compact, connected, and orientable.}}. 
We argue that the \textit{total topology}, characterized by $[B_v ,  \mathcal{C}^{\mathcal{G}_{\mathrm{AI}}}_v]$, is determined by a local topology and a global topology. The \textit{local topology} is given through the homotopy classes of every elementary component, i.e.~$[\boldsymbol{L}_{v}^{(i)} ,  \mathcal{C}^{\mathcal{G}_{\mathrm{AI}}}_v]$, which can be decomposed in terms of homotopy groups corresponding to the \textit{local topological charges} of each elementary nodal structure. 
The \textit{global topology} follows from the topology of the BZ itself, i.e.~of $\mathbb{T}^3$, which leads to constraints between the local topological charges of distinct elementary nodal structures. Generalizing from simple examples we derive the explicit formula of the decomposition of the total topology of any nodal valence bundle into its local topology and its global topology. 

\begin{figure}[htb]
\centering
\begin{tabular}{cc}
	\includegraphics[width=0.35\linewidth]{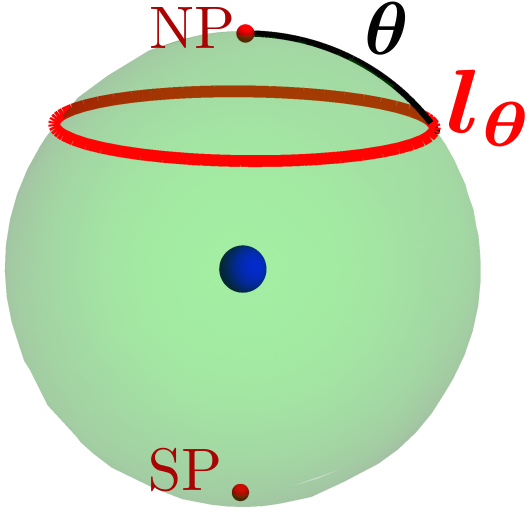} &
	\includegraphics[width=0.6\linewidth]{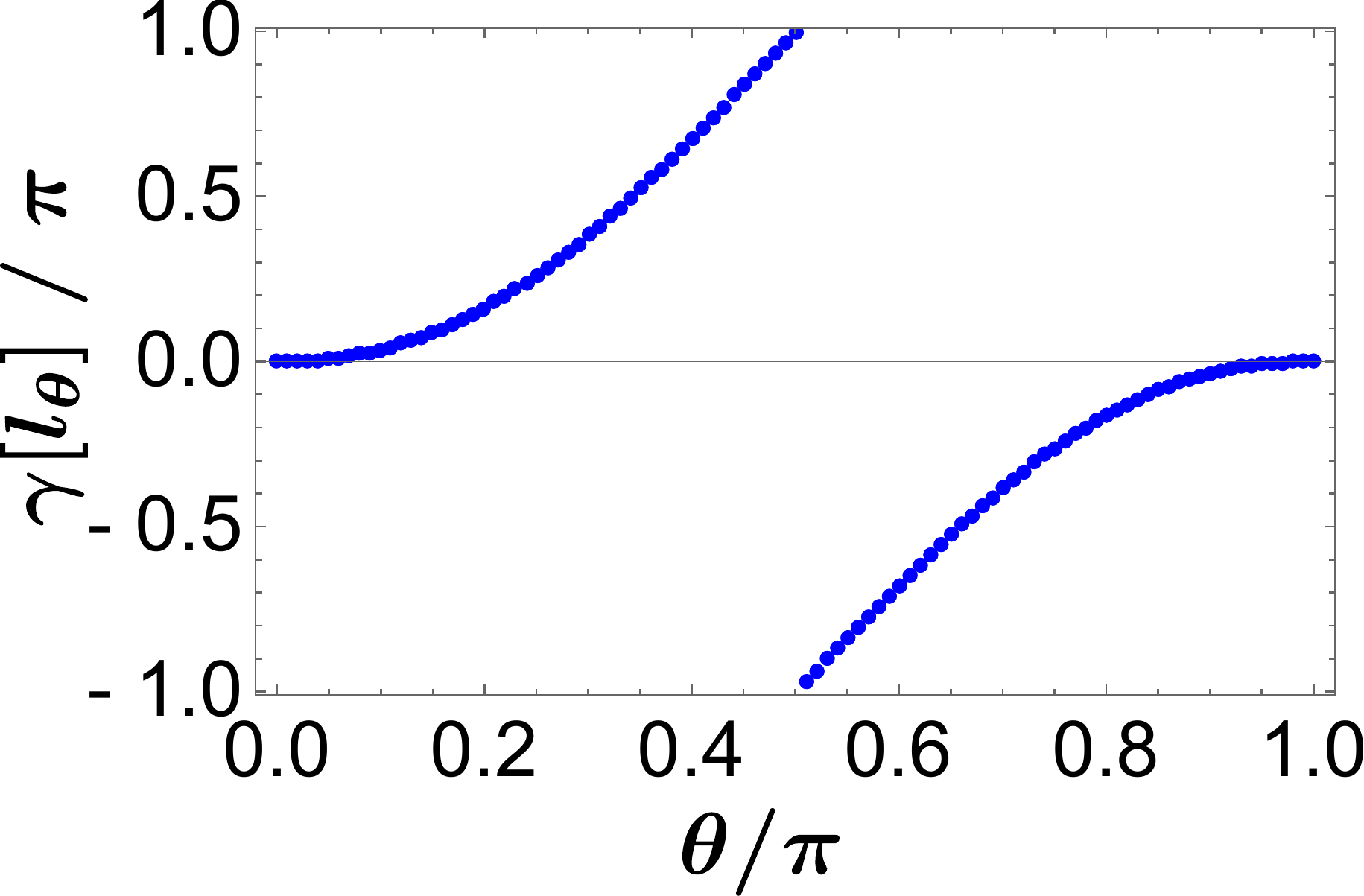}  \\
	(a) & (b)
\end{tabular} 
\caption{\label{fig_point_node} (a) Surrounding sphere bundle (green) of a point node (blue). The local topology is characterized by a Chern number $C_1 \in \mathbb{Z}$, that can be computed as the flow of Berry phase $\gamma[l_{\theta}]$ as we sweep the base loop $l_{\theta}$ (red) from the north pole to the south pole. (b) Numerical flow of Berry phase over a sphere surrounding one of the accidental point node in Fig.~\ref{fig_SG33_4B_lines}(d), giving $C_1=+1$. 
} 
\end{figure}
Let us first consider the case of a point node, as e.g.~found in Fig.~\ref{fig_SG33_4B_lines}(d). The surrounding surface of a single point is a sphere $\boldsymbol{L}_{v}^{(1)} \cong \mathbb{S}^2$, see Fig.~\ref{fig_point_node}(a), and its local topology is characterized through $[\mathbb{S}^2, Gr_{N_v}(\mathbb{C}^N) ] = \pi_{2}(Gr_{N_v}(\mathbb{C}^{N})) \cong \pi_1 (U(N_v)) \cong \pi_1 (SU(N_v))\times \pi_1 (U(1)) = \pi_1 (U(1)) = \mathbb{Z}$ with the topological invariant corresponding to the Chern number $C_1$. Note that while we have not assumed any symmetry ($Gr_{N_v}(\mathbb{C}^N)$ is the most general classifying space for complex vector bundles) including TRS does not change the result when we neglect spins and the point node is away from any HSPs \cite{ChiuSchnyder_reflection,Class_sym_review}. 
The above decomposition suggests that the Chern number $C_1$ of a point node can efficiently be computed through the flow of Berry phase $\gamma[l_{\theta}]$ as we sweep a base loop section $l_{\theta}$ on the sphere surrounding the point node, i.e.~$C_1 = \left(\gamma[l_{\theta=\pi}] - \gamma[l_{\theta=0}] \right)/2\pi$. Here we have parameterized the loop by the polar angle $\theta$ of the sphere, see Fig.~\ref{fig_point_node}(a). Indeed, the Berry phase defines a continuous mapping from base loop sections, $l_{\theta} \cong \mathbb{S}^1 \subset \mathbb{S}^2$, to the Berry phase factor $e^{i\gamma[l_{\theta}]} \in U(1)$, which hence belongs to the homotopy class $\pi_1(U(1))=\mathbb{Z}$. Fig.~\ref{fig_point_node}(b) shows the flow of Berry phase computed numerically around one of the accidental point nodes in Fig.~\ref{fig_SG33_4B_lines}(d) resulting in a Chern number $C_1=+1$.  

\begin{figure}[htb]
\centering
\begin{tabular}{cc} 
	\includegraphics[width=0.45\linewidth]{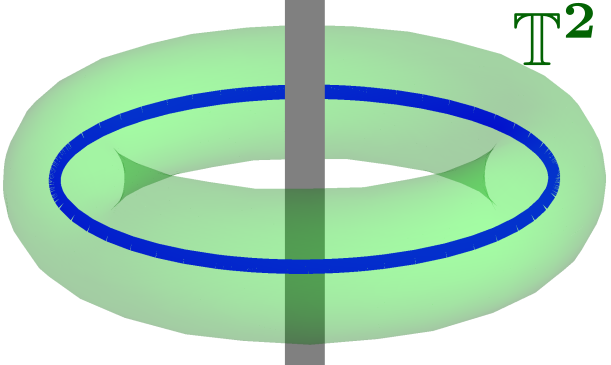} & 
	\includegraphics[width=0.5\linewidth]{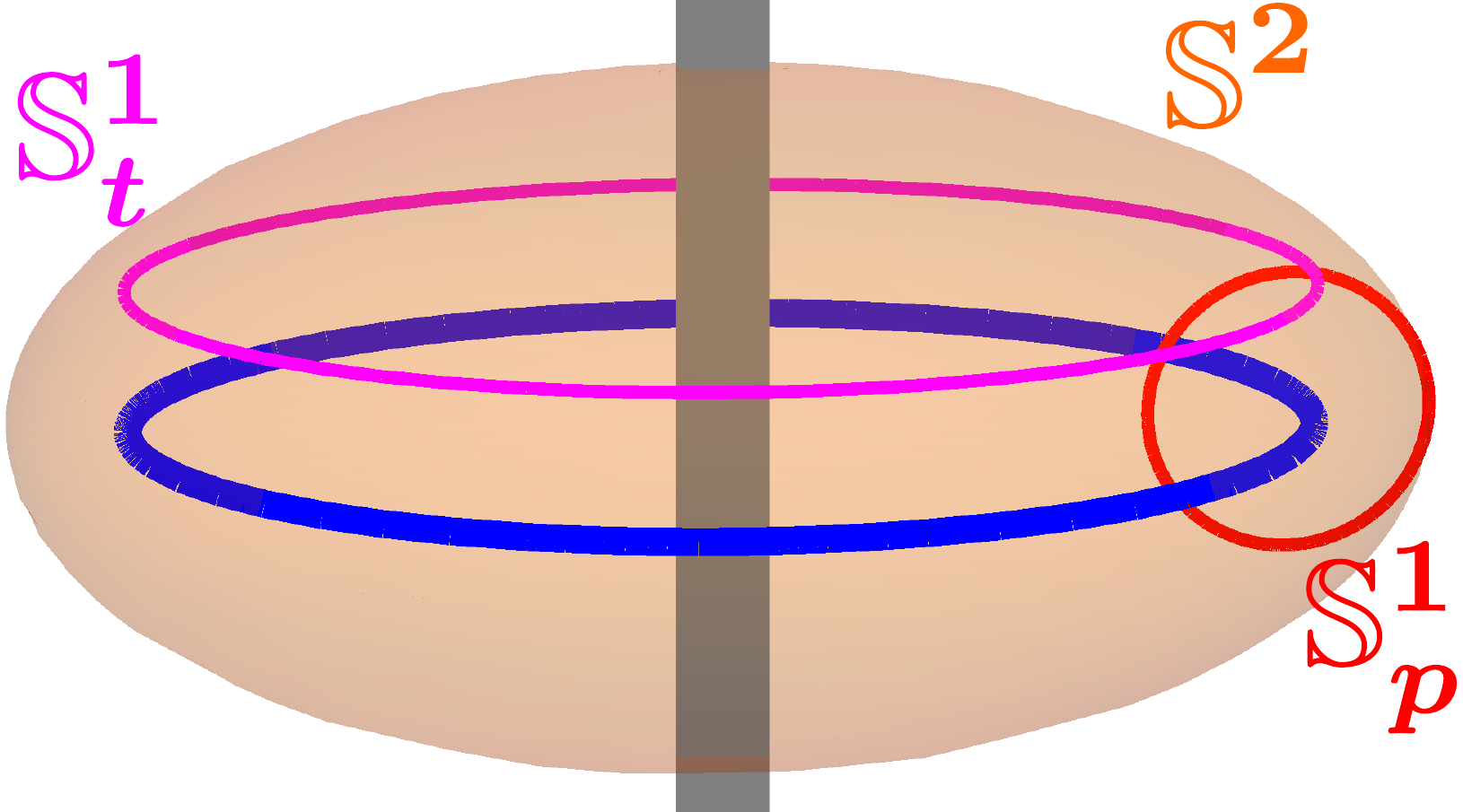} \\
	(a) & (b) 
\end{tabular}
\caption{\label{torus_plot} (a) Surrounding torus (green) of a NL (blue) with a possible obstruction (gray axis). (b) Poloidal loop $\mathbb{S}^1_p$ (red), toroidal loop $\mathbb{S}^1_t$ (purple), and sphere $\mathbb{S}^2$ (orange), corresponding to the base spaces of the local topological charges of a NL. The toroidal and the monopole bases are excluding each other depending on the obstruction represented by the gray axis. 
}
\end{figure}
Next we study a  NL that can be surrounded by a torus $\boldsymbol{L}_{v} \cong \mathbb{T}^2 = \mathbb{S}^1_p\times \mathbb{S}^1_t$, here decomposed into its poloidal and toroidal directions, see Fig.~\ref{torus_plot}. Thus the local topology is characterized by the homotopy classes $[ \mathbb{T}^2 , \mathcal{C}_v]  $. The torus homotopy group can be decomposed in the stable limit into a direct product of first and second homotopy groups according to \cite{ASSimon_83,Kitaev,Kennedy_homotopy}
\begin{equation}
\label{torus_local}
	[ \mathbb{T}^2 , \mathcal{C}_v] \cong  \pi_1(\mathcal{C}_v) \oplus \pi_1(\mathcal{C}_v) \oplus  \pi_2(\mathcal{C}_v) \;.
\end{equation}
Here one first homotopy group $\pi_1(\mathcal{C}_v)$ gives the set to which the local \textit{poloidal charge} belongs and the other $\pi_1(\mathcal{C}_v)$ is for the local \textit{toroidal charge}, i.e.~both charges are computed over a loop base space ($ \mathbb{S}^1_{p,t}$), while the second homotopy group $\pi_2(\mathcal{C}_v)$ gives the set to which belongs the \textit{monopole charge} computed over the sphere surrounding the whole NL ($ \mathbb{S}^2$), see Fig.~\ref{torus_plot}(b). In Fig.~\ref{torus_plot}(b) we intentionally left undecided whether or not the NL has an obstruction as represented by the gray axis. 

We now show how Eq.~(\ref{torus_local}) can be generalized in order to determine the local topology of any elementary nodal structure. For this we will artificially trivialize the BZ topology through the substitution $\mathbb{T}^3 \rightarrow \mathbb{R}^3$. First we need to identify the surrounding surface $\boldsymbol{L}_{v}^{(i)}$ of the $i$-th elementary nodal structure $L_{v}^{(i)}$. Then we contract every local poloidal and toroidal sections of $\boldsymbol{L}_{v}^{(i)}$ that are not obstructed. This gives the deformation retract of the punctured space $\mathbb{R}^3 \backslash L_{v}^{(i)} $ \cite{Hatcher_1}, which we write $r_{v}^{(i)}$. For any elementary nodal structure the deformation retract $r_{v}^{(i)}$ contains only one sphere, i.e.~the sphere surrounding the whole nodal structure. Let us also introduce the genus $\mathfrak{g}^{(i)} \in \mathbb{N}$ of the surrounding surface $\boldsymbol{L}_{v}^{(i)}$, which counts the number of handles. Then, we find that the deformation retract of the punctured space $\mathbb{R}^3 \backslash L_{v}^{(i)} $ is given by the wedge product
\begin{equation}
\label{def_retract}
	r_{v}^{(i)} =  \prod\limits_{l=1}^{\mathfrak{g}^{(i)}} \mathbb{S}^{1(i)}_{l}   \vee \mathbb{S}^{2(i)} \;, 
\end{equation}
between the Cartesian product of $\mathfrak{g}^{(i)}$ loops $\{\mathbb{S}^{1(i)}_{l}\}_{l=1,\dots,\mathfrak{g}^{(i)}}$, corresponding to all the non-contractible poloidal-toroidal loops of $r_{v}^{(i)}$, and the monopole sphere ($\mathbb{S}^{2(i)}$) surrounding the whole $i$-th elementary nodal structure $ L_{v}^{(i)}$. The generalized version of Eq.~(\ref{torus_local}) is then 
\begin{equation}
\label{loc_charges}
	[\boldsymbol{L}_{v}^{(i)}, \mathcal{C}_v] \cong  [ r_{v}^{(i)} , \mathcal{C}_v] \cong  \bigoplus\limits_{l=1}^{\mathfrak{g}^{(i)}} \pi^{(i)}_{1,l}( \mathcal{C}_v ) \oplus \pi^{(i)}_2(\mathcal{C}_v)	\;,
\end{equation}
where $\pi^{(i)}_{1,l}( \mathcal{C}_v )$ is the set for the $l$-th poloidal-toroidal charge over the $l$-th non-contractible loop section $\mathbb{S}^{1(i)}_{l}$ of the deformation retract $r_{v}^{(i)}$ obtained for the $i$-th elementary nodal structure, and $\pi^{(i)}_2(\mathcal{C}_v)$ is the set of the monopole charge.

\begin{figure}[htb]
\centering
\begin{tabular}{ccc} 
	\includegraphics[width=0.4\linewidth]{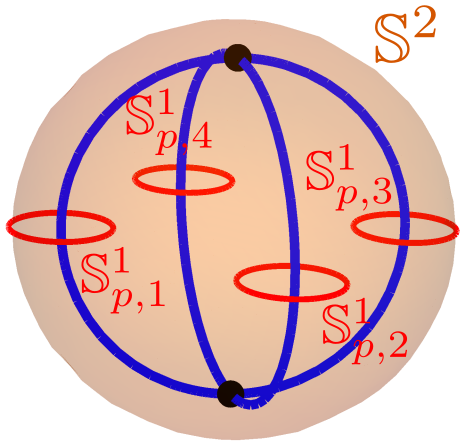} & \hspace{0.5cm} &
	\includegraphics[width=0.362\linewidth]{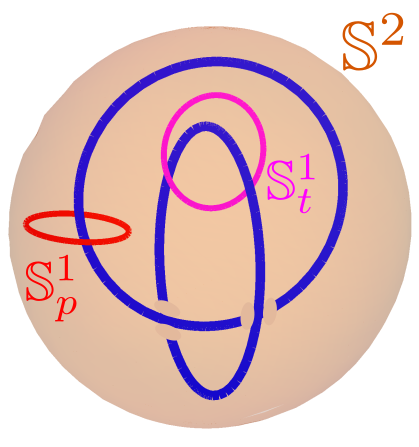}\\
	(a) & & (b)
\end{tabular}
\caption{\label{fig_ns_charges} (a) Elementary line-nodal structure consisting of two closed NLs connected at two points (a) and two closed NLs linked together (b). The local base spaces for every local topological charges expressed in Eq.~(\ref{loc_charges}) are shown. 
}
\end{figure}

Let us see how Eq.~(\ref{loc_charges}) works with a few simple cases. In the case of a closed and isolated NL, e.g.~in Fig.~\ref{fig_SG33_8B_lines}(a), there is no obstruction for contracting the toroidal loop into a point leaving a deformation retract composed of a poloidal loop and a monopole sphere, simply take Fig.~\ref{torus_plot}(b) and remove the obstructing gray axis. Therefore, a single NL is characterized fully by one poloidal charge, reflecting the genus $\mathfrak{g}=1$ of the original surrounding torus, and one monopole charge. 
Next consider the case of two closed NLs connected at two points, e.g.~as in Fig.~\ref{fig_SG33_8B_lines}(c). The surrounding surface is now more complicated with the genus $\mathfrak{g} = 4$. Each handle defines a local poloidal section, $\{\mathbb{S}^1_{p,l}\}_{l=1,\dots,4}$, but only one surrounding sphere $\mathbb{S}^2$ remains under the deformation retract of the punctured space, see Fig.~\ref{fig_ns_charges}(a). Therefore, the local topology of this nodal structure is characterized by four poloidal charges $\pi_{1,l}(\mathcal{C}_v)$ and one monopole charge $\pi_2(\mathcal{C}_v)$. 
We finally consider two closed NLs linked together as in Fig.~\ref{fig_ns_charges}(b). The deformation retract of the punctured space is now given by the wedge product of the torus surrounding one of the NLs and the sphere surrounding the whole structure \cite{Hatcher_1}. Then by Eq.~(\ref{loc_charges}), we get one poloidal charge, one toroidal charge and one monopole charge. Following this line of reasoning we can easily generalize the procedure to all the nodal structures present in this work.  

Importantly, we note that without any symmetry, $\pi_1(Gr_{N_v}(\mathbb{C}^N)) = \{e\}$ such that the NLs are not generically stable. A well known example of stable NLs is when the system is symmetric under the combination of TRS and inversion symmetry, $\mathcal{T}*I$, in class AI. In that case, the Hamiltonian can be chosen real and $\pi_1(Gr_{N_v,N}(\mathbb{R}))=\mathbb{Z}_2$. This corresponds to the $\mathbb{Z}_2$ quantization of the Berry phase (modulo $2\pi$): if the loop base space encircles an odd number of NLs the Berry phase of the poloidal direction is $\pi\mod2\pi$, otherwise it is $0\mod2\pi$. For SG33-AI we have already seen that instead crystalline symmetries protect NLs on high symmetry $m_x$- and $m_y$-invariant planes. We will in the following sections show that the poloidal-toroidal charges of symmetry protected nodal structures can be algebraically calculated in terms of quantized Wilson loop phases over symmetry-constrained momentum loops. The monopole charges are given through the spectral flow of Wilson loops as we sweep a symmetry-constrained momentum loop over a surrounding sphere centered on the nodal structure, in analogy to the computation of the Chern number in Fig.~\ref{fig_point_node}.

\begin{figure}[htb]
\centering
\begin{tabular}{cc} 
	\includegraphics[width=0.5\linewidth]{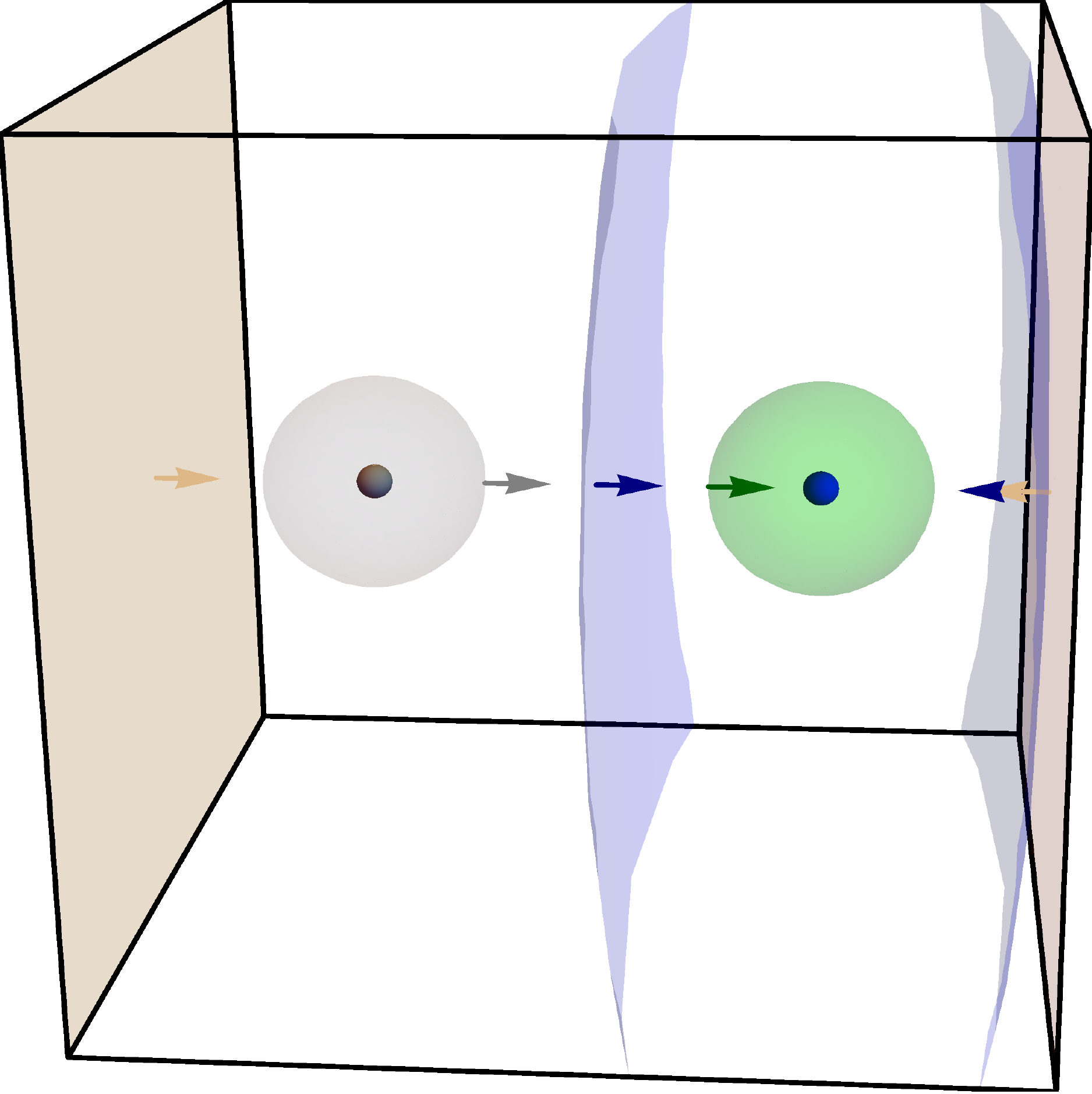} & 
	\includegraphics[width=0.5\linewidth]{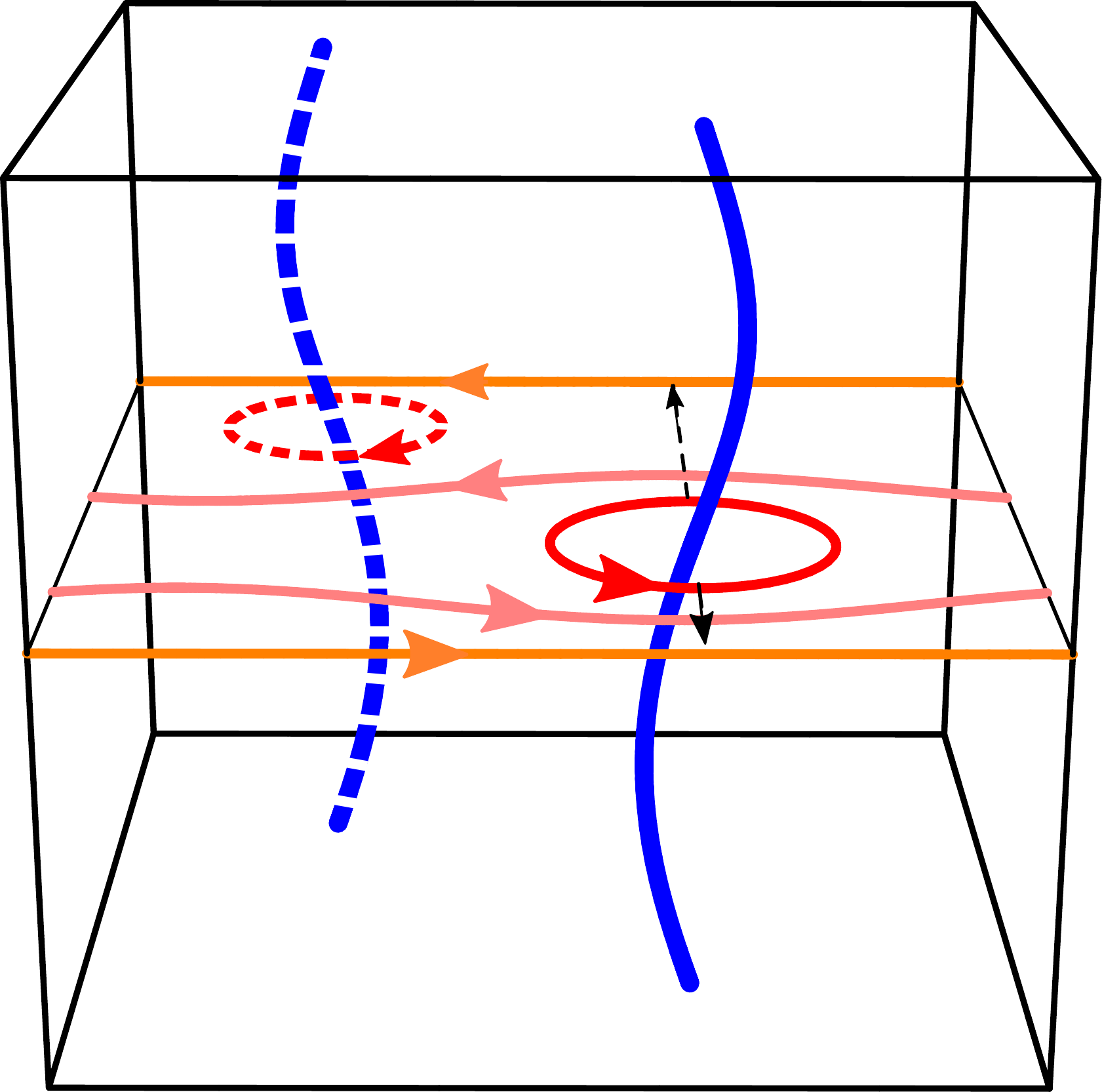}\\
	(a) & (b)
\end{tabular}
\caption{\label{fig_global_top} Global topology from geometry of the BZ. (a) Deformation of the oriented surrounding sphere (green, cyan, beige with increasing sizes) of a point node (blue) having a nontrivial monopole charge. Without the additional point node (gray) in the BZ$\cong\mathbb{T}^3$ there is no obstruction to deform the sphere and contract it to a point, which contradicts the nontrivial charge. (b) Deformation of the oriented poloidal loop (red, pink, orange with increasing sizes) encircling a NL thread (blue) threading the BZ, having a nontrivial poloidal Berry phase and thus a nontrivial thread charge. Without the additional threading NL (dashed blue) there is no obstruction to deform the poloidal loop into a point leading to a trivial Berry phase which contradicts the nontrivial charge.  
}
\end{figure}

We still have to characterize the global topology of nodal structures. We show that it is dictated by the geometry of the BZ itself. Let us in Fig.~\ref{fig_global_top}(a) again consider the case of one point node (blue) with a nontrivial monopole charge (e.g.~given by a Chern number $C_1 \neq0$ as derived above) but now imbedded in the BZ $\cong\mathbb{T}^3$. Without the additional point node (gray) in the BZ there would be no obstruction to progressively deforming the oriented surrounding sphere (green, purple, beige for increasing sizes) and contracting it to a point leading to a trivial monopole charge. This would however contradict the assumption of a nontrivial monopole charge. Hence, the global topology of $B_v$ imposes the presence of a second point node (gray) with the canceling monopole charge (e.g.~$-C_1$). In essence, this gives an heuristic explanation for the Nielsen-Ninomiya theorem \cite{Nielsen1, Nielsen1a, Nielsen2, Kiritsis, Witten_lect, MathaiThiang_Weyls_I}, which imposes that only pairs of point nodes with opposite charges be realized. We discuss further in Section \ref{four_top} the total topology of the point nodes in Fig.~\ref{fig_SG33_4B_lines}(d), taking into account the space group symmetries of the system and using Nielsen-Ninomiya theorem to argue that these are actually accidental. 

Here we show that a similar argument works in the case with one stable NL (i.e.~with a nontrivial poloidal Berry phase) that threads the BZ torus, shown in Fig.~\ref{fig_global_top}(b). Without the additional NL threading the torus BZ (dashed blue) there is no obstruction to deforming the oriented poloidal loop into a point, leading to a trivial Berry phase, again in contradiction with the assumption. Therefore, the global topology requires that such BZ-threading NLs and their nontrivial poloidal charges come in pairs. E.g.~take the two NLs threading the BZ of Fig.~\ref{fig_global_top}(b) each with a poloidal Berry phase $\pi \mod 2\pi$. This leads to the trivial global Berry phase factor $e^{i (\pi+\pi)\mathrm{mod}2\pi}=+1$ for a loop that encircles both NLs. We therefore call them \textit{NL threads} and say that these are characterized by nontrivial \textit{thread charges}. The above arguments can straightforwardly be extended from point nodes and NLs to general elementary line-nodal structures with nontrivial monopole and thread charges. Both types of nontrivial nodal objects are found in this work.  

Having unfolded the content of the local and global topologies of line-nodal structures we can now characterize the total topology of any nodal valence bundle. We have argued above that the total topology of the valence bundle it is given by $[B_v , \mathcal{C}_v]$. Let us write $b_v$ the deformation retract of the punctured BZ base space $B_v = \mathbb{T}^3\backslash L_v$. Contrary to the deformation retract $r_v^{(i)}$ (obtained from $\mathbb{R}^3\backslash L_v$) $b_v$ now includes the effects of the global topology discussed above. Therefore, the total topology of the nodal valence bundle is given together with Eq.~(\ref{loc_charges}) through 
\begin{equation}
\label{total_top}
	[B_v , \mathcal{C}_v] \cong  [b_v , \mathcal{C}_v] \cong \bigoplus\limits_{i}  [ r_{v}^{(i)} , \mathcal{C}_v] \;,
\end{equation}
in which only pairs of nontrivial monopole charges and pairs of nontrivial thread charges are allowed to appear.

While nontrivial local poloidal charges must always be realized in symmetry protected line-nodal structures, we find that nontrivial monopole, thread, and toroidal charges are much more rare. We show in Section \ref{monopoles} that the monopole charge depends on the centering of the nodal structure. When the surrounding surface of a symmetry-protected elementary nodal structure surrounds some HSPs, we find that it must have a trivial monopole charge. However, when the surrounding surface is topologically equivalent to a sphere (similarly to Fig.~\ref{fig_global_top}(a)) and it excludes all HSPs, the corresponding nodal structure$-$discarding accidental nodal structures which can be removed without any change of the valence IRREPs at HSPs$-$must have a nontrivial monopole charge. Also, when a symmetry-protected nodal structure threads the torus BZ (similarly to Fig.~\ref{fig_global_top}(b)) such that its surrounding surface, which now must be topologically equivalent to a two-torus, does not surround any HSP, then$-$again discarding accidental nodal structures which can be removed without any change of the valence IRREPs at HSPs$-$it must have a nontrivial thread charge. Concerning the toroidal charges it is clear that these can only appear when independent NLs are linked. We show one explicit example of this in Section \ref{Lifshitz}. 
This completes the formal characterization of the total topology of the nodal valence bundles combining local and global topology. We now turn to the problem of actually computing the local topological charges for which a pure algebraic approach is developed.

\section{Symmetry protected topological invariants for SG33-AI}\label{top_inv}
We have seen above that the topology of symmetry protected nodal structures is characterized through local topological charges classified by homotopy groups over poloidal-toroidal loops and monopole spheres. We will now show that every local charge is given by symmetry protected topological invariants corresponding to quantized Wilson loop phases over symmetry-constrained momentum loops. Here we present the algebraic algorithm that leads to all the symmetry protected topological invariants for SG33-AI, calculated using only the valence IRREPs at a the HSPs. In the sections following this we show how these can be directly used for characterization of the topology of all the symmetry protected nodal structures that we systematically identified through the valence IRREPs combinatorics in Section \ref{SG33}.   

We construct the momentum loops as closed loops $l^g$ in the BZ that connect direct neighboring HSPs acting as the vertices of the loop, and such that they are foldable under a point group symmetry $g\in C_{2v}$, possibly with the combined action of reciprocal lattice translations \footnote{Throughout this work we assume the periodic gauge, see Appendix \ref{sym_rep}.}. It is both the eigenvalues and the determinant (Berry phase factor) of a Wilson loop matrix over a closed symmetry constrained loop, $\mathcal{W}[l^g]$, that provide robust topological invariants \cite{Bernevig1, Bernevig_point_groups, Bernevig3, AlexBernevig_berryphase,Alex1,Alex2}. We present below the algebraic derivation of all symmetry protected topological invariants that can be defined from the Wilson loops over closed two- and four-point loops in the BZ.

\subsection{Two-point loop topological invariants}\label{line_inv}
\begin{figure}[htb]
\centering
\begin{tabular}{ccc}
	\includegraphics[width=0.28\linewidth]{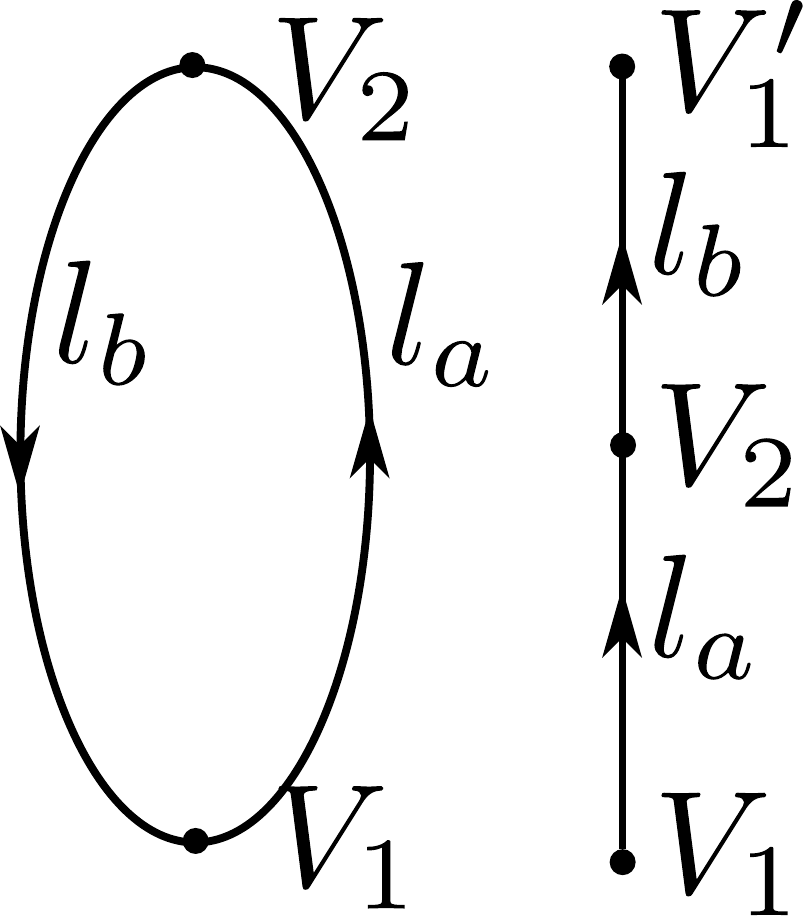} & &
	\includegraphics[width=0.55\linewidth]{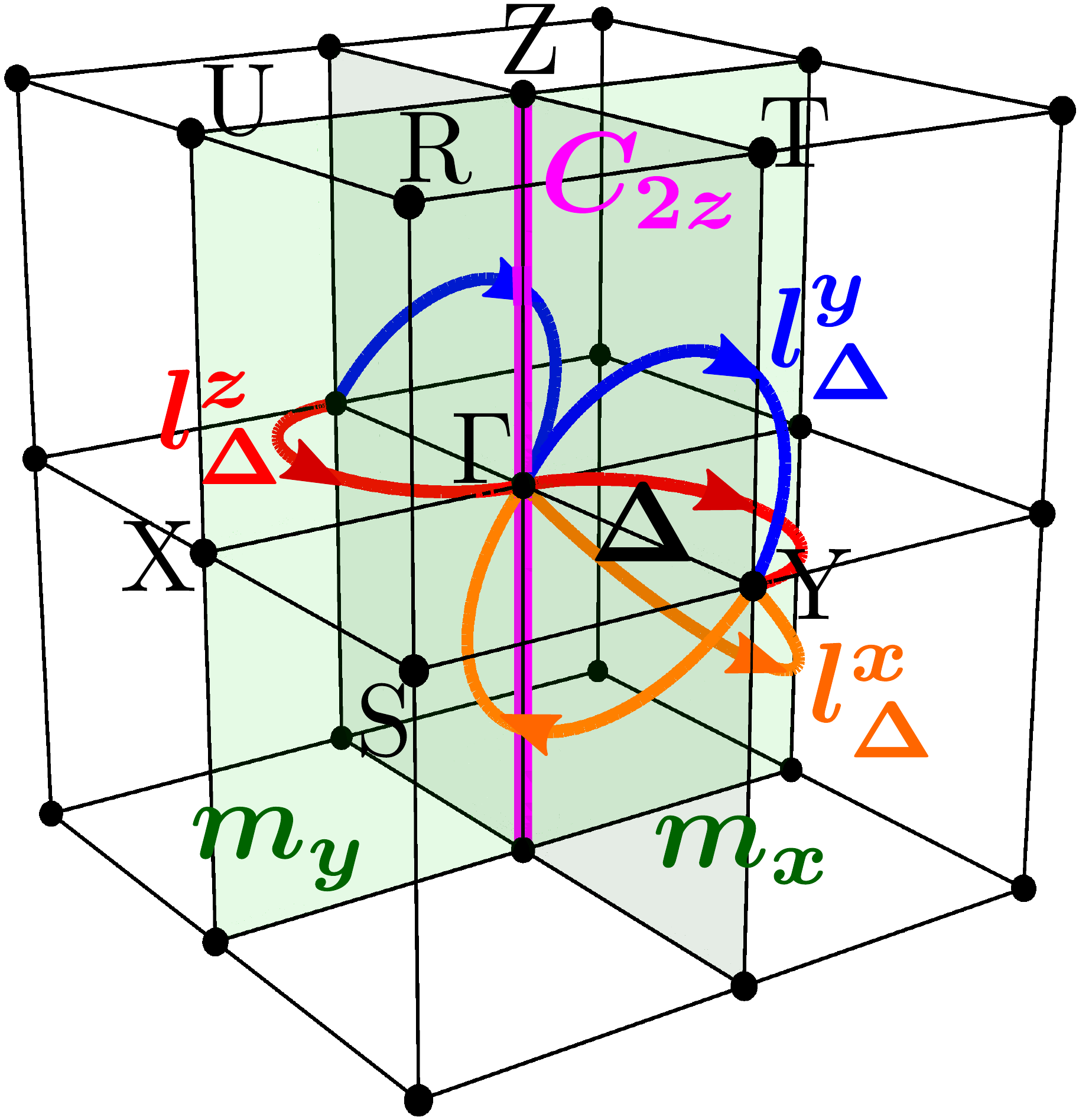}  \\
	(a) & & (b)
\end{tabular} 
\caption{\label{fig_loops} (a) Closed two-point loop with $l_L(0) = l_L(1) = V_1$ and $l_L(1/2)=V_2$, and quasi-closed two-point loop with $l_L(0) = V_1$, $l_L(1/2)=V_2$ and $l_L(1) = V_1'=V_1+\boldsymbol{K}$. (b) Examples of closed and quasi-closed two-point symmetry-constrained loops, $l^g_{\Delta}$ for $g=\{C_{2z},m_y,m_x\}$.
}
\end{figure}

The shortest loops we use connect only two nearest-neighbor HSPs, say $V_1$ and $V_2$, belonging to a common HSL $L\equiv \overline{V_1V_2}$, hence written $l_L$. Taking into account the periodicity in $k$-space we can either form closed or quasi-closed loops. A closed loop can be parameterized as $l : [0,1] \rightarrow \mathbb{S}^1,~t \mapsto l(t)$ which also implies an orientation for the loop. The derivation of the topological invariants is based on the symmetry folding of the Wilson loops. We therefore restrict ourselves to loops that can be decomposed into two ``anti-symmetric" segments $l_{L} = l_b \circ l_a$, i.e.~the composition of paths such that we first go through the oriented segment $l_a$ and then through $l_b$, where $g l_b = l_a^{-1}$ (here $l^{-1}$ is the loop $l$ with reversed orientation), with $g$ a point symmetry of the system, see Fig.~\ref{fig_loops}(a).
Hence the loop is foldable by the action of $g$ on one of the two segments. We write such a loop $l^g_{L}$ and call it a closed two-point symmetry-constrained loop. We plot an example of $l_{\Delta}^x$ (orange) in Fig.~\ref{fig_loops}(b). Fig.~\ref{fig_loops} also shows examples of quasi-closed loops. A quasi-closed loop is bounded by two vertices separated by a reciprocal lattice vector, i.e.~$V_1' = gV_1 = V_1+\boldsymbol{K}_g$, see Fig.~\ref{fig_loops}(a). The examples $l^{z}_{\Delta}$ (red) and $l^{y}_{\Delta}$ (blue) are shown in Fig.~\ref{fig_loops}(b). Within AI, only the following two-point loops are relevant for SG33: $l^g_L \in \{l_{\Sigma}^y,l_{\Delta}^x,l_{\Lambda}^{z,y,x}, l_{\text{A}}^y, l_{\text{P}}^x, l_{\text{Q}}^{z,y,x} \}$. For each $L$, the point symmetries available for the folding belong to the little co-group (stabilizer) of the line, i.e.~$g\in \overline{G}^{L}$. 

We now compute the Wilson loop over the closed two-point symmetry-constrained loops, $\mathcal{W}[l^g_{L}]$, defined in terms of the valence eigenvectors $\vert u_n, \boldsymbol{k} \rangle,~n=1,\cdots,N_v$. Note that the latter are analoguous to the cell-periodic part of Bloch eigenfunctions, see Appendix \ref{wilson} for definition and important properties of Wilson loops. It is convenient to have in mind that $\mathcal{W}[l^g_{L}]$ is a $N_v \times N_v$ unitary matrix. We show that the spectrum of the Wilson loop can be determined fully algebraically, directly leading to the symmetry-protected topological invariants. The approach we use below follows and generalizes Refs.~\cite{Alex1,Alex2}.%

It follows from the properties of Wilson loops that 
\begin{align}
	\mathcal{W}[l^g_{L}] = \mathcal{W}[l_b] \mathcal{W}[l_a] &= \breve{R}^{V_1}_g  \mathcal{W}[g l_b]   (\breve{R}^{V_2 }_g)^{-1} \mathcal{W}[l_a] \;, \nonumber\\
\label{wilson_folding}
	&=\breve{R}^{V_1}_g  \mathcal{W}[l_a]^{-1}   (\breve{R}^{V_2}_g)^{-1} \mathcal{W}[l_a] \;,
\end{align}
where at each vertex of the loop the folding symmetry $g$ is represented through a transformation matrix $[\breve{R}^{V_{1,2}}_g]_{mn} = \langle u_m, gV_{1,2} \vert ^{\{g\vert \boldsymbol{\tau}_g\}}\vert u_n , V_{1,2}\rangle $, $m,n=1,\dots,N_v$, see also Appendix \ref{sym_rep}. At invariant momenta $\bar{\boldsymbol{k}}$, i.e.~$g \bar{\boldsymbol{k}} = \bar{\boldsymbol{k}} +\boldsymbol{K}_n$ such that $g\in \overline{G}^{\bar{\boldsymbol{k}}}$, the matrix $\breve{R}^{\bar{\boldsymbol{k}}}_g$ has a block-diagonal form reflecting the IRREPs of the symmetry $\{g\vert \boldsymbol{\tau}_g\}$. Here we write $\breve{R}$ with a curled hat to symbolize that its explicit evaluation is basis dependent. In order to achieve gauge invariance we first perform the unitary transformation that diagonalizes the transformation matrices, i.e.~
\begin{align}
	\tilde{\mathcal{W}}[l^g_{L}] &= U^{V_1}_g \mathcal{W}[l^g_{L}] U^{V_1 \dagger}_g \nonumber\\
\label{wilson_folding_B}
	&= \breve{D}^{V_1}_g  \tilde{\mathcal{W}}[l_a]^{-1}   (\breve{D}^{V_2}_g)^{-1} \tilde{\mathcal{W}}[l_a] \;,
\end{align}
where $\breve{D}^{V_i}_g = U^{V_i \dagger}_g  \breve{R}^{V_i}_g U^{V_i}_g$ are now diagonal (hence gauge invariant) and $\tilde{\mathcal{W}}[l_a] = U^{V_{2} \dagger}_g \mathcal{W}[l_a] U^{V_{1} \dagger}_g$. Since the eigenvalues of a unitary matrix are invariant under unitary transformations, the spectrum of the Wilson loop matrix is not affected by this transformation. Also, by unitarity the Wilson loop spectrum is unimodular with eigenvalues of the form $\mathrm{eig}\{\mathcal{W}[l]\} = \{e^{i \varphi_i}\}_{i=1,\dots,N_v}$.
Moreover, since $g l^g_{L} = (l^g_{L})^{-1}$, the Wilson loop should also formally satisfy $\mathcal{W}[l^g_{L}]^{-1} =    (\breve{R}^{V_1}_g)^{-1} \mathcal{W}[l^g_{L}]   \breve{R}^{V_1}_g$ (and similarly for $\tilde{\mathcal{W}}[l^g_{L}]$).  Together this implies that $\mathrm{eig}\{\mathcal{W}[l^g_{L}]\} = \mathrm{eig}\{\mathcal{W}[l^g_{L}]\}^*$, i.e.~the Wilson loop spectrum must be symmetric under complex conjugation. Therefore, the Wilson loop eigenvalues can only be composed of $+1$, $-1$, and of complex conjugate pairs $\{e^{i \varphi_1},e^{-i \varphi_1}\}$. However, Eq.~(\ref{wilson_folding_B}) is not yet explicitly invariant under complex conjugation. Indeed, after diagonalization we have $\breve{D}^{V_i}_g = \mathrm{diag}(\lambda_{g,1},\dots,\lambda_{g,N_v})$, where each eigenvalue can be decomposed as $\lambda_{g,n} = s_{g,n} e^{i \theta_{g}} $ with $s_{g,n} = \pm1 $. Removing the global complex phase factor we define the \textit{bare IRREP}  $ R^{V_i}_g = e^{- i \theta_{g}} \breve{D}^{V_i}_g $ that is diagonal with only $\pm1$ elements, In  Appendix \ref{bare} we give the complete list of the bare IRREPs at HSPs for SG33-AI. 
Let us formally write the combined action of the diagonalization in Eq.~(\ref{wilson_folding_B}) and making the transformation matrices real as the pullback under the mapping $\Phi_{2}$, i.e.
\begin{equation}
\label{Phi_2}
	\Phi_{2}^{\star}\mathcal{W}[l^g_{L}]  =  R^{V_1}_g   \tilde{\mathcal{W}}[l_a]^{-1}  (R^{V_2}_g)^{-1}  \tilde{\mathcal{W}}[l_a]  \;.
\end{equation}
Here by construction the symmetry operation $g$ belongs to the little co-group of the vertices, i.e.~$g\in\overline{G}^{V_i}$. The Wilson loop spectrum, i.e.~$\mathrm{eig}\{\Phi_2^{\star}\mathcal{W}[l^g_{L}]\}$, is now explicitly symmetric under complex conjugation. 

It was shown in Refs.~\cite{Alex1,Alex2} that there exists a mapping $\mathcal{M}_2$ from the bare IRREP eigenvalues of all valence states of the two vertices of the loop ($V_1, V_2$) to the two-point symmetry-constrained Wilson loop spectrum, formally 
\begin{equation}
\label{M_2}
	\mathcal{M}_2  : (R^{V_1}_g,R^{V_2}_g) \mapsto \mathrm{eig}\{\Phi_2^{\star}\mathcal{W}[l^g_{L}]\} \;.
\end{equation}
This mapping is given in Appendix \ref{mappings} for two valence bands in Table~\ref{mapping_two} for four valence bands in Table~\ref{mapping_four}. Given the number of $\pm1$ bare IRREP eigenvalues at $V_i$, $N^{V_i}_{\pm}$, we define the minority eigenvalue $\xi \in \{+1,-1\}$, determined by $\min\limits_{\xi =\pm1} (N^{V_1}_{\xi}+N^{V_2}_{\xi} )$. Then the number of $\xi$-eigenvalues in the Wilson loop spectrum, which we write $\mathcal{N}^g_{L}$, is symmetry protected. The $N_v-\mathcal{N}^g_{L}$ remaining eigenvalues of the Wilson loop spectrum are then either $-\xi$ or unimodular complex conjugate pairs that can be mapped to $-\xi$ through an adiabatic and symmetry preserving transformation. 

By construction $\mathcal{N}^g_{L}$ is gauge invariant. It is also invariant under any continuous deformation of the loop $l^g_{L}$, as long as the vertices and the symmetry constraints are conserved and no nodes between the valence and conduction bands are crossed during the deformation, since this does not change the set of bare IRREP eigenvalues at the vertices of the loop. Therefore, we conclude that this is actually a symmetry-protected topological invariant that characterizes the bulk topology of the system. 
From this \textit{two-point loop topological invariant}
\begin{equation}
	\mathcal{N}^g_{L} \in \mathbb{N}\;,
\end{equation}
we readily get the quantized Berry phase factor from
\begin{align}
\label{Berry_phase_factor_2}
	e^{i \gamma^g_{L}} &= \mathrm{det}~\mathcal{W}[l^g_{L}]  = (-1)^{\mathcal{N}^g_{L}} \in \{ +1 , -1 \} \equiv \mathbb{Z}_2\;.
\end{align}

We will see that the Berry phase factor (Berry phase) is nontrivial, $e^{i \gamma^g_{L}}=-1$ ($\gamma^g_{L}\mod2\pi= \pi$), whenever the symmetry-constrained loop $l^g_{L}$ encircles an odd number of NLs, otherwise it is trivial, $e^{i \gamma^g_{L}}=+1$ ($\gamma^g_{L} \mod 2\pi = 0$), as it encircles an even number of NLs. We will also later see that the charges $\mathcal{N}^g_{L}$ refine the Berry phase. For example, on one hand, $\mathcal{N}^{x,y}_{L}$ counts the number of NLs belonging to a $m_{x,y}$-invariant plane of the BZ crossing the line $L$; on the other hand, the charges $\mathcal{N}^{z}_{L}$ give complementary information concerning the type of line-nodal structure. 
Since the above construction works for every point symmetry that belongs to the little co-group (stabilizer) $\overline{G}^{L}$ of $L$, we can also define 
\begin{equation}
	\mathcal{N}_{L} = \max\limits_{g\in \overline{G}^{L}} \mathcal{N}^g_{L} \in \mathbb{N} \;,
\end{equation}
that gives the number of symmetry protected NLs crossing the half of $L$.

The above construction works completely similarly for the quasi-closed two-point loops. While Wilson loops over the quasi-closed loops also provide important information, for instance in the bulk-boundary correspondence, their Wilson loop spectrum is not invariant under large gauge transformation and we therefore do not use them for characterization of the bulk topology. However, we show that they form the basis of the four-point loop topological invariants defined below. 
We also remark that the two-point loops can be further generalized by taking the vertices away from the HSPs. Indeed, any two-point loop $l^g_{X_1 X_2}$ with $X_1$ and $X_2$ taken within the same $g$-invariant region of the BZ can be made foldable under $g$ such that its symmetry protected Wilson loop eigenvalues can again be obtained algebraically through the same mapping $\mathcal{M}_2$, since the bare IRREPs $R^{X_1,X_2}_g$ are well defined all over any $g$-invariant region. This we use in Section \ref{eight_top} when discussing in details two illuminating numerical examples, and in Section \ref{Lifshitz} when considering local topological Lifshitz transitions.

\subsection{Four-point loop topological invariants}\label{plane_inv}
\begin{figure}[htb]
\centering
\begin{tabular}{cc}
	\includegraphics[width=0.28\linewidth]{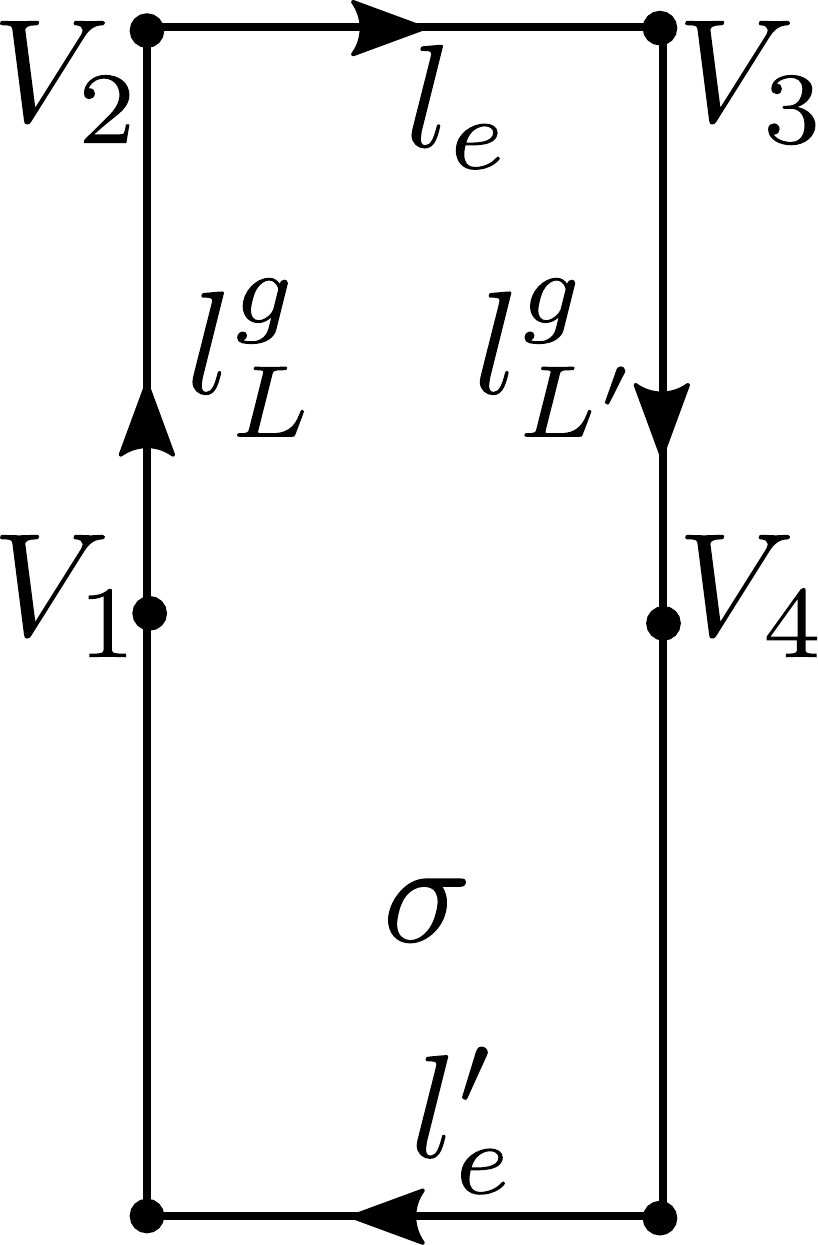}  &
	\includegraphics[width=0.6\linewidth]{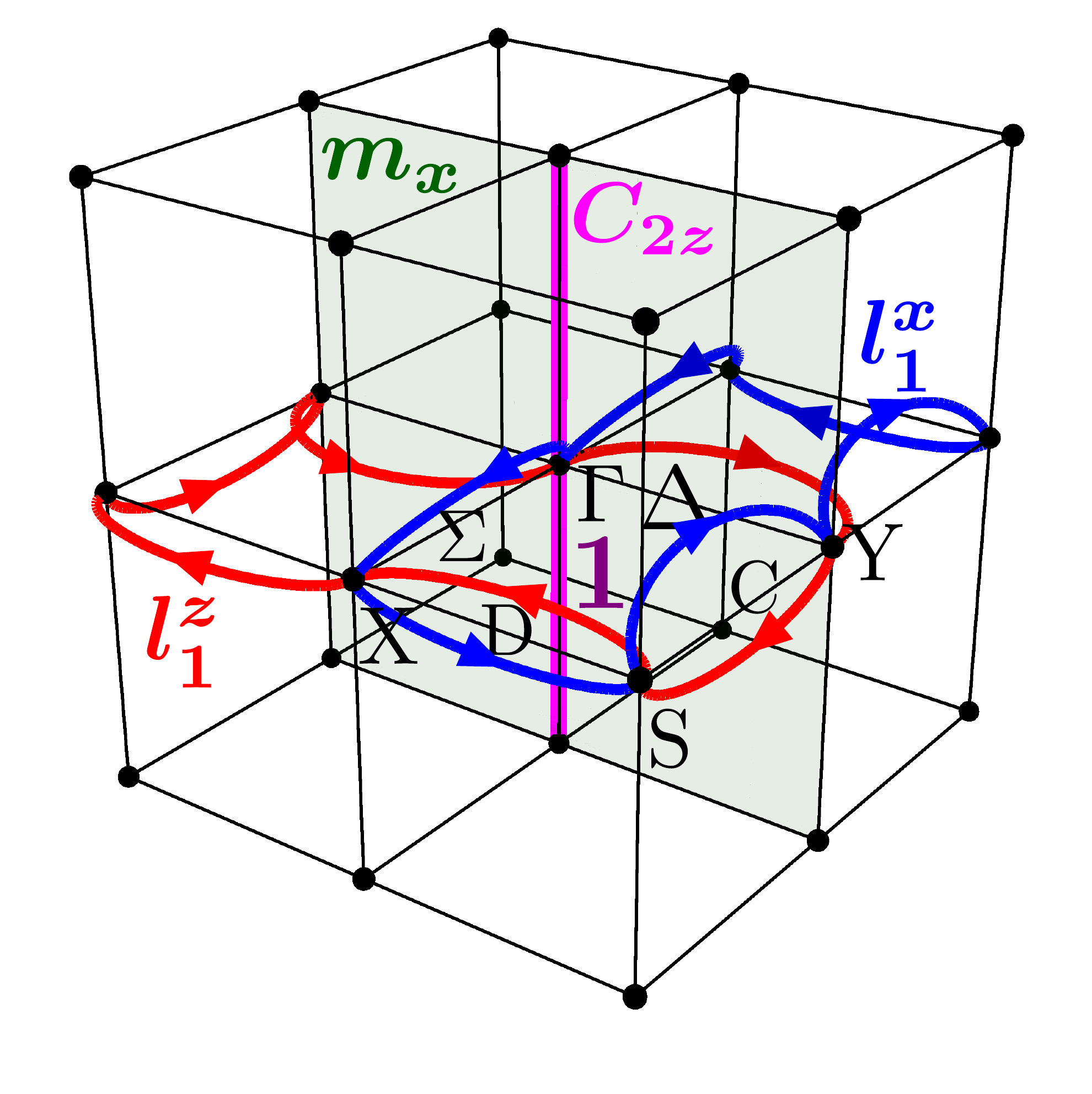} \\
	(a) & (b)
\end{tabular} 
\caption{\label{fig_plane_loops} (a) Closed four-point loop $l_{\sigma}$ built from four coplanar HSPs $V_1,V_2,V_3,V_4 \in \sigma$. It consists of two quasi-closed two-point loops, $l^g_{L}$ and $l^g_{L'}$, and two edge segments, $l_e$ and $l_e' = l^{-1}_e+\boldsymbol{K} $. (b) Examples of closed four-point symmetry-constrained loops, $l^{z}_{1}$ and $l^{x}_{1}$ for the plane $\sigma_1$. 
}
\end{figure}

We can extend the previous construction by now fixing four coplanar HSPs $\{V_1,V_2,V_3,V_4\}$ and build closed four-point symmetry-constrained loops by requiring them to be foldable under a single point group operation combined with reciprocal lattice translations. A four-point loop can be seen as the composition of two quasi-closed two-point loops, denoted $l^g_{L}$ and $l^g_{L'}$, connected by two edge segments, $l_e$ and $l'_e = l^{-1}_e+\boldsymbol{K}$, see Fig.~\ref{fig_plane_loops}(a). Taking any combination of four coplanar vertices among the eight inequivalent HSPs of the BZ, $\{\Gamma,\text{X},\text{Y},\text{Z},\text{S},\text{T},\text{U},\text{R}\}$, and discarding the diagonal planes, we generate the six $\sigma$-planes in Fig.~\ref{fig_SG33}(b). For each of the planes we can form a closed four-point symmetry-constrained loop $l^g_{\sigma}$ with $g\in \overline{C}^{\sigma}$, where $\overline{C}^{\sigma}$ is the centralizer of the plane $\sigma$. The closed four-point symmetry constrained loops allowed by SG33 are $ \{l_1^{z,y,x},l_2^{z,y,x},l_3^{z,x},l_4^{z,x},l_5^{z,y},l_6^{z,y}\}$. We show in Fig.~\ref{fig_plane_loops}(b) examples of $l_1^{z}$ and $l_1^x$ both formed from the four HSPs $\{\Gamma$,X,Y,S$\}$. In Fig.~\ref{fig_plane_loops}(b) $l_1^{z}$ is composed of the quasi-closed loops $ l_{\Delta}^z$ and $l^{z}_{\text{D}}$ and of the two edges $l_e=l_{\text{YS}}$ and $l'_e=l_{\text{S}-\boldsymbol{b}_2,\text{Y}-\boldsymbol{b}_2}\cong l^{-1}_{\text{YS}}-\boldsymbol{b}_2$, while $l_1^{x}$ is composed of the quasi-closed loops $ l_{\Sigma}^x$ and $l^{x}_{\text{C}}$ and of the two edges $l_e=l_{\text{XS}}$ and $l'_e=l_{\text{S}-\boldsymbol{b}_1\text{X}-\boldsymbol{b}_1}\cong l^{-1}_{\text{XS}}-\boldsymbol{b}_1$. 

Similarly as for the two-point loops, we want to derive algebraically the symmetry-protected topological invariants from the Wilson loop $\mathcal{W}[l^g_{\sigma}]$. The Wilson loop can in this case be decomposed as
\begin{align}
	\mathcal{W}[l^g_{\sigma}] = \mathcal{W}[l^g_{L'}]   \mathcal{W}[l_{e}]^{-1}   \mathcal{W}[l^g_{L}]   \mathcal{W}[l_e]  \;,
\end{align}
where the total oriented loop reads $l^g_{\sigma} =  l^g_{L'} \circ l'_e \circ l^g_{L} \circ l_e$, see Fig.~\ref{fig_plane_loops}(a), and where we have assumed the periodic gauge which trivializes the translation by $\boldsymbol{K}$ in $l'_e = l_e^{-1} + \boldsymbol{K}$. In analogy with the mapping $\mathcal{M}_2$ introduced in the previous section, we define $\mathcal{M}_4$ as the mapping from the Wilson loop spectra of the two quasi-closed two-point loops, $l^g_{L}$ and $l^g_{L'}$, to the Wilson loop spectrum of the four-point loop $l^g_{\sigma}$. We show below a construction that effectively reduces $\mathcal{M}_{4}$ to $\mathcal{M}_{2}$. It is based on the fact that we only need the symmetry protected topological invariants and not the Wilson loop spectrum \textit{per se}.  

We first make the two quasi-closed Wilson loops diagonal through the appropriate unitary transformation, i.e.~$\mathcal{D}[l^g_{L^{(')}}] = U^{\dagger}_{L^{(')}} \mathcal{W}[l^g_{L^{(')}}] U_{L^{(')}}$. This gives
\begin{align}
	\tilde{\mathcal{W}}[l^g_{\sigma}]  &= U^{\dagger}_{L'}\mathcal{W}[l^g_{\sigma}]  U_{L'} \;,\nonumber\\
	\label{wilson_folding_four}
	& =  \mathcal{D}[l^g_{L'}]  \tilde{\mathcal{W}}[l_{e}]^{-1}  \mathcal{D}[l^g_{L}]  \tilde{\mathcal{W}}[l_e]  \;,
\end{align}
where we have defined $ \tilde{\mathcal{W}}[l_e] = U^{\dagger}_{L}  \mathcal{W}[l_e] U_{L'}$. Note that the spectra of the two quasi-closed Wilson loops $\mathcal{D}[l^g_{L^{(')}}]$ are obtained through the symmetry folding and the mapping $\mathcal{M}_2 \circ \Phi_2$ in Eq.~(\ref{M_2}). 

We have seen in the previous section that the Wilson loop spectrum of closed two-point symmetry-constrained loops is composed of $\mathcal{N}_{\xi}$ symmetry protected $\xi$-eigenvalues and $N_v-\mathcal{N}_{\xi}$ remaining eigenvalues that can take the value $-\xi$ or form conjugate pairs $\{e^{i\varphi},e^{-i\varphi}\}$. Since only the $\xi$-eigenvalues are symmetry protected, we can adiabatically map all the complex conjugate pairs to $-\xi$. This transformation is adiabatic because it does not require a change of the band topology. The diagonalized quasi-closed two-point Wilson loops are then only composed of $\pm1$ eigenvalues, analogously to the bare IRREP eigenvalues, and we write them $\mathcal{R}[l^g_{L(')}]$. Let us formally write the combined action of the diagonalization of Eq.~(\ref{wilson_folding_four}) and making the Wilson loop spectra real as the pullback under the mapping $\Phi_4$
\begin{equation}
\label{Phi_4}
	\Phi_4^{\star}\mathcal{W}[l^g_{\sigma}] = \mathcal{R}[l^g_{L'}]   \tilde{\mathcal{W}}[l_{e}]^{-1}  \mathcal{R}[l^g_{L}]   \tilde{\mathcal{W}}[l_e] \;.
\end{equation}
This has exactly the same structure as Eq.~(\ref{Phi_2}) and the four-point Wilson loop spectrum, $\mathrm{eig}\{\Phi_4^{\star}\mathcal{W}[l^g_{\sigma}]\}$, is now explicitly symmetric under complex conjugation on top of being gauge invariant. Furthermore, through the equivalence of Eqs.~(\ref{Phi_2}) and (\ref{Phi_4}), we have formally reduced the problem of finding algebraically the symmetry protected spectrum of a four-point Wilson loop ($\mathcal{M}_4$) into the previous problem of finding the symmetry protected spectrum of a two-point Wilson loop ($\mathcal{M}_2$), i.e.~formally
\begin{equation}
\label{M_4}
	\mathcal{M}_4 \simeq \mathcal{M}_2  : 
	( \mathcal{R}[l^g_{L}], \mathcal{R}[l^g_{L'}]) \mapsto  
	\mathrm{eig}\{\Phi_4^{\star}\mathcal{W}[l^g_{\sigma}]\}	 \;,
\end{equation}
Thus we can use the mappings $\mathcal{M}_2$ tabulated in Appendix \ref{mappings}). We note that in the case of two valence bands the mapping $\mathcal{M}_4$ can actually be derived analytically without the need of $\Phi_4$ (see for instance Ref.~\cite{Alex1}), we give this result in Table \ref{mapping_Wilson_two} in Appendix \ref{mappings}. 

Then, given the minority eigenvalue $\xi$ over the four vertices of the loop ($V_1,V_2,V_3,V_4 \in \sigma$ ), determined by $\min\limits_{\xi=\pm1} (N^{V_1}_{\xi}+N^{V_2}_{\xi}+N^{V_3}_{\xi}+N^{V_4}_{\xi}) $, we get the number of symmetry protected $\xi$-eigenvalues of the total Wilson loop spectrum, which we write $\mathcal{N}^g_{\sigma}$. Similarly to the topological invariants of a two-point loop, $\mathcal{N}^g_{\sigma}$ constitutes a symmetry protected topological invariant that also characterizes the bulk topology of the system. 
Hence, we have defined the \textit{four-point loop topological invariant}
\begin{equation}
	\mathcal{N}^g_{\sigma} \in \mathbb{N} \;,
\end{equation}
from which we readily get the quantized Berry phase factor,
\begin{equation}
\label{Berry_phase_factor_4}
	e^{i \gamma^g_{\sigma}} = \mathrm{det}~\mathcal{W}[l^g_{\sigma}]  = (-1)^{\mathcal{N}^g_{\sigma}} \in \{ +1 , -1 \} \equiv \mathbb{Z}_2 \;.
\end{equation}
As mentioned above, we can use all the point symmetries that belong to the centralizer of a plane $\sigma$. 

Similarly to the two-point loop invariants, we will see in later sections that whenever a four-point loop encircles an odd number of NLs the Berry phase is nontrivial with $\gamma^g_{\sigma} = \pi \mod 2\pi$ but trivial ($\gamma^g_{\sigma} = 0 \mod 2\pi$) otherwise. Furthermore, we will see that $\mathcal{N}^{x,y}_{\sigma} $ counts the number of NLs belonging to $m_{x,y}$-invariant planes that cross the plane $\sigma$, while $\mathcal{N}^{z}_{\sigma} $ gives complementary informations that further characterize the type of line-nodal structure.

\subsection{Rosetta stone for SG33-AI}
\begin{figure}[htb]
\centering
\begin{tabular}{cc}
	\includegraphics[width=0.54\linewidth]{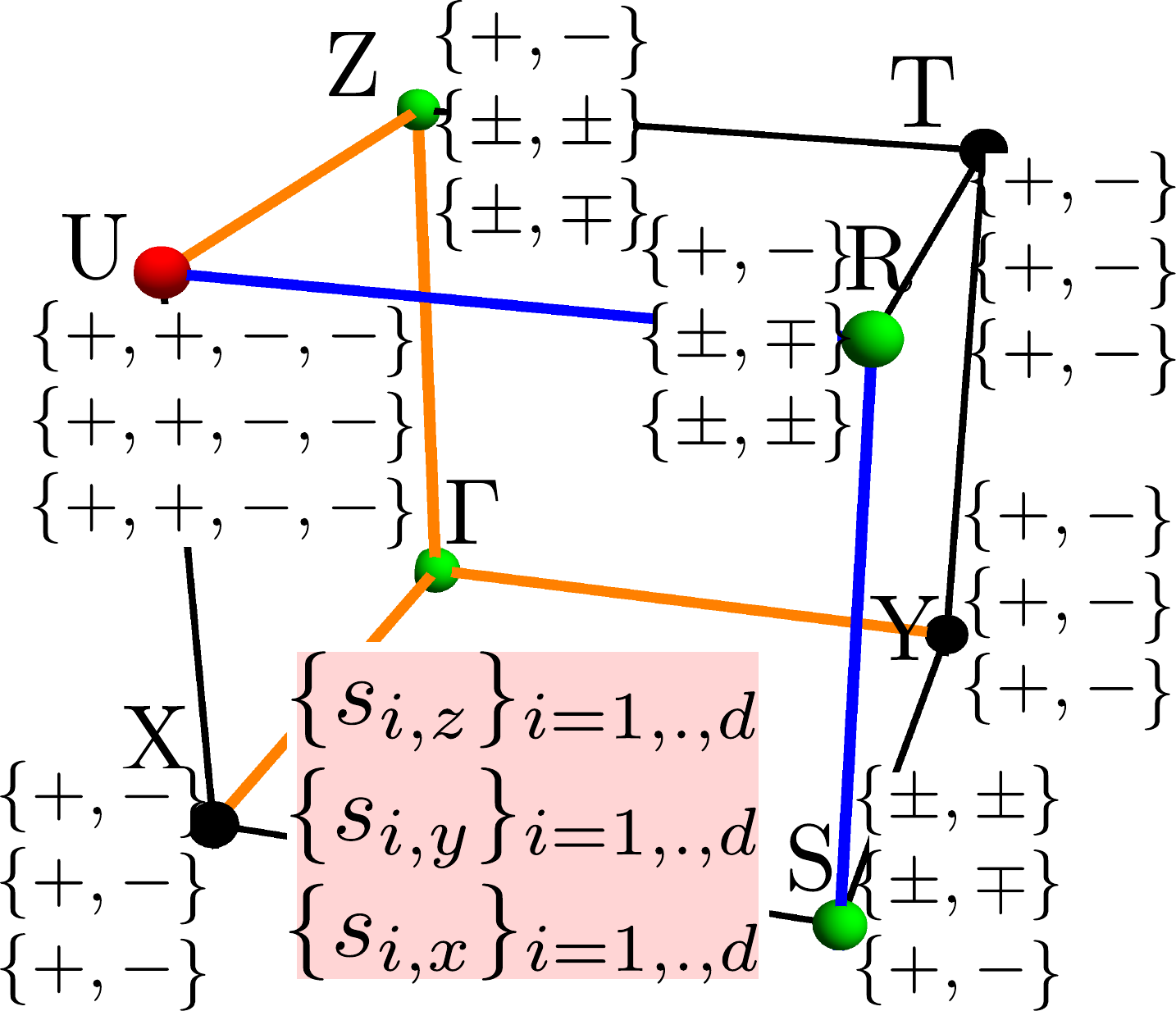} & 
	\includegraphics[width=0.46\linewidth]{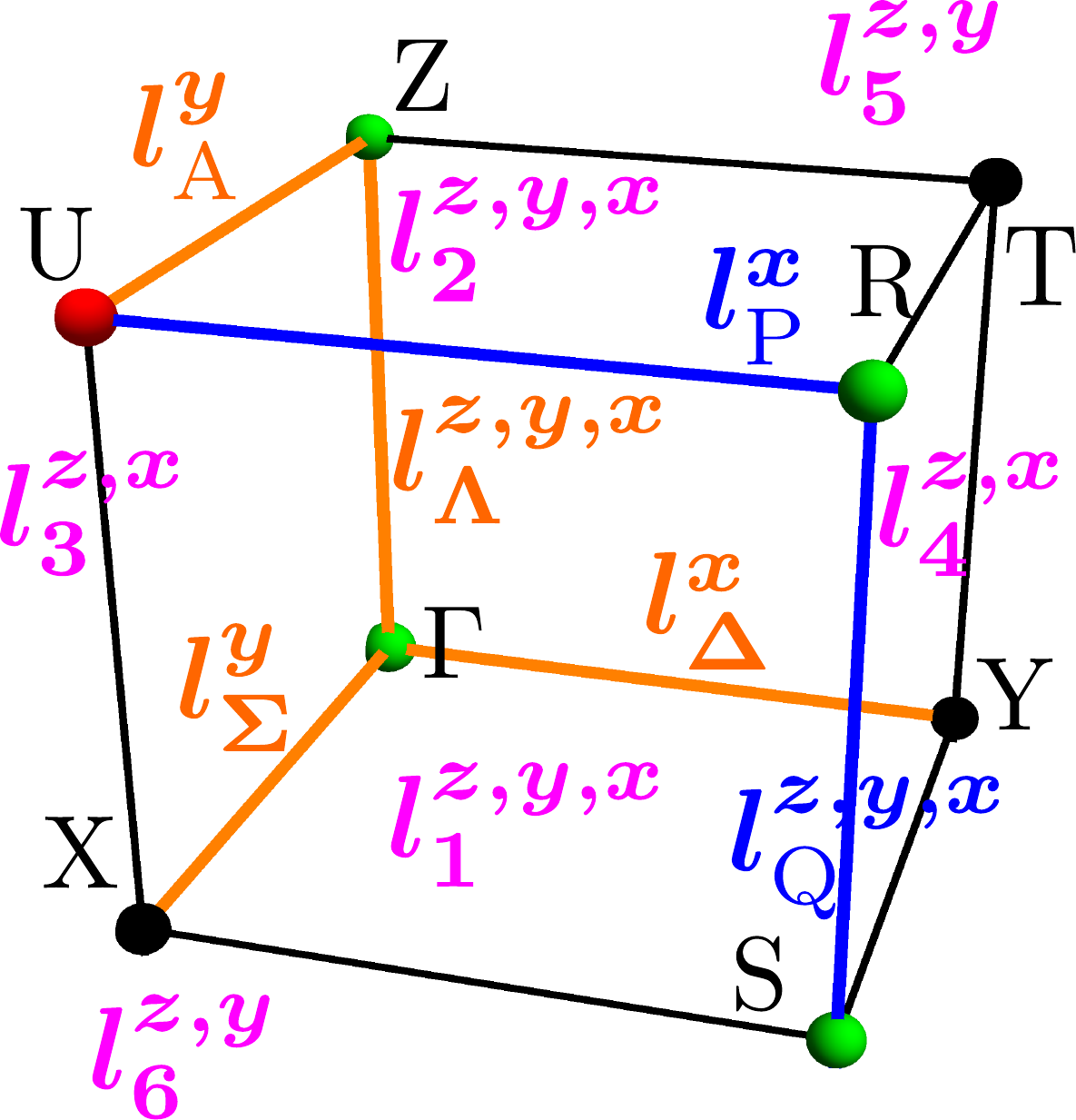} \\
	(a) & (b)
\end{tabular}
\caption{\label{rosetta_stone} Rosetta stone for SG33-AI to be used together with the character table at $\Gamma$ Fig.~\ref{fig_SG33}(a). (a) The bare IRREP eigenvalues at each HSP for $g=z,y,x$ listed in the order $(\{s_{i,z}\},\{s_{i,y}\},\{s_{i,x}\})_{i=1,\dots,d}^T$ with $d=2$ for 2D IRREPs and $d=4$ at U. Here $s_{i,z}=\pm$ at the HSP p means eigenvalue $+1$ for $p_5$ and $-1$ for $p_6$. (b) All closed two-point and four-point symmetry constrained loops for which we compute the symmetry protected topological invariants. 
} 
\end{figure}

All the two- and four-point loop symmetry protected topological invariants are obtained algebraically from the bare IRREP eigenvalues of the vertices of the closed symmetry-constrained loops and by applying the mappings $\mathcal{M}_2\circ \Phi_2$ in Eqs.~(\ref{Phi_2},\ref{M_2}) and $\mathcal{M}_4\circ \Phi_4$ in Eqs.~(\ref{Phi_4},\ref{M_4}). Therefore, all what is needed is the \textit{rosetta stone} shown in Fig.~\ref{rosetta_stone} together with the character table for the IRREPs at the $\Gamma$-point in Fig.~\ref{fig_SG33}(b). Fig.~\ref{rosetta_stone}(a) gives the bare IRREP eigenvalues at every HSP for $g=z,y,x$ listed in the order $(\{s_{i,z}\},\{s_{i,y}\},\{s_{i,x}\})_{i=1,\dots,d}^T$ with $d=2$ for 2D IRREPs and $d=4$ at U, and where $s_{i,z}=\pm$ at the HSP p means there are two allowed bare IRREPs with eigenvalues $+1$ for $p_5$ and $-1$ for $p_6$. Fig.~\ref{rosetta_stone}(b) gives all the closed two-point and four-point symmetry constrained loops for which we compute the symmetry protected topological invariants. Using this rosetta stone as input, the mappings $\mathcal{M}_2$ and $\mathcal{M}_4$ tabulated in Appendix \ref{mappings}) gives directly the topological invariants.

Let us illustrate the process works with a few illustrative examples. First we consider the situation with two valence states. Let us first compute the two-point invariant $\mathcal{N}^y_{\Sigma}$ for Fig.~\ref{fig_SG33_4B_lines}(a). The symmetry constrained two-point loop $l^y_{\Sigma}$ is made from the vertices $\Gamma$ and X, see Fig.~\ref{rosetta_stone}(b). The valence IRREPs at $\Gamma$ are $\{\Gamma_1,\Gamma_2\}$ with the bare $m_y$-eigenvalues $\{+,+\}$, see Fig.~\ref{fig_SG33}(c), and at X there is a single (projective) IRREP with the bare $m_y$-eigenvalues $\{+,-\}$, see Fig.~\ref{rosetta_stone}(a). The minority eigenvalue is thus $-1$. Then the mapping $\mathcal{M}_2$ gives $\{+,-\}$, therefore there is one Wilson loop phase of $ \pi \mod 2\pi$ that is protected by symmetry and we write $\mathcal{N}^y_{\Sigma}=1$. 

Next let us compute $\mathcal{N}^y_{\Lambda}$ in the case Fig.~\ref{fig_SG33_4B_lines}(d). The symmetry constrained two-point loop $l^y_{\Lambda}$ is made from the vertices $\Gamma$ and Z, see Fig.~\ref{rosetta_stone}(b). The IRREPs at $\Gamma$ gives the same bare $m_y$-eigenvalues as above, $\{+,+\}$, and the IRREP $Z_6$ at Z gives $\{-,-\}$. We can here choose the minority eigenvalue has $+1$ (or $-1$), and the mapping $\mathcal{M}_2$ gives $\{-,-\}$ ($\{+,+\}$). This gives two Wilson loop phases of $\pi \mod 2\pi$ that are protected by symmetry and we write $\mathcal{N}^y_{\Lambda} = 2$. 

Let us now consider the situation with four valence states. As an example, we compute $\mathcal{N}^y_{\Lambda}$ in the case Fig.~\ref{fig_SG33_8B_lines}(i). We need the bare $m_y$-eigenvalues of the IRREPs $\{\Gamma_1,\Gamma_1,\Gamma_2,\Gamma_3\}$ at $\Gamma$, and $\{Z_6,Z_6\}$ at Z. We find $\Gamma$:$\{+,+,+,-\}$ and Z:$\{-,-,-,-\}$. Let us rearrange them by multiplying both by $-1$, this will not change the result as this corresponds to simply  multiplying Eq.~(\ref{Phi_2}) by $+1$. We then have $\{-,-,-,+\}$ at $\Gamma$ and $\{+,+,+,+\}$ at Z. The minority eigenvalue is $-1$. Applying the mapping $\mathcal{M}_2$ we find $\{-,-,-,+\}$. Therefore there are three symmetry protected nontrivial Wilson loop phases that we write in the invariant $\mathcal{N}^y_{\Lambda}=3$. Note that had we chosen the IRREPs $\{Z_5,Z_5\}$ at Z instead, corresponding to the case Fig.~\ref{fig_SG33_8B_lines}(h), had we found $\mathcal{N}^y_{\Lambda} =1$. 

As a last example we compute $\mathcal{N}^y_{2}$ in the case Fig.~\ref{fig_SG33_8B_lines}(o). The four-point loop $l^y_2$ is made form the vertices Z, R, T, and U, with the IRREPs $\{Z_5,Z_5,R_5,R_6,T,T,U\}$. Reading the bare $m_y$-eigenvalues from Fig.~\ref{rosetta_stone}(a), we find $Z:\{+,+,+,+\}$, R:$\{+,-,-,+\}$, T:$\{+,-,+,-\}$, and U:$\{+,+,-,-\}$. The minority eigenvalue is $-1$. Applying the mapping $\mathcal{M}_2$ on the couples (Z,U) and (T,R), we find (Z,U)$\mapsto\{+,+,-,-\}$ and (T,R)$\mapsto\{\lambda_1,\lambda_1^*,\lambda_2,\lambda_2^*\}\simeq\{+,+,+,+\}$. Then applying the mapping $\mathcal{M}_4$, we find $\{+,+,-,-\}$. Therefore, there are two nontrivial Wilson loop phases over the four-point symmetry constrained loop $l^y_{2}$, which we write in the invariant $\mathcal{N}^y_{2} = 2$.

\subsection{From loop invariants to local charges of nodal structures}\label{sym_link}
Our aim is to characterize the topology of the nodal valence bundles obtained from the systematic classification of band structures for SG33-AI presented in Section \ref{SG33}. So far we have shown that the IRREPs at the HSPs determine every loop topological invariant for any fixed filling number $N_v$. Since the symmetry protected nodal structures are also determined by the set of valence IRREPs (through compatibility relations and band permutation rules), we naturally find that the loop invariants for the a given set of IRREPs directly reflect the corresponding nodal structure. Effectively, all the local topological charges of a nodal structure, i.e.~poloidal-toroidal and monopole charges, are given in terms of the invariants derived in this previous section. We show this explicitly for the four-band subspace in Section \ref{four_top} and for the eight-band subspace in Section \ref{eight_top}.

\section{Four-band topology}\label{four_top}
In this section we consider the case of an isolated four-band subspace at half-filling, i.e.~$N=4$, $N_v=2$. We identify all the symmetry protected topological invariants supported by SG33-AI and we show how they enter in the topological characterization of the symmetry protected nodal structures found in Fig.~\ref{fig_SG33_4B_lines}. It turns out that only poloidal charges are relevant for the line-nodal structure of the four-band subspace. For completeness, we also discuss the accidental point nodes of the case Fig.~\ref{fig_SG33_4B_lines}(d) characterized by their monopole charges. 

In the following we implicitly use the enlarged group of $k$-space point symmetries, $g\in C_{2v}\times \{E,I\} =D_{2h}$, of the band structures (due to TRS) and only count the symmetry independent NLs.

\subsection{Local topology of the line-nodal structures} 
Applying the algorithm derived in Section \ref{top_inv}, we find all the two-point and four-point invariants over all the loops in Fig.~\ref{rosetta_stone}(b) determined only in terms of the set of valence IRREPs at the HSPs. They are listed in Table~\ref{point_classes_fourband} for every case of Fig.~\ref{fig_SG33_4B_lines}. Note that all the loops and planes containing $\text{U}$ have been discarded since U is fourfold degenerate, thus preventing the distinction between valence and conduction eigenstates. While we focus on the half-filling case, it is straightforward to extend the analysis to arbitrary filling. We have confirmed that all the invariants of Table~\ref{point_classes_fourband} written in black correspond to base loops that encircle a single NL, i.e.~they are equivalent to a nontrivial Berry phase factor, Eqs.~(\ref{Berry_phase_factor_2},\ref{Berry_phase_factor_4}). The invariants written in blue instead lead to a trivial phase factor and count the even number of symmetry protected NLs within $m_x$- or $m_y$-planes encircled by the base loop. Below we give a detailed account of all line-nodal structures.
{\def\arraystretch{1.3}  
\begin{table*}[t!] 
\caption{\label{point_classes_fourband} Symmetry-protected topological invariants for each line-nodal structure in Fig.~\ref{fig_SG33_4B_lines} labeled from (a) to (e) of a four-band subspace at half-filling. First column lists the combinatoric sets of valence IRREPs. Second (third) set of columns gives the closed two-point (four-point) loop invariants. 
Invariants written in black correspond to a nontrivial Berry phase and mark the presence of single NL encircled by the base loop. Invariants written in blue give the (even) number of symmetry-protected nontrivial Wilson loop phases and count the multiple NLs within $m_x$- or $m_y$-invariant planes encircled by the base loop.}
\begin{tabular*}{\linewidth}{l @{\extracolsep{\fill}} | c  c c c | c c c}
\hline
\hline
	$\Gamma^{valence}$ & $\mathcal{N}^y_{\Sigma} $ & $\mathcal{N}^x_{\Delta} $ & $[\mathcal{N}^z_{\Lambda},\mathcal{N}^y_{\Lambda},\mathcal{N}^x_{\Lambda}]$ & $[\mathcal{N}^z_{\text{Q}},\mathcal{N}^y_{\text{Q}},\mathcal{N}^x_{\text{Q}}]$
	& $[\mathcal{N}^z_{1},\mathcal{N}^y_{1},\mathcal{N}^x_{1}]$  & $[\mathcal{N}^z_{4},\mathcal{N}^x_{4}]$ & $[\mathcal{N}^z_{5},\mathcal{N}^y_{5}]$ \\
	\hline
	$\{\Gamma_1,\Gamma_2,\text{Z}_5,\text{S}_{5(6)},\text{R}_{5(6)}\}_a$ & {\bf 1} & 0 & [0,0,0] & [{\bf 1},0,{\bf 1}] & [{\bf 1},{\bf 1},0] & [{\bf 1},{\bf 1}] & [0,0] \\
	$\{\Gamma_1,\Gamma_3,\text{Z}_{5(6)},\text{S}_{5(6)},\text{R}_{5(6)}\}_{b}$ & 0 & 0 & [{\bf 1},{\bf 1},0] &  [{\bf 1},0,{\bf 1}]  & [0,0,0] & [{\bf 1},{\bf 1}] & [{\bf 1},{\bf 1}] \\
	$\{\Gamma_1,\Gamma_4,\text{Z}_{5(6)},\text{S}_{5(6)},\text{R}_{5(6)}\}_c$ & 0 & {\bf 1} & [0,{\bf 1},{\bf 1}] & [{\bf 1},0,{\bf 1}] & [{\bf 1},0,{\bf 1}] & [{\bf 1},{\bf 1}] & [0,{\bf 1}] \\
	$\{\Gamma_1,\Gamma_2,\text{Z}_6,\text{S}_{5(6)},\text{R}_{5(6)}\}_{d,e}$ & {\bf 1} & 0 & [0,${\bf {\color{cyan} 2}}$,0] & [{\bf 1},0,{\bf 1}] & [{\bf 1},{\bf 1},0] & [{\bf 1},{\bf 1}]& [0,${\bf {\color{cyan} 2}}$]\\
 \hline
  \hline
\end{tabular*}
\end{table*} 
}

In the case Fig.~\ref{fig_SG33_4B_lines}(a), the NL belonging to the $m_y$-invariant plane $\sigma_3$ (actually the pair of NLs symmetric under $m_x$) and crossing the line $\Sigma$ is captured by the two-point loop invariant $\mathcal{N}^y_{\Sigma}=1$ and the four-point loop invariant $\mathcal{N}^y_{1}=1$ ($\sigma_1 \supset \Sigma$). Note also the four-point loop invariant $\mathcal{N}^z_1=1$. Since $l^{y}_{\Sigma}$, $l^{y}_{1}$ and $l^{z}_{1}$ all encircle a single NL, their loop invariants correspond to the nontrivial poloidal charge of the NL given through the Berry phase $\gamma^y_{\Sigma}=\gamma^y_1=\gamma^z_1 = \pi  \mod 2\pi$. Similarly, the NL belonging to the $m_x$-invariant plane $\sigma_6$ on the BZ boundary (again the pair of NLs symmetric under $m_z$) and crossing the line $\text{Q}$ is characterized by the two-point loop invariants $\mathcal{N}^x_{\text{Q}}=\mathcal{N}^z_{\text{Q}}=1$ and the four-point loop invariants $\mathcal{N}^x_{4}=\mathcal{N}^z_{4}=1$ ($\sigma_4 \supset \text{Q} $). Here too, $l^x_{\text{Q}}$, $l^z_{\text{Q}}$, $l^{x}_{4}$ and $l^{z}_{4}$ encircle a unique NL leading to the nontrivial poloidal Berry phase $\gamma^{x,z}_{\text{Q}} = \gamma^{x,z}_{4} = \pi \mod 2\pi$. In fact, this NL on the BZ boundary is present in all the cases of Fig.~\ref{fig_SG33_4B_lines}. 
The case Fig.~\ref{fig_SG33_4B_lines}(b) is directly analogous to Fig.~\ref{fig_SG33_4B_lines}(a) but with a NL on the $m_y$-invariant plane $\sigma_3$ crossing the line $\Lambda$ instead of $\Sigma$. It is characterized by the invariants $\mathcal{N}_{\Lambda}^{y,z}=1$ and $\mathcal{N}^{y,y}_{5}=1$ leading to a nontrivial poloidal Berry phase. 

The case Fig.~\ref{fig_SG33_4B_lines}(c) has two inequivalent NLs in the interior of the BZ, one belonging to the $m_y$-invariant plane $\sigma_3$ and characterized by the charges $\mathcal{N}^y_{\Lambda}=1$ and $\mathcal{N}^y_5=1$, and one belonging to the $m_x$-invariant plane $\sigma_5$ and characterized by the charges $\mathcal{N}^x_{\Lambda}=1$ and $\mathcal{N}^x_{1}=1$. Both lines hence have a nontrivial poloidal Berry phase. Although $l^z_{\Lambda}$ and $l^z_{5}$ encircles the two NLs their loop invariants are $\mathcal{N}^z_{\Lambda}=\mathcal{N}^z_{5}=0$. This is a consequence of there being no mismatch between the $C_{2z}$-bare IRREP eigenvalues at $\Gamma$, $\{s_{z}(\Gamma_1),s_{z}(\Gamma_4)\} = \{+1,-1\}$, and at Z, $\{s_{1,z},s_{2,z}\} = \{+1,-1\}$ for both $Z_{5(6)}$. We find that the $l^z$-loop invariants are always trivial like this when the loop encircles two NLs that cross at a simple point-node (twofold degenerate), here on $\Lambda$. 

Finally the cases of Fig.~\ref{fig_SG33_4B_lines}(d) and (e) exhibit two independent NLs on the same $m_y$-invariant plane $\sigma_3$ (not related by symmetry) with two independent crossings of the line $\Lambda$. They are characterized by the invariants $\mathcal{N}^{y,z}_{1}=1$ and $\mathcal{N}_{\Lambda}^{y}=\mathcal{N}^y_{5}=2$ ($\sigma_5 \supset \Lambda$). Since only one NL crosses $\Sigma$, the invariants of the four-point loops $l^{y,z}_1$ are directly analogous to the case Fig~\ref{fig_SG33_4B_lines}(a) and lead to the same nontrivial poloidal Berry phase corresponding to the odd parity of the number of NLs encircled. On the other hand, the invariants $\mathcal{N}_{\Lambda}^{y}=\mathcal{N}^y_{5}=2$, the only ones marked in blue in Table \ref{point_classes_fourband}, count the number of NLs within the $m_y$-invariant plane $\sigma_3 $ that are encircled by the loops $l^y_{\Lambda}$ and $l^y_{\sigma_5}$ crossing the HSL $\Lambda$ and crossing the high-symmetry plane $\sigma_5$, respectively. Hence they go beyond the (trivial) Berry phase $\gamma^y_{\Lambda}=\gamma^y_{5} = 0 \mod 2\pi$. We note that the case (e) differs from (d) by the presence of a small NL (and its partner at $-k_z$) within the perpendicular plane $\sigma_{5}$ connecting the two independent NLs of $\sigma_3$. This is achieved through a different relative ordering of the valence IRREPs at $\Gamma$, see Section \ref{four}, to which the invariants of Table \ref{point_classes_fourband} are insensitive. Nevertheless, the local topology in $k$-space can be further characterized with finer invariants. A detailed discussion for the eight-band case is given in Section \ref{Lifshitz}. We also note that neither toroidal nor monopole charges appear in the symmetry protected line-nodal structures of the four-band subspace.

\subsection{Accidental point nodes}\label{accidental_points}
Previously, we noted the eight accidental point-nodes of Fig.~\ref{fig_SG33_4B_lines}(d). As argued in Section \ref{top_meta}, a point node is locally characterized by a Chern number that can be computed numerically through the flow of Berry phase as we sweep a base loop over a sphere surrounding the point node, i.e.~$C_1 = \left(\gamma[l_{\theta=\pi}] - \gamma[l_{\theta=0}]\right)/2\pi$. In Fig.~\ref{fig_point_node} we show the numerical Berry phase flow around one of the point nodes in the case Fig.~\ref{fig_SG33_4B_lines}(d). From Fig.~\ref{fig_point_node}(b) we directly find $C_1 = +1$. As a point node at an arbitrary momentum is stable without any symmetry, the loops $l_{\theta}$ can be chosen without any symmetry constraint. 

For SG33-AI band structure is always symmetric under the enlarged group of $k$-space point symmetries $g\in C_{2v}\times \{E,I\} = D_{2h}$, and thus a point node at a general position must have seven other partners exhausting the orbit of $D_{2h}$. In fact, it can easily be shown that: (i) a pair of point nodes that are symmetric under a unitary rotation must have the same charge; (ii) a pair of point nodes that are symmetric under a unitary reflection must have opposite charges; and (iii) the charges of point node pairs relate reversely for the corresponding anti-unitary symmetries. Therefore, the point nodes are located symmetrically in the BZ according to $D_{2h}$ and with alternating signed charges according to (i), (ii) and (iii). We conclude that the nodal structure of Fig.~\ref{fig_SG33_4B_lines}(d) has four sources ($C_1=+1$) and four sinks ($C_1=-1$) of Berry curvature. This is consistent with Nielsen-Ninomiya theorem \cite{Nielsen1, Nielsen1a, Nielsen2, Kiritsis, Witten_lect}, which states that the total charge over the whole BZ must be zero. Furthermore, we can merge node and anti-node pairs onto each-other through a large adiabatic transformation, i.e.~without closing the band gap at the HSPs and HSLs and preserving all the symmetries of the system. Therefore, these point nodes are also accidental in the sense that they can be removed without changing the global band topology as fixed by the set of valence IRREPs at the HSPs. This discussion thus concludes the full topological classification of the nodal structure of the four-band subspace.

\section{Eight-band topology}\label{eight_top}
{\def\arraystretch{1.3}  
\begin{table*}[t]
\caption{\label{point_classes_eight} Symmetry protected topological invariants for each line-nodal structure in Fig.~\ref{fig_SG33_8B_lines} labeled from (a) to (v) of an eight-band subspace at half-filling. First column lists the high-symmetry point (HSP) classes defined in Table~\ref{point_classes} in terms of combinatoric sets of valence IRREPs.  Second (third) set of columns gives the two-point (four-point) loop invariants. Invariants that correspond to a nontrivial Berry phase are written in black, while other invariants are written in color (blue and red). Invariants $\mathcal{N}^{y(x)}_{\alpha}$ count the number of NLs within $m_y$($m_x$)-invariant planes surrounded by the loops $l^{y(x)}_{\alpha}$. Red invariants mark the presence of line-node monopole pairs. Rows (a-p) are for the HSP classes of the $\mathcal{B}_{\Gamma}$ domain, while rows (q-s) are for the HSP classes of the $\mathcal{B}_{\text{R}}$ domain.  Three last rows (t,u,v) correspond to the combined HSP classes of Fig.~\ref{fig_SG33_8B_lines}}
\begin{tabular*}{\linewidth}{@{\extracolsep{\fill}} l  | c  c c c c c | c c c c c c}
\hline
\hline
	$\Gamma^{val}$ & $\mathcal{N}^y_{\Sigma} $ & $\mathcal{N}^x_{\Delta} $ & $[\mathcal{N}^{z,y,x}_{\Lambda}]$ & $\mathcal{N}^y_{\text{A}} $ & $\mathcal{N}^x_{\text{P}} $ & $[\mathcal{N}^{z,y,x}_{\text{Q}}]$
	& $[\mathcal{N}^{z,y,x}_{1}]$& $[\mathcal{N}^{z,y,x}_{2}]$  & $[\mathcal{N}^z_{3},\mathcal{N}^x_{3}]$ & $[\mathcal{N}^z_{4},\mathcal{N}^x_{4}]$ & $[\mathcal{N}^z_{5},\mathcal{N}^y_{5}]$ & $[\mathcal{N}^z_{6},\mathcal{N}^y_{6}]$ \\
	\hline
	$(\Gamma_{II},\text{Z}_{I})_a$ & {\bf 1} & 0 & [{\bf 1},{\bf 1},0] & 0 & 0 & [0,0,0]  & [{\bf 1},{\bf 1},0] & [0,0,0] & [{\bf 1},0] & [0,0] & [{\bf 1},{\bf 1}] & [0,0] \\
	$(\Gamma_{II},\text{Z}_{I})_b$ & 0 & {\bf 1} & [{\bf 1},0,{\bf 1}] & 0 & 0 & [0,0,0]  & [{\bf 1},0,{\bf 1}] & [0,0,0] & [{\bf 1},{\bf 1}] & [0,0] & [{\bf 1},0] & [0,0] \\
	$(\Gamma_{II},\text{Z}_{I})_c$ & {\bf 1} & {\bf 1} & [0,{\bf 1},{\bf 1}] & 0 & 0 & [0,0,0]  & [0,{\bf 1},{\bf 1}] & [0,0,0] & [0,{\bf 1}] & [0,0] & [0,{\bf 1}] & [0,0] \\
	$(\Gamma_{III},\text{Z}_{I})_d$ & ${\bf {\color{cyan} 2}}$ & 0 & [0,${\bf {\color{cyan} 2}}$,0] & 0 & 0 & [0,0,0]  & [0,${\bf {\color{cyan} 2}}$,0] & [0,0,0] & [0,0] & [0,0] & [0,${\bf {\color{cyan} 2}}$] & [0,0] \\
	$(\Gamma_{III},\text{Z}_{I})_e$ & 0 & 0 & [${\bf {\color{red} 2}}$,0,0] & 0 & 0 & [0,0,0]  & [${\bf {\color{cyan} 2}}$,0,0] & [0,0,0] & [${\bf {\color{red} 2}}$,0] & [0,0] & [${\bf {\color{red} 2}}$,0] & [0,0] \\
	$(\Gamma_{III},\text{Z}_{I})_f$ & 0 & ${\bf {\color{cyan} 2}}$ & [0,0,${\bf {\color{cyan} 2}}$] & 0 & 0 & [0,0,0]  & [0,0,${\bf {\color{cyan} 2}}$] & [0,0,0] & [0,${\bf {\color{cyan} 2}}$] & [0,0] & [0,0] & [0,0] \\
	$(\Gamma_{I},\text{Z}_{II})_g$ & 0 & 0 & [0,${\bf {\color{cyan} 2}}$,0] & ${\bf {\color{cyan} 2}}$ & 0 & [0,0,0]  & [0,0,0] & [0,${\bf {\color{cyan} 2}}$,0] & [0,0] & [0,0] & [0,${\bf {\color{cyan} 2}}$] & [0,0] \\
	$(\Gamma_{II},\text{Z}_{II})_h$ & {\bf 1} & 0 & [{\bf 1},{\bf 1},0] & ${\bf {\color{cyan} 2}}$ & 0 & [0,0,0]  & [{\bf 1},{\bf 1},0] & [0,${\bf {\color{cyan} 2}}$,0] & [{\bf 1},0] & [0,0] & [{\bf 1},{\bf 1}] & [0,0] \\
	$(\Gamma_{II},\text{Z}_{II})_i$ & {\bf 1} & 0 & [{\bf 1},${\bf {\color{black} 3}}$,0] & ${\bf {\color{cyan} 2}}$ & 0 & [0,0,0]  & [{\bf 1},{\bf 1},0] & [0,${\bf {\color{cyan} 2}}$,0] & [{\bf 1},0] & [0,0] & [{\bf 1},${\bf {\color{black} 3}}$] & [0,0] \\
	$(\Gamma_{II},\text{Z}_{II})_j$ & 0 & {\bf 1} & [{\bf 1},${\bf {\color{cyan} 2}}$,{\bf 1}] & ${\bf {\color{cyan} 2}}$ & 0 & [0,0,0]  & [{\bf 1},0,{\bf 1}] & [0,${\bf {\color{cyan} 2}}$,0] & [{\bf 1},{\bf 1}] & [0,0] & [{\bf 1},${\bf {\color{cyan} 2}}$] & [0,0] \\
	$(\Gamma_{II},\text{Z}_{II})_k$ & {\bf 1} & {\bf 1} & [0,{\bf 1},{\bf 1}] & ${\bf {\color{cyan} 2}}$ & 0 & [0,0,0]  & [0,{\bf 1},{\bf 1}] & [0,${\bf {\color{cyan} 2}}$,0] & [0,{\bf 1}] & [0,0] & [0,{\bf 1}] & [0,0] \\
	$(\Gamma_{II},\text{Z}_{II})_l$ & {\bf 1} & {\bf 1} & [0,${\bf {\color{black} 3}}$,{\bf 1}] & ${\bf {\color{cyan} 2}}$ & 0 & [0,0,0]  & [0,{\bf 1},{\bf 1}] & [0,${\bf {\color{cyan} 2}}$,0] & [0,{\bf 1}] & [0,0] & [0,${\bf {\color{black} 3}}$] & [0,0] \\
	$(\Gamma_{III},\text{Z}_{II})_m$ & ${\bf {\color{cyan} 2}}$ & 0 & [0,0,0] & ${\bf {\color{cyan} 2}}$ & 0 & [0,0,0]  & [0,${\bf {\color{cyan} 2}}$,0] & [0,${\bf {\color{cyan} 2}}$,0] & [0,0] & [0,0] & [0,0] & [0,0] \\
	$(\Gamma_{III},\text{Z}_{II})_n$ & ${\bf {\color{cyan} 2}}$ & 0 & [0,${\bf {\color{cyan} 4}}$,0] & ${\bf {\color{cyan} 2}}$ & 0 & [0,0,0]  & [0,${\bf {\color{cyan} 2}}$,0] & [0,${\bf {\color{cyan} 2}}$,0] & [0,0] & [0,0] & [0,${\bf {\color{cyan} 4}}$] & [0,0] \\
	$(\Gamma_{III},\text{Z}_{II})_o$ & 0 & 0 & [${\bf {\color{cyan} 2}}$,${\bf {\color{cyan} 2}}$,0] & ${\bf {\color{cyan} 2}}$ & 0 & [0,0,0]  & [${\bf {\color{cyan} 2}}$,0,0] & [0,${\bf {\color{cyan} 2}}$,0] & [${\bf {\color{cyan} 2}}$,0] & [0,0] & [${\bf {\color{cyan} 2}}$,${\bf {\color{cyan} 2}}$] & [0,0] \\
	$(\Gamma_{III},\text{Z}_{II})_p$ & 0 & ${\bf {\color{cyan} 2}}$ & [0,${\bf {\color{cyan} 2}}$,${\bf {\color{cyan} 2}}$] & ${\bf {\color{cyan} 2}}$ & 0 & [0,0,0]  & [0,0,${\bf {\color{cyan} 2}}$] & [0,${\bf {\color{cyan} 2}}$,0] & [0,${\bf {\color{cyan} 2}}$] & [0,0] & [0,${\bf {\color{cyan} 2}}$] & [0,0] \\
\hline
	$(\text{S}_{II},\text{R}_{I})_q$ & 0 & 0 & [0,0,0] & 0 & 0 & [${\bf {\color{red} 2}}$,0,0]  & [${\bf {\color{cyan} 2}}$,0,0] & [0,0,0] & [0,0] & [${\bf {\color{red} 2}}$,0] & [0,0] & [${\bf {\color{red} 2}}$,0] \\
	$(\text{S}_{I},\text{R}_{II})_r$  & 0 & 0 & [0,0,0] & 0 & ${\bf {\color{cyan} 2}}$ & [0,0,${\bf {\color{cyan} 2}}$]  & [0,0,0] & [0,0,0] & [0,0] & [0,${\bf {\color{cyan} 2}}$] & [0,0] & [0,0] \\
	$(\text{S}_{II},\text{R}_{II})_s$ & 0 & 0 & [0,0,0] & 0 & ${\bf {\color{cyan} 2}}$ & [${\bf {\color{cyan} 2}}$,0,${\bf {\color{cyan} 2}}$]  & [${\bf {\color{cyan} 2}}$,0,0] & [0,0,${\bf {\color{cyan} 2}}$] & [0,0] & [${\bf {\color{cyan} 2}}$,${\bf {\color{cyan} 2}}$] & [0,0] & [${\bf {\color{cyan} 2}}$,0] \\
\hline
	$(\Gamma_{III},\text{S}_{II})_t$ & 0 & 0 & [${\bf {\color{red} 2}}$,0,0] & 0 & 0 & [${\bf {\color{red} 2}}$,0,0]  & [0,0,0] & [0,0,0] & [${\bf {\color{red} 2}}$,0] & [${\bf {\color{red} 2}}$,0] & [${\bf {\color{red} 2}}$,0] & [${\bf {\color{red} 2}}$,0] \\
	$(\Gamma_{II},\text{Z}_{II},\text{S}_{II})_u$ & 0 & {\bf 1} &  [{\bf 1},${\bf {\color{cyan} 2}}$,{\bf 1}] & ${\bf {\color{cyan} 2}}$ & 0 & [${\bf {\color{red} 2}}$,0,0]  & [${\bf1 }$,0,{\bf 1}] & [0,${\bf {\color{cyan} 2}}$,0] & [{\bf 1},{\bf 1}]  & [${\bf {\color{red} 2}}$,0] & [{\bf 1},${\bf {\color{cyan} 2}}$] & [${\bf {\color{red} 2}}$,0] \\
	$(\Gamma_{III},\text{Z}_{II},\text{S}_{II})_v$ & ${\bf {\color{cyan} 2}}$ & 0 & [0,0,0] & ${\bf {\color{cyan} 2}}$ & 0 & [${\bf {\color{red} 2}}$,0,0]  & [${\bf {\color{cyan} 2}}$,${\bf {\color{cyan} 2}}$,0] & [0,${\bf {\color{cyan} 2}}$,0] & [0,0] & [${\bf {\color{red} 2}}$,0] & [0,0] & [${\bf {\color{red} 2}}$,0] \\
 \hline
  \hline
\end{tabular*}
\end{table*} 
}

Moving on to the eight-band subspace, we list in Table~\ref{point_classes_eight} all the symmetry protected topological invariants for two-point and four-point symmetry-constrained loops for all the nodal structures of Fig.~\ref{fig_SG33_8B_lines} labelled from (a) to (v). The invariants of Table~\ref{point_classes_eight} written in black correspond to a nontrivial Berry phase factor, Eqs.~(\ref{Berry_phase_factor_2},\ref{Berry_phase_factor_4}), i.e.~their base loops encircle an odd number of NLs. The invariants written in color correspond to an even number of quantized Wilson phases which leads to a trivial Berry phase (consider for instance $\mathcal{N}=2$ and thus $\gamma = -i \log [(-1)^{\mathcal{N}}] \mod2\pi=0 $). We show that the invariants $\mathcal{N}^{y,x}_{\alpha}$ count the number of symmetry protected NLs within $m_x$- or $m_y$-planes encircled by the base loops. Finally, the red invariants mark the presence of line-node monopole pairs, which we discuss in detail in Section \ref{eight_top}. Similarly to Section \ref{four}, we here only count the symmetry independent NLs, keeping in mind the effective $D_{2h}$ symmetry of the band structure.

\subsection{Local topology of the line-nodal structures} 
The topological invariants for the cases in Fig.~\ref{fig_SG33_8B_lines}(a,b,c) are all $\mathcal{N}^{z,y,x}_{\alpha}\in\{0,1\}$, i.e.~they directly correspond to the Berry phase computed over the symmetry constrained loops. In particular, the case (a) has a single NL on the $m_y$-invariant plane $\sigma_3$ leading to the invariants $\mathcal{N}^y_{\Sigma,\Lambda,1,5}=1$. Furthermore, the invariants $\mathcal{N}^z_{\Lambda,1,3,5}=1$ are nontrivial since their corresponding symmetry constrained loops encircle the NL. 
The case (b) is directly analogous to (a) but now with a single NL on the $m_x$-invariant plane $\sigma_5$ leading to the invariants $\mathcal{N}^x_{\Delta,\Lambda,1,3}=1$. 
The case (c) exhibits two inequivalent NLs (one within $\sigma_3$ and one within $\sigma_5$) and combines the nontrivial invariants of the cases (a) and (b). The exception is for $\mathcal{N}^z_{\Lambda,1,3,5}=0$, where the corresponding loops encircle the two NLs crossing at a simple (twofold degenerate) point node on $\Lambda$, similarly to the four-band case Fig.~\ref{fig_SG33_4B_lines}(c) discussed previously. In all these cases the invariants determine the local poloidal charge of the NLs and there is no nontrivial monopole charge. The case (c) corresponds to the connected elementary nodal structure shown as an example in Fig.~\ref{fig_ns_charges}(a), which counted four poloidal charges without taking the symmetries into account. We have here identified the two poloidal invariants $\mathcal{N}^y_{\Sigma,\Lambda,1,5}=1$ and $\mathcal{N}^x_{\Delta,\Lambda,1,3}=1$, leading to four in total if we consider their images under $C_{2z}$.  

The cases (d,f,g,m,n,p,r) are all characterized by the topological invariants $\mathcal{N}^{x,y}_{\alpha}\in \{0,2,4\}$ (blue) and $\mathcal{N}^{z}_{\alpha}=0$ $\forall \alpha$. For instance, the case (d) has two inequivalent NLs within the $m_y$-invariant plane $\sigma_3$, leading to the invariants $\mathcal{N}^y_{\Sigma,1} = 2 $ giving the number of $m_y$-symmetry protected NLs crossing the line $\Sigma$ and the plane $\sigma_1\supset \Sigma$, and $\mathcal{N}^y_{\Lambda,5}=2$ giving the number of $m_y$-symmetry protected NLs crossing the line $\Lambda$ and the plane $\sigma_5 \supset \Lambda$. Thus, on top of the separate poloidal loops of each NL characterized by a nontrivial Berry phase, the loops that encircles two NLs all have a trivial Berry phase factor, i.e.~$e^{i\gamma} = e^{i(\pi\pm \pi)} = +1$, such that only the full Wilson loop spectrum characterizes the nontrivial local topology of the loop sub-bundles. In this case, $\mathcal{N}^y_{\Sigma,\Lambda,1,5} = 2$ indicates that the base loop encircles two NLs that each is characterized by a nontrivial poloidal Berry phase and such that the NLs do not cancel each other if they meet. We verify this interpretation with a detailed numerical computation of the Wilson loop in Section \ref{W_spectrum}.
The special case (n) even has four inequivalent NLs within the $m_y$-invariant plane $\sigma_3$ crossing the line $\Lambda$ and the plane $\sigma_5$ as captured by the invariants $\mathcal{N}^y_{\Lambda,5}=4$. Note that whenever the HSP class at Z is $Z_{II}$, the invariant $\mathcal{N}^{y}_{\text{A}}=2$ marks the double (fourfold degenerate) point node on $\text{A}$ where two NLs within $\sigma_3$ meet. We further note case (n) can actually be interpreted as the superposition of two disconnected nodal structures and we write (n)=(d)+(g). Here we assume that (d) has two small NLs on $\sigma_5$ as in the example Fig.~\ref{fig_transformed}(d'). While, there is again no nontrivial monopole charge for these cases, we further discuss in Section \ref{threads} the case (m) which exhibits a nontrivial thread charge. 

The cases (h,i,j,k,l) are characterized by two types of topological invariants $\mathcal{N}^{z,y,x}_{\alpha}\in\{0,1,3\}$ (black) and $\mathcal{N}^{x,y}_{\alpha}\in\{0,2,4\}$ (blue) $\forall \alpha$. Let us compare the cases (h) and (k). The two NLs within $\sigma_3$ are captured by the invariants $\mathcal{N}^y_{\Lambda,5}=2$ in both cases. The case (k), contrary to (h), has an unremovable NL within the $m_x$-invariant plane $\sigma_5$ and crossing the plane $\sigma_3$ leading to the invariants $\mathcal{N}^x_{\Lambda,1,3}=1$ which can be interpreted as nontrivial poloidal Berry phases. We note that these invariants reflect directly the fact that these cases are made of the combination of the HSP class $\Gamma_{II}$ with the HSP class Z$_{II}$. Indeed, seeing (h) as the combination of (a) and (g), we find $\mathcal{N}^z_{\Lambda,1,3,5}=1$ as for (a), seeing (k) as combination of (c) and (g), we find $\mathcal{N}^z_{\Lambda,1,3,5}=0$ as for (c), and similarly for the other cases of the class $(\Gamma_{III},\text{Z}_{II})$. Note that only in the cases (i), (j) and (l) are the line-nodal structures given as a mere superposition of disconnected nodal substructures, we write (i)=(a)+(g), (j)=(b)+(g) and (l)=(c)+(g). Also here there is no nontrivial monopole charge. 

The cases (e,o) are characterized by a set of valence states at $\Gamma$ that are all even (or odd) under $C_{2z}$, i.e.~either $s_{z}(\Gamma_j)=+1$ or $-1$ $\forall j= 1,2,3,4$. This directly leads to $\mathcal{N}^{z}_{\alpha}=2$ for all the loops $\alpha$ that has $\Gamma$ as a vertex. The cases (q,s) are similar but now with the vertex S playing the role of $\Gamma$. Note that all the cases (e,o,q,s) have $\mathcal{N}^{z}_{1}=2$ since the four-point loop $l_1$ has $\Gamma$ and S as vertices. We start with the case (o) that has two inequivalent NLs belonging to the $m_y$-invariant plane $\sigma_3$ characterized by $\mathcal{N}^{y}_{\Lambda,2,5}=2$. The strict evenness (or oddness) of the valence states at $\Gamma$ with respect to $C_{2z}$ leads to the charges $\mathcal{N}^{z}_{\Lambda,3,5}=2$. The case (s) is analogous to (o) but now with the NL on the $m_x$-variant plane $\sigma_6$ lying on the BZ boundary and characterized by $\mathcal{N}^{x}_{\text{Q},2,4}=2$. The pure evenness (or oddness) of the valence states at $\text{S}$ with respect to $C_{2z}$ leads to the nontrivial invariants $\mathcal{N}^{z}_{\text{Q},4,6}=2$. As noted above, in both cases $\mathcal{N}^z_{1}=2$. We show below that if we combine the HSP class of (o) with (s), i.e.~$(\Gamma_{III},\text{Z}_{II},\text{S}_{II},\text{R}_{II})$, then $\mathcal{N}^z_{1}=0$. Neither (o) nor (s) has a nontrivial monopole charge.

The cases (e) and (q) both exhibit a NL that is disconnected from its symmetry partner under $m_z$ (from the anti-unitary symmetry $C_{2z} \mathcal{T} = m_z \mathcal{K}$) and is not centered at a HSP. In particular, the NL can either be on the $m_y$- or $m_x$-invariant plane depending on the specific energy ordering of valence or conduction IRREPs. In (e), it crosses $\Lambda$ and is characterized by the invariants $\mathcal{N}^z_{\Lambda,3,5}=2$ (red). The case (q) is directly analogous to (e) with the only difference that the NL pair now belongs to the BZ boundary with the invariants $\mathcal{N}^z_{\text{Q},4,6}=2$. As we find in Section \ref{monopoles} these form monopole anti-monopole pairs explaining the red coloring. 

We already showed in Section \ref{additivity} that whenever a nontrivial HSP class of $\mathcal{B}_{\Gamma}$ is combined with a nontrivial HSP class of $\mathcal{B}_{\text{R}}$, the resulting line-nodal structure is a mere superposition of the respective line-nodal structures. This is true for the cases (t,u,v) with (t)=(e)+(q), (u)=(j)+(q) and (v)=(m)+(q). As a consequence, their topological invariants are directly given as a sum of the invariants of the separate cases composing them. The only exception is $\mathcal{N}^z_1$. Here $\mathcal{N}^z_1[(u)] = 1$, while $\mathcal{N}^z_1[(j)]+\mathcal{N}^z_1[(q)] = 2+1=3$. Still, $\mathcal{N}^z_1[(u)] = 1$ matches with the fact that a single NL is encircled by the base loop $l^z_1$ in the case (u).  

To summarize, all symmetry protected topological invariants have a very clear meaning in terms of local charges of the symmetry protected nodal structures but for the interpretation of the invariant $\mathcal{N}^z_1\in \{0,2\}$, which is still unclear to us. Indeed, it is not related to the presence of pairs of NLs encircled by the base loop $l^z_1$. Take for instance (d,e,f): two NLs are encircled by $l^z_1$ in (d,f) and we find $\mathcal{N}^z_1=0$, while no NL is encircled by $l^z_1$ in (e) and we still find $\mathcal{N}^z_1=2$. Similar remarks apply for $\mathcal{N}^z_1$ in the cases (m,n,o,p) and (q,r,s).

\subsection{Accidental point nodes}
Previously we noted the presence of point nodes in the cases Fig.~\ref{fig_SG33_8B_lines}(e,o,v). For the same reasons as discussed in Section \ref{accidental_points} these must be accidental and are stable with a Chern number $\vert C_1\vert =1$. First we note that they are all within the $k_z=0$ plane. Because of the $C_{2v}$ point symmetries there must be four points related by symmetry with alternating signs as imposed by the symmetry constraints (i), (ii) and (iii) discussed in Section \ref{accidental_points}. Therefore, we can bring opposite point nodes onto each other and annihilate them through a $C_{2v}$-symmetry preserving transformation without changing the set of valence and conduction IRREPs. This makes these point nodes accidental. It is worth remarking that two point nodes that are images under the anti-unitary symmetry $C_{2z}*\mathcal{T} = m_z \mathcal{K}$ must have the same charge. Therefore, the point nodes are locked on the $k_z=0$ plane since a shift in the $k_z$-direction would require a doubling of the charges of each point node which is forbidden. 

\subsection{Wilson loop spectrum}\label{W_spectrum}

To further illustrate the conclusions with regards to meaning of the topological invariants above, we give here a detailed comparison of numerical computation of the Wilson loops with the algebraic results of Section \ref{top_inv} for the case Fig.~\ref{fig_SG33_8B_lines}(d) showing a perfect agreement. 
\begin{figure}[htb]
\centering
\begin{tabular}{c}
	\includegraphics[width=0.95\linewidth]{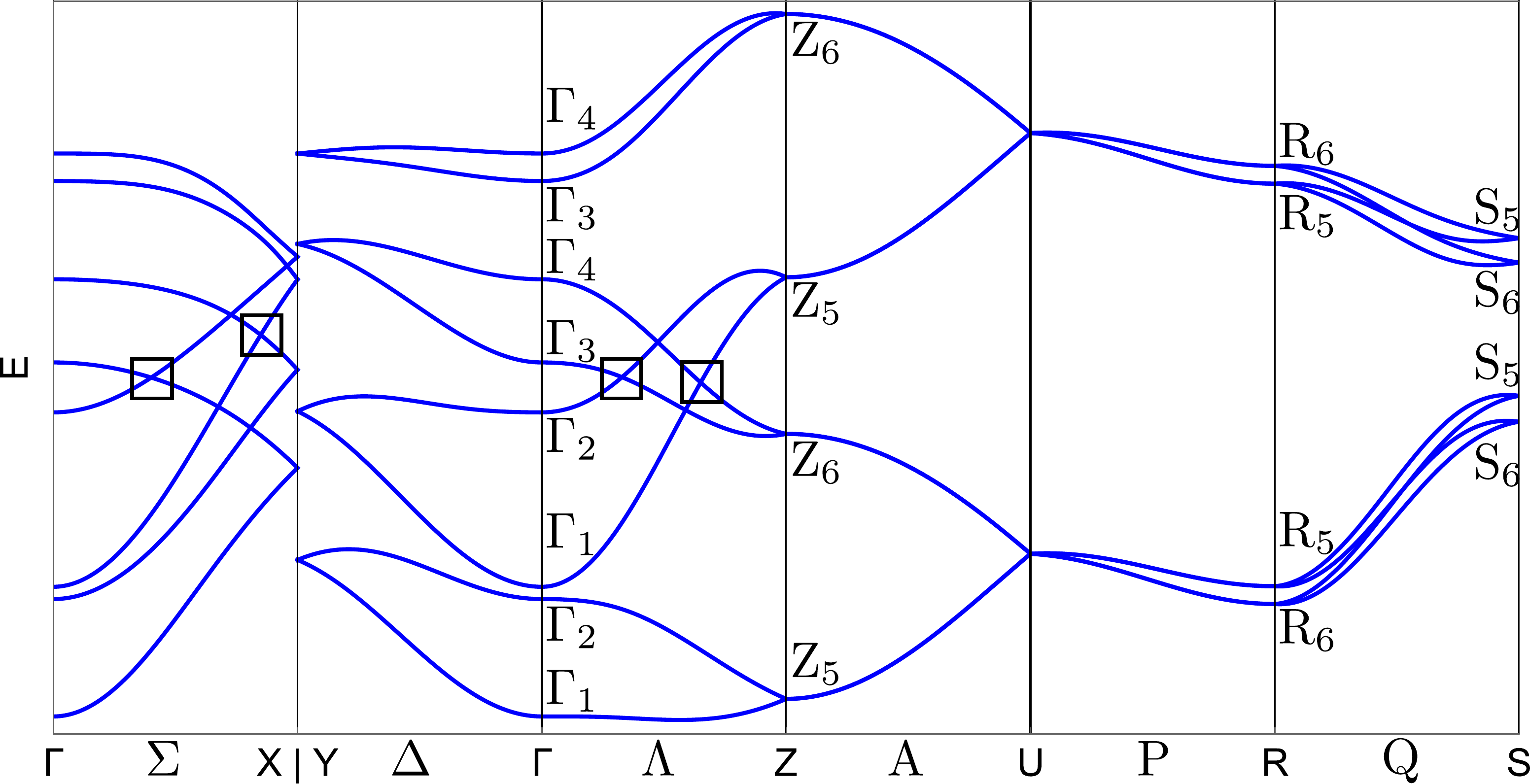} 

\end{tabular} 
\caption{\label{GIIIZI_d_bands} Example of band structure for the case Fig.~\ref{fig_SG33_8B_lines}(d). 
} 
\end{figure}
An example of band structure for the case Fig.~\ref{fig_SG33_8B_lines}(d) is shown in Fig.~\ref{GIIIZI_d_bands}. It has two independent NLs at half-filling within the plane $\sigma_3$ that cross the lines $\Sigma$ and $\Lambda$ at two distinct points (squares in Fig.~\ref{GIIIZI_d_bands}). We give a representation of these in Fig.~\ref{fig_Wilson_1122}(a). We here focus on the Wilson loop $\mathcal{W}[l^y_{\Sigma}]$. The detailed band structure along the $\Sigma$-line is shown in Fig.~\ref{fig_Wilson_1122}(b), with the momenta for the two symmetry-protected band crossings marked in green ($k_1$) and magenta ($k_2$). Also the IRREPs of the valence bands on the $\Sigma$-line are given, which is either even ($s_{n,y}(\Sigma_a)=+1$) or odd ($s_{n,y}(\Sigma_b)=-1$) under the glide symmetry $\{m_y\vert \boldsymbol{\tau}_y\}$.
\begin{figure}[htb]
\centering
\begin{tabular}{c}
	\includegraphics[width=0.6\linewidth]{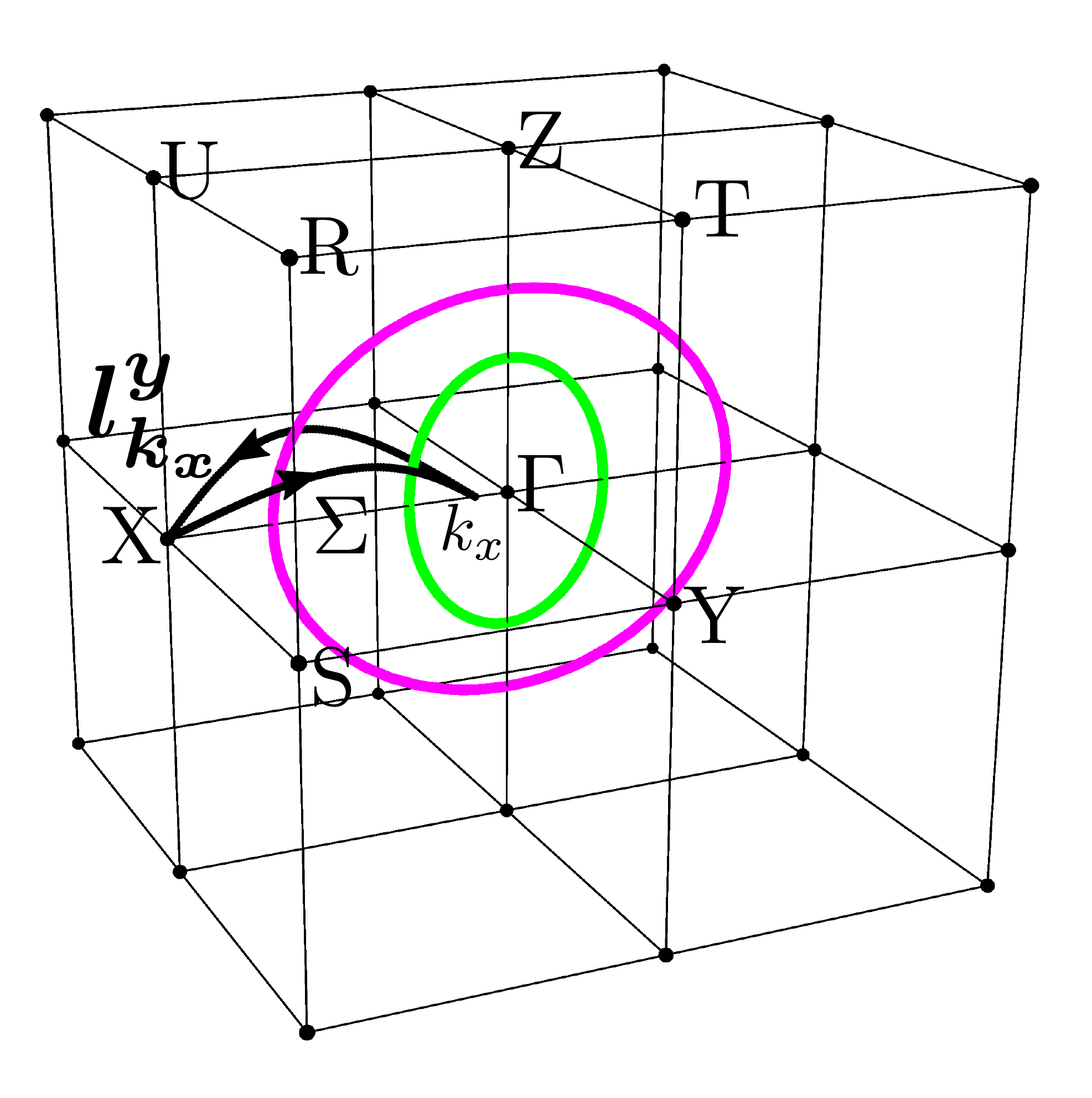}  \\
	(a) \\
	\includegraphics[width=0.64\linewidth]{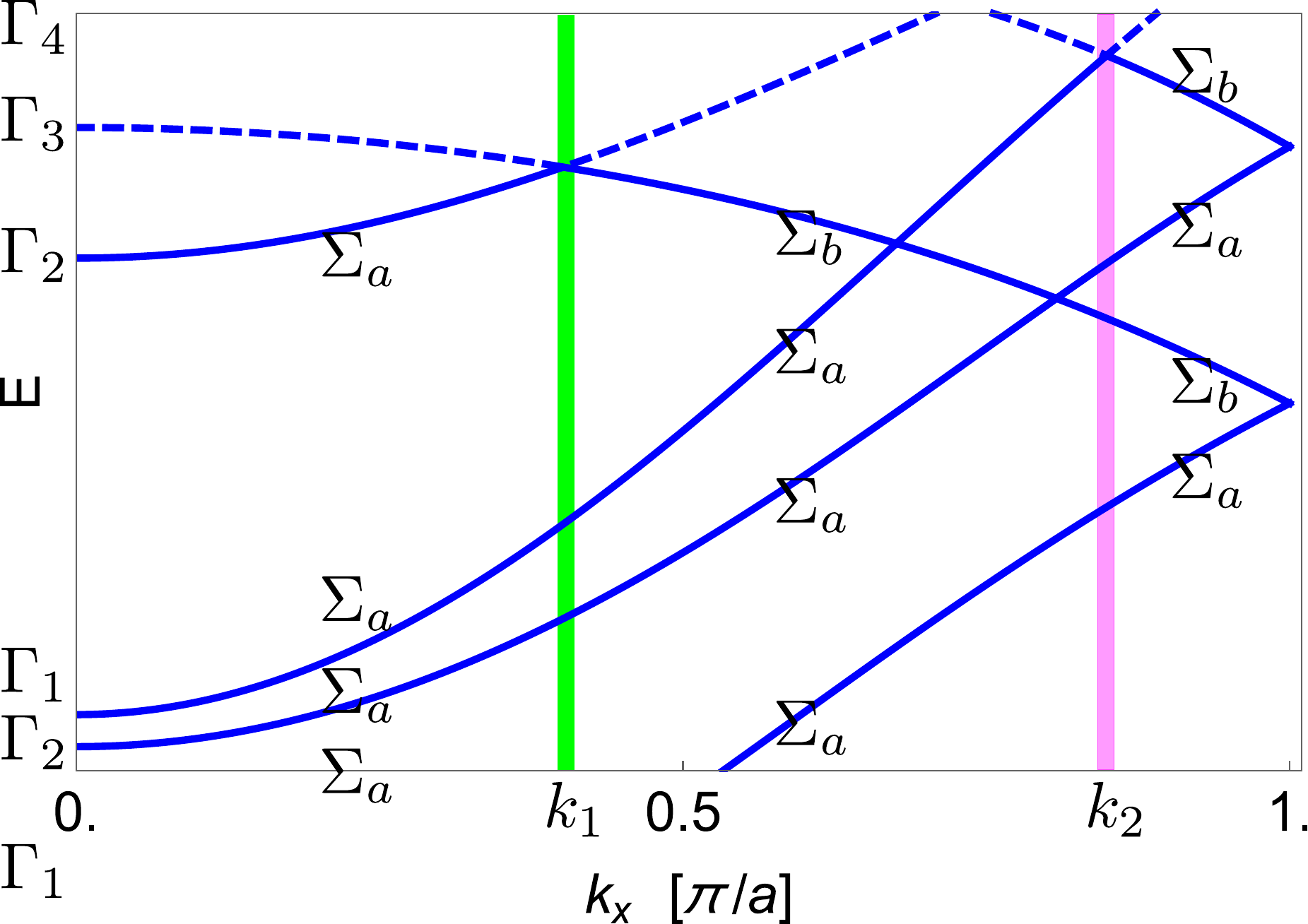} \\
	(b) \\
	\includegraphics[width=0.7\linewidth]{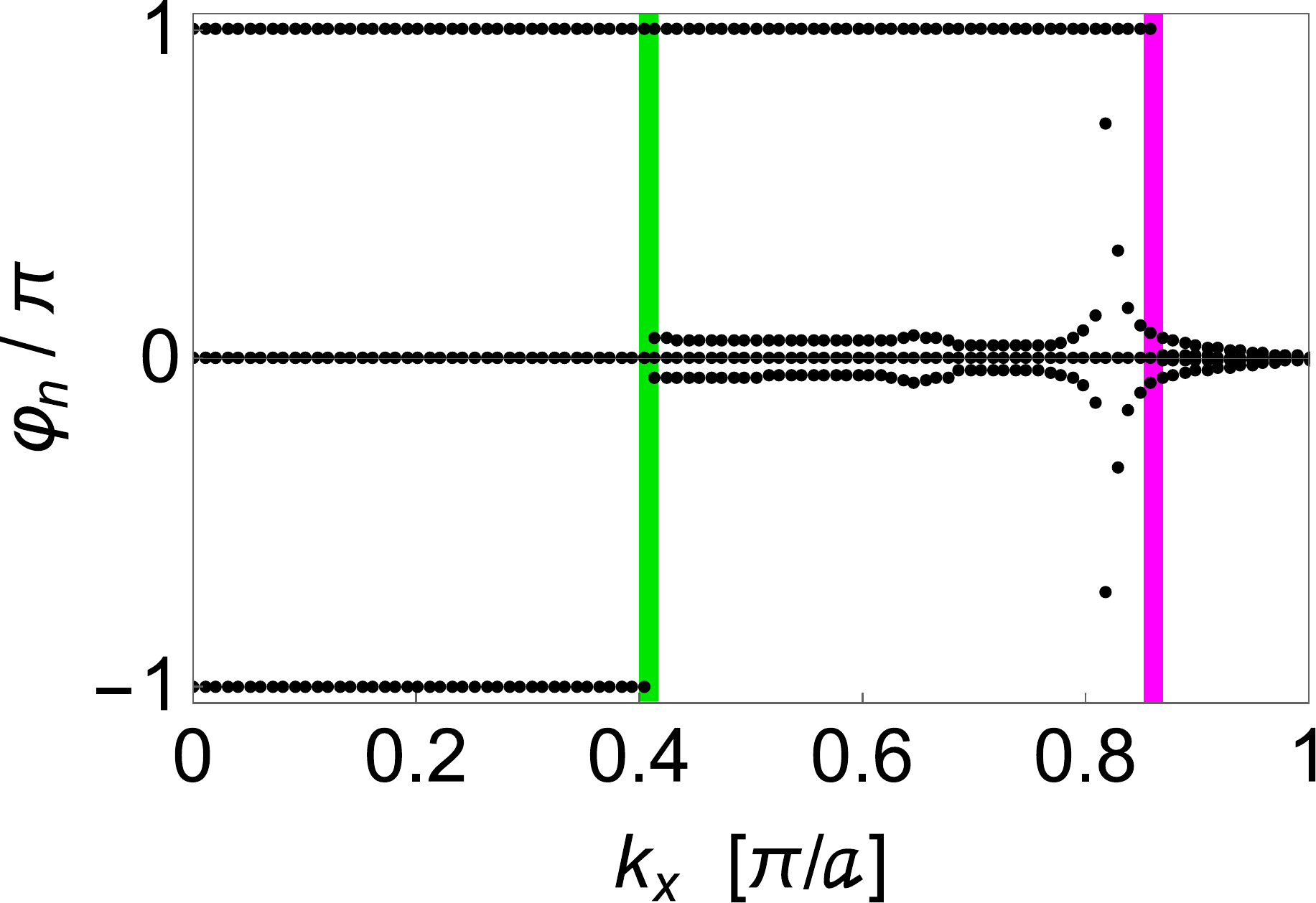} \\
	(c) 
\end{tabular} 
\caption{\label{fig_Wilson_1122} (a) NLs on the $\sigma_3$-plane (green and magneta) for the band structure Fig.~\ref{GIIIZI_d_bands} and the closed two-point symmetry-constrained loop, $l^y_{k_x}$, (black) over which the Wilson loops are evaluated, with one vertex at X and the other vertex on $(0,0,k_x) \in \Sigma$, i.e.~$k_x \in [0,\pi/a]$. (b) Details of the band structure along $\Sigma$ with the marked band crossings at half-filling at $k_1$ (green) and $k_2$ (magneta) and conduction bands with dashed lines. (c) Wilson loop spectrum for the two-point loop $l^y_{k_x}$ as a function of $k_x \in [0,\pi/a]$.
} 
\end{figure}

For a closed two-point loop $l^y_{V_1 V_2}$ foldable under $m_y$ with $V_1$ and $V_2$ taken on $\sigma_3$ (since $m_y \in \overline{G}^{\sigma_3}$, the little co-group of the plane), the Wilson loop spectrum indicates whether the loop encircles symmetry protected NLs or not. Fixing $V_2$ at X and $V_1$ at a point $k_x$ on the line $\Sigma$, we consider the $k_x$-dependent loop $l^y_{k_x}$ shown in black in Fig.~\ref{fig_Wilson_1122}(a). The Wilson loop spectrum computed numerically as a function of $k_x$ is shown in Fig.~\ref{fig_Wilson_1122}(c). Three distinct segments of $k_x\in \Sigma$ are found that are separated by the point nodes at $k_1$ and $k_2$: (i) for $k_x \in [X,k_2]$ we find $\mathrm{eig}\{\mathcal{W}[l^y_{k_x}]\} = \{e^{i\varphi_1},e^{-i\varphi_1},e^{i\varphi_2},e^{-i\varphi_2}\}$, i.e.~two pairs of complex conjugate Wilson loop eigenvalues, (ii) for $k_x\in[k_2,k_1]$ we have $\mathrm{eig}\{\mathcal{W}[l^y_{k_x}]\} = \{e^{i\varphi_1},e^{-i\varphi_1},+1,-1\}$ hence one zero Wilson loop phase and a Wilson loop phase of $\pi$, and (iii) for $k_x\in[k_1,0]$ we have $\mathrm{eig}\{\mathcal{W}[l^y_{k_x}]\} =\{+1,+1,-1,-1\}$ hence two zero and two $\pi$ Wilson loop phases. 

Alternatively, the Wilson loop spectrum can be obtained algebraically through the mapping $\mathcal{M}_2 \circ \Phi_2$ presented in Section \ref{line_inv}. For every $k_x$, taking the bare IRREPs of Fig.~\ref{fig_Wilson_1122}(a) as input for $\mathcal{M}_2$, we find the number of symmetry protected $-1$ Wilson loop eigenvalues $\mathcal{N}^z_{k_x}$, noting that in this case, the minority eigenvalue for every $k_x$ is $\xi=-1$. Then, taking all non-symmetry protected Wilson phases to zero, i.e.~$\varphi \rightarrow 0$ under the mapping $\Phi_4$ of Section \ref{plane_inv}, we find for the three segments introduced above: (i) $\mathcal{N}^y_{k_x>k_2 }=0$ reflecting the fact that no NL is encircled by the two-point loop; (ii) $\mathcal{N}^y_{k_1<k_x<k_2}=1$ capturing the fact that one NL is encircled, and (iii) $\mathcal{N}^y_{k_x<k_1}=2$ which marks the presence of two independent NLs encircled by the two-point loop. This matches exactly with the numerical result of Fig.~\ref{fig_Wilson_1122}(c). In particular, the nontrivial loop charge $\mathcal{N}^y_{k_x<k_1}=2$ marks the topological stability of the two independent NLs encircled by the base loop, i.e. they do not gap out if they touch (assuming the valence IRREP sets at HSPs are kept unchanged). This can be easily verified from the band structure Fig.~\ref{GIIIZI_d_bands} along the HSL $\Lambda$.

\subsection{Crystalline line-nodal monopoles}\label{monopoles}
In Section \ref{top_meta} we defined the monopole charge of a NL from the second homotopy group $\pi_2(\mathcal{C}_v)$ with the base space chosen as the sphere surrounding the whole NL, $\mathbb{S}^2_m$ in Fig.~\ref{torus_plot}(b), and with the classifying space of the valence sub-bundle taking all the symmetries into account, i.e.~$\mathcal{C}^{\mathcal{G}_{\mathrm{AI}}}_v \subsetneq \mathrm{Gr}_{N_v}(\mathbb{C}^N)$. We present here the heuristic computation of the monopole charge of a NL through the flow of the Wilson loop phases as we sweep a closed two-point symmetry-constrained loop over a symmetry-constrained surface surrounding the whole NL. This is in analogy with the definition of the Chern number from the flow of Berry phase except that the base space here needs to satisfy some symmetry constraints. Starting with a numerical computation of the Wilson loop, we find that the line-nodal monopole charge can also be determined algebraically in terms of the invariants derived in Section \ref{top_inv}. 

\begin{figure}[htb]
\centering
\begin{tabular}{c}
	\includegraphics[width=0.95\linewidth]{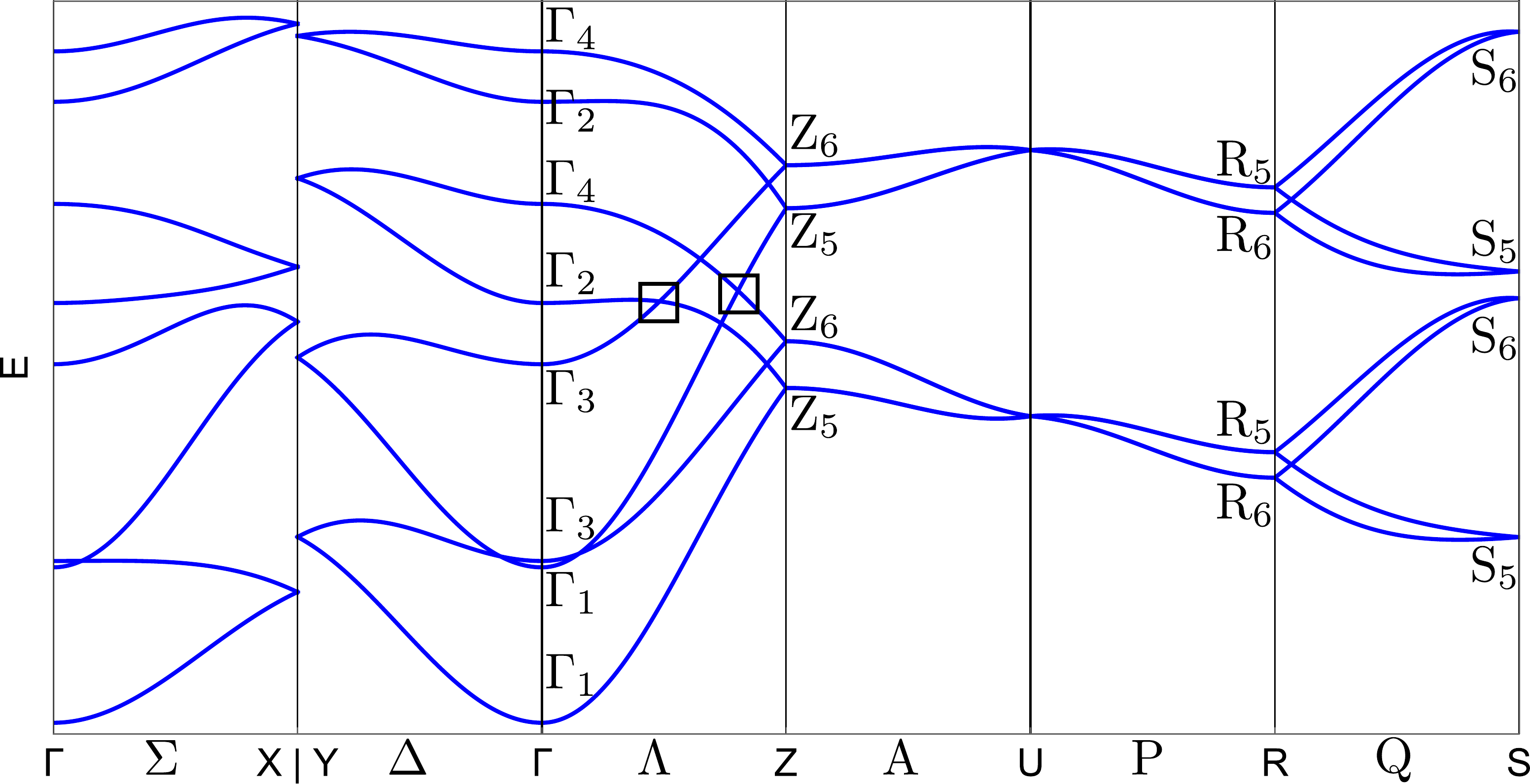}  
\end{tabular}  
\caption{\label{GIIIZI_e_bands} Example of band structure for the case Fig.~\ref{fig_SG33_8B_lines}(e).
} 
\end{figure}
\begin{figure}[htb]
\centering
\begin{tabular}{c}
	\includegraphics[width=0.6\linewidth]{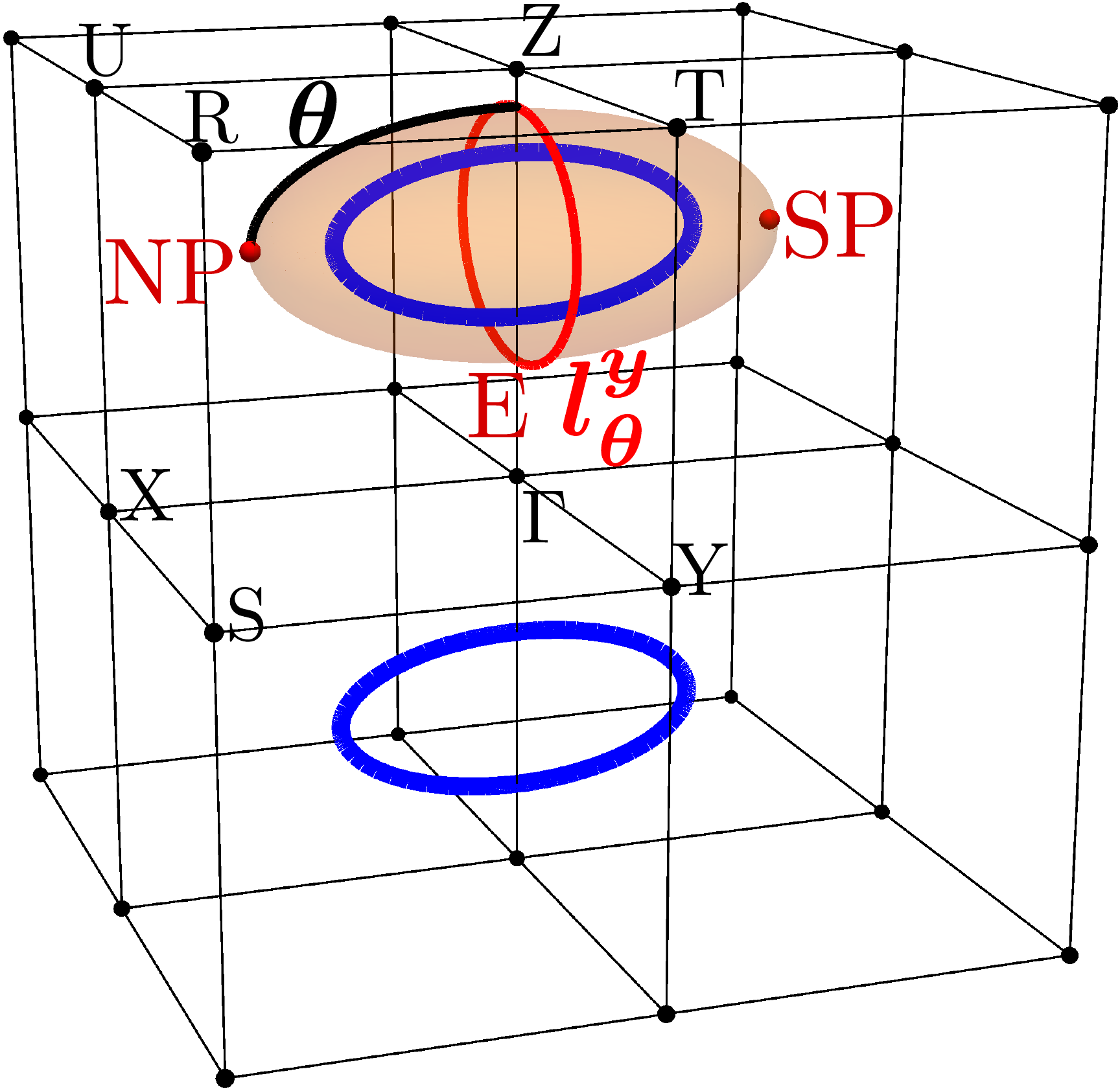}  \\
	(a)  
\end{tabular}
\begin{tabular}{cc}
	\includegraphics[width=0.5\linewidth]{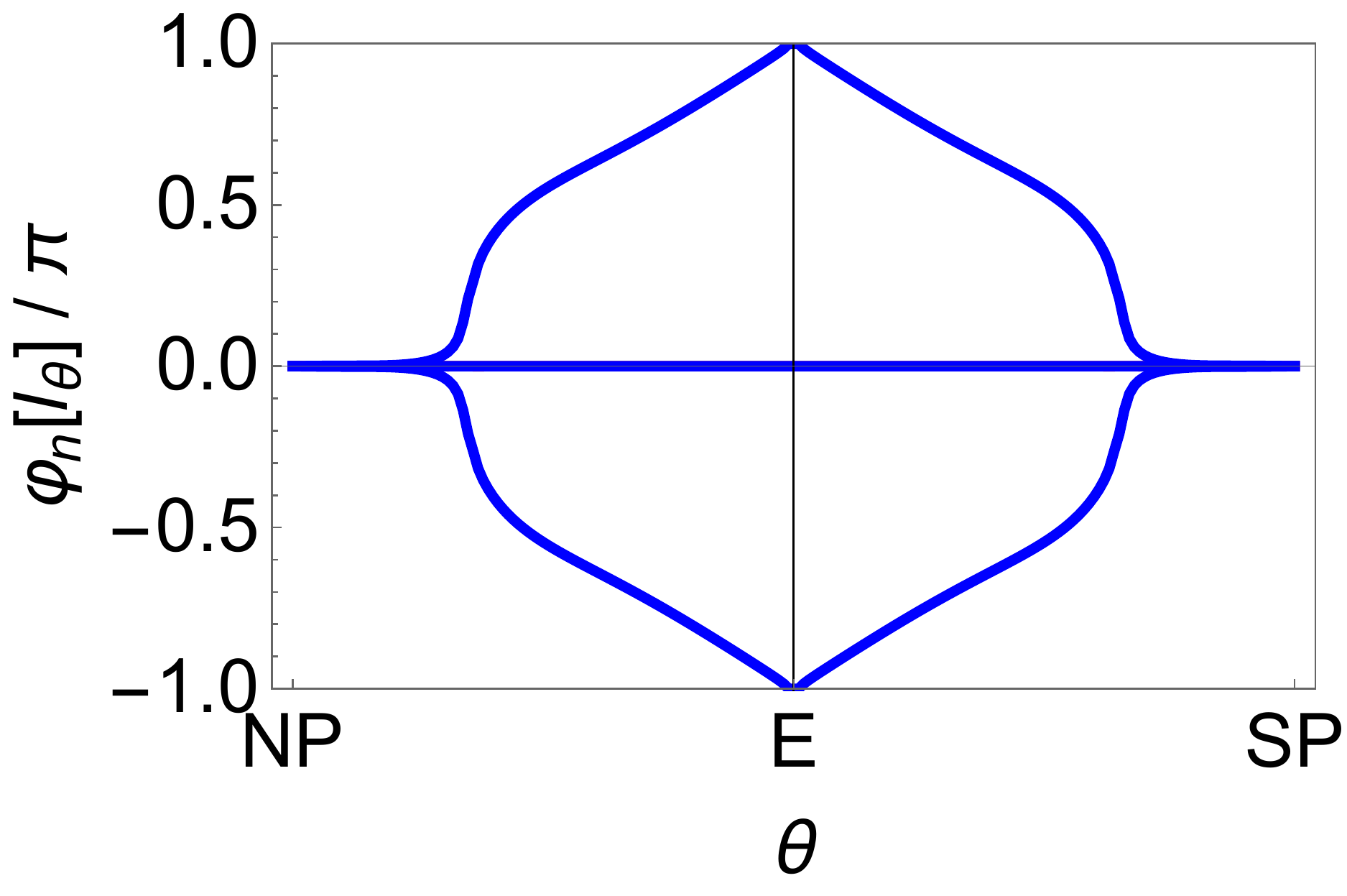}  &
	\includegraphics[width=0.5\linewidth]{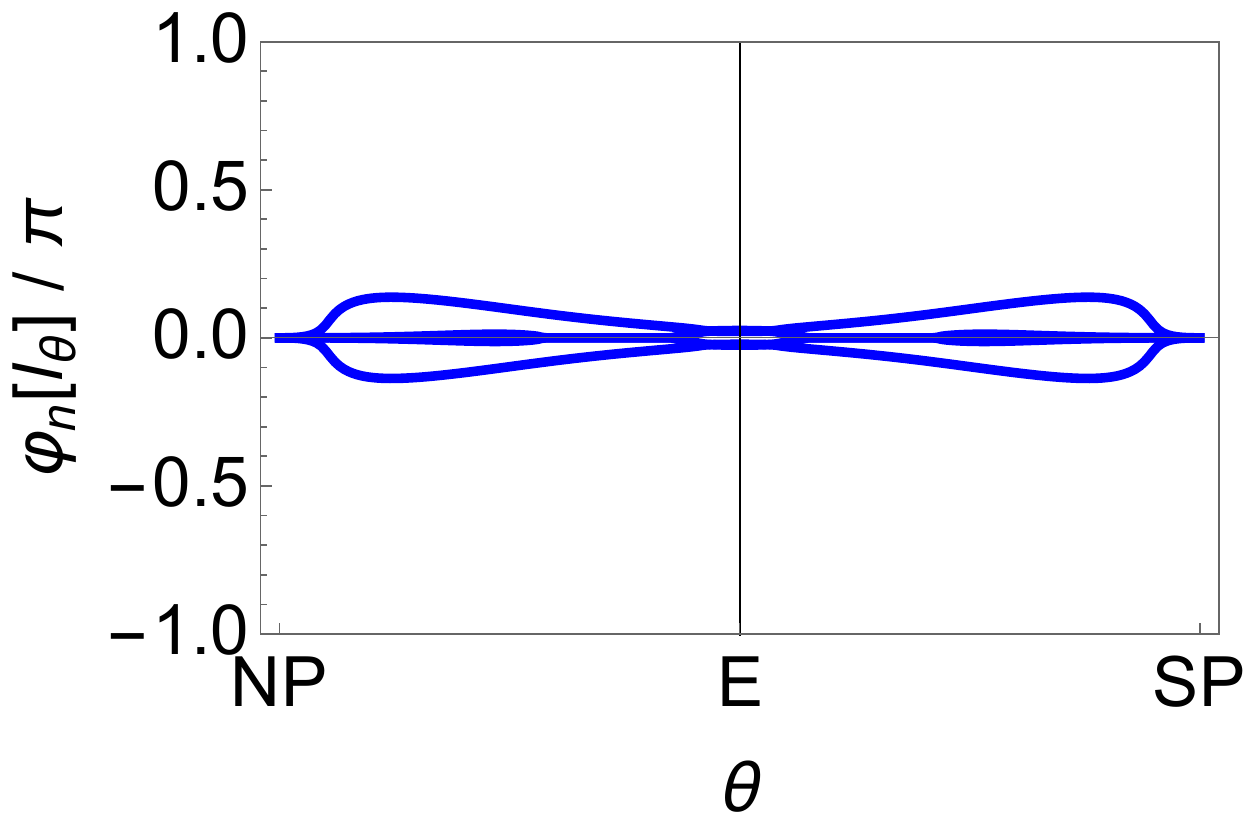}  \\
	(b) &	(c)
\end{tabular} 
\caption{\label{fig_Wilson_monop} (a) The pair of $m_z$-symmetric NLs (blue) in the case Fig.~\ref{fig_SG33_8B_lines}(e). The surrounding surface (orange) is parametrized with two-point loops $l^y_{\theta}$ for $\theta\in[0,\pi]$ and symmetric under $m_y$.  The loop at the equator ($\theta=\pi/2$) is also chosen symmetric under $C_{2z}$. (b) Spectral flow of the Wilson loops $\mathcal{W}[l^y_{\theta}]$ as we sweep the base loop over the sphere from the NP ($\theta=0$) to SP ($\theta=\pi$). The flow connects two topologically distinct symmetry protected spectra leading to a nontrivial monopole charge. (c) Spectral flow of the Wilson loop over a surface surrounding the two NLs in case Fig.~\ref{fig_SG33_8B_lines}(d). The flow is trivial leading to a trivial monopole charge. 
} 
\end{figure}

We focus on the case Fig.~\ref{fig_SG33_8B_lines}(e) that exhibits two NLs at half-filling within the $m_y$-invariant plane $\sigma_3$ and intersecting the line $\Lambda$. An example of the band structure is shown in Fig.~\ref{GIIIZI_e_bands}, where the two band crossings (along $\Lambda$) belonging to the same NL are marked by squares: one crossing between the bands $\{\Gamma_1,\Gamma_4\}$ and the other between $\{\Gamma_2,\Gamma_3\}$. Because of the effective $m_z$ symmetry of the band structure, there is a copy of the NL at $-k_z$, such that they form a pair of NLs. This is illustrated schematically in Fig.~\ref{fig_Wilson_monop}(a) where one NL is also surrounded by a closed surface that serves as the base space for the spectral flow of the Wilson loop. It is here important that this surrounding surface is chosen symmetric under $m_y$ and $C_{2z}$, and hence under the whole $C_{2v}$ point group.
Let us parametrize the sphere by $m_y$-symmetric loops $l^y_{\theta}$ (red) with $\theta$ being the polar angle with respect to a ``vertical'' axis chosen parallel to the $\overline{\Gamma \text{X}}$ line. At each $\theta$ we compute the Wilson loop spectrum $\mathrm{eig}\{\mathcal{W}[l^y_{\theta}]\}$ which must be symmetric under complex conjugation due to the $m_y$ symmetry of the loop $l^y_{\theta}$ (Section \ref{line_inv}). 
We show in Fig.~\ref{fig_Wilson_monop}(b) the Wilson loop flow as we sweep the loop from the north pole (NP, $\theta=0$) to the south pole (SP, $\theta=\pi$). At the NP and SP the Wilson loop phases must all be zero since the base loop is just a point. At the equator (E, $\theta=\pi/2$) the Wilson loop spectrum is $\{+1,+1,-1,-1\}$. Importantly, the equator $l^y_{\pi/2}$ is also symmetric under $C_{2z}$ by the $C_{2v}$ symmetry of the surrounding surface.

We have derived in Table~\ref{point_classes_eight} that the NL in the case Fig.~\ref{fig_SG33_8B_lines}(e) is characterized by the two-point loop topological invariant $\mathcal{N}^z_{\Lambda}=2$, counting the number of symmetry protected $-1$ Wilson loop eigenvalues. Notably,  this is in complete agreement with the numerical finding at the equator in Fig.~\ref{fig_Wilson_monop}(b). Therefore, the Wilson spectrum at the equator is symmetry protected. By $m_x$ symmetry, the Wilson loop flow from E to SP must also be the mirror symmetric of the flow from NP to E. We thus conclude that there is a nontrivial spectral flow of the Wilson loop protected by symmetry: as we cover the surface surrounding a line-nodal monopole it continuously connects the two topologically inequivalent sectors $\{+1,+1,+1,+1\}$ at NP and SP, and $\{+1,+1,-1,-1\}$ at E. We interpret this as the nontrivial monopole charge. Importantly, we note that this result is stable under breaking TRS, i.e.~the effective $m_z$ symmetry present with TRS is not needed. As a comparison, we computed the spectral flow of the Wilson loop over a surface surrounding the two NLs in the case Fig.~\ref{fig_SG33_8B_lines}(d) discussed in the previous section in Fig~\ref{fig_Wilson_monop}(c). Here we instead find a trivial flow over the surrounding surface indicating a trivial monopole charge. 

Contrary to all other NLs, a pair of line-nodal monopoles is made of two disconnected NLs that can only be removed if one collapses one onto the other. This is in analogy with the canceling of point node pairs as required by the Nielsen-Ninomiya theorem. As we also showed in Section \ref{top_meta}, it can be easily understood as a consequence of the geometry of the $\mathbb{T}^3$-BZ. This can be easily verified from the band structure along $\Lambda$, see Fig.~\ref{GIIIZI_e_bands}: no deformation of the bands will remove the NL as long as we keep the same set of valence IRREPs at $\Gamma$ and at Z. For instance, let us permute in energy the valence bands $\{\Gamma_1,Z_5\}$ with $\{\Gamma_3,Z_6\}$, in which case the band crossings are now between $\{\Gamma_1,\Gamma_2\}$ and between $\{\Gamma_3,\Gamma_4\}$, which leads to the crossing points now being part of a NL within the $m_x$-invariant plane $\sigma_5$. The fact that the line-nodal monopoles are not attached to a specific plane is captured by the topological invariants $\mathcal{N}^{x,y}_{\Lambda}=\mathcal{N}^{x}_3=\mathcal{N}^{y}_5=0$, which count the number of NLs constrained to lie on $m_x$($m_y$)-invariant planes.

In order to remove the pair of line-nodal monopoles we have to collapse the two line-nodal monopoles onto each-other. This is done through a topological Lifshitz transition that closes the band gap at $\Gamma$. Interestingly, while the monopole charges can be trivialized through a gap closing at Z (exchanging the valence IRREPs $\{Z_5,Z_6\}$ to $\{Z_i,Z_i\}_{i=5,6}$) the NLs do not go away and remain in a single connected elementary nodal structure, as can be seen in Fig.~\ref{fig_SG33_8B_lines}(o) where the two NLs are connected through double point nodes on the HSL A. This is captured by the invariant $\mathcal{N}^z_{\Lambda}=2$ in the cases (e) and (o). On the contrary, in the case Fig.~\ref{fig_SG33_8B_lines}(d) the two NLs are centered at $\Gamma$ such that the smaller NL can shrink on the $\Gamma$ point and disappear without touching the second NL. 

From this discussion we conclude that whenever there is pair of disconnected NLs that are not centered on an active HSP and with an invariant $\mathcal{N}^z_{\alpha}=2$, they must carry nontrivial monopole charges. This is true for the cases Fig.~\ref{fig_SG33_8B_lines}(e) and (q), where the line-nodal monopoles are lying on the plane $\sigma_4$ of the BZ boundary, and also the cases (t,u,v) since they contain the nodal structure of (q) by construction. In particular, the case (t) has two pairs of line-nodal monopoles as it is composed of both (e) and (q). All the other line-nodal structures of Fig.~\ref{fig_SG33_8B_lines} (with the exception of (m), see below) are centered on a HSP and thus have a trivial monopole charge. This can be directly generalized to all elementary nodal structures with a surrounding sphere that does not enclose active HSPs. Since then it is not possible to topologically change the valence IRREPs at HSPs without destroying the surrounding surface.
We finally remark that contrary to Ref.~\cite{Fu_line_node_monopole, Agterberg_BdGsurface, Bzdusek_mulitnodes, Moroz_nodalsurface}, the line-nodal monopoles discussed here are protected by the unitary lattice symmetries only. We particularly stress that the above construction of the monopole charge is based on the fact that a $C_{2v}$-symmetric surrounding surface can be defined. We therefore call them \textit{crystalline} line-nodal monopoles.

\subsection{Crystalline line-nodal threads}\label{threads}
\begin{figure}[htb]
\centering
\begin{tabular}{cc}
	\includegraphics[width=0.5\linewidth]{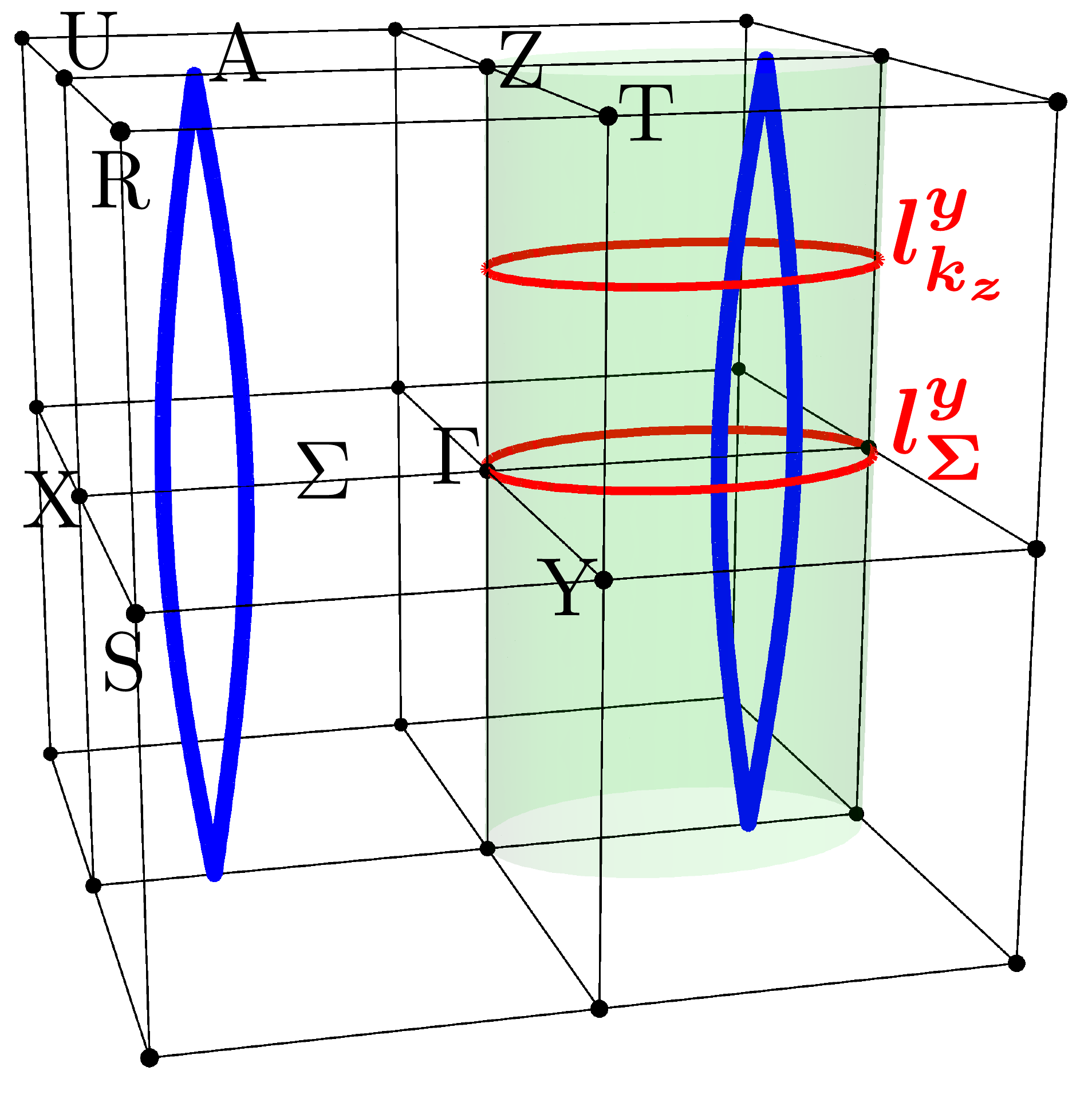}  &
	\includegraphics[width=0.5\linewidth]{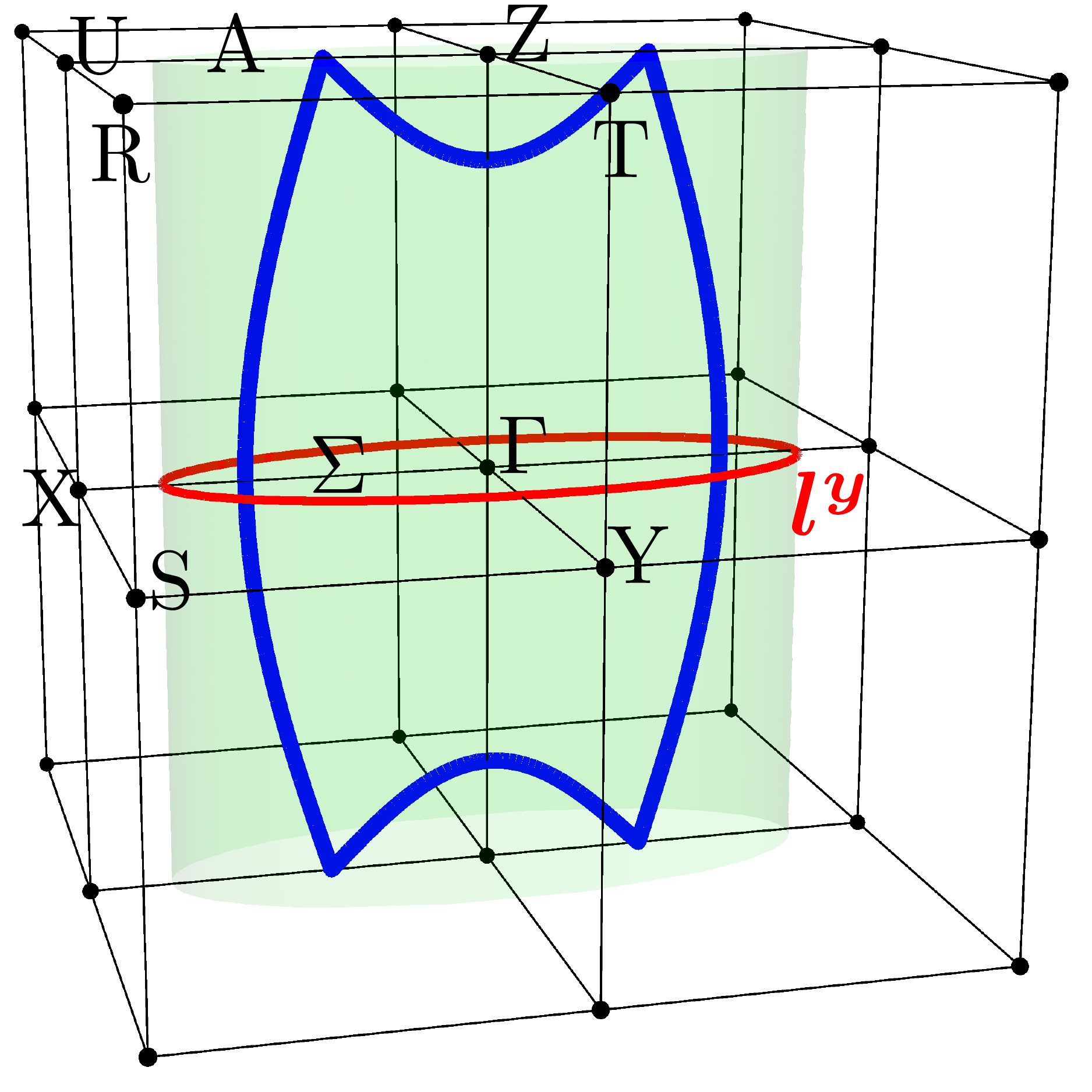}  \\
	(a) &	(b)
\end{tabular} 
\caption{\label{fig_threads} (a) Line-nodal threads (blue) with nontrivial thread charge $\mathcal{N}^y_{\Sigma} = \mathcal{N}^y_{k_z} =2 $~$\forall k_z$ (red) defined on surrounding cylinder (green) for nodal structure in Fig.~\ref{fig_SG33_8B_lines}(m). (b) Line-nodal thread with trivial thread charge $\mathcal{N}^y =0$ for nodal structure in Fig.~\ref{fig_SG33_8B_lines}(h). 
} 
\end{figure}

In Section \ref{top_meta} we defined a \textit{NL thread} as a NL that threads the BZ with a \textit{thread charge} that is then simply the poloidal charge of the NL. This can be generalized to any elementary nodal structure that threads the BZ. We here comparatively discuss the case Fig.~\ref{fig_SG33_8B_lines}(m) for which nontrivial thread charges can be defined and the case Fig.~\ref{fig_SG33_8B_lines}(h) that has a trivial thread charge. Fig.~\ref{fig_SG33_8bands} shows an example of band structure corresponding to the case (m). We pointed out (in Section \ref{eight}) the presence at half-filling of a pair of two NLs within the $m_y$-invariant plane $\sigma_3$ connected at double point nodes on the HSL A. We show a representation of the nodal structure in Fig.~\ref{fig_threads}(a), where one elementary line-nodal component threading the BZ is surrounded by a cylinder (green). A two-point loop section of the cylinder can be chosen as $l^y_{\Sigma}$ (red), or equivalently any other $m_y$-symmetric two-point loop section $l^y_{k_z}$ with $k_z\in[-\pi/c,\pi/c]$. The thread charge of the surrounded nodal structure is then nontrivial with $\mathcal{N}^y_{\Sigma} = \mathcal{N}^y_{k_z} =2 $ $\forall k_z$. When discussing global topology in Section \ref{top_meta} we concluded that a trivializing threading partner must be present, which directly explains the global nodal structure of the case (m) composed of a pair of nontrivial nodal threads. We point out that the surrounding cylinder of one nodal thread does not surround any HSP which makes the nodal thread nontrivial. Indeed, any topological change of valence IRREPs at HSPs that would remove the line-nodal thread would also remove the surrounding surface. Note that the monopole charge defined over the surrounding cylinder is trivial: precisely because $\mathcal{N}^y_{k_z} =2$ $\forall k_z$ there is no nontrivial flow of the Wilson loop over the cylinder. 

Let us also consider the case Fig.~\ref{fig_SG33_8B_lines}(h), with the threading nodal structure reproduced in Fig.~\ref{fig_threads}(b). Using the $m_x$ (or $C_{2z}$) symmetry of the band structure and applying the algebraic algorithm of Section \ref{line_inv}, we readily find the charge $\mathcal{N}^y = 0$ for any $m_y$-symmetric two-point base loop $l^y$ (red) that is a section of the surrounding cylinder (green). We thus conclude that the case (h) has a trivial thread charge. This can be interpreted as the consequence of the fact that the surrounding cylinder here encloses several active HSPs. Therefore, it is now possible to topologically change the valence IRREPs at those HSPs and make the nodal structure to disappear without removing the surrounding surface. We conclude this section by listing that (i,j,k,p,u) are all the other trivial nodal threads in Fig.~\ref{fig_SG33_8B_lines} while (v) has nontrivial thread charges similarly to (m).

\subsection{Local topological Lifshitz transitions}\label{Lifshitz}
\begin{figure*}[t]
\centering
\begin{tabular}{c|c|c|c} 
\hline
\hline
	$(\Gamma_{III},\text{Z}_{I})_{d'}$ & $(\Gamma_{II},\text{Z}_I)_{c'}$ & $(\Gamma_{III},\text{Z}_I)_{e'}$
	&  $(\Gamma_{II},\text{Z}_{II})_{j'}$ \\
	\hline
	\includegraphics[width=0.24\linewidth]{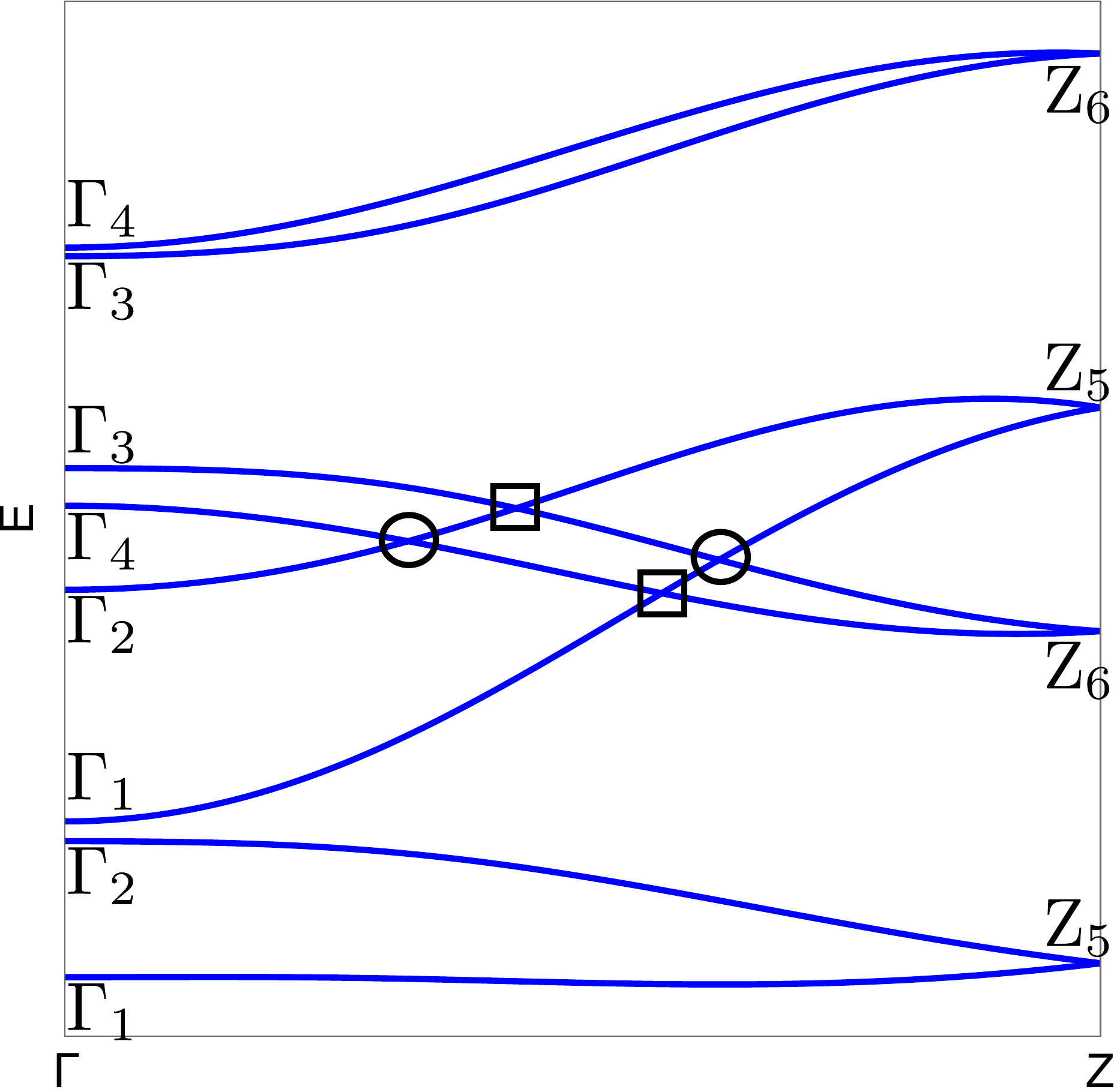} &
	\includegraphics[width=0.24\linewidth]{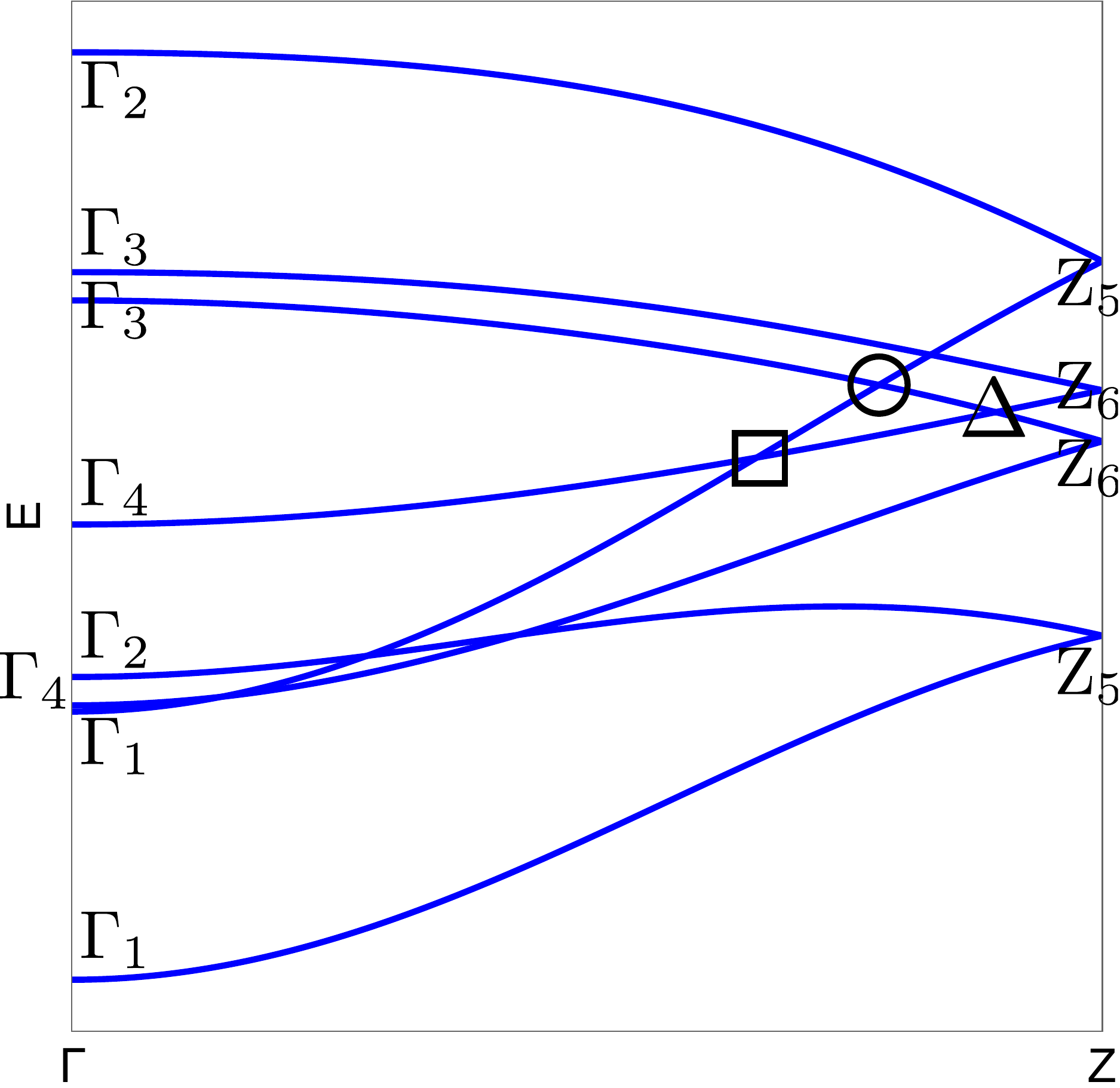} &
	\includegraphics[width=0.24\linewidth]{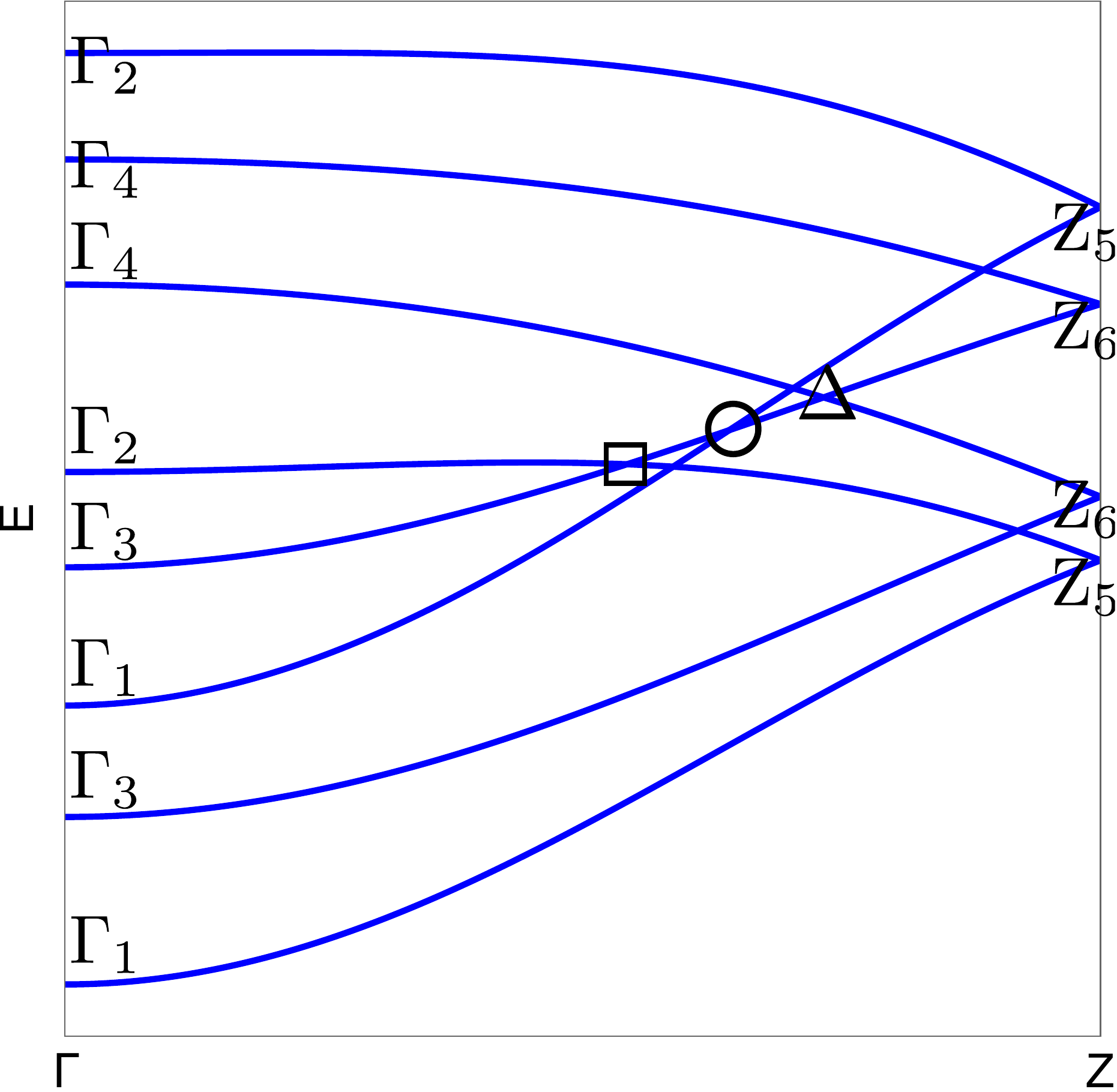} &
		\includegraphics[width=0.24\linewidth]{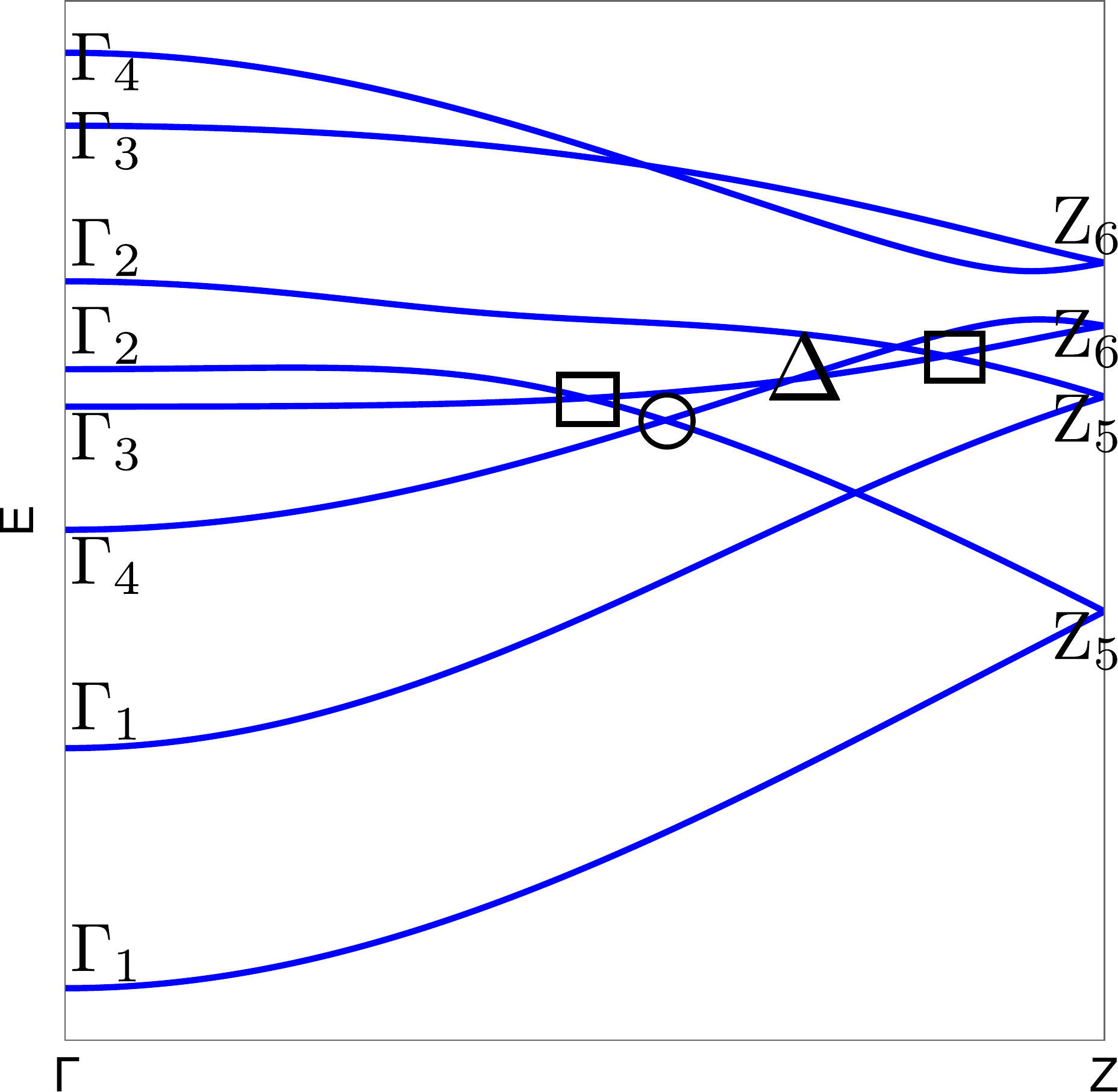} \\
	(d',1) & (c',1) & (e',1) 
	& (j',1) \\
	\includegraphics[width=0.2\linewidth]{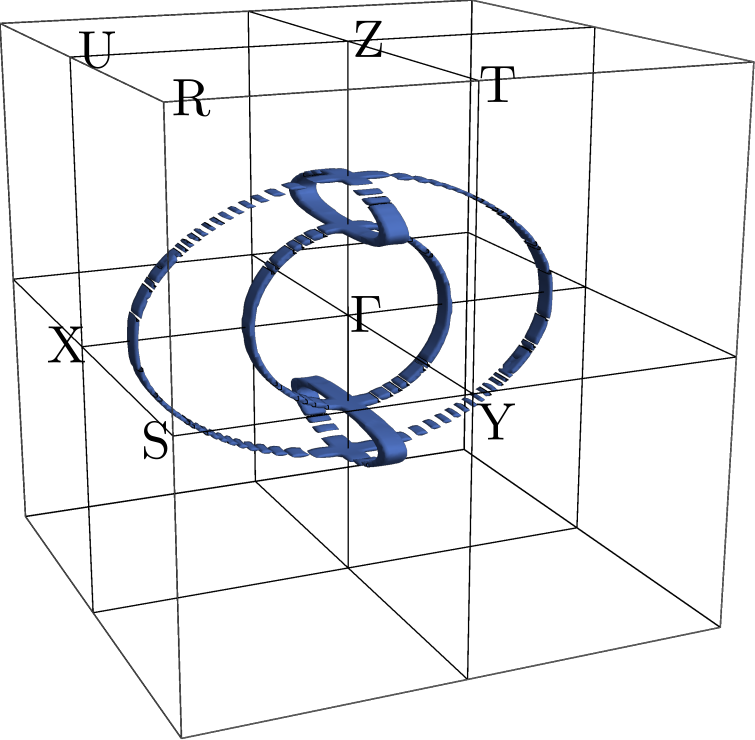} 	&
	\includegraphics[width=0.2\linewidth]{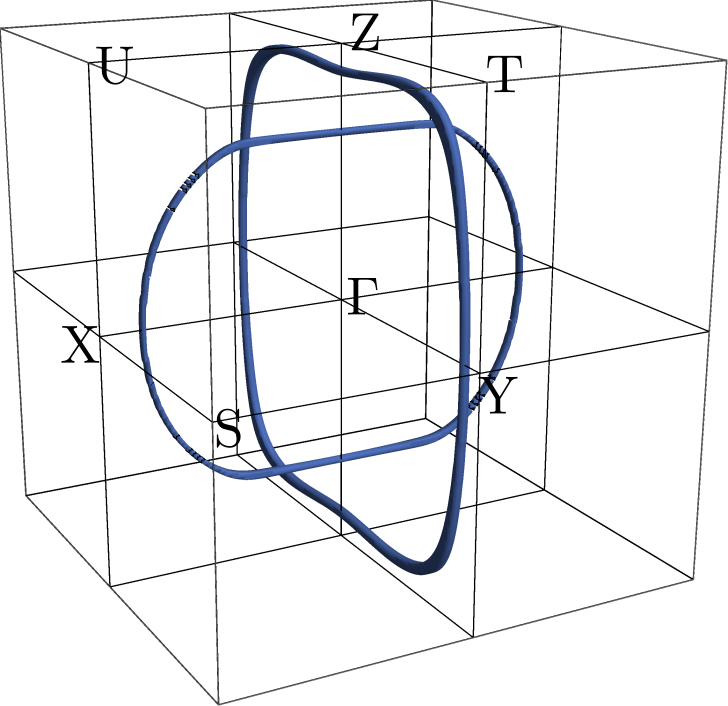} &
	\includegraphics[width=0.2\linewidth]{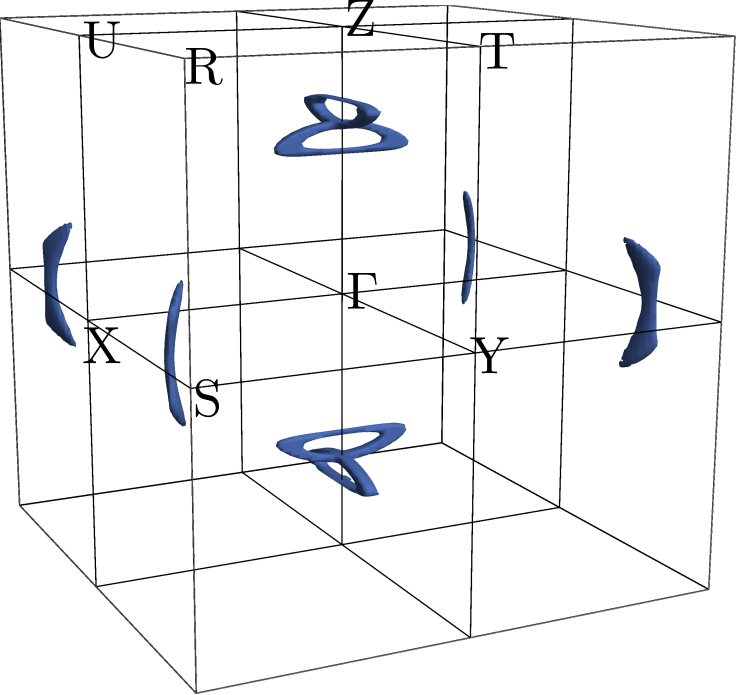} &
	\includegraphics[width=0.2\linewidth]{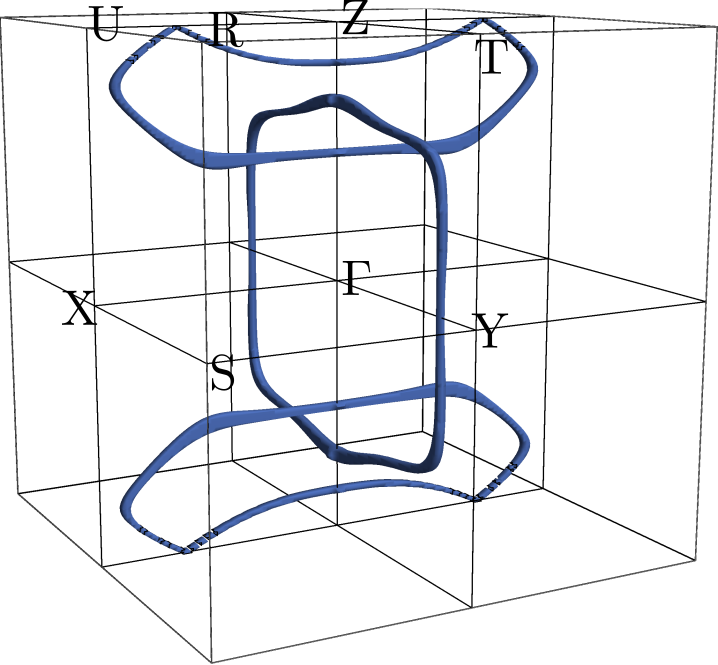} \\
	(d',2) & (c',2) & (e',2) 
	& (j',2) \\
	\includegraphics[width=0.24\linewidth]{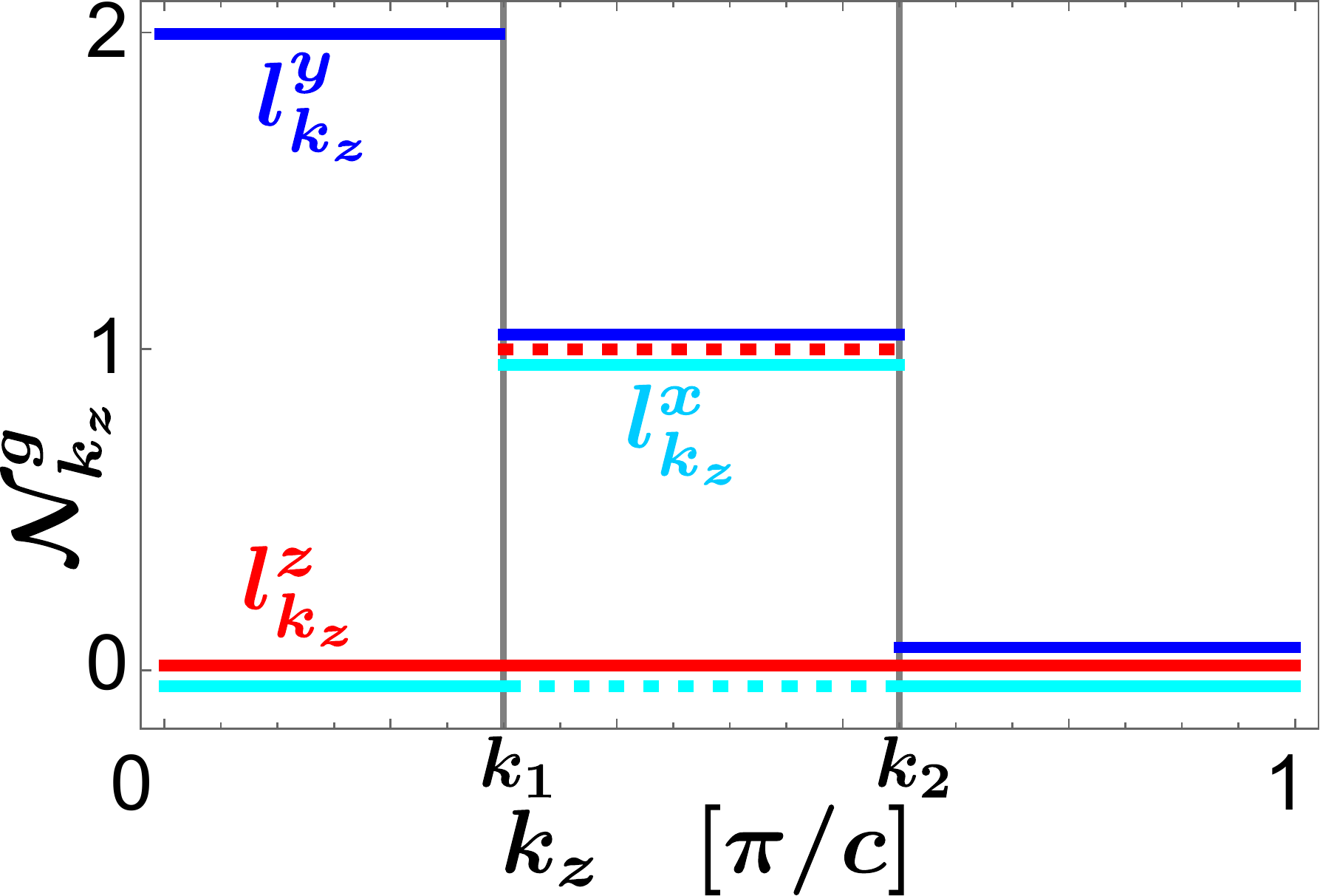}  &
	\includegraphics[width=0.24\linewidth]{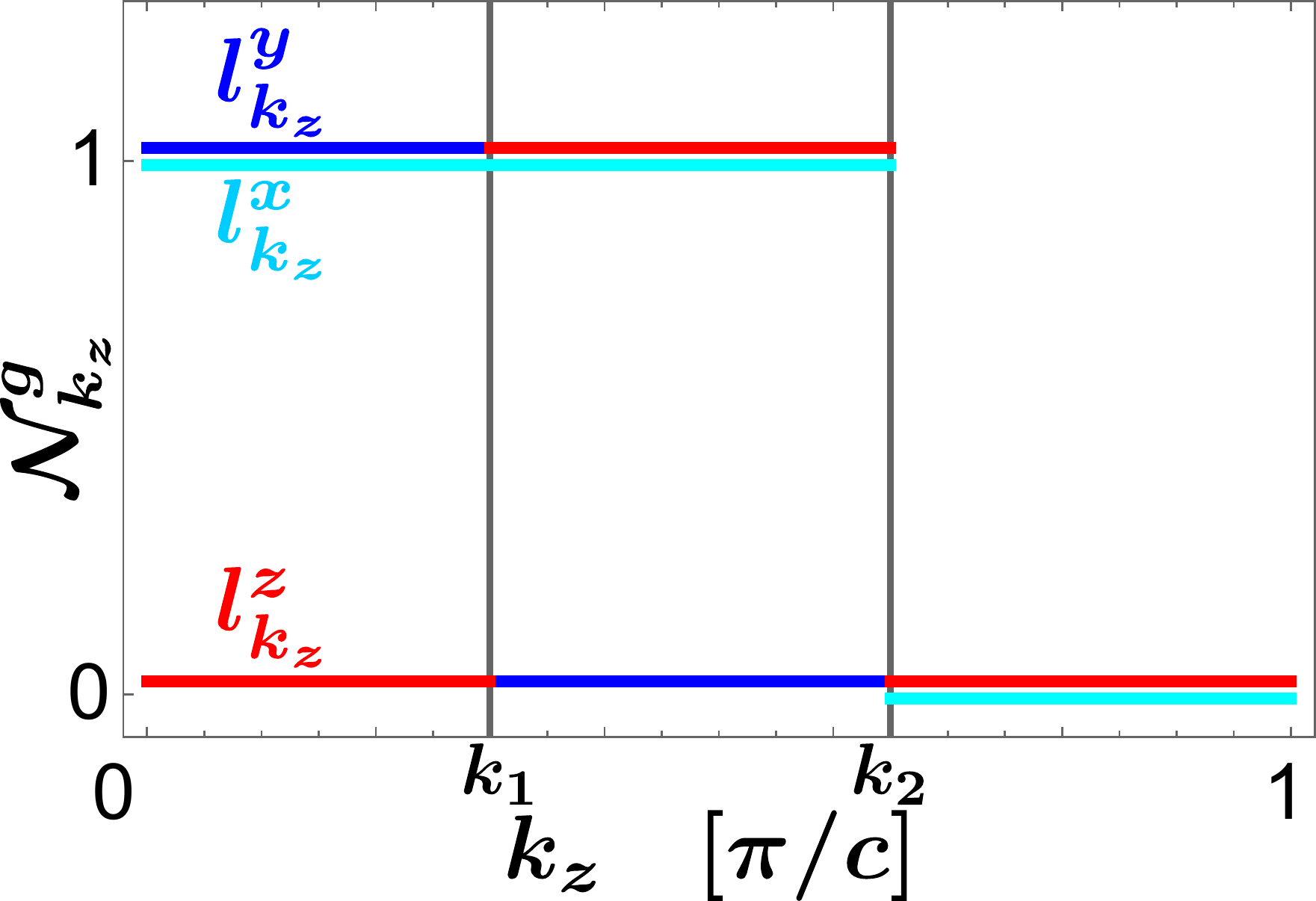} & 
	\includegraphics[width=0.24\linewidth]{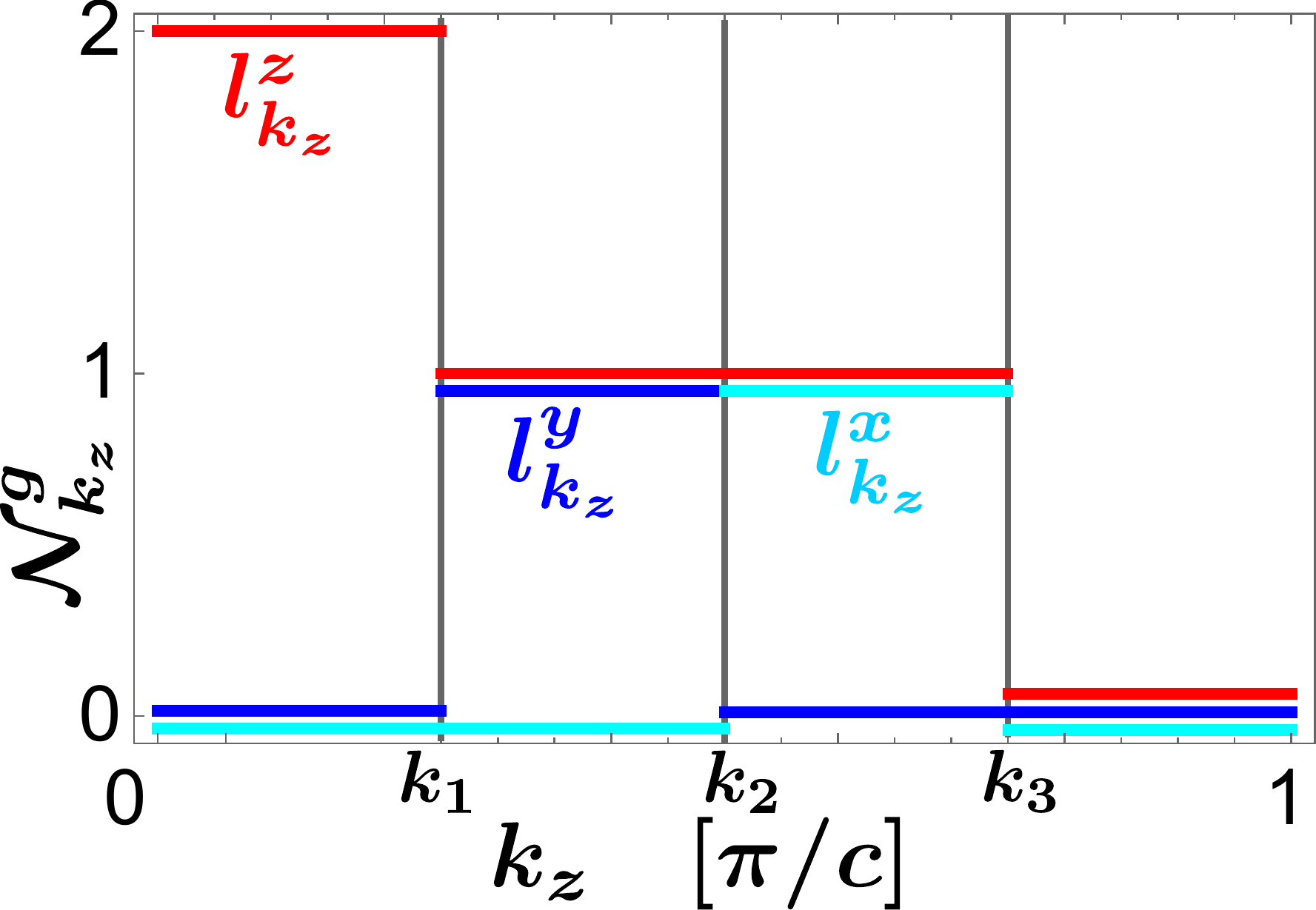} & 
	\includegraphics[width=0.24\linewidth]{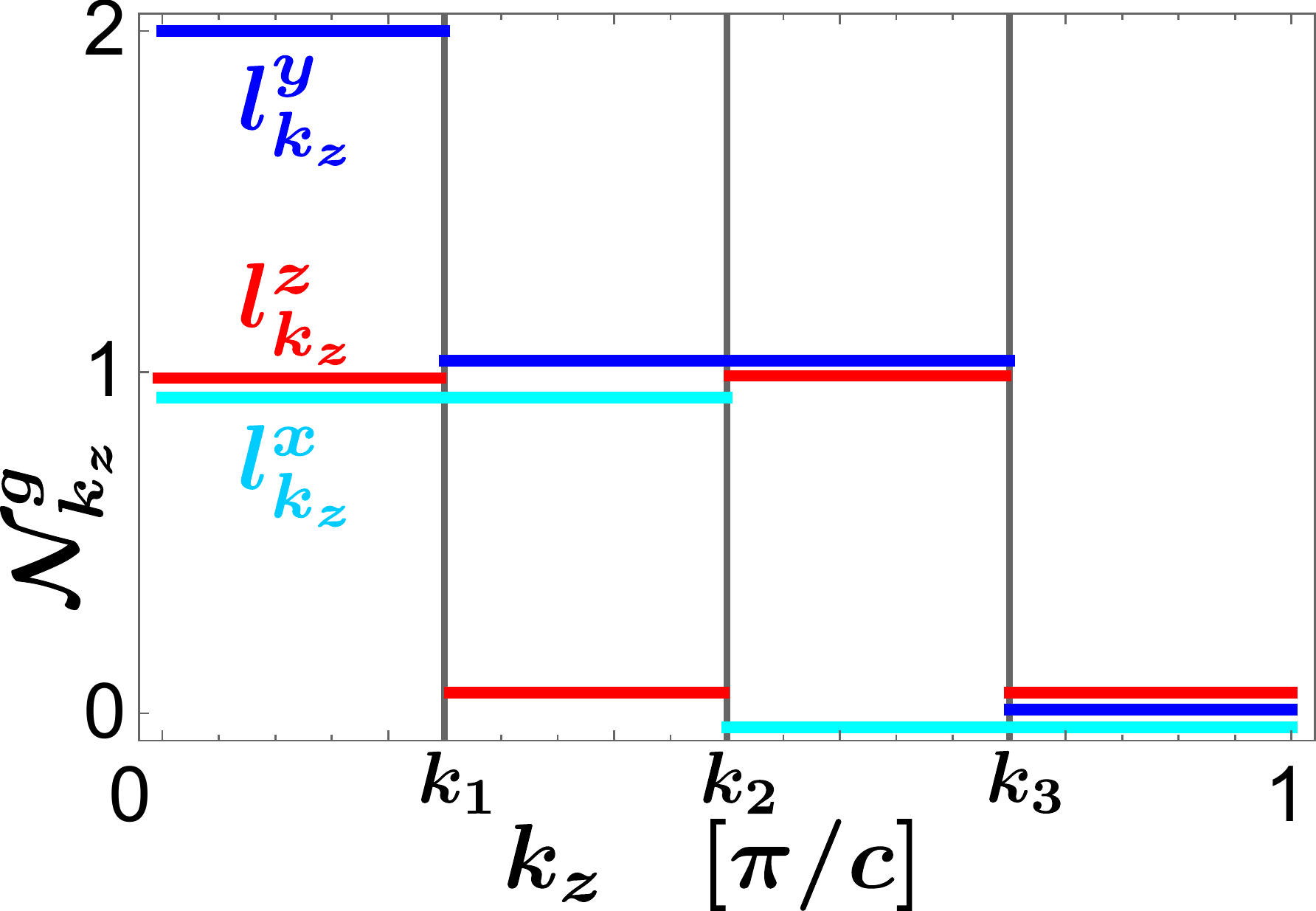} \\
	(d',3) & (c',3) & (e',3) 
	& (j',3) \\
\hline
\hline
\end{tabular}
\caption{\label{fig_transformed} Examples of line-nodal structures obtained from the cases (d,c,e,j) of Fig.~\ref{fig_SG33_8B_lines} through local topological Lifshitz transitions in $k$-space. (\textunderscore,1) band structure along $\Lambda$; (\textunderscore,2) global nodal structure at half-filling; (\textunderscore,3) two-point loop invariants $[\mathcal{N}^{z,y,x}_{k_z}]$ for a loop $l^{g}_{k_z}$ with as vertices Z and $k_z \in \Lambda \equiv \overline{\Gamma \mathrm{Z}}$.
}
\end{figure*}
The exchange of valence IRREPs with conduction IRREPs, while keeping the filling number fixed, involves the closing of the band gap and can lead to inequivalent global band topologies. We call this a topological Lifshitz transition. In our case, this is easily realized through the tuning of the microscopic tight-binding parameters. We distinguish between local transitions in $k$-space that do not affect the set of valence IRREPs at the HSPs and global transitions accompanied by a change of valence IRREPs' sets. Global topological Lifshitz transitions thus correspond to the transitions between the different cases of Fig.~\ref{fig_SG33_4B_lines} or Fig.~\ref{fig_SG33_8B_lines} and the remaining cases that can be formed from the combinatorics of Table \ref{point_classes}. In this section we instead discuss the local transitions that allow qualitatively different nodal structures but still with fixed sets of IRREPs at the active HSPs $\{\Gamma,\text{Z},\text{S},\text{R}\}$. 

\subsubsection{Connecting and disconnecting NLs}
We first consider the cases Fig.~\ref{fig_SG33_8B_lines}(d,f,g,i,l,n,r), that can all conditionally exhibit a small NL connecting two independent and unremovable coplanar NLs. This small NL is present in Fig.~\ref{fig_SG33_8B_lines}(f,g,i,l,n,r), but for example absent in (d). The presence of the small NL depends on the specific dispersion along the $\Lambda$-line. Taking for instance the case $(\Gamma_{II},\text{Z}_{I})_d$ in Fig.~\ref{fig_SG33_8B_lines}(d), the band structure in Fig.~\ref{GIIIZI_d_bands} marks the presence of two crossing points on $\Lambda$ (squares) that continue as two independent NLs on the plane $\sigma_3$. An alternative  band structure shown in Fig.~\ref{fig_transformed}(d',1) has the same set of valence IRRPEPs as $(\Gamma_{II},\text{Z}_{I})_d$, but now exhibits two additional crossing points on $\Lambda$ (circles), which are each part of two NLs, one NL on the plane $\sigma_3$ and one NL on $\sigma_5$. Hence, each crossing point is now the connecting point of two distinct NLs belonging to perpendicular planes. Noting that the global class requires two crossing points on each half of the $\Sigma$-line, while there should be none on the $\Delta$-line, we conclude that there are two distinct symmetry protected NLs on the $m_y$-invariant plane ($\sigma_3$) and, connecting them, one extra small NL on the $m_x$-invariant plane $\sigma_5$ as clearly seen in Fig.~\ref{fig_transformed}(d',2). 

We are free to move the relative positions of the two crossing points on $\Lambda$. Starting from the previous configuration we bring the two circle-type crossing points closer and eventually reverse their order along $\Lambda$, hence exchanging the two circle-type crossings into the square-type crossings. As a result, only the two independent NLs on the $m_y$-invariant plane remain as in Fig.~\ref{fig_SG33_8B_lines}(d). Notably, this is done without closing the energy gap between the valence and conduction bands at any HSP, i.e.~$(\Gamma_{II},\text{Z}_{I})_d \cong (\Gamma_{II},\text{Z}_{I})_{d'}$.

Forming the two-point symmetry-constrained loops from the HSP $\text{Z}$ and $k_z\in \Lambda$, $l^g_{k_z\in \Lambda}$, we plot in Fig.~\ref{fig_transformed}(d',3) the $k_z$-dependent symmetry protected topological invariants $\mathcal{N}^g_{k_z\in \Lambda}$ with (full lines) and without (dashed lines) the connecting NL on the $\sigma_5$-plane. We find that the above transformation is characterized by the change of the two-point invariants within the range $k_z\in [k_{1},k_{2}]\subset \Lambda$, where $k_1<k_2$ are the point nodes on $\Lambda$. More specifically, we find $[\mathcal{N}^{z}_{k_z},\mathcal{N}^{x}_{k_z} ]= [0,1]$ with the connecting NL (full line), but $[\mathcal{N}^{z}_{k_z},\mathcal{N}^{x}_{k_z} ] = [1,0]$ without it (dashed line). This defines a local topological transition in $k$-space, which we call a \textit{local} topological Lifshitz transition because it clearly does not change the global band topology fixed by the sets of valence IRREPs at the HSPs.

We also show the effect of a local topological Lifshitz transition in the case Fig.~\ref{fig_SG33_8B_lines}(c), where the two connected NLs can instead be disconnected, see Fig.~\ref{fig_transformed}(c',2). This happens by pushing down the branch $\Gamma_4$ below the circle-type crossing point marked in Fig.~\ref{fig_transformed}(c',1). This can for instance be achieved by inverting the energy ordering the $Z_5$ and $Z_6$ conduction states. As a consequence there is an intermediary region $k_z\in[k_1,k_2]$ on $\Lambda$ with $[\mathcal{N}^{z,y,x}_{k_z}] =[1,0,1]$ that differs from $[\mathcal{N}^{z,y,x}_{k_z=0}] =[0,1,1]$ and is absent when the two NLs are connected, see Fig.~\ref{fig_transformed}(c',3). 

Finally, in Fig.~\ref{fig_transformed}(e') we show the effect of a local topological Lifshitz transition on the monopole case Fig.~\ref{fig_SG33_8B_lines}(e). Here the nodal monopoles are now each composed of two perpendicular NLs connected at one point on $\Lambda$, see Fig.~\ref{fig_transformed}(e',2) (the four accidental point nodes within the plane $\sigma_1$ are still also present). This is also achieved by inverting the $Z_5$ and $Z_6$ conduction states, as seen in Fig.~\ref{fig_transformed}(e',1). This transformation is accompanied by the presence of an intermediary region on $\Lambda$ with $[\mathcal{N}^{z,y,x}_{k_1<k_z<k_2}] =[1,1,0]$ and $[\mathcal{N}^{z,y,x}_{k_2<k_z<k_3}] =[1,0,1]$ which is absent in the untransformed case.

\subsubsection{Linked NLs and toroidal charge} 
As a last example, we show in Fig.~\ref{fig_transformed}(j') the effect of, starting from Fig.~\ref{fig_SG33_8B_lines}(j), pushing the branch $\Gamma_3$ above the circle-type crossing shown in Fig.~\ref{fig_transformed}(j',1). As a consequence, the (connected) NLs on $\sigma_3$ are now disconnected from the NL on $\sigma_5$ such that they are linked all together, see Fig.~\ref{fig_transformed}(j',2). This leads to an intermediary region on $\Lambda$ with $[\mathcal{N}^{z,y,x}_{k_1<k_z<k_2}] =[0,1,1]$. Here $\mathcal{N}^{y}_{k_1<k_z<k_2}=1$ gives the nontrivial poloidal Berry phase of the upper branch of the connected NLs on $\sigma_3$, while $\mathcal{N}^{x}_{k_1<k_z<k_2}=1$ gives the nontrivial toroidal Berry phase of the NL on $\sigma_5$. This nodal structure is the only one in the work with a toroidal charge. Many more linked nodal structures can be achieved by allowing a larger number of bands, but since they do not lead to any new qualitative features, i.e.~all of them can be described in terms of the invariants of Section \ref{top_inv} and the local charges of Section \ref{top_meta}, we can still restrict the discussion to the four- and eight-band subspaces. 
Finally, we remark that this great variety of locally topologically distinct line-nodal structures appears as a consequence of the rich little co-group (stabilizer) of the line $\Lambda$ $C_{2v}$ that results from the nonsymmorphic symmetries of SG33.

\section{Concluding remarks}\label{discussion}
Before concluding we here offer a few clarifying remarks. These concern the relation between the line-nodal structures so far discussed and the actual Fermi surfaces as well as the full topological classification of SG33-A1. 

\begin{figure}[htb]
\centering
\begin{tabular}{c}
	\includegraphics[width=0.6\linewidth]{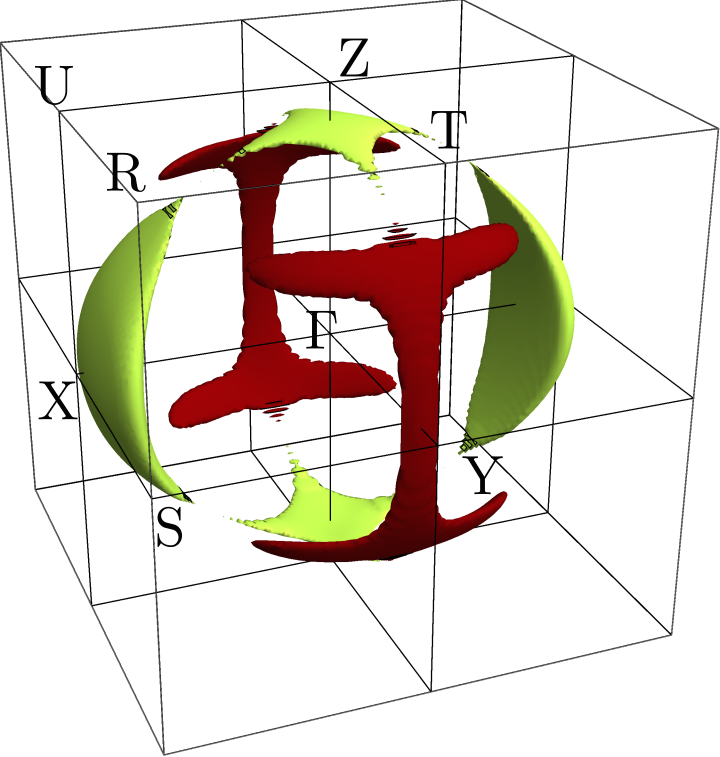} 
\end{tabular} 
\caption{\label{fig_Fermi_surface} Example of Fermi surface (iso-energy surface) obtained for the case Fig.~\ref{fig_SG33_8B_lines}(c) at physical ``half-filling''. Colors indicate the contributions from distinct energy ordered bands. The Fermi surface is topologically equivalent to the inflated nodal structure, i.e.~the surface $\boldsymbol{L}_v$ surrounding the nodal structure $L_v$, with bottlenecks point nodes marking the crossing of the line-nodal structure with the Fermi level. 
} 
\end{figure}

We have already discussed that the nodal structures we have classified and characterized in this work, defined by a fixed number of valence states over the whole BZ, are in general not to be thought as Fermi surfaces. Indeed, nothing prevents a nontrivial energy dependence of the nodal structure and its coexistence with electron or hole pockets, such that the number of valence states varies over the BZ, which affects the topology of the Fermi surface. However, it is safe to consider that the Fermi surface (almost) always separates a valence subspace from a conduction subspace at every HSPs. Hence, the method we have introduced in this work always determines, knowing the set of valence IRREPs at the HSPs, whether the Fermi surface must contain nontrivial band crossings or not. Typically, whenever the averaged filling number enforces a line-nodal structure to cross the Fermi energy level, the corresponding qualitative Fermi surface can straightforwardly be obtained as the following: take the surrounding surface $\boldsymbol{L}_v$ of the nodal structure $L_v$ defined in Section \ref{top_meta} and shrink it into bottleneck point nodes whenever the line-nodal structure crosses the Fermi level. As an example, we show the computed Fermi surface in Fig.~\ref{fig_Fermi_surface} when the Fermi level crosses the line-nodal structure of Fig.~\ref{fig_SG33_8B_lines}(c). As it has been shown very recently from minimal models \cite{Mele_response_linenodes}, we expect the physical properties of the quasiparticles at low energy to be qualitatively affected by the nontrivial topology of the underlying nodal structure. This will be further explored in the future.  

As for the full topological classification, following the approach of Refs.~\cite{TopQuantChem,Cano_EBR}, it is straightforward to establish that every band structure of an insulator with SG33 in class AI must be topologically trivial, since we can always adiabatically map it onto an atomic insulator without breaking any symmetry. This follows from the fact that the compatibility relations impose every valence structure isolated by a band gap from above and from below to be composed of the following IRREPs at the active HSPs, at $\Gamma$: $N\times \left( \Gamma_1 \oplus \Gamma_2\oplus \Gamma_3\oplus \Gamma_4 \right)$, and at p: $N\times \left( p_5 \oplus p_6 \right)$ for p=Z, S, R, for $N\in \mathbb{N}$. Also any realization of the unique Wyckoff's position of SG33 can be mapped adiabatically to an other position realization without breaking the symmetries. As a consequence, every isolated valence band structure is compatible with a decomposition into elementary band representations \cite{TopQuantChem,Cano_EBR} of an atomic insulator. Thus, the space group of lattice symmetries strongly constrains the combinatoric space of band structures preventing the formation of any nontrivial topological insulating phases. We can therefore conclude that our characterization line-nodal structures in SG33-AI is exhaustive. 

To summarize, we have in this work presented a complete and systematic topological classification and characterization of all symmetry protected line-nodal structures realized in four-band and eight-band subspaces with SG33 in class AI, i.e.~preserving TRS and with no spin-orbit coupling, at half-filling.
We achieved this through a two step process. We started with the topological classification of all line-nodal structures. This was accomplished by only knowing the valence IRREPs combinatorics at the active HSPs and the constraints by the compatibility relations and the band permutation rules due to the nonsymmorphic symmetries in SG33. We were able to show that a great variety of nodal structures can be realized. In particular, we found highly connected nodal structures, composed nodal structures, as well as nodal monopole pairs and nodal thread pairs. This procedure can straightforwardly be extended in the case of a larger number of bands and arbitrary filling conditions. 

As a second step, we solved the complementary task of topological characterization of any nodal structure in terms of its local topology and global topology. For this purpose we derived a general algorithm leading to all the local poloidal-toroidal and monopole and thread charges of any elementary nodal structure. These were expressed as homotopy groups over every local loop and the unique surrounding sphere or cylinder resulting from the deformation retract of the punctured BZ (i.e.~after subtracting the nodal structure). 
We then derived and extracted the necessary symmetry protected topological invariants for the charges from the quantized Wilson loop phases computed over two-point (formed by two HSP vertices) and four-point (formed by four HSP vertices) symmetry constrained loops. We showed that these can all be determined fully algebraically from the \textit{bare} valence IRREPs. To that end, we proposed the drawing of a \textit{rosetta stone}, namely the compact presentation on the BZ of the bare IRREPs at all HSPs and all possible two and four-point symmetry constrained loops, from which every topological invariants can be readily obtained. 
We discussed in detail how these two- and four-point invariants determine all the local topological charges of every nodal structure. We also illustrated a complete agreement between the numerical computation of the Wilson loop and this algebraic approach. 
We finally demonstrated the existence of local and symmetry preserving topological Lifshitz transitions through which independent NLs can be connected, disconnected, and also linked. 
Taken together, this work constitute an heuristic answer to the exhaustive classification and characterization of the bulk topology of lattice systems with SG33 in class AI, which is very easily extendable to other systems. Furthermore, since the only inputs needed for this definitive topological characterization of semimetallic phases are the valence band IRREPs at the HSPs, our method is optimally suited to be at the core of future intelligent data mining schemes based on {\it ab-initio} calculations searching for new topological materials, and thus naturally extends such existing proposals \cite{BorysovGeilhufeBalatsky_mining,TopQuantChem,Vishwanath_mining}.

\begin{acknowledgments}
We thank A.~Alexandradinata for useful comments and R.-J.~Slager for insightful discussions. This work was supported by the Swedish Research Council (Vetenskapsr\aa det) and the Knut and Alice Wallenberg Foundation.
\end{acknowledgments}

\appendix 
 
\section{Bloch basis, symmetry representations, and periodic gauge}\label{sym_rep}
This appendix briefly introduces the Bloch basis used and the corresponding representation of space group symmetries. It also defines the periodic gauge used in the symmetry reduction of Wilson loops.

In this work we use the Bloch-L\"owdin-Zak basis based on site-localized Wannier functions,
\begin{equation*}
	\vert \varphi_i,\boldsymbol{k}\rangle = \dfrac{1}{\sqrt{N}} \sum\limits_{\boldsymbol{R}_n} \mathrm{e}^{i \boldsymbol{k}\cdot (\boldsymbol{R}_n+\boldsymbol{r}_i)}  \vert w,\boldsymbol{R}_n+\boldsymbol{r}_i\rangle \;,
\end{equation*}
where $\boldsymbol{R}_n$ points to the symmetry center of each unit cell of the Bravais lattice and $\boldsymbol{r}_i$ points to a sub-lattice site $i$ within the unit cell. This choice of basis set for the Fourier transform of the Hamiltonian differential operator gives a trivialization of the total Bloch bundle, as needed in Section \ref{top_meta}, and leads to a physically relevant definition of the Berry connection and the corresponding holonomy \cite{FruchartCarpentier1,Berry_connection_Moore}. 

We can then form the symmetrized Bloch basis
\begin{equation*}
	\vert \boldsymbol{\phi},\boldsymbol{k}\rangle = \vert \boldsymbol{\varphi}_i,\boldsymbol{k}\rangle \cdot \hat{U}^{S}\;,
\end{equation*}
that transforms under the space group symmetry operation $\{g\vert \boldsymbol{\tau}_g\} \in \mathcal{G}$ as
\begin{align*}
	^{\{g\vert \boldsymbol{\tau}_g\}}  \vert \boldsymbol{\phi},\boldsymbol{k}\rangle &= \vert \boldsymbol{\phi},g\boldsymbol{k}\rangle \cdot \hat{U}_{g}(\boldsymbol{k})\;, \\
	\hat{U}_g(\boldsymbol{k}) &= \mathrm{e}^{-i g\boldsymbol{k} \cdot \boldsymbol{\tau}_g } \bigoplus\limits_{j} \chi^{\Gamma_j}(g)\;.
\end{align*}
The Hamiltonian in this basis reads 
\begin{equation*}
	\mathcal{H} = \sum\limits_{\boldsymbol{k}\in\mathrm{BZ}} \vert \boldsymbol{\phi},\boldsymbol{k}\rangle H(\boldsymbol{k}) \langle \boldsymbol{\phi},\boldsymbol{k}\vert\;,
\end{equation*}
with $H(\boldsymbol{k})$ being a $N\times N$ Hermitian matrix. Then the Bloch eigenfunctions transform under $\{g\vert \boldsymbol{\tau}_g\}$ according to
\begin{align*}
	^{\{g\vert \boldsymbol{\tau}_g\}}  \vert \boldsymbol{\psi},\boldsymbol{k}\rangle &= \vert \boldsymbol{\psi},g\boldsymbol{k}\rangle \cdot \breve{S}^{\boldsymbol{k}}_{g}\;, \\
	\breve{S}^{\boldsymbol{k}}_{g} &= \breve{U}^{\dagger}(g\boldsymbol{k})\cdot \hat{U}_g(\boldsymbol{k}) \cdot \breve{U}(\boldsymbol{k})\;,
\end{align*}
where $\breve{U}(\boldsymbol{k})$ is the diagonalizing matrix of the Hamiltonian, i.e.~$\vert \boldsymbol{\psi},\boldsymbol{k}\rangle = \vert \boldsymbol{\phi},\boldsymbol{k}\rangle \breve{U}(\boldsymbol{k})$ and $\breve{U}^{\dagger}(\boldsymbol{k})H(\boldsymbol{k})\breve{U}(\boldsymbol{k}) = \mathrm{diag}(E_1(\boldsymbol{k}),\dots, E_N(\boldsymbol{k}))$. The transformation matrix $\breve{S}^{\boldsymbol{k}}_{g}$ must have a block-diagonal structure corresponding to the decomposition of the eigenstates at $\boldsymbol{k}$ into the IRREPs of the little group $\overline{G}^{\boldsymbol{k}}$ \cite{BradCrack}. Therefore, $\breve{S}^{\boldsymbol{k}}_{g}$ gives an explicit representation of the symmetry operation $\{g\vert \boldsymbol{\tau}_g\}$ that depends on the basis set in which we have written the Hamiltonian. In this work, we only need the eigenvalues of $\breve{S}^{\boldsymbol{k}}_{g}$, so we can use any tabulated IRREP matrices, given for instance in Refs.~\cite{BradCrack} or \cite{Bilbao}. 

We obtain a first useful representation of symmetry operations acting on a Bloch eigenvector $[\breve{U}(\boldsymbol{k})]_n \equiv \vert U_n , \boldsymbol{k} \rangle $, i.e.
\begin{equation*}
	\hat{U}_g(\boldsymbol{k})  \vert U_n , \boldsymbol{k} \rangle  = \vert U_m , g\boldsymbol{k} \rangle \breve{S}^{\boldsymbol{k}}_{mn,g}	\;.
\end{equation*}
It is here convenient to separate the phase factor present in nonsymmorphic symmetries and write 
\begin{equation*}
	\breve{S}^{\boldsymbol{k}}_{g} = \mathrm{e}^{-i g\boldsymbol{k} \cdot \boldsymbol{\tau}_g } \breve{R}^{\boldsymbol{k}}_{g} \;.
\end{equation*}
The cell periodic part of the Bloch eigenfunction is defined as $\vert \boldsymbol{u},\boldsymbol{k}\rangle = \mathrm{e}^{-i \boldsymbol{k} \cdot \hat{\boldsymbol{r}}}  \vert \boldsymbol{\psi},\boldsymbol{k}\rangle $ (with $\hat{\boldsymbol{r}}$ is the position operator), such that it transforms according to
\begin{align*}
	^{\{g\vert \boldsymbol{\tau}_g\}}  \vert \boldsymbol{u},\boldsymbol{k}\rangle &= \vert \boldsymbol{u},g\boldsymbol{k}\rangle \cdot \breve{R}^{\boldsymbol{k}}_{g}\;, 
\end{align*}
i.e.~the phase factor $\mathrm{e}^{-i g\boldsymbol{k} \cdot \boldsymbol{\tau}_g } $ is now absent. This is convenient since in general the phase depends on the choice of origin with respect to the Bravais lattice in real space. 

Because of the nonsymmorphycity of the space group, Bloch eigenfunctions transform nontrivially under a reciprocal lattice translation,
\begin{align*}
	 \vert \boldsymbol{\psi},\boldsymbol{k}+\boldsymbol{K}\rangle &= \vert \boldsymbol{\psi},\boldsymbol{k}\rangle \cdot \breve{T}(\boldsymbol{K})\;,\\
	  \breve{T}(\boldsymbol{K}) &=  \breve{U}^{\dagger}(\boldsymbol{k}) \hat{T}(\boldsymbol{K})\breve{U}(\boldsymbol{k}+\boldsymbol{K})\;,
\end{align*}
where, for a four-band system, the translation matrix in the symmetrized Bloch-L\"owdin-Zak basis is explicitly given by 
\begin{equation*}
	 \hat{T}(\boldsymbol{K}) = U^{S\dagger} \mathrm{diag}\left[\mathrm{e}^{i \boldsymbol{K} \cdot \boldsymbol{r}_1},\mathrm{e}^{i \boldsymbol{K} \cdot \boldsymbol{r}_2},\mathrm{e}^{i \boldsymbol{K} \cdot \boldsymbol{r}_3},\mathrm{e}^{i \boldsymbol{K} \cdot \boldsymbol{r}_4} \right] U^S\;.
\end{equation*}
The periodic gauge is defined by $\breve{U}^p(\boldsymbol{k}+\boldsymbol{K}) = \hat{T}^{\dagger}(\boldsymbol{K}) \breve{U}^p(\boldsymbol{k})$, such that $\vert \boldsymbol{\psi}^p,\boldsymbol{k}+\boldsymbol{K}\rangle = \vert \boldsymbol{\psi}^p,\boldsymbol{k}\rangle$. Therefore the cell periodic Bloch eigenfunctions transform under reciprocal translation as
\begin{align*}
	 \vert \boldsymbol{u}^p,\boldsymbol{k}+\boldsymbol{K}\rangle &= \mathrm{e}^{-i \boldsymbol{K}\cdot \hat{\boldsymbol{r}}}  \vert \boldsymbol{u}^p,\boldsymbol{k}\rangle\;,
\end{align*}
which leads to the multiplication of each Bloch-L\"owdin-Zak basis function $\vert \varphi_i,\boldsymbol{k}\rangle$ by a phase factor $\mathrm{e}^{-i \boldsymbol{K}\cdot (\boldsymbol{R}_n+\boldsymbol{r}_i) } = \mathrm{e}^{-i \boldsymbol{K}\cdot \boldsymbol{r}_i }$, i.e.~they transform like the eigenvectors $[\breve{U}^p(\boldsymbol{k})]_n$.

\section{IRREPs and bare IRREPs for SG33-AI}\label{bare}
Since it is essential four our analysis, we give in this Appendix in Table~\ref{table_IRREP} the IRREP matrices at the HSPs $\bar{\boldsymbol{k}}\in \{\text{X},\text{Y},\text{Z},\text{S},\text{T},\text{U},\text{R}\}$ for each symmetry operation of their little co-group ($\overline{G}^{\bar{\boldsymbol{k}}} = C_{2v}$) for SG33-AI, as calculated in Ref.~\cite{Bilbao}. At the $\Gamma$-point there are only 1D IRREPs given by the character table of Fig.~\ref{fig_SG33}(c). Only a single 2D IRREP is allowed at $ \{\text{X},\text{Y},\text{T}\}$, a single 4D IRREP allowed at U, and two 2D IRREPs allowed at $\{\text{Z},\text{S},\text{R}\}$, which we write $X_{5,6}$ for $X=Z,S,R$. 
{\def\arraystretch{1.3}  
\begin{table}[thb]
\caption{\label{table_IRREP} IRREP matrices $\breve{S}^{\bar{\boldsymbol{k}}}_g$ for SG33-AI at the HSPs, $\bar{\boldsymbol{k}} \in \{\text{X},\text{Y},\text{Z},\text{S},\text{R},\text{T},\text{U}\}$, for each symmetry of the of the little co-groups $g\in \overline{G}^{\bar{\boldsymbol{k}}} \cong \{E,C_{2z},m_y,m_x\}$, as given by Ref.~\cite{Bilbao}.}
\begin{tabular*}{\linewidth}{@{\extracolsep{\fill}} l ccc}
\hline 
\hline 
	$\mathrm{IRREP}$ & $C_{2z}$ & $m_y$  & $m_x$  \\
	\hline 
	$\breve{S}^\text{X}_{g}$ & $\left(\begin{array}{cc} 0 & 1\\1 & 0 \end{array}\right)$  & 
		$\left(\begin{array}{cc} 0 & 1\\-1 & 0 \end{array}\right)$ & 
			 $\left(\begin{array}{cc} 1 & 0\\0 & -1 \end{array}\right)$ \\
	$\breve{S}^\text{Y}_{g}$ & $\left(\begin{array}{cc} 0 & 1\\1 & 0 \end{array}\right)$  & 
		$\left(\begin{array}{cc} 0 & -i\\i & 0 \end{array}\right)$ & 
			 $\left(\begin{array}{cc} -i & 0\\0 & i \end{array}\right)$ \\
	$\breve{S}^\text{Z}_{g,\left(\substack{5 \\6}\right)}$ & $\left(\begin{array}{cc} 0 & 1\\-1 & 0 \end{array}\right)$  & 
		$\left(\begin{array}{cc} \pm1 & 0\\0 & \pm 1 \end{array}\right)$ & 
			 $\left(\begin{array}{cc} 0 & \pm1 \\ \mp 1 & 0 \end{array}\right)$ \\
	$\breve{S}^\text{S}_{g,\left(\substack{5 \\6}\right)}$ & $\left(\begin{array}{cc} \pm1 & 0\\0 & \pm1 \end{array}\right)$  & 
		$\left(\begin{array}{cc} 0 & \pm1 \\\mp1 & 0 \end{array}\right)$ & 
			 $\left(\begin{array}{cc} 0 & 1\\-1 & 0 \end{array}\right)$ \\
	$\breve{S}^\text{R}_{g,\left(\substack{5 \\6}\right)}$ & $\left(\begin{array}{cc} 0 & 1\\-1 & 0 \end{array}\right)$  & 
		$\left(\begin{array}{cc} 0 & \mp1\\\pm1  & 0 \end{array}\right)$ & 
			 $\left(\begin{array}{cc} \pm1 & 0\\0 & \pm1 \end{array}\right)$ \\
	$\breve{S}^\text{T}_{g}$ &$ \left(\begin{array}{cc} 0 & -1\\1 & 0 \end{array}\right)$  & 
		$\left(\begin{array}{cc} 0 & 1 \\ 1 & 0 \end{array}\right)$ & 
			 $\left(\begin{array}{cc} 1 & 0\\0 & -1 \end{array}\right)$ \\
	$\breve{S}^\text{U}_{g,1}$ &$ \left(\begin{array}{cc} 0 & -1\\1 & 0 \end{array}\right)$  & 
		$\left(\begin{array}{cc} 0 & -i\\-i & 0 \end{array}\right)$ & 
			 $\left(\begin{array}{cc} -i & 0\\0 & i \end{array}\right)$ \\
	\hline
	\hline
\end{tabular*}
\end{table} 
}

From the IRREP matrices $\breve{S}^{\bar{\boldsymbol{k}}}_g$ written in the Bloch eigenfunctions we get the IRREP matrices in the cell periodic Bloch eigenfunctions from $\breve{S}^{\bar{\boldsymbol{k}}}_{g} = \mathrm{e}^{-i g\bar{\boldsymbol{k}} \cdot \boldsymbol{\tau}_g } \breve{R}^{\bar{\boldsymbol{k}}}_{g}$. Then, diagonalizing and removing any complex phase factor, as described in Section \ref{line_inv}, we get the \textit{bare} IRREP eigenvalue matrices $R^{\bar{\boldsymbol{k}}}_g$ composed only of $\pm1$. We list these in Table \ref{table_IRREP_bare} and show them also in the rosetta stone Fig.~\ref{rosetta_stone}(a). 
{\def\arraystretch{1.3}  
\begin{table}[thb]
\caption{\label{table_IRREP_bare} Bare IRREP eigenvalues in the cell periodic Bloch basis derived from Table \ref{table_IRREP}.}
\begin{tabular*}{\linewidth}{@{\extracolsep{\fill}} l ccc}
\hline 
\hline 
	$\mathrm{bare}$ & $C_{2z}$ & $m_y$  & $m_x$  \\
	\hline 
	$R^\text{X}_{g}$ & $\left(\begin{array}{cc} 1 & 0\\0 & -1 \end{array}\right)$  & 
		$\left(\begin{array}{cc} 1 & 0\\0 & -1 \end{array}\right)$ & 
			 $\left(\begin{array}{cc} 1 & 0\\0 & -1 \end{array}\right)$ \\
	$R^\text{Y}_{g}$ & $\left(\begin{array}{cc} 1 & 0\\0 & -1 \end{array}\right)$  & 
		$\left(\begin{array}{cc} 1 & 0\\0 & -1 \end{array}\right)$ & 
			 $\left(\begin{array}{cc} 1 & 0\\0 & -1 \end{array}\right)$ \\
	$R^\text{Z}_{g,\left(\substack{5 \\6}\right)}$ & $\left(\begin{array}{cc} 1 & 0\\0 & -1 \end{array}\right)$  & 
		$\left(\begin{array}{cc} \pm1 & 0\\0 & \pm 1 \end{array}\right)$ & 
			 $\left(\begin{array}{cc} \pm 1 & 0\\0 & \mp 1 \end{array}\right)$ \\
	$R^\text{S}_{g,\left(\substack{5 \\6}\right)}$ & $\left(\begin{array}{cc} \pm1 & 0\\0 & \pm1 \end{array}\right)$  & 
		$\left(\begin{array}{cc} \pm 1 & 0\\0 & \mp 1 \end{array}\right)$ & 
			 $\left(\begin{array}{cc} 1 & 0\\0 & -1 \end{array}\right)$ \\
	$R^\text{R}_{g,\left(\substack{5 \\6}\right)}$ & $\left(\begin{array}{cc} 1 & 0\\0 & -1 \end{array}\right)$  & 
		$\left(\begin{array}{cc} \pm 1 & 0\\0 & \mp 1 \end{array}\right)$ & 
			 $\left(\begin{array}{cc} \pm1 & 0\\0 & \pm1 \end{array}\right)$ \\
	$R^\text{T}_{g}$ &$ \left(\begin{array}{cc} 1 & 0\\0 & -1 \end{array}\right)$  & 
		$\left(\begin{array}{cc} 1 & 0\\0 & -1  \end{array}\right)$ & 
			 $\left(\begin{array}{cc} 1 & 0\\0 & -1 \end{array}\right)$ \\
	$R^\text{U}_{g,1}$ & $\left(\begin{array}{cc} 1 & 0\\0 & -1 \end{array}\right)$  & 
		$\left(\begin{array}{cc} 1 & 0\\0 & -1 \end{array}\right)$ & 
			 $\left(\begin{array}{cc} 1 & 0\\0 & -1 \end{array}\right)$ \\
	\hline
	\hline
\end{tabular*}
\end{table} 
}

\section{Wilson loop}\label{wilson}
In this work we use the Wilson loop approach for the computation of topological invariants as developed in \cite{Bernevig1, Bernevig_point_groups,Alex1,Alex_noSOC_noTRS,Kane_line,AlexBernevig_berryphase,Bernevig3,Alex0,Alex2}. In this Appendix we briefly define the continuum Wilson loop and its symmetry properties and then introduce the tight-binding Wilson loop. The continuum Wilson loop is used in the algebraic derivation of symmetry protected topological invariants. The tight-binding Wilson loop is used for the numerical results.  

First we define the Berry-Wilczek-Zee connection defined in terms of the valence cell periodic Bloch eigenfunctions $\vert u_{n},\boldsymbol{k} \rangle $ with $n = 1,\dots, N_v$, i.e.~
\begin{align*}
	\mathcal{A}_{mn,\mu} &= \langle u_{m},\boldsymbol{k} \vert \partial_{k^{\mu}} \vert u_{n},\boldsymbol{k} \rangle \;.
\end{align*}
Then the Wilson loop over a loop $l$ in $k$-space, is defined as
\begin{align*}
	\mathcal{W}[l] &= P \exp\left[- \int_{l} d\boldsymbol{k}\cdot \boldsymbol{\mathcal{A}} \right]\;,
\end{align*}
where $P\int_{l}$ means the path-ordered integration. The total $U(1)$ Berry phase over a loop $l$ is then obtained through
\begin{align*}
	\mathrm{e}^{i \gamma[l]} &= P \exp\left[- \int_{l} d\boldsymbol{k}\cdot \mathrm{Tr}~\boldsymbol{\mathcal{A}} \right]\; , \\
	&= \mathrm{det}~\mathcal{W}[l] \;.
\end{align*}

The Wilson loop can also be expressed in terms of a Wilson loop operator given as a product of valence projectors at successively infinitesimally close points of the loop \cite{Bernevig0}. Choosing $\boldsymbol{k}_0$ as the base point of the loop $l$, it reads
\begin{align*}
\label{WL_op}
	\mathcal{W}_{mn}[l] &= \langle u_m , \boldsymbol{k}_0 \vert \hat{W}[l] \vert   u_n , \boldsymbol{k}_0 \rangle \;, \\
	\hat{W}[l] &= P \prod\limits_{\boldsymbol{k}\in l} \mathcal{P}_{v}(\boldsymbol{k}) \;, 
\end{align*}
with the projector operator $\mathcal{P}_{v}(\boldsymbol{k}) = \sum\limits_{n=1}^{N_v} \vert   u_n , \boldsymbol{k} \rangle \langle    u_n , \boldsymbol{k} \vert$.
The unitarity of the Wilson loop then follows since $\mathcal{W}^{-1}[l] = \mathcal{W}^{\dagger}[l] = \mathcal{W}[l^{-1}]$.

Let us write $\breve{R}_g(g\boldsymbol{k},\boldsymbol{k}) \equiv \breve{R}^{\boldsymbol{k}}_{g} $. Then from
\begin{align*}
	^{\{g\vert \boldsymbol{\tau}_g\}}\vert \boldsymbol{u},\boldsymbol{k}\rangle &= \vert \boldsymbol{u},g\boldsymbol{k}\rangle \cdot \breve{R}_g(g\boldsymbol{k},\boldsymbol{k})	\\
	\Leftrightarrow\quad \vert \boldsymbol{u},g\boldsymbol{k}\rangle &= \left.^{\{g\vert \boldsymbol{\tau}_g\}}\vert \boldsymbol{u},\boldsymbol{k}\rangle\right. \cdot \breve{R}_{g^{-1}}(\boldsymbol{k},g\boldsymbol{k})\;, \\
	\langle \boldsymbol{u},g\boldsymbol{k}\vert^{\{g\vert \boldsymbol{\tau}_g\}} &= \breve{R}_g(g\boldsymbol{k},\boldsymbol{k}) \cdot \langle \boldsymbol{u},\boldsymbol{k}\vert 	\\
	\Leftrightarrow\quad \langle \boldsymbol{u},g\boldsymbol{k}\vert &= \breve{R}_g(g\boldsymbol{k},\boldsymbol{k}) \cdot \langle \boldsymbol{u},\boldsymbol{k}\vert^{\{g^{-1}\vert -g^{-1}\boldsymbol{\tau}_g\}}\;,
\end{align*}
where we have used $ \breve{R}_{g^{-1}}(\boldsymbol{k},g\boldsymbol{k}) = (\breve{R}^{\boldsymbol{k}}_g)^{-1} = (\breve{R}^{\boldsymbol{k}}_g)^{\dagger} $, we derive the symmetry transformation of the Wilson loop,
\begin{align*}
	\mathcal{W}[g\boldsymbol{k}_2 \leftarrow g\boldsymbol{k}_1] &= \breve{R}^{\boldsymbol{k}_2}_{g} \cdot \mathcal{W}[\boldsymbol{k}_2 \leftarrow \boldsymbol{k}_1] \cdot (\breve{R}^{\boldsymbol{k}_1}_{g})^{\dagger}\;.
\end{align*}
Then if $gl=l^{-1}$ and choosing an invariant base point $g\boldsymbol{k}_0=\boldsymbol{k}_0$, we have
\begin{align*}
		\mathcal{W}[gl] &= \breve{R}^{\boldsymbol{k}_0}_{g} \cdot \mathcal{W}[l] \cdot (\breve{R}^{\boldsymbol{k}_0}_{g})^{\dagger} \\
		&=  \mathcal{W}[l^{-1}]= \mathcal{W}^{\dagger}[l]\;,
\end{align*}
which leads to $\mathrm{eig}\{\mathcal{W}[l]\} = \mathrm{eig}\{\mathcal{W}[l]\}^{*}$ and, since $\mathcal{W}[l]$ is unitary, $\mathrm{det} \mathcal{W}[l] \in \{+ 1,-1\}$.

For numerical computations it is convenient to use the discretized tight-binding Wilson loop, $\bar{\mathcal{W}}$, expressed in terms of the valence Bloch eigenvectors $[\breve{U}(\boldsymbol{k})]_n$ with $n=1,\dots,N_v$ \cite{Bernevig0,Alex1}. Discretizing the loop in $k$-space $l\cong \{\boldsymbol{k}_i\}_{i=1,\dots,N_l}$, the tight-binding Wilson loop is given by
\begin{align*}
	\bar{\mathcal{W}}_{mn}[l] &= \langle U_m , \boldsymbol{k}_{N_l} \vert \hat{\bar{W}}[l] \vert   U_n , \boldsymbol{k}_1 \rangle \;, \\
	\hat{\bar{W}}[l] &= \prod\limits_{\boldsymbol{k}_i}^{\boldsymbol{k}_{N_l} \leftarrow \boldsymbol{k}_1} \bar{\mathcal{P}}_v(\boldsymbol{k}_i) \;, \\
	\bar{\mathcal{P}}_v(\boldsymbol{k}_i) &= \sum\limits_{n=1}^{N_v} \vert   U_n , \boldsymbol{k}_i \rangle \langle    U_n , \boldsymbol{k}_1 \vert\;.
\end{align*}
This latest form is fully gauge invariant when $\boldsymbol{k}_{N_l} = \boldsymbol{k}_1$, i.e.~when the path is identically closed. For quasi-closed loops, i.e.~$\boldsymbol{k}_{N_l} = \boldsymbol{k}_1 + \boldsymbol{K}$ where $\boldsymbol{K}$ is a reciprocal lattice vector, we use the periodic gauge which gives,
\begin{align*}
	\bar{\mathcal{W}}_{mn}[l] &= \langle U_m , \boldsymbol{k}_1 \vert \hat{T}(\boldsymbol{K}) \hat{W}[l]\vert   U_n , \boldsymbol{k}_1 \rangle \;.
\end{align*}
The translation matrix $\hat{T}(\boldsymbol{K})$ depends on the choice of the lattice origin. Shifting the origin by $\boldsymbol{\delta}_0$ leads to $\hat{T}_{\boldsymbol{\delta}_0}(\boldsymbol{K}) = \mathrm{e}^{-i \boldsymbol{K}\cdot \boldsymbol{\delta}_0} \hat{T}(\boldsymbol{K})$. Therefore there is a global $U(1)$ gauge ambiguity of the quasi-closed Wilson loop that comes from the arbitrariness of the definition of the real space origin with respect to the Bravais lattice. This phase can be fixed by requiring that the spectrum of the Wilson loop be symmetric under complex conjugation whenever the loop is foldable by a point symmetry, i.e.~whenever $gl = l^{-1}$.

\section{Mapping from the bare IRREP eigenvalues to Wilson loop spectra}\label{mappings}
The explicit mapping $\mathcal{M}_2$ used in the definition of the two-point loop symmetry protected topological invariants of Section \ref{line_inv} is given in Table \ref{mapping_two} for two valence states and in Table \ref{mapping_four} for four valence states, following Refs.~\cite{Alex1,Alex2}. The explicit mapping $\mathcal{M}_4$ used in the definition of the four-point loop symmetry protected topological invariants of Section \ref{plane_inv} is given in Table \ref{mapping_Wilson_two} for two valence states and in Table \ref{mapping_four} for four valence states. 

{\def\arraystretch{1.3}  
\begin{table}[htb]
\caption{\label{mapping_two} Mapping $\mathcal{M}_2$ from the bare IRREP eigenvalues (obtained under the action of $\Phi_2$, see Section \ref{line_inv}) at the vertices $V_1$ and $V_2$ to the Wilson loop spectrum over a the closed two-point symmetry-constrained loop $l^g_L$ for two valence states, following Refs.~\cite{Alex1,Alex2}. }
\begin{tabular*}{\linewidth}{c @{\extracolsep{\fill}} X c c c }
\hline
\hline
	 $R^{V_1}_g$ & $R^{V_2}_g$ & $ \rightarrow$ & $\mathrm{eig}\{\mathcal{W}[l^g_{L}]\}$  \\
\hline
	$\{+1,+1\}$ & $\{+1,+1\}$ & & $\{+1,+1\} $   \\
	$\{+1,-1\}$ & $\{+1,-1\}$ & & $\{\mathrm{e}^{i \varphi},\mathrm{e}^{-i \varphi}\}$ \\
	$\{+1,+1\}$ & $\{-1,-1\}$ & & $\{-1,-1\}$  \\
	$\{+1,+1\}$ & $\{+1,-1\}$ & & $\{+1,-1\}$   \\
\hline
\hline
\end{tabular*}
\end{table} 
}
{\def\arraystretch{1.3}  
\begin{table}[htb]
\caption{\label{mapping_Wilson_two} Mapping $\mathcal{M}_4$ from the Wilson loop spectra of two quasi-closed two-point symmetry-constrained loops $l^g_L$ and $l^g_{L'}$ to the Wilson loop spectrum of the closed four-point loop $l^g_{\sigma}$ ($\supset \{l^g_L, l^g_{L'}\}$) for two valence states, as discussed in Section \ref{plane_inv}.}
\begin{tabular*}{\linewidth}{ c @{\extracolsep{\fill}} c c c  }
\hline
\hline
	 $\mathrm{eig}\{\mathcal{W}[l^g_{L}]\}$ & $\mathrm{eig}\{\mathcal{W}[l^g_{L'}]\}$ & $\rightarrow$ & $\mathrm{eig}\{\mathcal{W}[l^g_{\sigma}]\}$   \\
\hline
	$\{+1,+1\}$ & $\{+1,+1\}$ & & $\{+1,+1\}$   \\
	$\{-1,-1\}$ & $\{-1,-1\}$ & & $\{+1,+1\}$    \\
	$\{\mathrm{e}^{i \varphi_L},\mathrm{e}^{-i \varphi_L}\}$ & $\{\mathrm{e}^{i \varphi_{L'}},\mathrm{e}^{-i \varphi_{L'}}\}$ & & $\{\mathrm{e}^{i \varphi},\mathrm{e}^{-i \varphi}\}$  \\
	$\{+1,+1\}$ &  $\{\mathrm{e}^{i \varphi_{L'}},\mathrm{e}^{-i \varphi_{L'}}\}$ & & $\{\mathrm{e}^{i \varphi},\mathrm{e}^{-i \varphi}\}$ \\
	$\{+1,-1\}$ & $\{+1,-1\}$ & & $\{\mathrm{e}^{i \varphi},\mathrm{e}^{-i \varphi}\}$ \\
	$\{+1,+1\}$ & $\{-1,-1\}$ & & $\{-1,-1\}$    \\
	$\{\mathrm{e}^{i \varphi_L},\mathrm{e}^{-i \varphi_L}\}$ & $\{ -1, -1\}$ & & $\{-\mathrm{e}^{i \varphi},-\mathrm{e}^{-i \varphi}\}$  \\
	$\{\pm1,\pm1\}$ & $\{+1,-1\}$ & & $\{+1,-1\}$    \\
	$\{\mathrm{e}^{i \varphi_L},\mathrm{e}^{-i \varphi_L}\}$ & $\{+1,-1\}$ & & $\{\mathrm{e}^{i \varphi},-\mathrm{e}^{-i \varphi}\}$ \\
\hline
\hline
\end{tabular*}
\end{table} 
}
{\def\arraystretch{1.3}  
\begin{table}[htb]
\caption{\label{mapping_four} Mapping $\mathcal{M}_2$ and $\mathcal{M}_4$ (preceded by the action of $\Phi_4$, see Section \ref{plane_inv}) for four valence states, following Refs.~\cite{Alex1,Alex2}. We  use the contracted notation $+=+1$, $-=-1$, and $\lambda_1 = e^{i \varphi_1}$.
}
\begin{tabular*}{\linewidth}{ c @{\extracolsep{\fill}} X c c c  }
\hline
\hline
	  $R^{V_1}_g,\mathcal{R}[l^g_{L}]$ & $R^{V_2}_g,\mathcal{R}[l^g_{L'}]$ & $\rightarrow$ & $\mathrm{eig}\{\mathcal{W}[l^g_{L}]\},\mathrm{eig}\{\mathcal{W}[l^g_{\sigma}]\}$   \\
\hline
	$\{+,+,+,+\}$ & $\{+,+,+,+\}$ &  & $\{+,+,+,+\}$  \\
	$\{+,+,+,-\}$ & $\{+,+,+,-\}$ & & $\{\lambda_1,\lambda_1^*,+,+\}$ \\
	$\{+,+,-,-\}$ & $\{+,+,-,-\}$ & & $\{\lambda_1,\lambda_1^*,\lambda_2,\lambda_2^*\}$  \\
	$\{+,+,+,+\}$ & $\{+,+,+,-\}$ & & $\{+,+,+,-\}$    \\
	$\{+,+,+,-\}$ & $\{+,+,-,-\}$ & & $\{\lambda_1,\lambda_1^*,+,-\}$ \\
	$\{+,+,+,+\}$ & $\{+,+,-,-\}$ & & $\{+,+,-,-\}$   \\
	$\{+,+,+,-\}$ & $\{+,-,-,-\}$ & & $\{\lambda_1,\lambda_1^*,-,-\}$   \\
	$\{+,+,+,+\}$ & $\{+,-,-,-\}$ & & $\{+,-,-,-\}$   \\
	$\{+,+,+,+\}$ & $\{-,-,-,-\}$ & & $\{-,-,-,-\}$   \\
\hline
\hline
\end{tabular*}
\end{table} 
}
\pagebreak
\bibliography{mybib}

\begin{thebibliography}{90}%
\makeatletter
\providecommand \@ifxundefined [1]{%
 \@ifx{#1\undefined}
}%
\providecommand \@ifnum [1]{%
 \ifnum #1\expandafter \@firstoftwo
 \else \expandafter \@secondoftwo
 \fi
}%
\providecommand \@ifx [1]{%
 \ifx #1\expandafter \@firstoftwo
 \else \expandafter \@secondoftwo
 \fi
}%
\providecommand \natexlab [1]{#1}%
\providecommand \enquote  [1]{``#1''}%
\providecommand \bibnamefont  [1]{#1}%
\providecommand \bibfnamefont [1]{#1}%
\providecommand \citenamefont [1]{#1}%
\providecommand \href@noop [0]{\@secondoftwo}%
\providecommand \href [0]{\begingroup \@sanitize@url \@href}%
\providecommand \@href[1]{\@@startlink{#1}\@@href}%
\providecommand \@@href[1]{\endgroup#1\@@endlink}%
\providecommand \@sanitize@url [0]{\catcode `\\12\catcode `\$12\catcode
  `\&12\catcode `\#12\catcode `\^12\catcode `\_12\catcode `\%12\relax}%
\providecommand \@@startlink[1]{}%
\providecommand \@@endlink[0]{}%
\providecommand \url  [0]{\begingroup\@sanitize@url \@url }%
\providecommand \@url [1]{\endgroup\@href {#1}{\urlprefix }}%
\providecommand \urlprefix  [0]{URL }%
\providecommand \Eprint [0]{\href }%
\providecommand \doibase [0]{http://dx.doi.org/}%
\providecommand \selectlanguage [0]{\@gobble}%
\providecommand \bibinfo  [0]{\@secondoftwo}%
\providecommand \bibfield  [0]{\@secondoftwo}%
\providecommand \translation [1]{[#1]}%
\providecommand \BibitemOpen [0]{}%
\providecommand \bibitemStop [0]{}%
\providecommand \bibitemNoStop [0]{.\EOS\space}%
\providecommand \EOS [0]{\spacefactor3000\relax}%
\providecommand \BibitemShut  [1]{\csname bibitem#1\endcsname}%
\let\auto@bib@innerbib\@empty
\bibitem [{\citenamefont {Kane}\ and\ \citenamefont
  {Mele}(2005)}]{KaneMele_Z2}%
  \BibitemOpen
  \bibfield  {author} {\bibinfo {author} {\bibfnamefont {C.~L.}\ \bibnamefont
  {Kane}}\ and\ \bibinfo {author} {\bibfnamefont {E.~J.}\ \bibnamefont
  {Mele}},\ }\href@noop {} {\bibfield  {journal} {\bibinfo  {journal} {Phys.
  Rev. Lett.}\ }\textbf {\bibinfo {volume} {95}},\ \bibinfo {pages} {146802}
  (\bibinfo {year} {2005})}\BibitemShut {NoStop}%
\bibitem [{\citenamefont {Fu}\ and\ \citenamefont {Kane}(2006)}]{FuKane_Z2}%
  \BibitemOpen
  \bibfield  {author} {\bibinfo {author} {\bibfnamefont {L.}~\bibnamefont
  {Fu}}\ and\ \bibinfo {author} {\bibfnamefont {C.~L.}\ \bibnamefont {Kane}},\
  }\href@noop {} {\bibfield  {journal} {\bibinfo  {journal} {Phys. Rev. B}\
  }\textbf {\bibinfo {volume} {74}},\ \bibinfo {pages} {195312} (\bibinfo
  {year} {2006})}\BibitemShut {NoStop}%
\bibitem [{\citenamefont {Fu}\ \emph {et~al.}(2007)\citenamefont {Fu},
  \citenamefont {Kane},\ and\ \citenamefont {Mele}}]{FuKaneMele_Z2}%
  \BibitemOpen
  \bibfield  {author} {\bibinfo {author} {\bibfnamefont {L.}~\bibnamefont
  {Fu}}, \bibinfo {author} {\bibfnamefont {C.~L.}\ \bibnamefont {Kane}}, \ and\
  \bibinfo {author} {\bibfnamefont {E.~J.}\ \bibnamefont {Mele}},\ }\href@noop
  {} {\bibfield  {journal} {\bibinfo  {journal} {Phys. Rev. Lett.}\ }\textbf
  {\bibinfo {volume} {98}},\ \bibinfo {pages} {106803} (\bibinfo {year}
  {2007})}\BibitemShut {NoStop}%
\bibitem [{\citenamefont {Wen}(2016)}]{Wen_zoo_top}%
  \BibitemOpen
  \bibfield  {author} {\bibinfo {author} {\bibfnamefont {X.-G.}\ \bibnamefont
  {Wen}},\ }\href@noop {} {} (\bibinfo {year} {2016}),\ \Eprint
  {http://arxiv.org/abs/arXiv:1610.03911} {arXiv:1610.03911} \BibitemShut
  {NoStop}%
\bibitem [{\citenamefont {Schnyder}\ \emph {et~al.}(2008)\citenamefont
  {Schnyder}, \citenamefont {Ryu}, \citenamefont {Furusaki},\ and\
  \citenamefont {Ludwig}}]{Schnyder08}%
  \BibitemOpen
  \bibfield  {author} {\bibinfo {author} {\bibfnamefont {A.}~\bibnamefont
  {Schnyder}}, \bibinfo {author} {\bibfnamefont {S.}~\bibnamefont {Ryu}},
  \bibinfo {author} {\bibfnamefont {A.}~\bibnamefont {Furusaki}}, \ and\
  \bibinfo {author} {\bibfnamefont {W.}~\bibnamefont {Ludwig}},\ }\href@noop {}
  {\bibfield  {journal} {\bibinfo  {journal} {Phys. Rev. B}\ }\textbf {\bibinfo
  {volume} {78}},\ \bibinfo {pages} {195125} (\bibinfo {year}
  {2008})}\BibitemShut {NoStop}%
\bibitem [{\citenamefont {Kitaev}(2009)}]{Kitaev}%
  \BibitemOpen
  \bibfield  {author} {\bibinfo {author} {\bibfnamefont {A.}~\bibnamefont
  {Kitaev}},\ }\href@noop {} {\bibfield  {journal} {\bibinfo  {journal} {AIP
  Conference Proceedings}\ }\textbf {\bibinfo {volume} {1134}},\ \bibinfo
  {pages} {22} (\bibinfo {year} {2009})}\BibitemShut {NoStop}%
\bibitem [{\citenamefont {Ryu}\ \emph {et~al.}(2010)\citenamefont {Ryu},
  \citenamefont {Schnyder}, \citenamefont {Furusaki},\ and\ \citenamefont
  {Ludwig}}]{RyuSchnyder_10ways}%
  \BibitemOpen
  \bibfield  {author} {\bibinfo {author} {\bibfnamefont {S.}~\bibnamefont
  {Ryu}}, \bibinfo {author} {\bibfnamefont {A.}~\bibnamefont {Schnyder}},
  \bibinfo {author} {\bibfnamefont {A.}~\bibnamefont {Furusaki}}, \ and\
  \bibinfo {author} {\bibfnamefont {A.}~\bibnamefont {Ludwig}},\ }\href@noop {}
  {\bibfield  {journal} {\bibinfo  {journal} {New J. Phys.}\ }\textbf {\bibinfo
  {volume} {12}},\ \bibinfo {pages} {065010} (\bibinfo {year}
  {2010})}\BibitemShut {NoStop}%
\bibitem [{\citenamefont {Volovik}(2003)}]{Volovik}%
  \BibitemOpen
  \bibfield  {author} {\bibinfo {author} {\bibfnamefont {G.~E.}\ \bibnamefont
  {Volovik}},\ }\href@noop {} {\emph {\bibinfo {title} {The Universe in a
  Helium Droplet}}},\ \bibinfo {edition} {oxford university press}\ ed.\
  (\bibinfo {year} {2003})\BibitemShut {NoStop}%
\bibitem [{\citenamefont {Teo}\ and\ \citenamefont
  {Kane}(2010)}]{Top_defect_tenfold}%
  \BibitemOpen
  \bibfield  {author} {\bibinfo {author} {\bibfnamefont {J.~C.~Y.}\
  \bibnamefont {Teo}}\ and\ \bibinfo {author} {\bibfnamefont {C.~L.}\
  \bibnamefont {Kane}},\ }\href@noop {} {\bibfield  {journal} {\bibinfo
  {journal} {Phys. Rev. B}\ }\textbf {\bibinfo {volume} {82}},\ \bibinfo
  {pages} {115120} (\bibinfo {year} {2010})}\BibitemShut {NoStop}%
\bibitem [{\citenamefont {Zhao}\ and\ \citenamefont
  {Wang}(2013)}]{ZhaoWang_FS}%
  \BibitemOpen
  \bibfield  {author} {\bibinfo {author} {\bibfnamefont {Y.~X.}\ \bibnamefont
  {Zhao}}\ and\ \bibinfo {author} {\bibfnamefont {Z.~D.}\ \bibnamefont
  {Wang}},\ }\href@noop {} {\bibfield  {journal} {\bibinfo  {journal} {Phys.
  Rev. Lett.}\ }\textbf {\bibinfo {volume} {110}},\ \bibinfo {pages} {240404}
  (\bibinfo {year} {2013})}\BibitemShut {NoStop}%
\bibitem [{\citenamefont {Chiu}\ and\ \citenamefont
  {Schnyder}(2014)}]{ChiuSchnyder_reflection}%
  \BibitemOpen
  \bibfield  {author} {\bibinfo {author} {\bibfnamefont {C.-K.}\ \bibnamefont
  {Chiu}}\ and\ \bibinfo {author} {\bibfnamefont {A.~P.}\ \bibnamefont
  {Schnyder}},\ }\href@noop {} {\bibfield  {journal} {\bibinfo  {journal}
  {Phys. Rev. B}\ }\textbf {\bibinfo {volume} {90}},\ \bibinfo {pages} {205136}
  (\bibinfo {year} {2014})}\BibitemShut {NoStop}%
\bibitem [{\citenamefont {Michel}\ and\ \citenamefont {Zak}(1999)}]{Zak1}%
  \BibitemOpen
  \bibfield  {author} {\bibinfo {author} {\bibfnamefont {L.}~\bibnamefont
  {Michel}}\ and\ \bibinfo {author} {\bibfnamefont {J.}~\bibnamefont {Zak}},\
  }\href@noop {} {\bibfield  {journal} {\bibinfo  {journal} {Phys. Rev. B}\
  }\textbf {\bibinfo {volume} {59}},\ \bibinfo {pages} {5998} (\bibinfo {year}
  {1999})}\BibitemShut {NoStop}%
\bibitem [{\citenamefont {Michel}\ and\ \citenamefont {Zak}(2000)}]{Zak2}%
  \BibitemOpen
  \bibfield  {author} {\bibinfo {author} {\bibfnamefont {L.}~\bibnamefont
  {Michel}}\ and\ \bibinfo {author} {\bibfnamefont {J.}~\bibnamefont {Zak}},\
  }\href@noop {} {\bibfield  {journal} {\bibinfo  {journal} {Europhys. Lett.}\
  }\textbf {\bibinfo {volume} {50}},\ \bibinfo {pages} {519} (\bibinfo {year}
  {2000})}\BibitemShut {NoStop}%
\bibitem [{\citenamefont {Zak}(2002)}]{Zak3}%
  \BibitemOpen
  \bibfield  {author} {\bibinfo {author} {\bibfnamefont {J.}~\bibnamefont
  {Zak}},\ }\href@noop {} {\bibfield  {journal} {\bibinfo  {journal} {J. Phys.
  A: Math. Gen.}\ }\textbf {\bibinfo {volume} {35}},\ \bibinfo {pages} {6509}
  (\bibinfo {year} {2002})}\BibitemShut {NoStop}%
\bibitem [{\citenamefont {Lee}\ \emph {et~al.}(2003)\citenamefont {Lee},
  \citenamefont {Kim},\ and\ \citenamefont {Zak}}]{Zak4}%
  \BibitemOpen
  \bibfield  {author} {\bibinfo {author} {\bibfnamefont {G.}~\bibnamefont
  {Lee}}, \bibinfo {author} {\bibfnamefont {J.~S.}\ \bibnamefont {Kim}}, \ and\
  \bibinfo {author} {\bibfnamefont {J.}~\bibnamefont {Zak}},\ }\href@noop {}
  {\bibfield  {journal} {\bibinfo  {journal} {J. Phys.: Cond. Matter}\ }\textbf
  {\bibinfo {volume} {15}},\ \bibinfo {pages} {2005} (\bibinfo {year}
  {2003})}\BibitemShut {NoStop}%
\bibitem [{\citenamefont {Bradley}\ and\ \citenamefont
  {Cracknell}(1972)}]{BradCrack}%
  \BibitemOpen
  \bibfield  {author} {\bibinfo {author} {\bibfnamefont {C.}~\bibnamefont
  {Bradley}}\ and\ \bibinfo {author} {\bibfnamefont {A.}~\bibnamefont
  {Cracknell}},\ }\href@noop {} {\emph {\bibinfo {title} {The Mathematical
  Theory of Symmetry in Solids}}},\ edited by\ \bibinfo {editor} {\bibfnamefont
  {O.~U.}\ \bibnamefont {Press}}\ (\bibinfo {year} {1972})\BibitemShut
  {NoStop}%
\bibitem [{\citenamefont {Parameswaran}\ \emph {et~al.}(2013)\citenamefont
  {Parameswaran}, \citenamefont {Turner}, \citenamefont {Arovas}, \citenamefont
  {Watanabe},\ and\ \citenamefont {Vishwanath}}]{Vishwanath1}%
  \BibitemOpen
  \bibfield  {author} {\bibinfo {author} {\bibfnamefont {S.~A.}\ \bibnamefont
  {Parameswaran}}, \bibinfo {author} {\bibfnamefont {A.~M.}\ \bibnamefont
  {Turner}}, \bibinfo {author} {\bibfnamefont {D.~P.}\ \bibnamefont {Arovas}},
  \bibinfo {author} {\bibfnamefont {H.}~\bibnamefont {Watanabe}}, \ and\
  \bibinfo {author} {\bibfnamefont {A.}~\bibnamefont {Vishwanath}},\
  }\href@noop {} {\bibfield  {journal} {\bibinfo  {journal} {Nat. Phys.}\
  }\textbf {\bibinfo {volume} {9}},\ \bibinfo {pages} {299} (\bibinfo {year}
  {2013})}\BibitemShut {NoStop}%
\bibitem [{\citenamefont {Young}\ \emph {et~al.}(2012)\citenamefont {Young},
  \citenamefont {S.~Zaheer}, \citenamefont {Kane}, \citenamefont {Mele},\ and\
  \citenamefont {Rappe}}]{YoungKane_simple}%
  \BibitemOpen
  \bibfield  {author} {\bibinfo {author} {\bibfnamefont {S.}~\bibnamefont
  {Young}}, \bibinfo {author} {\bibfnamefont {J.~T.}\ \bibnamefont
  {S.~Zaheer}}, \bibinfo {author} {\bibfnamefont {C.}~\bibnamefont {Kane}},
  \bibinfo {author} {\bibfnamefont {E.}~\bibnamefont {Mele}}, \ and\ \bibinfo
  {author} {\bibfnamefont {A.}~\bibnamefont {Rappe}},\ }\href@noop {}
  {\bibfield  {journal} {\bibinfo  {journal} {Phys. Rev. Lett.}\ }\textbf
  {\bibinfo {volume} {108}},\ \bibinfo {pages} {140405} (\bibinfo {year}
  {2012})}\BibitemShut {NoStop}%
\bibitem [{\citenamefont {Young}\ and\ \citenamefont
  {Kane}(2015)}]{Kane_nonsym}%
  \BibitemOpen
  \bibfield  {author} {\bibinfo {author} {\bibfnamefont {S.~M.}\ \bibnamefont
  {Young}}\ and\ \bibinfo {author} {\bibfnamefont {C.~L.}\ \bibnamefont
  {Kane}},\ }\href@noop {} {\bibfield  {journal} {\bibinfo  {journal} {Phys.
  Rev. Lett.}\ }\textbf {\bibinfo {volume} {115}},\ \bibinfo {pages} {126803}
  (\bibinfo {year} {2015})}\BibitemShut {NoStop}%
\bibitem [{\citenamefont {Watanabe}\ \emph {et~al.}(2015)\citenamefont
  {Watanabe}, \citenamefont {Po}, \citenamefont {Michael},\ and\ \citenamefont
  {Zaletel}}]{Vishwanath2}%
  \BibitemOpen
  \bibfield  {author} {\bibinfo {author} {\bibfnamefont {H.}~\bibnamefont
  {Watanabe}}, \bibinfo {author} {\bibfnamefont {H.~C.}\ \bibnamefont {Po}},
  \bibinfo {author} {\bibfnamefont {A.~V.}\ \bibnamefont {Michael}}, \ and\
  \bibinfo {author} {\bibfnamefont {M.~P.}\ \bibnamefont {Zaletel}},\
  }\href@noop {} {\bibfield  {journal} {\bibinfo  {journal} {Proc. Natl. Acad.
  Sci. U.S.A.}\ }\textbf {\bibinfo {volume} {112}},\ \bibinfo {pages} {14551}
  (\bibinfo {year} {2015})}\BibitemShut {NoStop}%
\bibitem [{\citenamefont {Wieder}\ \emph {et~al.}(2016)\citenamefont {Wieder},
  \citenamefont {Kim}, \citenamefont {Rappe},\ and\ \citenamefont
  {Kane}}]{WiederKane_double}%
  \BibitemOpen
  \bibfield  {author} {\bibinfo {author} {\bibfnamefont {B.~J.}\ \bibnamefont
  {Wieder}}, \bibinfo {author} {\bibfnamefont {Y.}~\bibnamefont {Kim}},
  \bibinfo {author} {\bibfnamefont {A.~M.}\ \bibnamefont {Rappe}}, \ and\
  \bibinfo {author} {\bibfnamefont {C.~L.}\ \bibnamefont {Kane}},\ }\href@noop
  {} {\bibfield  {journal} {\bibinfo  {journal} {Phys. Rev. Lett.}\ }\textbf
  {\bibinfo {volume} {116}},\ \bibinfo {pages} {186402} (\bibinfo {year}
  {2016})}\BibitemShut {NoStop}%
\bibitem [{\citenamefont {Watanabe}\ \emph {et~al.}(2016)\citenamefont
  {Watanabe}, \citenamefont {Po}, \citenamefont {Zaletel},\ and\ \citenamefont
  {Vishwanath}}]{Vishwanath3}%
  \BibitemOpen
  \bibfield  {author} {\bibinfo {author} {\bibfnamefont {H.}~\bibnamefont
  {Watanabe}}, \bibinfo {author} {\bibfnamefont {H.~C.}\ \bibnamefont {Po}},
  \bibinfo {author} {\bibfnamefont {M.~P.}\ \bibnamefont {Zaletel}}, \ and\
  \bibinfo {author} {\bibfnamefont {A.}~\bibnamefont {Vishwanath}},\
  }\href@noop {} {\bibfield  {journal} {\bibinfo  {journal} {Phys. Rev. Lett.}\
  }\textbf {\bibinfo {volume} {117}},\ \bibinfo {pages} {096404} (\bibinfo
  {year} {2016})}\BibitemShut {NoStop}%
\bibitem [{\citenamefont {Bzdusek}\ \emph {et~al.}(2016)\citenamefont
  {Bzdusek}, \citenamefont {Wu}, \citenamefont {R\"uegg}, \citenamefont
  {Sigrist},\ and\ \citenamefont {Soluyanov}}]{Thomas_line}%
  \BibitemOpen
  \bibfield  {author} {\bibinfo {author} {\bibfnamefont {T.}~\bibnamefont
  {Bzdusek}}, \bibinfo {author} {\bibfnamefont {Q.~S.}\ \bibnamefont {Wu}},
  \bibinfo {author} {\bibfnamefont {A.}~\bibnamefont {R\"uegg}}, \bibinfo
  {author} {\bibfnamefont {M.}~\bibnamefont {Sigrist}}, \ and\ \bibinfo
  {author} {\bibfnamefont {A.~A.}\ \bibnamefont {Soluyanov}},\ }\href@noop {}
  {\bibfield  {journal} {\bibinfo  {journal} {Nature}\ }\textbf {\bibinfo
  {volume} {538}},\ \bibinfo {pages} {75} (\bibinfo {year} {2016})}\BibitemShut
  {NoStop}%
\bibitem [{\citenamefont {Zhu}\ \emph {et~al.}(2016)\citenamefont {Zhu},
  \citenamefont {Winkler}, \citenamefont {Wu}, \citenamefont {Li},\ and\
  \citenamefont {Soluyanov}}]{Solouyanov_triplepoints}%
  \BibitemOpen
  \bibfield  {author} {\bibinfo {author} {\bibfnamefont {Z.}~\bibnamefont
  {Zhu}}, \bibinfo {author} {\bibfnamefont {G.~W.}\ \bibnamefont {Winkler}},
  \bibinfo {author} {\bibfnamefont {Q.}~\bibnamefont {Wu}}, \bibinfo {author}
  {\bibfnamefont {J.}~\bibnamefont {Li}}, \ and\ \bibinfo {author}
  {\bibfnamefont {A.~A.}\ \bibnamefont {Soluyanov}},\ }\href@noop {} {\bibfield
   {journal} {\bibinfo  {journal} {Phys. Rev. X}\ }\textbf {\bibinfo {volume}
  {6}},\ \bibinfo {pages} {031003} (\bibinfo {year} {2016})}\BibitemShut
  {NoStop}%
\bibitem [{\citenamefont {Wieder}\ and\ \citenamefont
  {Kane}(2016)}]{Kane_2D_SOC}%
  \BibitemOpen
  \bibfield  {author} {\bibinfo {author} {\bibfnamefont {B.~J.}\ \bibnamefont
  {Wieder}}\ and\ \bibinfo {author} {\bibfnamefont {C.~L.}\ \bibnamefont
  {Kane}},\ }\href@noop {} {\bibfield  {journal} {\bibinfo  {journal} {Phys.
  Rev. B}\ }\textbf {\bibinfo {volume} {94}},\ \bibinfo {pages} {155108}
  (\bibinfo {year} {2016})}\BibitemShut {NoStop}%
\bibitem [{\citenamefont {Zhao}\ and\ \citenamefont
  {Schnyder}(2016)}]{ZhaoSchnyder_1D}%
  \BibitemOpen
  \bibfield  {author} {\bibinfo {author} {\bibfnamefont {Y.}~\bibnamefont
  {Zhao}}\ and\ \bibinfo {author} {\bibfnamefont {A.}~\bibnamefont
  {Schnyder}},\ }\href@noop {} {\bibfield  {journal} {\bibinfo  {journal}
  {Phys. Rev. B}\ }\textbf {\bibinfo {volume} {94}},\ \bibinfo {pages} {195109}
  (\bibinfo {year} {2016})}\BibitemShut {NoStop}%
\bibitem [{\citenamefont {Yang}\ \emph {et~al.}(2017)\citenamefont {Yang},
  \citenamefont {Bojesen}, \citenamefont {Morimoto},\ and\ \citenamefont
  {Furusaki}}]{Furusaki_line}%
  \BibitemOpen
  \bibfield  {author} {\bibinfo {author} {\bibfnamefont {B.-J.}\ \bibnamefont
  {Yang}}, \bibinfo {author} {\bibfnamefont {T.~A.}\ \bibnamefont {Bojesen}},
  \bibinfo {author} {\bibfnamefont {T.}~\bibnamefont {Morimoto}}, \ and\
  \bibinfo {author} {\bibfnamefont {A.}~\bibnamefont {Furusaki}},\ }\href@noop
  {} {\bibfield  {journal} {\bibinfo  {journal} {Phys. Rev. B}\ }\textbf
  {\bibinfo {volume} {95}},\ \bibinfo {pages} {075135} (\bibinfo {year}
  {2017})}\BibitemShut {NoStop}%
\bibitem [{\citenamefont {Geilhufe}\ \emph
  {et~al.}(2017{\natexlab{a}})\citenamefont {Geilhufe}, \citenamefont {Bouhon},
  \citenamefont {Borysov},\ and\ \citenamefont {Balatsky}}]{GB_line}%
  \BibitemOpen
  \bibfield  {author} {\bibinfo {author} {\bibfnamefont {R.~M.}\ \bibnamefont
  {Geilhufe}}, \bibinfo {author} {\bibfnamefont {A.}~\bibnamefont {Bouhon}},
  \bibinfo {author} {\bibfnamefont {S.~S.}\ \bibnamefont {Borysov}}, \ and\
  \bibinfo {author} {\bibfnamefont {A.~V.}\ \bibnamefont {Balatsky}},\
  }\href@noop {} {\bibfield  {journal} {\bibinfo  {journal} {Phys. Rev. B}\
  }\textbf {\bibinfo {volume} {95}},\ \bibinfo {pages} {041103} (\bibinfo
  {year} {2017}{\natexlab{a}})}\BibitemShut {NoStop}%
\bibitem [{\citenamefont {Geilhufe}\ \emph
  {et~al.}(2017{\natexlab{b}})\citenamefont {Geilhufe}, \citenamefont
  {Borysov}, \citenamefont {Bouhon},\ and\ \citenamefont
  {Balatsky}}]{GelBoBoBa_II}%
  \BibitemOpen
  \bibfield  {author} {\bibinfo {author} {\bibfnamefont {R.~M.}\ \bibnamefont
  {Geilhufe}}, \bibinfo {author} {\bibfnamefont {S.~S.}\ \bibnamefont
  {Borysov}}, \bibinfo {author} {\bibfnamefont {A.}~\bibnamefont {Bouhon}}, \
  and\ \bibinfo {author} {\bibfnamefont {A.~V.}\ \bibnamefont {Balatsky}},\
  }\href@noop {} {\bibfield  {journal} {\bibinfo  {journal} {Scientific
  Reports}\ }\textbf {\bibinfo {volume} {7}},\ \bibinfo {pages} {7298}
  (\bibinfo {year} {2017}{\natexlab{b}})}\BibitemShut {NoStop}%
\bibitem [{\citenamefont {Furusaki}(2017)}]{Furusaki_multiplescrews}%
  \BibitemOpen
  \bibfield  {author} {\bibinfo {author} {\bibfnamefont {A.}~\bibnamefont
  {Furusaki}},\ }\href@noop {} {\bibfield  {journal} {\bibinfo  {journal}
  {Science Bulletin}\ }\textbf {\bibinfo {volume} {62}},\ \bibinfo {pages}
  {788} (\bibinfo {year} {2017})}\BibitemShut {NoStop}%
\bibitem [{\citenamefont {Takahashi}\ \emph {et~al.}(2017)\citenamefont
  {Takahashi}, \citenamefont {Hirayama},\ and\ \citenamefont
  {Murakami}}]{Murakami_NLsym}%
  \BibitemOpen
  \bibfield  {author} {\bibinfo {author} {\bibfnamefont {R.}~\bibnamefont
  {Takahashi}}, \bibinfo {author} {\bibfnamefont {M.}~\bibnamefont {Hirayama}},
  \ and\ \bibinfo {author} {\bibfnamefont {S.}~\bibnamefont {Murakami}},\
  }\href@noop {} {} (\bibinfo {year} {2017}),\ \Eprint
  {http://arxiv.org/abs/arXiv:1704.02151} {arXiv:1704.02151} \BibitemShut
  {NoStop}%
\bibitem [{\citenamefont {Tsirkin}\ \emph {et~al.}(2017)\citenamefont
  {Tsirkin}, \citenamefont {Souza}, ,\ and\ \citenamefont
  {Vanderbilt}}]{Vanderbilt_WeylPoints}%
  \BibitemOpen
  \bibfield  {author} {\bibinfo {author} {\bibfnamefont {S.~S.}\ \bibnamefont
  {Tsirkin}}, \bibinfo {author} {\bibfnamefont {I.}~\bibnamefont {Souza}}, , \
  and\ \bibinfo {author} {\bibfnamefont {D.}~\bibnamefont {Vanderbilt}},\
  }\href@noop {} {\bibfield  {journal} {\bibinfo  {journal} {Phys. Rev. B}\
  }\textbf {\bibinfo {volume} {96}},\ \bibinfo {pages} {045102} (\bibinfo
  {year} {2017})}\BibitemShut {NoStop}%
\bibitem [{\citenamefont {Shiozaki}\ \emph {et~al.}(2016)\citenamefont
  {Shiozaki}, \citenamefont {Sato},\ and\ \citenamefont {Gomi}}]{Sato_3}%
  \BibitemOpen
  \bibfield  {author} {\bibinfo {author} {\bibfnamefont {K.}~\bibnamefont
  {Shiozaki}}, \bibinfo {author} {\bibfnamefont {M.}~\bibnamefont {Sato}}, \
  and\ \bibinfo {author} {\bibfnamefont {K.}~\bibnamefont {Gomi}},\ }\href@noop
  {} {\bibfield  {journal} {\bibinfo  {journal} {Phys. Rev. B}\ }\textbf
  {\bibinfo {volume} {93}},\ \bibinfo {pages} {195413} (\bibinfo {year}
  {2016})}\BibitemShut {NoStop}%
\bibitem [{\citenamefont {Chang}\ \emph {et~al.}(2016)\citenamefont {Chang},
  \citenamefont {Sanchez}, \citenamefont {Wieder}, \citenamefont {Xu},
  \citenamefont {Schindler}, \citenamefont {Belopolski}, \citenamefont {Huang},
  \citenamefont {Singh}, \citenamefont {Wu}, \citenamefont {Neupert},
  \citenamefont {Chang}, \citenamefont {Lin},\ and\ \citenamefont
  {Hasan}}]{1611.07925}%
  \BibitemOpen
  \bibfield  {author} {\bibinfo {author} {\bibfnamefont {G.}~\bibnamefont
  {Chang}}, \bibinfo {author} {\bibfnamefont {D.~S.}\ \bibnamefont {Sanchez}},
  \bibinfo {author} {\bibfnamefont {B.~J.}\ \bibnamefont {Wieder}}, \bibinfo
  {author} {\bibfnamefont {S.-Y.}\ \bibnamefont {Xu}}, \bibinfo {author}
  {\bibfnamefont {F.}~\bibnamefont {Schindler}}, \bibinfo {author}
  {\bibfnamefont {I.}~\bibnamefont {Belopolski}}, \bibinfo {author}
  {\bibfnamefont {S.-M.}\ \bibnamefont {Huang}}, \bibinfo {author}
  {\bibfnamefont {B.}~\bibnamefont {Singh}}, \bibinfo {author} {\bibfnamefont
  {D.}~\bibnamefont {Wu}}, \bibinfo {author} {\bibfnamefont {T.}~\bibnamefont
  {Neupert}}, \bibinfo {author} {\bibfnamefont {T.-R.}\ \bibnamefont {Chang}},
  \bibinfo {author} {\bibfnamefont {H.}~\bibnamefont {Lin}}, \ and\ \bibinfo
  {author} {\bibfnamefont {M.~Z.}\ \bibnamefont {Hasan}},\ }\href@noop {} {}
  (\bibinfo {year} {2016}),\ \Eprint {http://arxiv.org/abs/arXiv:1611.07925}
  {arXiv:1611.07925} \BibitemShut {NoStop}%
\bibitem [{\citenamefont {Aroyo}\ \emph {et~al.}(2006)\citenamefont {Aroyo},
  \citenamefont {Kirov}, \citenamefont {Capillas}, \citenamefont {Perez-Mato},\
  and\ \citenamefont {Wondratschek}}]{Bilbao}%
  \BibitemOpen
  \bibfield  {author} {\bibinfo {author} {\bibfnamefont {M.~I.}\ \bibnamefont
  {Aroyo}}, \bibinfo {author} {\bibfnamefont {A.}~\bibnamefont {Kirov}},
  \bibinfo {author} {\bibfnamefont {C.}~\bibnamefont {Capillas}}, \bibinfo
  {author} {\bibfnamefont {J.~M.}\ \bibnamefont {Perez-Mato}}, \ and\ \bibinfo
  {author} {\bibfnamefont {H.}~\bibnamefont {Wondratschek}},\ }\href@noop {}
  {\bibfield  {journal} {\bibinfo  {journal} {Acta Cryst. A}\ }\textbf
  {\bibinfo {volume} {62}},\ \bibinfo {pages} {115} (\bibinfo {year}
  {2006})}\BibitemShut {NoStop}%
\bibitem [{\citenamefont {Hughes}\ \emph {et~al.}(2011)\citenamefont {Hughes},
  \citenamefont {Prodan},\ and\ \citenamefont {Bernevig}}]{Bernevig1}%
  \BibitemOpen
  \bibfield  {author} {\bibinfo {author} {\bibfnamefont {T.}~\bibnamefont
  {Hughes}}, \bibinfo {author} {\bibfnamefont {E.}~\bibnamefont {Prodan}}, \
  and\ \bibinfo {author} {\bibfnamefont {B.~A.}\ \bibnamefont {Bernevig}},\
  }\href@noop {} {\bibfield  {journal} {\bibinfo  {journal} {Phys. Rev. B}\
  }\textbf {\bibinfo {volume} {83}},\ \bibinfo {pages} {245132} (\bibinfo
  {year} {2011})}\BibitemShut {NoStop}%
\bibitem [{\citenamefont {Fang}\ \emph {et~al.}(2012)\citenamefont {Fang},
  \citenamefont {Gilbert},\ and\ \citenamefont
  {Bernevig}}]{Bernevig_point_groups}%
  \BibitemOpen
  \bibfield  {author} {\bibinfo {author} {\bibfnamefont {C.}~\bibnamefont
  {Fang}}, \bibinfo {author} {\bibfnamefont {M.~J.}\ \bibnamefont {Gilbert}}, \
  and\ \bibinfo {author} {\bibfnamefont {B.~A.}\ \bibnamefont {Bernevig}},\
  }\href@noop {} {\bibfield  {journal} {\bibinfo  {journal} {Phys. Rev. B}\
  }\textbf {\bibinfo {volume} {86}},\ \bibinfo {pages} {115112} (\bibinfo
  {year} {2012})}\BibitemShut {NoStop}%
\bibitem [{\citenamefont {Alexandradinata}\ \emph
  {et~al.}(2014{\natexlab{a}})\citenamefont {Alexandradinata}, \citenamefont
  {Dai},\ and\ \citenamefont {Bernevig}}]{Alex1}%
  \BibitemOpen
  \bibfield  {author} {\bibinfo {author} {\bibfnamefont {A.}~\bibnamefont
  {Alexandradinata}}, \bibinfo {author} {\bibfnamefont {X.}~\bibnamefont
  {Dai}}, \ and\ \bibinfo {author} {\bibfnamefont {B.~A.}\ \bibnamefont
  {Bernevig}},\ }\href@noop {} {\bibfield  {journal} {\bibinfo  {journal}
  {Phys. Rev. B}\ }\textbf {\bibinfo {volume} {89}},\ \bibinfo {pages} {155114}
  (\bibinfo {year} {2014}{\natexlab{a}})}\BibitemShut {NoStop}%
\bibitem [{\citenamefont {Alexandradinata}\ \emph
  {et~al.}(2014{\natexlab{b}})\citenamefont {Alexandradinata}, \citenamefont
  {Fang}, \citenamefont {Gilbert},\ and\ \citenamefont
  {Bernevig}}]{Alex_noSOC_noTRS}%
  \BibitemOpen
  \bibfield  {author} {\bibinfo {author} {\bibfnamefont {A.}~\bibnamefont
  {Alexandradinata}}, \bibinfo {author} {\bibfnamefont {C.}~\bibnamefont
  {Fang}}, \bibinfo {author} {\bibfnamefont {M.~J.}\ \bibnamefont {Gilbert}}, \
  and\ \bibinfo {author} {\bibfnamefont {B.~A.}\ \bibnamefont {Bernevig}},\
  }\href@noop {} {\bibfield  {journal} {\bibinfo  {journal} {Phys. Rev. Lett.}\
  }\textbf {\bibinfo {volume} {113}},\ \bibinfo {pages} {116403} (\bibinfo
  {year} {2014}{\natexlab{b}})}\BibitemShut {NoStop}%
\bibitem [{\citenamefont {Kim}\ \emph {et~al.}(2015)\citenamefont {Kim},
  \citenamefont {Wieder}, \citenamefont {Kane},\ and\ \citenamefont
  {Rappe}}]{Kane_line}%
  \BibitemOpen
  \bibfield  {author} {\bibinfo {author} {\bibfnamefont {Y.}~\bibnamefont
  {Kim}}, \bibinfo {author} {\bibfnamefont {B.~J.}\ \bibnamefont {Wieder}},
  \bibinfo {author} {\bibfnamefont {C.~L.}\ \bibnamefont {Kane}}, \ and\
  \bibinfo {author} {\bibfnamefont {A.~M.}\ \bibnamefont {Rappe}},\ }\href@noop
  {} {\bibfield  {journal} {\bibinfo  {journal} {Phys. Rev. Lett.}\ }\textbf
  {\bibinfo {volume} {115}},\ \bibinfo {pages} {036806} (\bibinfo {year}
  {2015})}\BibitemShut {NoStop}%
\bibitem [{\citenamefont {Alexandradinata}\ and\ \citenamefont
  {Bernevig}(2016)}]{AlexBernevig_berryphase}%
  \BibitemOpen
  \bibfield  {author} {\bibinfo {author} {\bibfnamefont {A.}~\bibnamefont
  {Alexandradinata}}\ and\ \bibinfo {author} {\bibfnamefont {B.~A.}\
  \bibnamefont {Bernevig}},\ }\href@noop {} {\bibfield  {journal} {\bibinfo
  {journal} {Phys. Rev. B}\ }\textbf {\bibinfo {volume} {93}},\ \bibinfo
  {pages} {205104} (\bibinfo {year} {2016})}\BibitemShut {NoStop}%
\bibitem [{\citenamefont {Wang}\ \emph {et~al.}(2016)\citenamefont {Wang},
  \citenamefont {Alexandradinata}, \citenamefont {Cava},\ and\ \citenamefont
  {Bernevig}}]{Bernevig3}%
  \BibitemOpen
  \bibfield  {author} {\bibinfo {author} {\bibfnamefont {Z.}~\bibnamefont
  {Wang}}, \bibinfo {author} {\bibfnamefont {A.}~\bibnamefont
  {Alexandradinata}}, \bibinfo {author} {\bibfnamefont {R.}~\bibnamefont
  {Cava}}, \ and\ \bibinfo {author} {\bibfnamefont {B.~A.}\ \bibnamefont
  {Bernevig}},\ }\href@noop {} {\bibfield  {journal} {\bibinfo  {journal}
  {Nature}\ }\textbf {\bibinfo {volume} {532}},\ \bibinfo {pages} {189}
  (\bibinfo {year} {2016})}\BibitemShut {NoStop}%
\bibitem [{\citenamefont {Alexandradinata}\ \emph {et~al.}(2016)\citenamefont
  {Alexandradinata}, \citenamefont {Wang},\ and\ \citenamefont
  {Bernevig}}]{Alex0}%
  \BibitemOpen
  \bibfield  {author} {\bibinfo {author} {\bibfnamefont {A.}~\bibnamefont
  {Alexandradinata}}, \bibinfo {author} {\bibfnamefont {Z.}~\bibnamefont
  {Wang}}, \ and\ \bibinfo {author} {\bibfnamefont {B.~A.}\ \bibnamefont
  {Bernevig}},\ }\href@noop {} {\bibfield  {journal} {\bibinfo  {journal}
  {Phys. Rev. X}\ }\textbf {\bibinfo {volume} {6}},\ \bibinfo {pages} {021008}
  (\bibinfo {year} {2016})}\BibitemShut {NoStop}%
\bibitem [{\citenamefont {Muechler}\ \emph {et~al.}(2016)\citenamefont
  {Muechler}, \citenamefont {Alexandradinata}, \citenamefont {Neupert},\ and\
  \citenamefont {Car}}]{Alex2}%
  \BibitemOpen
  \bibfield  {author} {\bibinfo {author} {\bibfnamefont {L.}~\bibnamefont
  {Muechler}}, \bibinfo {author} {\bibfnamefont {A.}~\bibnamefont
  {Alexandradinata}}, \bibinfo {author} {\bibfnamefont {T.}~\bibnamefont
  {Neupert}}, \ and\ \bibinfo {author} {\bibfnamefont {R.}~\bibnamefont
  {Car}},\ }\href@noop {} {\bibfield  {journal} {\bibinfo  {journal} {Phys.
  Rev. X}\ }\textbf {\bibinfo {volume} {6}},\ \bibinfo {pages} {041069}
  (\bibinfo {year} {2016})}\BibitemShut {NoStop}%
\bibitem [{\citenamefont {Fang}\ \emph {et~al.}(2015)\citenamefont {Fang},
  \citenamefont {Chen}, \citenamefont {Kee},\ and\ \citenamefont
  {Fu}}]{Fu_line_node_monopole}%
  \BibitemOpen
  \bibfield  {author} {\bibinfo {author} {\bibfnamefont {C.}~\bibnamefont
  {Fang}}, \bibinfo {author} {\bibfnamefont {Y.}~\bibnamefont {Chen}}, \bibinfo
  {author} {\bibfnamefont {H.-Y.}\ \bibnamefont {Kee}}, \ and\ \bibinfo
  {author} {\bibfnamefont {L.}~\bibnamefont {Fu}},\ }\href@noop {} {\bibfield
  {journal} {\bibinfo  {journal} {Phys. Rev. B}\ }\textbf {\bibinfo {volume}
  {92}},\ \bibinfo {pages} {081201(R)} (\bibinfo {year} {2015})}\BibitemShut
  {NoStop}%
\bibitem [{\citenamefont {Agterberg}\ \emph {et~al.}(2017)\citenamefont
  {Agterberg}, \citenamefont {Brydon},\ and\ \citenamefont
  {Timm}}]{Agterberg_BdGsurface}%
  \BibitemOpen
  \bibfield  {author} {\bibinfo {author} {\bibfnamefont {D.}~\bibnamefont
  {Agterberg}}, \bibinfo {author} {\bibfnamefont {P.}~\bibnamefont {Brydon}}, \
  and\ \bibinfo {author} {\bibfnamefont {C.}~\bibnamefont {Timm}},\ }\href@noop
  {} {\bibfield  {journal} {\bibinfo  {journal} {Phys. Rev. Lett.}\ }\textbf
  {\bibinfo {volume} {118}},\ \bibinfo {pages} {127001} (\bibinfo {year}
  {2017})}\BibitemShut {NoStop}%
\bibitem [{\citenamefont {Bzdusek}\ and\ \citenamefont
  {Sigrist}(2017)}]{Bzdusek_mulitnodes}%
  \BibitemOpen
  \bibfield  {author} {\bibinfo {author} {\bibfnamefont {T.}~\bibnamefont
  {Bzdusek}}\ and\ \bibinfo {author} {\bibfnamefont {M.}~\bibnamefont
  {Sigrist}},\ }\href@noop {} {} (\bibinfo {year} {2017}),\ \Eprint
  {http://arxiv.org/abs/arXiv:1705.07126} {arXiv:1705.07126} \BibitemShut
  {NoStop}%
\bibitem [{\citenamefont {T\"urker}\ and\ \citenamefont
  {Moroz}(2017)}]{Moroz_nodalsurface}%
  \BibitemOpen
  \bibfield  {author} {\bibinfo {author} {\bibfnamefont {O.}~\bibnamefont
  {T\"urker}}\ and\ \bibinfo {author} {\bibfnamefont {S.}~\bibnamefont
  {Moroz}},\ }\href@noop {} {} (\bibinfo {year} {2017}),\ \Eprint
  {http://arxiv.org/abs/arXiv:1709.01561} {arXiv:1709.01561} \BibitemShut
  {NoStop}%
\bibitem [{\citenamefont {Slager}\ \emph {et~al.}(2013)\citenamefont {Slager},
  \citenamefont {Mesaros}, \citenamefont {Juricic},\ and\ \citenamefont
  {Zaanen}}]{Slager_SG_pfaffian}%
  \BibitemOpen
  \bibfield  {author} {\bibinfo {author} {\bibfnamefont {R.-J.}\ \bibnamefont
  {Slager}}, \bibinfo {author} {\bibfnamefont {A.}~\bibnamefont {Mesaros}},
  \bibinfo {author} {\bibfnamefont {V.}~\bibnamefont {Juricic}}, \ and\
  \bibinfo {author} {\bibfnamefont {J.}~\bibnamefont {Zaanen}},\ }\href@noop {}
  {\bibfield  {journal} {\bibinfo  {journal} {Nature Physics}\ }\textbf
  {\bibinfo {volume} {9}},\ \bibinfo {pages} {98} (\bibinfo {year}
  {2013})}\BibitemShut {NoStop}%
\bibitem [{\citenamefont {Zhang}\ and\ \citenamefont {Liu}(2015)}]{Liu_coreps}%
  \BibitemOpen
  \bibfield  {author} {\bibinfo {author} {\bibfnamefont {R.-X.}\ \bibnamefont
  {Zhang}}\ and\ \bibinfo {author} {\bibfnamefont {C.-X.}\ \bibnamefont
  {Liu}},\ }\href@noop {} {\bibfield  {journal} {\bibinfo  {journal} {Phys.
  Rev. B}\ }\textbf {\bibinfo {volume} {91}},\ \bibinfo {pages} {115317}
  (\bibinfo {year} {2015})}\BibitemShut {NoStop}%
\bibitem [{\citenamefont {Dong}\ and\ \citenamefont {Liu}(2016)}]{Liu_reps}%
  \BibitemOpen
  \bibfield  {author} {\bibinfo {author} {\bibfnamefont {X.-Y.}\ \bibnamefont
  {Dong}}\ and\ \bibinfo {author} {\bibfnamefont {C.-X.}\ \bibnamefont {Liu}},\
  }\href@noop {} {\bibfield  {journal} {\bibinfo  {journal} {Phys. Rev. B}\
  }\textbf {\bibinfo {volume} {93}},\ \bibinfo {pages} {045429} (\bibinfo
  {year} {2016})}\BibitemShut {NoStop}%
\bibitem [{\citenamefont {Kruthoff}\ \emph {et~al.}(2016)\citenamefont
  {Kruthoff}, \citenamefont {de~Boer}, \citenamefont {van Wezel}, \citenamefont
  {Kane},\ and\ \citenamefont {Slager}}]{Slager_K}%
  \BibitemOpen
  \bibfield  {author} {\bibinfo {author} {\bibfnamefont {J.}~\bibnamefont
  {Kruthoff}}, \bibinfo {author} {\bibfnamefont {J.}~\bibnamefont {de~Boer}},
  \bibinfo {author} {\bibfnamefont {J.}~\bibnamefont {van Wezel}}, \bibinfo
  {author} {\bibfnamefont {C.~L.}\ \bibnamefont {Kane}}, \ and\ \bibinfo
  {author} {\bibfnamefont {R.-J.}\ \bibnamefont {Slager}},\ }\href@noop {} {}
  (\bibinfo {year} {2016}),\ \Eprint {http://arxiv.org/abs/arXiv:1612.02007}
  {arXiv:1612.02007} \BibitemShut {NoStop}%
\bibitem [{\citenamefont {Bouhon}\ and\ \citenamefont
  {Black-Schaffer}(2017)}]{BBS_1}%
  \BibitemOpen
  \bibfield  {author} {\bibinfo {author} {\bibfnamefont {A.}~\bibnamefont
  {Bouhon}}\ and\ \bibinfo {author} {\bibfnamefont {A.~M.}\ \bibnamefont
  {Black-Schaffer}},\ }\href@noop {} {\bibfield  {journal} {\bibinfo  {journal}
  {Phys. Rev. B}\ }\textbf {\bibinfo {volume} {95}},\ \bibinfo {pages}
  {241101(R)} (\bibinfo {year} {2017})}\BibitemShut {NoStop}%
\bibitem [{\citenamefont {Shiozaki}\ \emph {et~al.}(2017)\citenamefont
  {Shiozaki}, \citenamefont {Sato},\ and\ \citenamefont
  {Gomi}}]{ShiozakiWallPaper_K}%
  \BibitemOpen
  \bibfield  {author} {\bibinfo {author} {\bibfnamefont {K.}~\bibnamefont
  {Shiozaki}}, \bibinfo {author} {\bibfnamefont {M.}~\bibnamefont {Sato}}, \
  and\ \bibinfo {author} {\bibfnamefont {K.}~\bibnamefont {Gomi}},\ }\href@noop
  {} {} (\bibinfo {year} {2017}),\ \Eprint
  {http://arxiv.org/abs/arXiv:1701.08725} {arXiv:1701.08725} \BibitemShut
  {NoStop}%
\bibitem [{\citenamefont {Po}\ \emph {et~al.}(2017{\natexlab{a}})\citenamefont
  {Po}, \citenamefont {Vishwanath},\ and\ \citenamefont
  {Watanabe}}]{PoVishWata_symmetry_indic}%
  \BibitemOpen
  \bibfield  {author} {\bibinfo {author} {\bibfnamefont {H.~C.}\ \bibnamefont
  {Po}}, \bibinfo {author} {\bibfnamefont {A.}~\bibnamefont {Vishwanath}}, \
  and\ \bibinfo {author} {\bibfnamefont {H.}~\bibnamefont {Watanabe}},\
  }\href@noop {} {\bibfield  {journal} {\bibinfo  {journal} {Nat. Commun.}\
  }\textbf {\bibinfo {volume} {8}},\ \bibinfo {pages} {50} (\bibinfo {year}
  {2017}{\natexlab{a}})}\BibitemShut {NoStop}%
\bibitem [{\citenamefont {Vergniory}\ \emph {et~al.}(2017)\citenamefont
  {Vergniory}, \citenamefont {Elcoro}, \citenamefont {Wang}, \citenamefont
  {Cano}, \citenamefont {Felser}, \citenamefont {Aroyo}, \citenamefont
  {Bernevig},\ and\ \citenamefont {Bradlyn}}]{Graph_theory}%
  \BibitemOpen
  \bibfield  {author} {\bibinfo {author} {\bibfnamefont {M.~G.}\ \bibnamefont
  {Vergniory}}, \bibinfo {author} {\bibfnamefont {L.}~\bibnamefont {Elcoro}},
  \bibinfo {author} {\bibfnamefont {Z.}~\bibnamefont {Wang}}, \bibinfo {author}
  {\bibfnamefont {J.}~\bibnamefont {Cano}}, \bibinfo {author} {\bibfnamefont
  {C.}~\bibnamefont {Felser}}, \bibinfo {author} {\bibfnamefont {M.~I.}\
  \bibnamefont {Aroyo}}, \bibinfo {author} {\bibfnamefont {B.~A.}\ \bibnamefont
  {Bernevig}}, \ and\ \bibinfo {author} {\bibfnamefont {B.}~\bibnamefont
  {Bradlyn}},\ }\href@noop {} {\bibfield  {journal} {\bibinfo  {journal} {Phys.
  Rev. E}\ }\textbf {\bibinfo {volume} {96}},\ \bibinfo {pages} {023310}
  (\bibinfo {year} {2017})}\BibitemShut {NoStop}%
\bibitem [{\citenamefont {Bradlyn}\ \emph
  {et~al.}(2017{\natexlab{a}})\citenamefont {Bradlyn}, \citenamefont {Elcoro},
  \citenamefont {Vergniory}, \citenamefont {Cano}, \citenamefont {Wang},
  \citenamefont {Felser}, \citenamefont {Aroyo},\ and\ \citenamefont
  {Bernevig}}]{Band_connectivity}%
  \BibitemOpen
  \bibfield  {author} {\bibinfo {author} {\bibfnamefont {B.}~\bibnamefont
  {Bradlyn}}, \bibinfo {author} {\bibfnamefont {L.}~\bibnamefont {Elcoro}},
  \bibinfo {author} {\bibfnamefont {M.~G.}\ \bibnamefont {Vergniory}}, \bibinfo
  {author} {\bibfnamefont {J.}~\bibnamefont {Cano}}, \bibinfo {author}
  {\bibfnamefont {Z.}~\bibnamefont {Wang}}, \bibinfo {author} {\bibfnamefont
  {C.}~\bibnamefont {Felser}}, \bibinfo {author} {\bibfnamefont {M.~I.}\
  \bibnamefont {Aroyo}}, \ and\ \bibinfo {author} {\bibfnamefont {B.~A.}\
  \bibnamefont {Bernevig}},\ }\href@noop {} {} (\bibinfo {year}
  {2017}{\natexlab{a}}),\ \Eprint {http://arxiv.org/abs/arXiv:1709.01937}
  {arXiv:1709.01937} \BibitemShut {NoStop}%
\bibitem [{\citenamefont {Bradlyn}\ \emph
  {et~al.}(2017{\natexlab{b}})\citenamefont {Bradlyn}, \citenamefont {Elcoro},
  \citenamefont {Cano}, \citenamefont {Vergniory}, \citenamefont {Wang},
  \citenamefont {Felser}, \citenamefont {Aroyo},\ and\ \citenamefont
  {Bernevig}}]{TopQuantChem}%
  \BibitemOpen
  \bibfield  {author} {\bibinfo {author} {\bibfnamefont {B.}~\bibnamefont
  {Bradlyn}}, \bibinfo {author} {\bibfnamefont {L.}~\bibnamefont {Elcoro}},
  \bibinfo {author} {\bibfnamefont {J.}~\bibnamefont {Cano}}, \bibinfo {author}
  {\bibfnamefont {M.~G.}\ \bibnamefont {Vergniory}}, \bibinfo {author}
  {\bibfnamefont {Z.}~\bibnamefont {Wang}}, \bibinfo {author} {\bibfnamefont
  {C.}~\bibnamefont {Felser}}, \bibinfo {author} {\bibfnamefont {M.~I.}\
  \bibnamefont {Aroyo}}, \ and\ \bibinfo {author} {\bibfnamefont {B.~A.}\
  \bibnamefont {Bernevig}},\ }\href@noop {} {\bibfield  {journal} {\bibinfo
  {journal} {Nature}\ }\textbf {\bibinfo {volume} {547}},\ \bibinfo {pages}
  {298} (\bibinfo {year} {2017}{\natexlab{b}})}\BibitemShut {NoStop}%
\bibitem [{\citenamefont {H\"oller}\ and\ \citenamefont
  {Alexandradinata}(2017)}]{Alex_BlochOsc}%
  \BibitemOpen
  \bibfield  {author} {\bibinfo {author} {\bibfnamefont {J.}~\bibnamefont
  {H\"oller}}\ and\ \bibinfo {author} {\bibfnamefont {A.}~\bibnamefont
  {Alexandradinata}},\ }\href@noop {} {} (\bibinfo {year} {2017}),\ \Eprint
  {http://arxiv.org/abs/arXiv:1708.02943} {arXiv:1708.02943} \BibitemShut
  {NoStop}%
\bibitem [{\citenamefont {Cano}\ \emph {et~al.}(2017)\citenamefont {Cano},
  \citenamefont {Bradlyn}, \citenamefont {Wang}, \citenamefont {Elcoro},
  \citenamefont {Vergniory}, \citenamefont {Felser}, \citenamefont {Aroyo},\
  and\ \citenamefont {Bernevig}}]{Cano_EBR}%
  \BibitemOpen
  \bibfield  {author} {\bibinfo {author} {\bibfnamefont {J.}~\bibnamefont
  {Cano}}, \bibinfo {author} {\bibfnamefont {B.}~\bibnamefont {Bradlyn}},
  \bibinfo {author} {\bibfnamefont {Z.}~\bibnamefont {Wang}}, \bibinfo {author}
  {\bibfnamefont {L.}~\bibnamefont {Elcoro}}, \bibinfo {author} {\bibfnamefont
  {M.~G.}\ \bibnamefont {Vergniory}}, \bibinfo {author} {\bibfnamefont
  {C.}~\bibnamefont {Felser}}, \bibinfo {author} {\bibfnamefont {M.~I.}\
  \bibnamefont {Aroyo}}, \ and\ \bibinfo {author} {\bibfnamefont {B.~A.}\
  \bibnamefont {Bernevig}},\ }\href@noop {} {} (\bibinfo {year} {2017}),\
  \Eprint {http://arxiv.org/abs/arXiv:1709.01935} {arXiv:1709.01935}
  \BibitemShut {NoStop}%
\bibitem [{\citenamefont {Po}\ \emph {et~al.}(2017{\natexlab{b}})\citenamefont
  {Po}, \citenamefont {Watanabe},\ and\ \citenamefont
  {Vishwanath}}]{Po_FragileTop}%
  \BibitemOpen
  \bibfield  {author} {\bibinfo {author} {\bibfnamefont {H.~C.}\ \bibnamefont
  {Po}}, \bibinfo {author} {\bibfnamefont {H.}~\bibnamefont {Watanabe}}, \ and\
  \bibinfo {author} {\bibfnamefont {A.}~\bibnamefont {Vishwanath}},\
  }\href@noop {} {} (\bibinfo {year} {2017}{\natexlab{b}}),\ \Eprint
  {http://arxiv.org/abs/arXiv:1709.06551} {arXiv:1709.06551} \BibitemShut
  {NoStop}%
\bibitem [{\citenamefont {Fu}\ and\ \citenamefont {Kane}(2007)}]{FuKane_inv}%
  \BibitemOpen
  \bibfield  {author} {\bibinfo {author} {\bibfnamefont {L.}~\bibnamefont
  {Fu}}\ and\ \bibinfo {author} {\bibfnamefont {C.~L.}\ \bibnamefont {Kane}},\
  }\href@noop {} {\bibfield  {journal} {\bibinfo  {journal} {Phys. Rev. B}\
  }\textbf {\bibinfo {volume} {76}},\ \bibinfo {pages} {045302} (\bibinfo
  {year} {2007})}\BibitemShut {NoStop}%
\bibitem [{\citenamefont {Teo}\ \emph {et~al.}(2008)\citenamefont {Teo},
  \citenamefont {Fu}, ,\ and\ \citenamefont {Kane}}]{TeoFuKane_mirror_Z2}%
  \BibitemOpen
  \bibfield  {author} {\bibinfo {author} {\bibfnamefont {J.~C.~Y.}\
  \bibnamefont {Teo}}, \bibinfo {author} {\bibfnamefont {L.}~\bibnamefont
  {Fu}}, , \ and\ \bibinfo {author} {\bibfnamefont {C.~L.}\ \bibnamefont
  {Kane}},\ }\href@noop {} {\bibfield  {journal} {\bibinfo  {journal} {Phys.
  Rev. B}\ }\textbf {\bibinfo {volume} {78}},\ \bibinfo {pages} {045426}
  (\bibinfo {year} {2008})}\BibitemShut {NoStop}%
\bibitem [{Note1()}]{Note1}%
  \BibitemOpen
  \bibinfo {note} {We use the parameterization of the International Tables for
  Crystallography \cite {ITC}.}\BibitemShut {Stop}%
\bibitem [{Note2()}]{Note2}%
  \BibitemOpen
  \bibinfo {note} {An essential degeneracy happens on the whole connected
  subspace of the BZ that is characterized by the same little co-group.
  Therefore, these include band crossing points at HSPs, band crossing lines at
  HSLs, and band crossing planes at high-symmetry planes.}\BibitemShut {Stop}%
\bibitem [{Note3()}]{Note3}%
  \BibitemOpen
  \bibinfo {note} {All nodal structures of Fig.~\ref {fig_SG33_4B_lines}
  actually come from an eight-band tight-binding model that splits into two
  four-band subspaces separated by an energy band gap. It is easy to show that
  any four-by-four tight-binding model with SG33-AI (corresponding to four
  sub-lattice sites with a single electronic orbital per site) has an
  accidental fourfold degenerate NL along the whole HSL $\protect \text {P}$.
  We therefore introduce more bands in order to avoid this artificial
  degeneracy.}\BibitemShut {Stop}%
\bibitem [{Note4()}]{Note4}%
  \BibitemOpen
  \bibinfo {note} {Most of Fig.~\ref {fig_SG33_8B_lines} is obtained
  numerically from a generic eight-band tight-binding model, but a few cases
  (i,l,n,r,s) are drawn by hand as they would require an eight-band subspace
  imbedded in additional bands in order to lift artificial
  symmetries.}\BibitemShut {Stop}%
\bibitem [{Note5()}]{Note5}%
  \BibitemOpen
  \bibinfo {note} {This is obtained through the Fourier transform of an
  appropriate basis set modeling the physical degrees of freedom. In
  particular, we assume real space functions that have the periodicity of the
  sub-lattice sites belonging to the same Wyckoff position and not only the
  periodicity of the Bravais lattice. This choice of a trivializing reference
  section of the total Bloch bundle has been shown to be the more physically
  relevant for studying parallel transports, see Refs.~\cite
  {FruchartCarpentier1, Berry_connection_Moore}.}\BibitemShut {Stop}%
\bibitem [{\citenamefont {Fruchart}\ \emph {et~al.}(2014)\citenamefont
  {Fruchart}, \citenamefont {Carpentier},\ and\ \citenamefont
  {Gawedzki}}]{FruchartCarpentier1}%
  \BibitemOpen
  \bibfield  {author} {\bibinfo {author} {\bibfnamefont {M.}~\bibnamefont
  {Fruchart}}, \bibinfo {author} {\bibfnamefont {D.}~\bibnamefont
  {Carpentier}}, \ and\ \bibinfo {author} {\bibfnamefont {K.}~\bibnamefont
  {Gawedzki}},\ }\href@noop {} {\bibfield  {journal} {\bibinfo  {journal} {Eur.
  Phys. Lett.}\ }\textbf {\bibinfo {volume} {106}},\ \bibinfo {pages} {60002}
  (\bibinfo {year} {2014})}\BibitemShut {NoStop}%
\bibitem [{\citenamefont {Moore}(2017)}]{Berry_connection_Moore}%
  \BibitemOpen
  \bibfield  {author} {\bibinfo {author} {\bibfnamefont {G.~W.}\ \bibnamefont
  {Moore}},\ }\href@noop {} {\enquote {\bibinfo {title} {A comment on berry
  connections},}\ } (\bibinfo {year} {2017}),\ \Eprint
  {http://arxiv.org/abs/arXiv:1706.01149} {arXiv:1706.01149} \BibitemShut
  {NoStop}%
\bibitem [{Note6()}]{Note6}%
  \BibitemOpen
  \bibinfo {note} {The Grassmanian can also be defined as the space of valence
  projector matrices $P_{v,\boldsymbol {k}} = \sum _{n=1}^{N_v} \vert \breve
  {U}_n, \boldsymbol {k} \rangle \langle \breve {U}_n, \boldsymbol {k} \vert $,
  which takes a vector in $V_{\boldsymbol {k}}$ and gives a vector in
  $V_{v,\boldsymbol {k}}$ such that $\mathrm {Ran}~P_{v,\boldsymbol {k}} =
  V_{v,\boldsymbol {k}}$.}\BibitemShut {Stop}%
\bibitem [{Note7()}]{Note7}%
  \BibitemOpen
  \bibinfo {note} {This follows from the valence bundle always be obtainable as
  the pullback of the tautological valence bundle $\protect \mathcal {F}_v =
  \DOTSB \bigcup@ \slimits@ _{P_v \in Gr_{N_{v}}(\protect \mathbb {C}^N)}
  \protect \mathrm {Ran}~P_v $ by the continuous map $\Phi :B_v \rightarrow
  Gr_{N_{v}}(\protect \mathbb {C}^N)$, i.e.~$\protect \mathcal {E}_v = \Phi
  ^*\protect \mathcal {F}_v$, and noting the invariance of bundle isomorphism
  classes under homotopy \cite {NittisGomiAI}.}\BibitemShut {Stop}%
\bibitem [{Note8()}]{Note8}%
  \BibitemOpen
  \bibinfo {note} {Strictly speaking $\boldsymbol {L}_{v}^{(i)}$ is the
  manifold obtained as the retract of $(\mathbb {T}^3 \backslash L_{v}^{(i)})
  \cap D_{v,i}$, where $D_{v,i}$ is an open disk covering $L_{v}^{(i)}$ \cite
  {MathaiThiang_Weyls_I}. It is compact, connected, and
  orientable.}\BibitemShut {Stop}%
\bibitem [{\citenamefont {Chiu}\ \emph {et~al.}(2016)\citenamefont {Chiu},
  \citenamefont {Teo}, \citenamefont {Schnyder},\ and\ \citenamefont
  {Ryu}}]{Class_sym_review}%
  \BibitemOpen
  \bibfield  {author} {\bibinfo {author} {\bibfnamefont {C.-K.}\ \bibnamefont
  {Chiu}}, \bibinfo {author} {\bibfnamefont {J.~C.}\ \bibnamefont {Teo}},
  \bibinfo {author} {\bibfnamefont {A.~P.}\ \bibnamefont {Schnyder}}, \ and\
  \bibinfo {author} {\bibfnamefont {S.}~\bibnamefont {Ryu}},\ }\href@noop {}
  {\bibfield  {journal} {\bibinfo  {journal} {Rev. Mod. Phys.}\ }\textbf
  {\bibinfo {volume} {88}},\ \bibinfo {pages} {035005} (\bibinfo {year}
  {2016})}\BibitemShut {NoStop}%
\bibitem [{\citenamefont {Avran}\ \emph {et~al.}(1983)\citenamefont {Avran},
  \citenamefont {Seiler},\ and\ \citenamefont {Simon}}]{ASSimon_83}%
  \BibitemOpen
  \bibfield  {author} {\bibinfo {author} {\bibfnamefont {J.}~\bibnamefont
  {Avran}}, \bibinfo {author} {\bibfnamefont {R.}~\bibnamefont {Seiler}}, \
  and\ \bibinfo {author} {\bibfnamefont {B.}~\bibnamefont {Simon}},\
  }\href@noop {} {\bibfield  {journal} {\bibinfo  {journal} {Phys. Rev. Lett.}\
  }\textbf {\bibinfo {volume} {51}},\ \bibinfo {pages} {51} (\bibinfo {year}
  {1983})}\BibitemShut {NoStop}%
\bibitem [{\citenamefont {Kennedy}\ and\ \citenamefont
  {Guggenheim}(2015)}]{Kennedy_homotopy}%
  \BibitemOpen
  \bibfield  {author} {\bibinfo {author} {\bibfnamefont {R.}~\bibnamefont
  {Kennedy}}\ and\ \bibinfo {author} {\bibfnamefont {C.}~\bibnamefont
  {Guggenheim}},\ }\href@noop {} {\bibfield  {journal} {\bibinfo  {journal}
  {Phys. Rev. B}\ }\textbf {\bibinfo {volume} {91}},\ \bibinfo {pages} {245148}
  (\bibinfo {year} {2015})}\BibitemShut {NoStop}%
\bibitem [{\citenamefont {Hatcher}(2001)}]{Hatcher_1}%
  \BibitemOpen
  \bibfield  {author} {\bibinfo {author} {\bibfnamefont {A.}~\bibnamefont
  {Hatcher}},\ }\href@noop {} {\emph {\bibinfo {title} {Algebraic Topology}}},\
  \bibinfo {edition} {cambridge university press}\ ed.\ (\bibinfo {year}
  {2001})\BibitemShut {NoStop}%
\bibitem [{\citenamefont {Nielsen}\ and\ \citenamefont
  {Ninomiya}(1981{\natexlab{a}})}]{Nielsen1}%
  \BibitemOpen
  \bibfield  {author} {\bibinfo {author} {\bibfnamefont {H.}~\bibnamefont
  {Nielsen}}\ and\ \bibinfo {author} {\bibfnamefont {M.}~\bibnamefont
  {Ninomiya}},\ }\href@noop {} {\bibfield  {journal} {\bibinfo  {journal}
  {Nucl. Phys. B}\ }\textbf {\bibinfo {volume} {185}},\ \bibinfo {pages} {20}
  (\bibinfo {year} {1981}{\natexlab{a}})}\BibitemShut {NoStop}%
\bibitem [{\citenamefont {Nielsen}\ and\ \citenamefont
  {Ninomiya}(1982)}]{Nielsen1a}%
  \BibitemOpen
  \bibfield  {author} {\bibinfo {author} {\bibfnamefont {H.}~\bibnamefont
  {Nielsen}}\ and\ \bibinfo {author} {\bibfnamefont {M.}~\bibnamefont
  {Ninomiya}},\ }\href@noop {} {\bibfield  {journal} {\bibinfo  {journal} {E:
  Nucl. Phys. B}\ }\textbf {\bibinfo {volume} {195}},\ \bibinfo {pages} {541}
  (\bibinfo {year} {1982})}\BibitemShut {NoStop}%
\bibitem [{\citenamefont {Nielsen}\ and\ \citenamefont
  {Ninomiya}(1981{\natexlab{b}})}]{Nielsen2}%
  \BibitemOpen
  \bibfield  {author} {\bibinfo {author} {\bibfnamefont {H.}~\bibnamefont
  {Nielsen}}\ and\ \bibinfo {author} {\bibfnamefont {M.}~\bibnamefont
  {Ninomiya}},\ }\href@noop {} {\bibfield  {journal} {\bibinfo  {journal}
  {Nucl. Phys. B}\ }\textbf {\bibinfo {volume} {193}},\ \bibinfo {pages} {173}
  (\bibinfo {year} {1981}{\natexlab{b}})}\BibitemShut {NoStop}%
\bibitem [{\citenamefont {Kiritsis}(1987)}]{Kiritsis}%
  \BibitemOpen
  \bibfield  {author} {\bibinfo {author} {\bibfnamefont {E.}~\bibnamefont
  {Kiritsis}},\ }\href@noop {} {\bibfield  {journal} {\bibinfo  {journal}
  {Commun. Math. Phys.}\ }\textbf {\bibinfo {volume} {111}},\ \bibinfo {pages}
  {417} (\bibinfo {year} {1987})}\BibitemShut {NoStop}%
\bibitem [{\citenamefont {Witten}(2015)}]{Witten_lect}%
  \BibitemOpen
  \bibfield  {author} {\bibinfo {author} {\bibfnamefont {E.}~\bibnamefont
  {Witten}},\ }\href {\doibase 10.1393/ncr/i2016-10125-3} {\  (\bibinfo {year}
  {2015}),\ 10.1393/ncr/i2016-10125-3},\ \Eprint
  {http://arxiv.org/abs/arXiv:1510.07698} {arXiv:1510.07698} \BibitemShut
  {NoStop}%
\bibitem [{\citenamefont {Mathai}\ and\ \citenamefont
  {Thiang}(2017)}]{MathaiThiang_Weyls_I}%
  \BibitemOpen
  \bibfield  {author} {\bibinfo {author} {\bibfnamefont {V.}~\bibnamefont
  {Mathai}}\ and\ \bibinfo {author} {\bibfnamefont {G.~C.}\ \bibnamefont
  {Thiang}},\ }\href@noop {} {\bibfield  {journal} {\bibinfo  {journal} {J.
  Phys. A: Math. Theor.}\ }\textbf {\bibinfo {volume} {50}},\ \bibinfo {pages}
  {1} (\bibinfo {year} {2017})}\BibitemShut {NoStop}%
\bibitem [{Note9()}]{Note9}%
  \BibitemOpen
  \bibinfo {note} {Throughout this work we assume the periodic gauge, see
  Appendix \ref {sym_rep}.}\BibitemShut {Stop}%
\bibitem [{\citenamefont {Ahn}\ \emph {et~al.}(2017)\citenamefont {Ahn},
  \citenamefont {Mele},\ and\ \citenamefont {Min}}]{Mele_response_linenodes}%
  \BibitemOpen
  \bibfield  {author} {\bibinfo {author} {\bibfnamefont {S.}~\bibnamefont
  {Ahn}}, \bibinfo {author} {\bibfnamefont {E.~J.}\ \bibnamefont {Mele}}, \
  and\ \bibinfo {author} {\bibfnamefont {H.}~\bibnamefont {Min}},\ }\href@noop
  {} {\bibfield  {journal} {\bibinfo  {journal} {Phys. Rev. Lett.}\ }\textbf
  {\bibinfo {volume} {119}},\ \bibinfo {pages} {147402} (\bibinfo {year}
  {2017})}\BibitemShut {NoStop}%
\bibitem [{\citenamefont {Borysov}\ \emph {et~al.}(2017)\citenamefont
  {Borysov}, \citenamefont {Geilhufe},\ and\ \citenamefont
  {Balatsky}}]{BorysovGeilhufeBalatsky_mining}%
  \BibitemOpen
  \bibfield  {author} {\bibinfo {author} {\bibfnamefont {S.~S.}\ \bibnamefont
  {Borysov}}, \bibinfo {author} {\bibfnamefont {R.~M.}\ \bibnamefont
  {Geilhufe}}, \ and\ \bibinfo {author} {\bibfnamefont {A.~V.}\ \bibnamefont
  {Balatsky}},\ }\href@noop {} {\bibfield  {journal} {\bibinfo  {journal} {PLoS
  ONE}\ }\textbf {\bibinfo {volume} {12}},\ \bibinfo {pages} {e0171501}
  (\bibinfo {year} {2017})}\BibitemShut {NoStop}%
\bibitem [{\citenamefont {Chen}\ \emph {et~al.}(2017)\citenamefont {Chen},
  \citenamefont {Po}, \citenamefont {Neaton},\ and\ \citenamefont
  {Vishwanath}}]{Vishwanath_mining}%
  \BibitemOpen
  \bibfield  {author} {\bibinfo {author} {\bibfnamefont {R.}~\bibnamefont
  {Chen}}, \bibinfo {author} {\bibfnamefont {H.~C.}\ \bibnamefont {Po}},
  \bibinfo {author} {\bibfnamefont {J.~B.}\ \bibnamefont {Neaton}}, \ and\
  \bibinfo {author} {\bibfnamefont {A.}~\bibnamefont {Vishwanath}},\ }\href
  {http://dx.doi.org/10.1038/nphys4277} {\bibfield  {journal} {\bibinfo
  {journal} {Nat Phys}\ } (\bibinfo {year} {2017})}\BibitemShut {NoStop}%
\bibitem [{\citenamefont {Yu}\ \emph {et~al.}(2011)\citenamefont {Yu},
  \citenamefont {Qi}, \citenamefont {Bernevig}, \citenamefont {Fang},\ and\
  \citenamefont {Dai}}]{Bernevig0}%
  \BibitemOpen
  \bibfield  {author} {\bibinfo {author} {\bibfnamefont {R.}~\bibnamefont
  {Yu}}, \bibinfo {author} {\bibfnamefont {X.~L.}\ \bibnamefont {Qi}}, \bibinfo
  {author} {\bibfnamefont {A.}~\bibnamefont {Bernevig}}, \bibinfo {author}
  {\bibfnamefont {Z.}~\bibnamefont {Fang}}, \ and\ \bibinfo {author}
  {\bibfnamefont {X.}~\bibnamefont {Dai}},\ }\href@noop {} {\bibfield
  {journal} {\bibinfo  {journal} {Phys. Rev. B}\ }\textbf {\bibinfo {volume}
  {84}},\ \bibinfo {pages} {075119} (\bibinfo {year} {2011})}\BibitemShut
  {NoStop}%
\bibitem [{\citenamefont {(ed.)}(2006)}]{ITC}%
  \BibitemOpen
  \bibfield  {author} {\bibinfo {author} {\bibfnamefont {T.~H.}\ \bibnamefont
  {(ed.)}},\ }\href {http://it.iucr.org/Ac/} {\emph {\bibinfo {title}
  {International Tables for Crystallography. Volume A, Space-group
  symmetry}}},\ \bibinfo {edition} {online}\ ed.\ (\bibinfo {year}
  {2006})\BibitemShut {NoStop}%
\bibitem [{\citenamefont {Nittis}\ and\ \citenamefont
  {Gomi}(2014)}]{NittisGomiAI}%
  \BibitemOpen
  \bibfield  {author} {\bibinfo {author} {\bibfnamefont {G.~D.}\ \bibnamefont
  {Nittis}}\ and\ \bibinfo {author} {\bibfnamefont {K.}~\bibnamefont {Gomi}},\
  }\href@noop {} {\bibfield  {journal} {\bibinfo  {journal} {J. Geom. Phys.}\
  }\textbf {\bibinfo {volume} {86}},\ \bibinfo {pages} {303} (\bibinfo {year}
  {2014})}\BibitemShut {NoStop}%
\end{thebibliography}%

\end{document}